\documentclass[10pt,aps,prd,noshowpacs,nofootinbib,superscriptaddress,notitlepage,eqsecnum,preprintnumbers]{revtex4-1}

\usepackage{xspace}
\usepackage{graphicx}
\usepackage[top=2cm,bottom=3cm]{geometry}
\usepackage[svgnames]{xcolor}
\usepackage[bookmarks=false,colorlinks=true,linkcolor=DarkBlue,citecolor=DarkBlue]{hyperref}
\usepackage[caption=false]{subfig}
\usepackage{rotating}
\usepackage{units}
\usepackage{amsmath}
\usepackage{bm}     % nice math bold italics
\usepackage{lineno}
\usepackage{listings} 
\usepackage[normalem]{ulem}
\usepackage[section]{placeins}
\usepackage{SIunits}
\usepackage{hepunits}
\usepackage{hepparticles}
\usepackage{cancel}
\usepackage{hepnames}
\usepackage{epstopdf}
\usepackage{mathtools}
\usepackage[capitalise]{cleveref}
\usepackage{braket}
\usepackage{slashed}
\usepackage{xcolor}
\usepackage{multirow}
\definecolor{NALblue}{rgb}{0,0.2,0.6}
\definecolor{NALgreen}{rgb}{0.317647,0.603922,0.141176}
\definecolor{NALorange}{rgb}{0.858824,0.447059,0.0470588}

\newcommand{\cut}[1]{}                          % suppress undesired text
\newcommand{\st}[1]{}
\newcommand{\ask}[1]{\textcolor{black}{#1}}
\newcommand{\jgm}[1]{\textcolor{black}{#1}}  
\newcommand{\lar}[1]{\textcolor{black}{#1}}
\newcommand{\oscask}[1]{\textcolor{black}{#1}}  % Accepting Andreas' changes in osc section

\newcommand{\ie}{i.e., }
\newcommand{\eg}{e.g., }

\newcommand{\numu}{\ensuremath{\nu_\mu}\xspace}
\newcommand{\nue}{\ensuremath{\nu_e}\xspace}
\newcommand{\numubar}{\ensuremath{\overline{\nu}_\mu}\xspace}
\newcommand{\nuebar}{\ensuremath{\overline{\nu}_e}\xspace}
\newcommand{\dcp}{\ensuremath{\delta_{CP}}\xspace}
\newcommand{\thtt}{\ensuremath{\theta_{23}}\xspace}
\newcommand{\dmsq}{\ensuremath{|\Delta m^2_{32}|}\xspace}

\def\evsqc    {\ensuremath{\textrm{eV}^{2} / c^{4}}}

\newcommand{\nova}{NOvA\xspace}
\newcommand{\mb}{MiniBooNE\xspace}
\newcommand{\minerva}{MINER$\nu$A\xspace}
\newcommand{\HK}{T2HK\xspace}

\newcommand{\Enu}{\ensuremath{E_{\nu}}\xspace}

\newcommand{\n}[1]{\ensuremath{|\bm{#1}|}}

\newcommand{\Ekp}{\ensuremath{E_{\ell}}}

\newcommand{\Eppi}{\ensuremath{E_{\bm{p}_i'}}}

\newcommand{\tcoh}{\ensuremath{\left|t\right|\xspace}}
\newcommand{\Etrue}{\ensuremath{E_\text{true}}\xspace}
\newcommand{\Ereco}{\ensuremath{E_\text{reco}}\xspace}
\newcommand{\ptrue}{\ensuremath{\bm{p}_\text{true}}\xspace} % 4 vector, KM
\newcommand{\preco}{\ensuremath{\bm{p}_\text{reco}}\xspace} % 4 vector, KM
\newcommand{\pthreco}{\ensuremath{\vec{p}_\text{reco}}\xspace} % 3 vector, KM

\begin{document}
\preprint{\bf FERMILAB-PUB-17-195-ND-T}
\preprint{\bf INT-PUB-17-020}
%\preprint{ }

% Remove FNAL pre-print number for now - we will add a bunch later.
%\hspace{4.75in} \mbox{FERMILAB-PUB-17-195-ND-T}
%INT-PUB-17-020
\title{\vspace*{+1.0 cm} NuSTEC\footnote{Neutrino Scattering Theory Experiment Collaboration \url{http://nustec.fnal.gov} } White Paper: Status and Challenges of \\ Neutrino-Nucleus Scattering}
%{http://nustec.fnal.gov}
\date{\today}

\author{L. Alvarez-Ruso}\affiliation{Instituto de F\'isica Corpuscular (IFIC), Centro Mixto CSIC-Universidad de Valencia, E-46071 Valencia, Spain}
\author{M. \surname{Sajjad Athar}}\affiliation{Department of Physics, Aligarh Muslim University,
Aligarh-202 002, India}
\author{M.~B.~Barbaro}\affiliation{Dipartimento di Fisica, Universit\`{a} di Torino and INFN, Sezione di Torino, \cut{Via P. Giuria 1,} 10125 Torino, Italy}
\author{D.~Cherdack}\affiliation{Department of Physics, Colorado State University, Fort Collins, CO 80523, USA}
\author{M.~E.~Christy}\affiliation{Department of Physics, Hampton University, Hampton, Virginia, 23668 USA}
\author{P.~Coloma}\affiliation{Fermi National Accelerator Laboratory, Batavia, IL 60510, USA}
\author{T.~W.~Donnelly}\affiliation{Center for Theoretical Physics, Laboratory for Nuclear Science, and Department of Physics, Massachusetts Institute of Technology, Cambridge, MA 02139, USA}
\author{S. Dytman}\affiliation{Department of Physics and Astronomy, University of Pittsburgh, Pittsburgh PA 15260, USA}
\author{A.~de~Gouv\^ea}\affiliation{Northwestern University, Evanston, IL 60208, USA}
\author{R.~J.~Hill}\affiliation{Perimeter Institute for Theoretical Physics, Waterloo, ON, N2L 2Y5 Canada}
\affiliation{Fermi National Accelerator Laboratory, Batavia, IL 60510, USA}
\author{P.~Huber}\affiliation{Center of Neutrino Physics, Virginia Tech, Blacksburg, VA 24061, USA}
\author{N.~Jachowicz}\affiliation{Department of Physics and Astronomy, Ghent University, B-9000 Gent, Belgium }
\author{T.~Katori}\affiliation{Queen Mary University of London, London, UK}
\author{A.~S.~Kronfeld}\affiliation{Fermi National Accelerator Laboratory, Batavia, IL 60510, USA}
\author{K.~Mahn}\affiliation{Department of Physics and Astronomy, Michigan State University, East Lansing, MI 48824, USA}
\author{M.~Martini}\affiliation{ESNT, CEA, IRFU, Service de Physique Nucl\'eaire, Universit\'e de Paris-Saclay, F-91191 Gif-sur-Yvette, France}
\author{J.~G. Morf\'{i}n}\affiliation{Fermi National Accelerator Laboratory, Batavia, IL 60510, USA}
\author{J.~Nieves}\affiliation{Instituto de F\'isica Corpuscular (IFIC), Centro Mixto CSIC-Universidad de Valencia, E-46071 Valencia, Spain}
\author{G.~Perdue}\affiliation{Fermi National Accelerator Laboratory, Batavia, IL 60510, USA}
\author{R.~Petti}\affiliation{Department of Physics and Astronomy, University of South Carolina, Columbia SC 29208, USA}
\author{D.~G.~Richards}\affiliation{Jefferson Laboratory,  Newport News, VA 23606, USA}
\author{F.~S\'anchez}\affiliation{IFAE, Barcelona Institute Science and Technology,
Barcelona, Spain}
\author{T. Sato}\affiliation{Department of Physics, Osaka University, Toyonaka, Osaka 560-0043, Japan}
\affiliation{J-PARC Branch, KEK Theory Center, KEK, Tokai, 319-1106, Japan}
\author{J.~T. Sobczyk}\affiliation{Institute of Theoretical Physics, University of Wroc\l aw, Wroc\l aw, Poland}
\author{G.~P. Zeller}\affiliation{Fermi National Accelerator Laboratory, Batavia, IL 60510, USA}
 
\begin{abstract}
The precise measurement of neutrino properties is among the highest priorities in fundamental particle physics, involving many experiments worldwide. Since the experiments rely on the interactions of neutrinos with bound nucleons inside atomic nuclei, the planned advances in the scope and precision of these experiments requires a commensurate effort in the understanding and modeling of the hadronic and nuclear physics of these interactions, which is incorporated as a nuclear model in neutrino event generators. This model is essential to every phase of experimental analyses and its theoretical uncertainties play an important role in interpreting every result.

In this White Paper we discuss in detail the impact of neutrino-nucleus interactions, especially the nuclear effects, on the measurement of neutrino properties using the determination of oscillation parameters as a central example. After an Executive Summary and a concise Overview of the issues, we explain how the neutrino event generators work, what can be learned from electron-nucleus interactions and how each underlying physics process---from quasi-elastic to deep inelastic scattering---is understood today. We then emphasize how our understanding must improve to meet the demands of future experiments. With every topic we find that the challenges can be met only with the active support and collaboration among specialists in strong interactions and electroweak physics that include theorists and experimentalists from both the nuclear and high energy physics communities.

\end{abstract}

\maketitle
%\linenumbers

\newpage
\tableofcontents

\newpage
\section{Executive Summary }

\label{exec}

The precise measurement of neutrino properties and interactions is among the highest priorities in fundamental particle physics. The discovery of nonzero neutrino masses \jgm {at} the end of the twentieth century remains one of the very few hints regarding the nature of physics beyond the standard model of particle physics and \jgm{the ability to fully explore} 
this new physics points to high-statistics, high-precision, neutrino oscillation experiments. Indeed, pursuing the physics responsible for neutrino masses was identified as one of the science drivers for particle physics by the 2014 Strategic Plan for U.S. Particle Physics (P5) and a beam-based, long-baseline, neutrino oscillation experiment was identified as the highest priority intermediate-future effort by the U.S. community. This effort has taken the form of the Deep Underground Neutrino Experiment (DUNE), an international project to be hosted by Fermilab. Similar sentiments were expressed in the European Strategy for Particle Physics in 2013, while the particle physics community in Japan has identified the Tokai-to-Hyper-Kamiokande project (\HK) as one of its highest particle physics priorities for the next decade.

Qualitative improvement on the measurement of neutrino properties in oscillation experiments, including the thorough exploration of CP-invariance violation in the lepton sector and nontrivial tests of the three-massive-neutrinos paradigm, requires percent-level control of systematic uncertainties. This unprecedented level of precision translates into novel challenges and opportunities for our understanding of the scattering of neutrinos with a variety of complex nuclei, including argon and oxygen.

The exploitation of the physics capabilities of the neutrino facilities currently being planned for the next decade, with both near and far detectors, requires improving our ability to describe neutrino--nucleus scattering.    
The current state of the art for interaction systematic uncertainties is in the neighborhood of (5-10)\%, and even a modest improvement could, for example, dramatically shorten the required running time for five-sigma coverage of at least half of the allowed values of the Dirac CP-odd phase in the leptonic mixing matrix (a useful benchmark for experimental reach). These current systematic uncertainties associated with neutrino--nucleus interactions already play a significant role even \cut{once one takes}\lar{after taking} into account essential information from the near detectors.  It is important to appreciate that while near-detector facilities play a useful role in understanding \jgm{the neutrino flux}, 
they  are not sufficient to solve the problem of neutrino-nucleus interaction uncertainties.
\lar{To address them,} strengthening investments \lar{are required} in both theoretical and experimental aspects of this complex phenomenon.

A defining challenge for neutrino experiments is that neither the incoming neutrino \jgm{energy nor the particle configuration and} kinematics \jgm{of the interaction within the nucleus} are known. This means one must work with ensembles of events and rely on Monte Carlo simulation (event generators) to produce probability-weighted maps that connect observations in the detector to distributions of possible true kinematics. Inaccuracies or biases in the construction of these maps can lead to problems in neutrino energy reconstruction that distort the spectrum to an unacceptable degree, even in a near-detector complex.  Therefore, measurements of neutrino oscillation probabilities as a function of the incoming neutrino energy, often using a specific reaction channel, are highly dependent on fundamentally accurate models of neutrino-nucleus interactions that must also be extensive.
\jgm{That is, one must know the energy-dependent cross section of every initial interaction that, through nuclear effects, could contribute to an observed final state in the detector. And this for multiple nuclei, should there be a suite of diverse nuclear targets in the detector.} \lar{To properly inform these
theoretical models and the state-of-the-art event generators that employ them, it is crucial that there exist a diversity of experiments covering a variety of targets and beam energies, along with excellent communication between theorists, experimentalists and Monte Carlo simulation experts} 
  
\cut{\jgm{To properly inform these
theoretical models and the state-of-the-art event generators that employ them, it is crucial that a diversity of experiments covering a variety of targets and beam energies along with excellent communication between theorists, experimentalists and Monte Carlo model builders 
be available.  
The overall problem is complex enough that no one measurement, and no one theory contribution, can solve everything at once. Many pieces working together are required and coordination between  \lar{particle and nuclear theory}
\cut{the nuclear theory and particle theory}, along with scattering experiments is mandatory.}  }

Neutrino--nucleus scattering is a multi-scale problem, especially at the energy \cut{scales} \lar{region} of interest to long-baseline neutrino-oscillation experiments (hundreds to thousands of MeV). At these energy scales, it is convenient to describe neutrino interactions as the scattering of neutrinos off nucleons that are bound inside \cut{complex} nuclei. The physics of neutrino--nucleon scattering is in the realm of \cut{traditional} theoretical particle physics\cut{, and}: precision calculations are required in order to meet the stringent requirements of next-generation experiments. Contributions from lattice QCD, for example, are necessary to fill important gaps in the understanding \cut{neutrino--nucleus scattering} \lar{of nucleon structure}. Moreover, the proper treatment of radiative corrections is also a requirement, especially for experiments that plan to use the more numerous $\nu_{\mu}$ events in their near detector complex to constrain features of the $\nu_{e}$ cross section. \jgm{On the other hand, further} computations in the realm of theoretical nuclear physics are necessary in order to properly characterize the target \lar{bound} nucleons, \cut{which are bound inside their respective nuclei} allow for different multi-component initial and final states, take into account final state interactions, and properly describe the propagation of the products of the \jgm{bound-}nucleon level scattering inside the nuclear medium\cut{, etc}. This necessary close cooperation of nuclear physics (NP) and high-energy physics (HEP) highlights a problem facing neutrino-nucleus scattering that is rooted in the boundaries erected between \cut{high energy physics and nuclear physics} \lar{these subjects} by important overseeing agencies.  This separation results in more difficult collaboration and cooperation between groups that are natural stakeholders in a CP-violation measurement at a long-baseline experiment, or in a sterile neutrino search at short baselines.  Nature does not respect this division of knowledge and we need to be flexible enough to utilize the organizing structures in our field to make tasks easier.

Finally, \jgm{it is important to emphasize that} neutrino-nucleus scattering is also interesting in its own right. Neutrinos provide very useful and complementary information on nuclear \jgm{and bound-nucleon} structure that is not easily available in charged-lepton- or photon-nucleus scattering. The large data samples expected at different near detector facilities will also allow for the search of new neutrino--matter interactions and may provide invaluable information concerning new fundamental particles and interactions. 

\newpage

\section{Introduction and Overview of the Current Challenges}
\label{Overview}

\subsection{Introduction: General Challenges}

The recent increased interest in neutrino-nucleus interactions is mostly due to its importance in neutrino oscillation studies. 
The next generation of oscillation experiments with a goal to measure CP violation phase (DUNE in the United States and Hyper-Kamiokande in Japan) are costly enterprises requiring international level coordination and cooperation.
It \cut{is not always}\ask{must be} recognized that their success may depend on a significant effort in \ask{understanding, quantifying and} reducing the systematic error coming from modeling neutrino-nucleus interactions. 
Apart from a critical importance in neutrino oscillation studies, neutrino interaction research \ask{supplements electron- and photon-scattering studies of}\cut{is also relevant to} hadronic physics, \cut{providing supplementary information to that attainable in electron and photon scattering studies through the inclusion of}\ask{by including the} axial-vector \cut{current}\ask{interactions}.
Both perspectives are discussed in this paper.

\begin{figure}[btp]
	\includegraphics[width=0.66\textwidth,trim={50pt 90pt 50pt 90pt},clip]{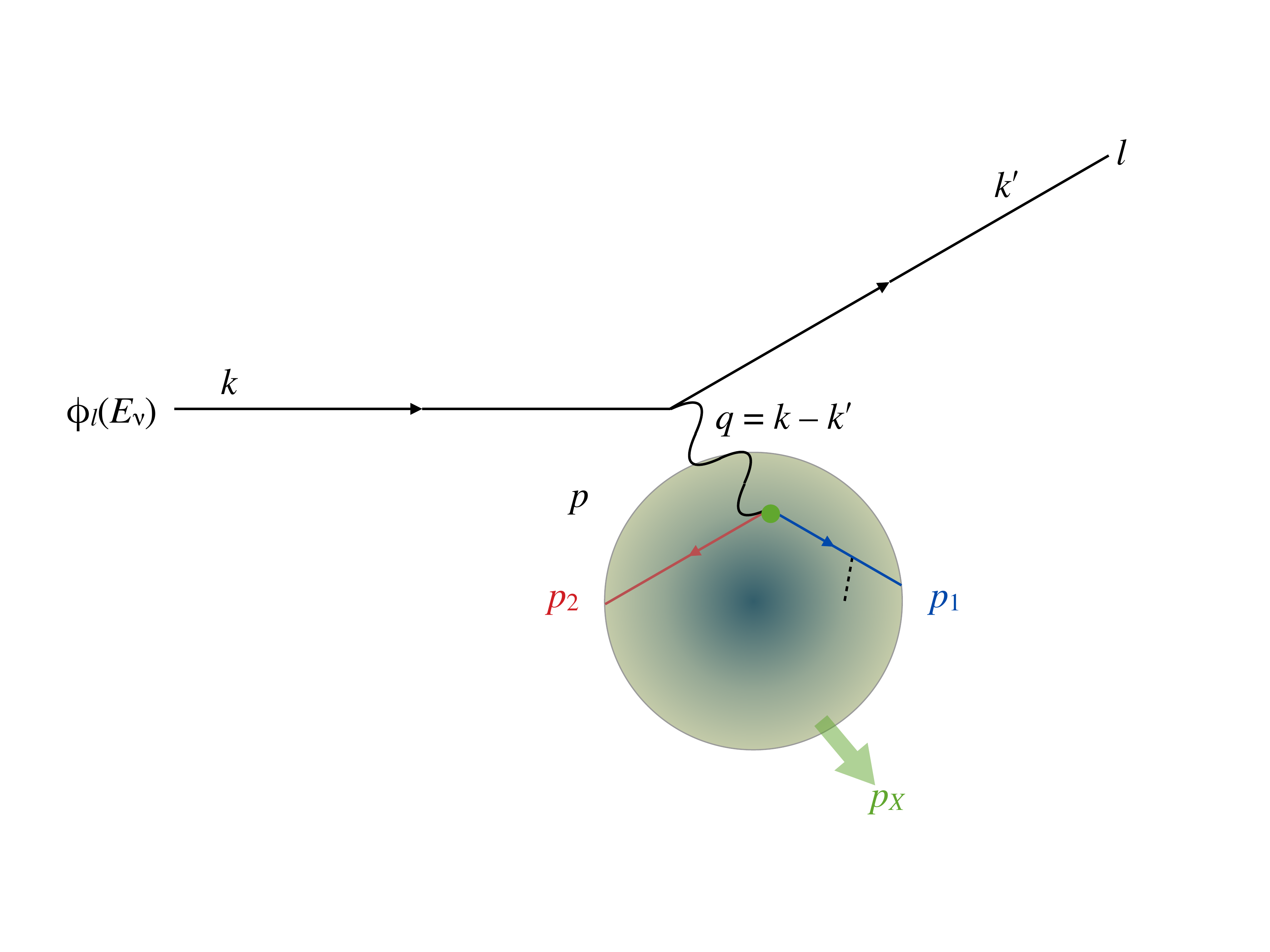}
    \caption{\color{black} In neutrino-nucleus scattering a neutrino of energy $E_\nu$ and flavor $l$ within a beam with energy spectrum~$\phi_l(E_\nu)$, strikes a nucleus of atomic number~$A$. In charged (neutral) current interaction the associated charged lepton~$l$ (neutrino of same flavor) emerges. Hadrons emerge from the initial interaction vertex as well that include one or more nucleons and, typically, pions (black dashed lines).}
    \label{fig:overview:scatter}
\end{figure}
\ask{The basic setup of a lepton-nucleus scattering experiment is shown in Fig.~\ref{fig:overview:scatter}.} \jgm{A neutrino of unknown energy enters the detector made of heavier nuclei and interacts.
In charged-current neutrino scattering, the final-state lepton is the charged partner of the incoming flavor while in neutral-current scattering the final state lepton is a neutrino of the same flavor as the incoming neutrino.
Typically, the exchanged $W$ or $Z$ boson interacts with a bound nucleon, moving with Fermi momentum $p_F$ within the nucleus, producing an outgoing nucleon of four-momentum~$p_1$ and, if the neutrino energy is high enough, additional hadrons, mostly pions.  Occasionally the exchanged boson interacts with a pair of correlated nucleons and a second nucleon is released in the initial interaction: these ``two-particle-two-hole'' events are fascinating from the perspective of nuclear physics and, it turns out, of quantitative importance in measuring neutrino-oscillation parameters. These nuclear effects of the initial interaction; including the Fermi momentum of the bound nucleon and the existence of correlated multi-nucleon ensembles, affect the initial kinematic distribution of both the outgoing lepton and hadronic shower. }

\begin{figure}
	\includegraphics[width=0.66\textwidth] {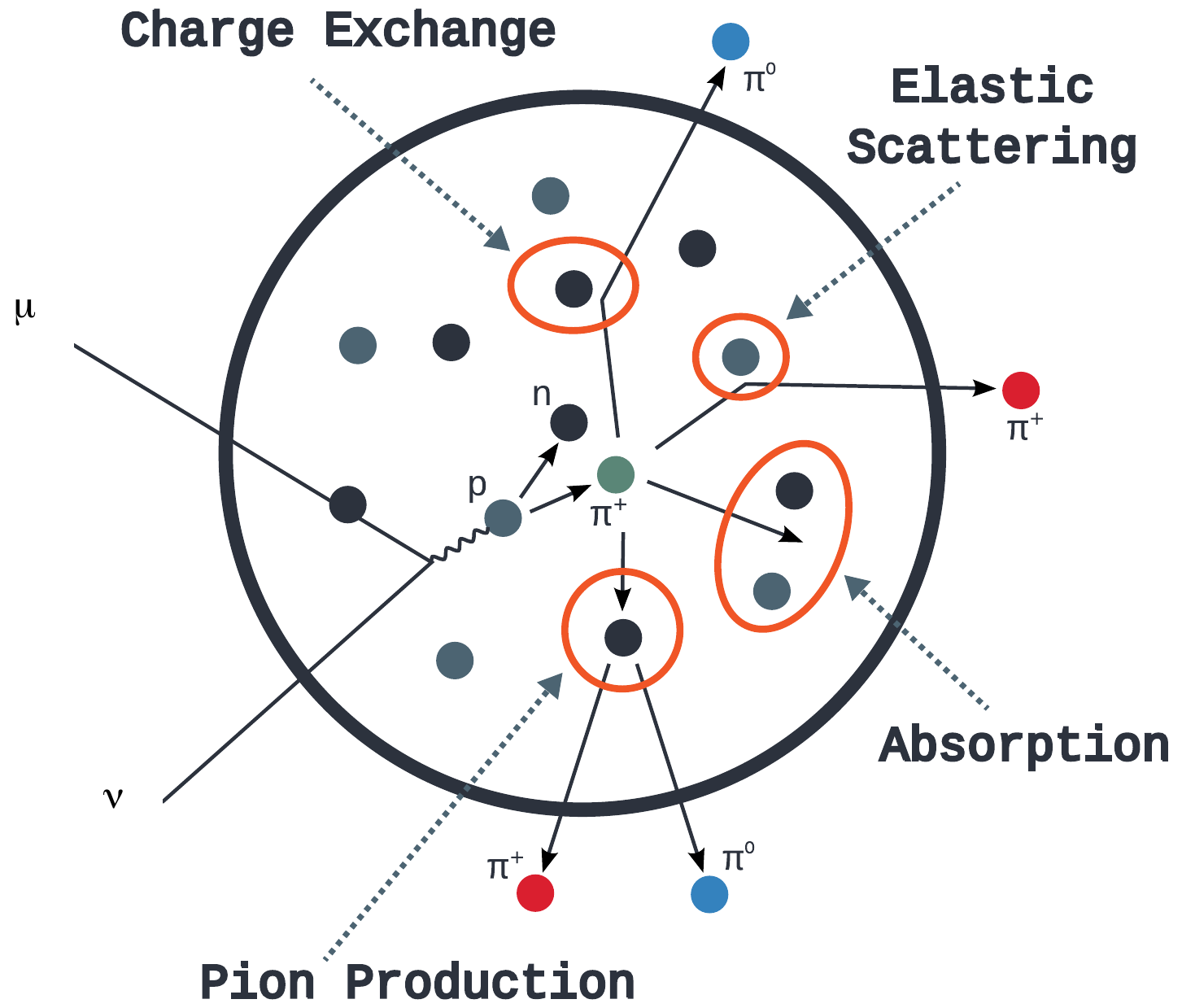}
 %   ,trim={50pt 90pt 50pt 90pt},clip]{FSI}
    \caption{The hadronic shower produced in the initial interaction must still traverse the dense nuclear matter and is then subject to Final State Interactions (FSI) before appearing in the detector.  These FSI include nucleon-nucleon interactions as well as pion-nucleon interactions as illustrated. Figure from Tomasz Golan.}
    \label{fig:overview:FSI}
\end{figure}

\jgm{ The final state lepton escapes the nucleus, however the initially produced hadronic shower undergoes significant further nuclear effects as it proceeds through the dense nuclear matter within the nucleus. As illustrated in Fig.~\ref{fig:overview:FSI}} these final state interactions (FSI) can change the energy, angle and even charge state of the originally produced hadrons with the pions having reasonable probability of even being totally absorbed within the nucleus and not emerging in the detector. The above picture of course assumes that processes can be factorized (interactions occur on individual bound nucleons), though it seems to be justified only for large enough values of momentum transfer.

\ask{It cannot be stressed enough that the incident neutrino energy is not a~priori known. This situation differs dramatically from electron or muon scattering studies where the amounts of energy and momentum that are transfered to the nucleus is known precisely on event-by-event basis.}
\jgm{For neutrino nucleus scattering the incoming neutrino energy and initially produced hadronic particles, which have been subject to the above mentioned nuclear effects, can only be estimated from what is observed in the detector.}

\jgm{Since it is the initial neutrino energy spectrum as well as signal and background topologies that have to be used in the extraction of oscillation parameters, the strong dependence of the unbiased extraction of neutrino-oscillation parameters on neutrino-interaction physics can best be summarized by noting that the energy and configuration of interactions observed in experimental detectors are, aside from detector effects, the convolution of the energy-dependent neutrino flux, the energy-dependent neutrino-nucleon cross section, and these significant energy-dependent nuclear effects.}
% *** TEXT Stops here and continues where indicated below
\cut {As stated above, these energy-dependent nuclear effects include the motion of the struck nucleon within the nucleus, the probability that the interaction is taking place off correlated multi-nucleon states, and the possible interaction of the hadron shower, produced at the interaction point, while passing through the dense hadronic components of the nucleus denoted as final-state interactions (FSI)}

\cut{Interacting neutrino energy is not a~priori known and must be estimated, based on a careful study and reconstruction of the particles in the final state.
Good understanding of all exclusive reaction channels is required.
The situation is very distinct from electron scattering studies where amounts of energy and momentum that is transfered to the nucleus is known precisely on event by event basis.}

\jgm{\cut {In neutrino oscillation  
experiments, the nucleus can distort the neutrino energy and channel of the initial interaction into what the experiments reconstruct in their detectors.
\jgm{\sl It is literally true. If the (low-energy) experiment uses the muon kinematics to estimate $E_nu$ for QE-like events they can underestimate the real energy when 2p-2h or pion absorption fools them.  For calorimetric detectors an initial proton can interact giving much energy to a neutron that then escapes detection or a pion can be absorbed on a di-nucleon and these nucleons cascade through the nucleus yielding only a reduced detectable fraction in the detector... }
Since it is the initial neutrino energy spectrum that has to be used in the extraction of oscillation parameters, the experimenters need to simulate the effects of the nucleus in their event generators.}}
% *** TEXT CONTINUES HERE
Practically, experimenters combine information about the energy dependence of all exclusive cross sections as well as nuclear effects into a \emph{nuclear model}.
This model along with the best estimate of the spectrum of incoming neutrino energies then enters the Monte Carlo predictions of target nucleus response and topology of final states and is a critical component of oscillation analyses.

To illustrate how oscillation experiments depend on this nuclear model, consider the following  illustrative conceptual outline of a two-detector, long-baseline oscillation analysis:
\begin{enumerate}
\item Reconstruct the observed event topology and energy (final state particles identification and their momenta) in the near detector  (ND).
\item Use the \emph{nuclear model} to take the reconstructed event topology and energy
back through the nucleus to infer the neutrino interaction energy $E^\text{nd}_\nu$.
\item Using information on geometric differences between near and far detector fluxes and perturbed via an oscillation hypothesis, project the resulting initial interaction neutrino energy spectrum $\phi(E^\text{nd}_\nu)$, into the predicted spectrum $\phi'(E^\text{fd}_\nu)$ at the far detector.
\item Following an \st{initial} interaction in the far detector, use the \emph{nuclear model} to take the initial $E^\text{fd}_\nu$ through the nucleus to an estimate of the reconstructed neutrino energy and topology in the far detector. 
\item Compare this estimated far neutrino energy spectrum (flux) in the far detector with the reconstructed neutrino energy spectrum to test the assumed oscillation parameters.
\end{enumerate}

\jgm{Contrary to what is sometimes assumed, the use of a near detector although extremely useful does not reduce the oscillation analysis to a simple rescaling.
Differences, both geometric and oscillation-induced, between near and far fluxes 
make the precise modeling of neutrino-nucleus  interactions a necessary and critical element of an experimental setup.}
A large and growing body of work over the past several years highlights how mis-modeling of the nucleus
(the nuclear model) could lead to unacceptably large systematic uncertainties or, worse, biased measurements in current and future oscillation experiments~\cite{Ankowski:2016bji,Mosel:2016cwa,Coloma:2013tba,Coloma:2013rqa}.
This suggests that since, for example, 
the discovery of CP violation at DUNE/LBNF will require as-yet unachieved percent-level control over the appearance signals, the understanding of the nuclear model has to be critically examined refined, and quantified.  

It is important to realize that the neutrino-nucleus interaction is the least understood component of a detector's response to neutrinos.
Understanding the subtleties of the nuclear model and its effects on what neutrino experimentalists measure in their detectors can only be accurately performed with the input of theorists specializing in this topic.

To be more specific, the following is a list of general challenges facing the community.
For some of them a strategy of how to address them seems already clear, while others require a wider discussion among experts in the field.
\begin{itemize}
\item Significant improvements of nuclear models by theorists are essential and should include: 
\begin{enumerate}
\item The development of a unified model of nuclear structure giving the initial kinematics and dynamics of nucleons bound in the nucleus.

\item Modeling neutrino--bound-nucleon cross sections not only at the lepton semi-inclusive cross section level, but also in the full phase space for all the exclusive channels that are kinematically allowed.

\item Improving our understanding of the role played by nucleon-nucleon correlations in interactions and implementing this understanding in MC generators, in order to avoid double counting.

\item Improving models of final state interactions,
which may call for further experimental input from other communities such as pion-nucleus scattering.

\item Expressing these improvements of the nuclear model in terms that can be successfully incorporated in the simulation of neutrino events by neutrino event generators.

\end{enumerate}

These steps can most efficiently be accomplished with additional support of theorists working in this area 
\ask{in a well-coordinated international program}. It is then vital to have an established  procedure that promotes nuclear and high energy theorists joining neutrino interaction generator experts and neutrino experimentalists in working toward this goal.
\cut{to}\ask{The aim of this program should be to} provide more robust models to meet the requirements of the oscillation \cut{program}\ask{experiments,} and \ask{to deepen the engagement}\cut{improving the conversation} between theorists and generator builders so as to speed the implementation of improved models in generators.
\cut{This must be an international well coordinated program aiming to involve nuclear physics groups in many countries.}

\item To establish priorities for necessary improvements to the\cut{ implemented} nuclear model requires identifying in an unambiguous quantitative way which ingredients of nuclear model currently implemented in Monte Carlo generators are most critical for the success of future neutrino oscillation experiments. Rapidly incorporating these improvements in event generators is equally important and requires a collaborative effort of the HEP and NP communities.

\item The critical role of neutrino nucleus event generators needs to be emphasized and more community resources devoted to keeping them widely available, accurate, transparent, and current.
Involvement of  the dedicated resources of leading laboratories like Fermilab and CERN is essential.

\item It is critical to benchmark improved nuclear models and the generators that employ them against both accelerator-based precision neutrino-nucleus interaction measurements and, via a collaborative HEP and NP effort, electron-nucleus interaction measurements. 
For example, expanded use of the existing Jefferson Laboratory data could bring significant insight.

\item The current experimental neutrino interaction program (\minerva, \nova Near Detector, MicroBooNE, T2K Near Detector) continues to provide important data and should be supported to its conclusion. This should include efforts to improve the precision with which the neutrino flux is known and an agreement on guidelines for a community-wide data format to enable more effective and efficient comparisons between experiments. 

\medskip
\item Future high-precision neutrino interaction experiments are needed to extend the current program of GeV-scale neutrino interactions and should include:
\begin{enumerate} 
\item A feasibility study of a high-statistics hydrogen or deuterium scattering experiment to supplement the currently poorly known (anti)neutrino-nucleon cross sections.
\item The need for (anti)neutrino Ar scattering data in the energy range relevant for the DUNE experiment.
\item The possibility of muon-based neutrino beams providing extremely accurate knowledge of the neutrino flux and an intense electron neutrino beam.  
\end{enumerate}

\item Current and future long- and short-baseline neutrino oscillation programs should evaluate and articulate what additional neutrino-nucleus interaction data is required to meet their ambitious goals and support experiments that provide this data. 

\end{itemize}

\cut{In addition to these general challenges facing the community, there are significant challenges for specific topics and particular interaction channels that often take the form of observations, problems and/or recommendations. Following are condensed summaries of these challenges. However, to best understand the current status of these topics and channels and the resulting challenges, the reader is encouraged to consult the subsequent sections of this paper. }
\lar{In addition to these general challenges facing the community, there are more specific concerns for particular topics and interaction channels. These are summarized below in the form of  observations, problem description or recommendations. For a deeper insight, the reader is encouraged to consult the subsequent sections of this paper. }

\subsection{Challenges: The Determination of Neutrino Oscillation Parameters and Neutrino-Nucleus Interaction Physics (Section ~\ref{expt_motive})}

Several initial processes can contribute to each observable topology in our detectors due to nuclear effects and the significant energy spread of neutrino and antineutrino beams. It is clear that nuclear effects are a major issue for current and future experiments. To achieve the future program, we need to tackle the following challenges:

\begin{itemize}
\item Current and future long- and short-baseline neutrino oscillation programs should evaluate and articulate what additional neutrino-nucleus interaction data or support measurements are required to meet their ambitious goals. This can be done with a combination of phenomenological and direct theoretical estimations.
 
\item Near detectors are powerful in oscillation analyses, but do have fundamental and practical limitations in the near-to-far extrapolation of event rates. New experimental methods such as NuPRISM ~\cite{Bhadra:2014oma}, which enables variable neutrino energy fluxes to enter the near detector, could circumvent the problem of different fluxes at the near and far detector. 
 
\item  Of specific interest is  precise knowledge of electron/muon neutrino cross section differences which historically have been difficult to measure in near detectors. Increased theoretical effort is necessary to determine if there are any unexpected differences. It is important to understand the level at which this quantity will be known by the proposed future experimental programs.   
\item Neutrino energy estimators are sensitive to threshold effects and model-based particle composition and kinematics. As neutrino-antineutrino event-rate comparisons are important for \dcp measurements, the relative neutron composition of final hadronic states is significant.  It is important to understand the prospects for semi-inclusive theoretical models that can predict this neutron composition. Experimentally, programs to detect neutrons are essential. Electron scattering data may also provide insights to the hadronic state.

\item The calculated detector efficiency often depends on the nuclear model. Whether or not current uncertainties cover this issue needs to be studied. If this is an important effect for current and future experiments, systematic errors must reflect the range of nuclear models used in calculating this efficiency. It may be that experimental and computational approaches will be necessary.

\end{itemize}
 
\subsection{Challenges: Generators (Section ~\ref{generators})}

Monte Carlo (MC) generators serve as a bridge between theor\cut{ists}\ask{etical models} and experimental\cut{ists communities} \jgm{measurements}.
For future neutrino oscillation experiments it is of critical importance that they contain the best knowledge of neutrino-\ask{bound-nucleus} cross sections and nuclear effects.

Because of practical importance, many MC related challenges are listed above, as particular interaction modes are discussed.
\cut{Below}\ask{Here} we present more general MC problems:
\begin{itemize}
\item \cut{How to fully engage the theory community in the design and implementation of event generators?}
\ask{The design and implementation of event generators must fully engage the relevant theory community.}
\ask{Indeed, s}\cut{S}uperior, more modern theory and models are available, but the current mechanisms for improving MC generators  \cut{been slow}\ask{have not led to rapid deployment in the codes}.
\cut{Can we consider a scheme with d}\ask{D}irect collaboration of nuclear theorists in \cut{the }generator development\cut{?}\ask{, for example, via standardized code interfaces, would hasten implementation.} 

\item Individual channels neutrino cross sections are known with a precision not exceeding 20--30\%.
There is a hope, however, that a joint global fit to the existing data could reduce the uncertainties.
When tuning generators in this kind of global fits, \cut{how should we handle}\ask{a mechanism for examining} ``tensions'' in datasets\cut{?} \ask{should be established.}  A useful goal would be  a “universal” or global tune as achieved by QCD global fits of parton distribution functions. 

\item \cut{How do we coordinate efforts between}\ask{It will be beneficial to coordinate among} generator groups to minimize duplication of effort, \cut{but still}\ask{while} preserv\ask{ing}\cut{e} the advantages of independent approaches and ideas\cut{?}\ask{.}
\cut{Should we think about a universal MC generator supported by one of the leading high energy laboratories like CERN or Fermilab?} 
\ask{For example, with support from a suitable source of funding, a universal MC generator \emph{framework, allowing users to unify}\cut{ing} the strengths of the existing tools, should be created.} 
\cut{Will experimental groups be ready in time to use external MCs rather than developing their own?}%
\ask{Similar efforts were supported for LHC experiments and proved very successful.}
\end{itemize}

\subsection{Challenges: Electron-nucleus Scattering (Section ~\ref{eA})}

Any nuclear model used to describe neutrino-nucleus scattering should first be validated against these data. Since the vector part of the weak response is related to the electro-magnetic response through CVC, such a test is necessary, but not sufficient, to ensure the validity of a model for given kinematics, namely given values of the transferred energy $\omega  (= \nu$ for neutrinos) and momentum q.  The main challenges in connecting electron and neutrino reactions:

\begin{itemize}
\item matching models used to predict neutrino-nucleus observables to
electron scattering data
\item expanding theory to include more semi-inclusive predictions
\item provide semi-inclusive neutron, proton and pion data sets with as
broad an angular range as possible
 \end{itemize}

\subsection{Challenges: Quasielastic Peak Region (Section ~\ref{CCQE})}
\label{over:CCQE}
 
The charged current quasielastic (CCQE) reactions
$$
\nu_\mu n\to \mu^- p,\qquad \bar\nu_\mu p\to \mu^+ n
$$
are the most important \cut{in}\ask{when the neutrino flux is} predominantly sub-GeV\ask{, such as in the}\cut{neutrino flux experiments like} T2K or MicroBooNE \ask{experiments.
However CCQE remains}\cut{and} significant even \cut{for}\ask{at} higher neutrino energ\cut{y}\ask{ies, such as in the}\cut{ experiments like} NOvA and DUNE \ask{experiments}.
While the CCQE reaction is uniquely defined in the case of \ask{a} free nucleon target\cut{ reactions},
in the case of neutrino-nucleus scattering it usually refers to a neutrino bound-nucleon interaction in \ask{which} 
the intermediate \ask{vector} boson is absorbed by \ask{only} one nucleon\cut{ only}.

Its experimental identification \cut{is not obvious}\ask{can be ambiguous} due to hadronic final-state interaction effects.
\ask{Even so, u}\cut{U}nbiased reconstruction of the interacting neutrino energy is simpler for CCQE than for any other reaction channel, so its systematic error should be the smallest \ask{and most robust}.
\cut{Also a} Apart from the significant nuclear effects, the theory of CCQE scattering is \cut{rather simple}\ask{straightforward} and is reduced to a knowledge of several vector and axial form factors \ask{of the nucleon}.

%\ask{\P}
The  major challenges for this reaction channel are
\begin{itemize}
\item \cut{How to }improv\cut{e}\ask{ing} our knowledge of the axial part of the \jgm{nucleon\cut{\ask{neutrino}}-nucleon} transition matrix elements via
\begin{enumerate}
\item \cut{Do we need a new high-statistics hydrogen and/or deuterium bubble chamber cross section experiment?}
\ask{a new high-statistics hydrogen and/or deuterium \cut{bubble chamber} cross section experiment; or}
\item 
\ask{lattice-QCD calculations of the nucleon form factors at the same level of quality and precision as for  meson form factors used in quark-flavor physics;}
\end{enumerate}

\item \cut{How important are radiative corrections? }
The inclusion of radiative corrections is critical for required precision cross sections.
Radiative corrections impact theoretical predictions for absolute cross section normalizations, kinematic distributions, and $\nu_\mu/\nu_e$ cross section ratios.

\item \cut{What is the best way to include two-body current contribution in Monte Carlo generators?
How should one account correctly for one- and two-body current interference?
Currently, there are many theoretical proposals with large differences in their prediction.}
\ask{refining the theoretical description of correlated nucleon effects, especially in view of large differences in \jgm{predictions}, and then implement the best description(s) in Monte Carlo generators;}

\item \cut{How to use results from ab initio computations in Monte Carlo generators?
Limitations come from the fact that the available results are non-relativistic, confined to light nuclei and a limited phase space.}
\ask{extending the reach of \emph{ab~initio} computations of nuclear structure beyond nonrelativistic kinematics in light nuclei and to a greater portion of phase space.}

\item \cut{Keeping in mind increasing}\ask{The} interest in final-state proton studies \ask{is increasing, raising the profile of final-state-interaction models and their implementation in generators.}\cut{: how reliable are nucleon final state interaction models used in Monte Carlo generators?}

\item 
\ask{Superscaling---i.e., the empirical observation that electron-scattering \jgm{experimental results} can be brought into a form relying on a single kinematic variable---should be extended to and tested in neutrino scattering.}
\end{itemize}

\subsection{Challenges: The Resonance Region (Section ~\ref{1pi})}
\label{over:RES}

The resonance region is characterized by transfers of energy larger than in QE peak region corresponding to larger hadronic invariant mass.
The most important contribution is from the $\Delta(1232)$ resonance\ask{:} \cut{excitation region}
%\hfill \ask{\sl typeset antineutrino analog (as in QE)?}
$$
	\nu_\mu p\to \mu^- \Delta^{++},\quad \Delta^{++}\to p \pi^+ 
$$
and
$$
   \bar\nu_\mu n\to \mu^+ \Delta^{-},\quad \Delta^{-}\to n \pi^-,
$$
However better knowledge of contributions from heavier resonances is also important for higher energy experiments like NOvA and DUNE and seriously lacking.

The most important challenges are
\begin{itemize}
\item \cut{How to }improv\cut{e}\ask{ing} our knowledge of the axial part of \cut{the }nucleon-$\Delta$ transition matrix elements\cut{?}\ask{, either via a new hydrogen and/or deuterium experiment or via lattice-QCD calculations;}

\item describing nonresonant contributions to pion production channels. Understanding the range of applicability of models based on chiral perturbation theory particularly for higher mass states where no calculations currently exist;
 
 \item \cut{How to }incorporat\cut{e}\ask{ing} more modern models of pion production in the $\Delta$ region and 2-pion production channels in current neutrino event generators\cut{.}\ask{;}
 
 \item \cut{How}\ask{evaluating the} importan\cut{t}\ask{ce of nucleon-nucleon correlated pairs in}\cut{is a contribution to} pion production\cut{ coming from nucleon-nucleon correlated pairs.}\ask{;}
 
 \item \cut{What is}\ask{understanding} the origin of \cut{widely discussed}\ask{the} tensions between MiniBooNE and \minerva pion production measurements on (mostly) carbon target\ask{s} in the $\Delta\ask{(1232)}$ region\ask{.}\cut{? How to resolve the tension in the most effective way?}
\end{itemize}

\subsection{Challenges: Shallow and Deep-Inelastic Scattering Region (Section~\ref{sisdis})}

The description of inclusive lepton scattering in the transition region between resonance excitation and deep-inelastic scattering (DIS) is a subject of continuing study. This region, sometimes referred to as shallow inelastic scattering (SIS), can contribute significantly to the determination of neutrino oscillation parameters through feed-down via nuclear effect into both signal and background estimates.
%\ask{Sec.~\ref{over:RES}} and }is 
\cut{The q}\ask In electro-production experiments Quark-hadron (QH) duality\cut{, first introduced by Bloom and Gilman to explain electron-proton scattering,} has been shown to provide a connection between the average value of interaction strengths in the quark-gluon description of the DIS formalism \cut {at high $Q^2$}and the average value of interaction strengths in the pion-nucleon description in the region of resonance excitation \cut{at lower~$Q^2$}. However, the application of QH duality in neutrino scattering is still being investigated.

\ask{At even higher \jgm {hadronic mass} and four-momentum transfer the reaction is described by the interaction with partons and perturbative QCD successfully describes this reqion.}
$$ 	
	\nu_l/\bar \nu_l + N \rightarrow \mu^{\mp} + X
$$
In the studies of charged lepton nucleus DIS there is evidence from experimental measurements as well as theoretical studies that the quark parton distribution function for the nucleons bound in nuclei (nPDF) differs from the quark PDFs in the free nucleon. These partonic nuclear effects demonstrate themselves even down into the SIS region. In addition, non-perturbative High Twist (HT) effects 
also play a significant role in the SIS/DIS region for the typical kinematics of modern (anti)neutrino experiments. Both the HT and nuclear 
corrections in (anti)neutrino scattering are still characterized by large uncertainties which require more experimental and theoretical efforts.
It is worth noting that the existing data from (anti)neutrino SIS/DIS indicate some discrepancies and have limited precision. Various analyses 
of the nuclear effects in (anti)neutrino-nucleus scattering suggest possible differences in the behavior of nuclear effects observed in the 
case of the charged lepton-nucleus scattering. These differences may have implications while doing a combined analyses using neutrino 
and charged-lepton data sets for the extraction of nuclear and proton PDFs

Further study of these kinematic regions require
\begin{itemize}
\item \cut{O}\ask{o}ptimiz\ask{ation of}\cut{e} the description of the transition region from DIS to resonance production and definition of the kinematic limits of applicability of the DIS formalism for structure functions and cross sections;

\item study of the interplay of various nuclear effects (Fermi motion, nuclear binding, meson exchange currents, nuclear shadowing, off-shell effects, etc.) in different regions of $x_\text{Bj}$ and $Q^2$ for neutrino and anti-neutrino interactions with bound nucleons;

\item study of the impact on cross sections of higher-twist contributions, the $F_L$ structure function, and radiative corrections;

\item \cut{Perform}\ask{carrying out} new precise measurements with neutrinos and antineutrinos of differential and total cross sections on a variety of nuclear targets in the same experiment with wide $x_\text{Bj}$ and $Q^2$ coverage to compare nucleus-dependent extracted structure functions and their ratios.

\item making model independent measurements of nuclear effects on structure functions with neutrinos and antineutrinos by comparing measurements on nuclear targets to new precise measurements on free proton and deuteron targets in the same experiment across $x_{Bj}$ and $Q^2$;

\item understanding the differences in the nuclear effects for electromagnetic and weak DIS structure functions and cross sections and consequent extraction of nuclear parton distributions; 

\item clarifying existing discrepancies among existing measurements and between (anti) neutrinos and charged leptons across $x_\text{Bj}$;

\item Improve hadronization models in modern generators in order to describe exclusive hadron production at all W values;

\item obtaining a consistent description of SIS/DIS (anti)neutrino cross sections with respect to recent models and other developments. 
\end{itemize}

\subsection{Challenges: Coherent Meson Production (Section~\ref{cohdiff})}

A proper understanding of the coherent and diffractive processes is very important in the analysis of neutrino $\nu_\mu$ oscillation experiments. 
\ask{These processes take the form}
$$
\nu_l + A \rightarrow l^- + m^+ + A , \quad \bar{\nu}_l + A \rightarrow l^+ + m^- + A
$$
with $m^\pm = \pi^\pm, K^\pm, \rho^\pm, \ldots$, while in the NC case, one has
$$
\nu_l + A \rightarrow \nu_l + m^0 + A , \quad \bar{\nu}_l + A \rightarrow \bar{\nu}_l + m^0 + A
$$
with $m^0 = \gamma, \pi^0, \rho^0, \ldots$.
In particular, \ask{neutral-current production of} $\pi^0$ or $\gamma$ \cut{production by neutral current (NC) are important $\nu_\mu$ induced backgrounds to $\nu_\mu\to\nu_e$ oscillations because $\gamma$ or $\pi^0$ events }can mimic \cut{$\nu_e$ signal events}\ask{final-state electrons}.
\ask{Thus, their production results in important backgrounds to $\nu_\mu\to\nu_e$ oscillations.}
Furthermore, in many experiments, \ask{coherent photon events}\cut{single showers induced by NC coherent $\gamma$ emission} can hardly be distinguished from those coming from \ask{the reference process of} $\nu$-$e$ elastic scattering\cut{, which is a reference process in neutrino physics}.

Specific challenges are
\begin{itemize}
\item Ambiguities in the predictions of \jgm{coherent} \cut{PCAC \ask{\sl not defined in this \S}} pion production models implemented in different neutrino event generators should be resolved.
A validation criterion could be the ability to describe pion nucleus scattering.

\item For pion and kaon coherent production, it is important to understand if the accuracy goals justify the need for models better than the simple and fast coherent production models. 

\item Microscopic models must be more efficiently implemented and extended beyond the $\Delta(1232)$ region. 

\item Microscopic models must be validated with other reactions such as coherent meson photo-and-electro-production, meson-nucleus scattering.

\item Address coherent gamma production both theoretically and experimentally in the neutrino energy range of interest for DUNE, HK, and short-baseline (SBN) experiments. 

\item Other coherent meson production channels such as Coherent $\rho$ production should be studied both theoretically and experimentally.
\cut{\st{Coherent $\rho$ production is coming soon in GENIE.}
\ask{\sl grayed out bit is not an action item}}

\item Address theoretically (isolate from inclusive pion production) neutrino-nucleon diffractive pion production at low~hadronic mass. 

\item Measure the nucleus $A$~dependence of coherent scattering off a range of nuclei and compare data to theoretical predictions.

\item Perform new measurements of coherent and diffractive scattering to complement MINERvA measurements.
\end{itemize}

\newpage 

\section{The Impact of Neutrino Nucleus Interaction Physics on Oscillation Physics Analyses}
\label{expt_motive}
 
\subsection{Neutrino oscillations and the extraction of oscillation parameters}
\label{sec:basics}

The basic phenomenology of any oscillation experiment can be understood from the two-flavor limit.
For two families in vacuum, the probability that a neutrino of flavor $\alpha$ oscillates into\cut{ a neutrino of} flavor $\beta$,
after propagating through a distance $L$, can be written as
\begin{equation}
    P (\nu_\alpha \to \nu_\beta) \simeq \sin^22\theta \sin^2\left( \frac{\Delta m^2 L}{4 E}\right)  ,
    \label{eq:Ptwofam}
\end{equation}
where $\Delta m^2\cut{ \equiv m^2_\text{heavy} - m^2_\text{light}}$ is the mass-squared splitting between the two mass eigenstates
of the system, $\theta$ is the mixing angle which changes between the flavor and mass bases, and $E$ is the neutrino energy.
As can be seen from Eq.~(\ref{eq:Ptwofam}), the oscillation probability is maximized for values of $L$ and $E$ such that
$\Delta m^2 L/4E \sim (n + \frac{1}{2})\pi$, $n$ being an integer. The neutrino energy at which the maximum of the oscillation takes place tells us the value of the mass splitting (\ie the frequency
of the oscillation), while the amplitude of the oscillation tells us the value of $\sin^22\theta$.

A unitary mixing matrix, $U$, for three Dirac neutrinos can be parametrized with three mixing angles, $\theta_{12}$, $\theta_{23}$, and $\theta_{13}$, plus a CP-violating phase, $\dcp$.%
\footnote{In the case of Majorana neutrinos, two additional CP-violating phases enter the mixing matrix; oscillation experiments
are, however, insensitive to these phases.}
The customary way the angles and phase parametrize $U$ is the same as in the quark-mixing matrix~\cite{Olive:2016xmw}.%
\footnote{The elements of $U$ are denoted $U_{\ell i}$, $\ell=e,\mu,\tau$, $i=1,2,3$.}
With three flavors, the oscillation pattern is governed by two different oscillation frequencies:
these are given by the two mass-squared differences $\Delta m^2_{21} $ and $\Delta m^2_{31}$, usually referred to as the solar and
the atmospheric mass-squared splittings, from the observations that first established them as nonzero.
The most recently updated values for the neutrino mixing parameters obtained from a global fit to neutrino oscillation data can be found, for instance, in Refs.~\cite{Esteban:2016qun,nufit,Capozzi:2016rtj}, and are summarized in Table~\ref{tab:oscparams}.
%%%%%%%%%%%%%%%%%%%%%%%%%%%%%

\begin{table}[bp]
\setlength{\tabcolsep}{7pt}
\centering
\renewcommand{\arraystretch}{1.6}
\begin{tabular}{l  c c c c c c}
\hline\hline
&  $\theta_{12}$ &  $\theta_{13}$ &  $\theta_{23}$ &  $\Delta m^2_{21}/10^{-5}$ &  $\Delta m^2_{3j}/10^{-3}$ & $\dcp$ \\ \hline
 Normal Ordering &  
 $33.56^{+0.77}_{-0.75}$ &   $8.46^{+0.15}_{-0.15}$ &  $41.6^{+1.5}_{-1.2}$ & $7.50^{+0.19}_{-0.17}$ & $2.524^{+0.039}_{-0.040}$ & $261^{+51}_{-59}$ \\
Inverted Ordering &  
 $33.56^{+0.77}_{-0.75}$ &   $8.49^{+0.15}_{-0.15}$ &  $50.0^{+1.1}_{-1.4}$ & $7.50^{+0.19}_{-0.17}$ & $-2.514^{+0.038}_{-0.041}$ &  $277^{+40}_{-46}$ \\ \hline \hline
\end{tabular}
\caption{Experimentally allowed ranges for the oscillation parameters from a global fit to neutrino oscillation data, taken from Ref.~\cite{Esteban:2016qun}. All mixing angles and the CP-phase are given in degrees, while the $\Delta m^2_{ij}$ are given in eV$^2$. The values in the table indicate the current best-fit and the edges of the allowed confidence regions at $1\sigma$, for the two possible neutrino mass orderings, normal ($m_a < m_2 < m_3$) and inverted ($m_3 < m_1 < m_2$). The value given for $\Delta m^2_{3j}$ corresponds to $\Delta m^2_{31}$ for normal ordering and $\Delta m^2_{32}$ for inverted ordering.} 
\label{tab:oscparams}
\end{table}
%%%%%%%%%%%%%%%%%%%%%%%%%%%%

As can be seen from Table~\ref{tab:oscparams}, the first hints for CP violation and the octant of $\theta_{23}$ are slowly emerging at $1\sigma$. However, this preference takes place at low statistical significance and completely disappears at $3\sigma$; see Refs.~\cite{Esteban:2016qun,Capozzi:2016rtj} for details. Furthermore, current neutrino data show only a very mild preference for normal ordering. For instance, in Ref.~\cite{Esteban:2016qun} the authors find $\Delta \chi^2 = 0.83$ for the inverted ordering hypothesis. The current and future generation of oscillation experiments will aim for the following three main goals:
\begin{enumerate}
    \item establish whether nature violates CP in the lepton sector and, if so, measure $\dcp$;
    \item improve the accuracy on $\theta_{23}$ and, if not maximal, a determination of the octant it belongs to: $\theta_{23}<\pi/4$ 
        vs.\ $\theta_{23}>\pi/4$;
    \item determine the neutrino mass ordering at high confidence level: $m_1 < m_2 < m_3$ vs.\ $m_3 < m_1 < m_2$.
\end{enumerate} 
These goals will all require an unprecedented level of accuracy in oscillation experiments,
in order to pin down subleading effects.
A reliable and accurate estimation of the incoming neutrino energy, discussed in Sec.~\ref{sec:reco-neut}, will also be 
crucial to lift\cut{ potentially harmful} parametric degeneracies.
In the following, we will focus on the determination of $\theta_{23}$ and $\dcp$, as these are 
especially subject to the impact of systematic uncertainties and reconstruction effects.

The possibility of CP violation in neutrino oscillations relies on the interference between the two
contributions to the oscillation amplitude from $\Delta m^2_{21} $ and $\Delta m^2_{31}$: it is a genuine three-flavor effect.
At the first atmospheric oscillation maximum, $L/E\sim 500~\text{km/GeV}$, the interference is
already observable.
Under the well-justified assumption of CPT conservation, CP violation can only be observed if the initial and final neutrino flavors
are different.
It is very difficult to create $\nu_\tau$ beams or efficiently detect $\nu_\tau$.
Therefore, searches for CP violation at long-baseline experiments measure
oscillations in the appearance channels $\nu_\mu\to\nu_e$ and $\bar\nu_\mu\to\bar\nu_e$.%
\footnote{From a purely physical point of view, the time-reversed channels $\nu_e \to \nu_\mu$ and $\bar\nu_e \to \bar\nu_\mu$,
thanks to CPT invariance, contain the same information.
However, $\nu_e/\bar\nu_e$ beams are technically more difficult to obtain than $\nu_\mu /\bar\nu_\mu$ beams.}
  
For long-baseline experiments, analytical expressions for the oscillation probabilities in this channel can be obtained by expanding in the small quantities $\theta_{13}$,
$\Delta_{21}/\Delta_{32}$, $\Delta_{21}/A$, where $\Delta_{ij} \equiv \Delta m^2_{ij} L /2E $.
Here, $A \equiv \sqrt{2}G_FN_e$ is the matter potential felt by the neutrinos as they travel through the Earth,
with  the Fermi constant $G_F$ and  the density of electrons $N_e$.
At second order, the oscillation probability $\nu_\mu\to\nu_e$ reads%
\footnote{Because $\theta_{13}$ is not small, additional terms should be included in 
the expansion to increase its level of accuracy.
More accurate expressions of the oscillation probabilities can be found in Refs.~\cite{Asano:2011nj,Denton:2016wmg}.}%
~\cite{Cervera:2000kp}:
\begin{align}
    P_{\mu e} &= s_{23}^2 \sin^2 2 \theta_{13} \left( \frac{ \Delta_{31} }{ \tilde B_\mp } \right)^2
        \sin^2 \left( \frac{\tilde B_\mp L}{2} \right) +
        c_{23}^2 \sin^2 2\theta_{12} \left( \frac{ \Delta_{21} }{A} \right)^2 \sin^2 \left( \frac{A  L}{2} \right ) \nonumber  \\
    &\hspace{2em} + 
    \tilde{J}\, \frac{\Delta_{21} }{A} \frac{\Delta_{31}}{\tilde B_\mp} \sin \left( \frac{ A L}{2}\right) 
        \sin \left( \frac{\tilde B_{\mp} L}{2}\right) \cos \left( \mp \dcp - \frac{ \Delta_{31}  L}{2} \right )  , 
\label{eq:Pmue}
\end{align}
where $s_{ij}\equiv \sin\theta_{ij}$, $c_{ij}\equiv \cos\theta_{ij}$,
$\tilde{J}\equiv c_{13}\sin^22\theta_{13}\sin^22\theta_{12}\sin^22\theta_{23}$, and
$\tilde B_\mp \equiv |A \mp \Delta_{31}|$.
The upper (lower) sign correspond to the neutrino~(antineutrino,
$\bar\nu_\mu\to\bar\nu_e$) channel.
Thus, CP-violation searches can be performed by combining measurements of
$P(\nu_\mu\to\nu_e)$ and $P(\bar{\nu}_\mu\to\bar{\nu}_e)$, trying to observe a different behavior for particles and antiparticles.
The information gathered at different neutrino energies also generally helps to reduce the size of the allowed confidence regions,
which overall results in a better determination of the value of $\dcp$.

As can be seen from Eq.~(\ref{eq:Pmue}), however, every term entering the oscillation probability
is suppressed either with the value of $\theta_{13}$, the value of $\Delta_{21}$, or the product of the
two (in the case of the interference term), making this measurement very challenging from the start.
Moreover, matter effects also violate CP, because the matter potential takes a different
sign for neutrinos and antineutrinos, \cut{which can further}\oscask{thereby} hinder\oscask{ing} a signal of intrinsic CP violation.
To measure the $P(\nu_\mu\to\nu_e)$ and its $L/E$ dependence\oscask{, it is necessary to identify}\cut{requires identification of} the
neutrino flavor and \oscask{reconstruct the} neutrino energy\oscask{.
The reconstruction requires a solid}\cut{, as well as} knowledge of the interaction rate of a $\nu_e$ of given energy on the
target nucleus, as well as the reconstruction efficiency.
Furthermore, as both $P(\nu_\mu\to\nu_e)$ and $P(\bar{\nu}_\mu\to\bar{\nu}_e)$ are \cut{relevant for}\oscask{key to} CP violation
searches, the \oscask{separate interaction and reconstruction} behavior of both neutrinos and antineutrinos \cut{needs to be
characterized}\oscask{must be understood}.

\begin{figure}[tbp]
    \centering
    \includegraphics[scale=0.5]{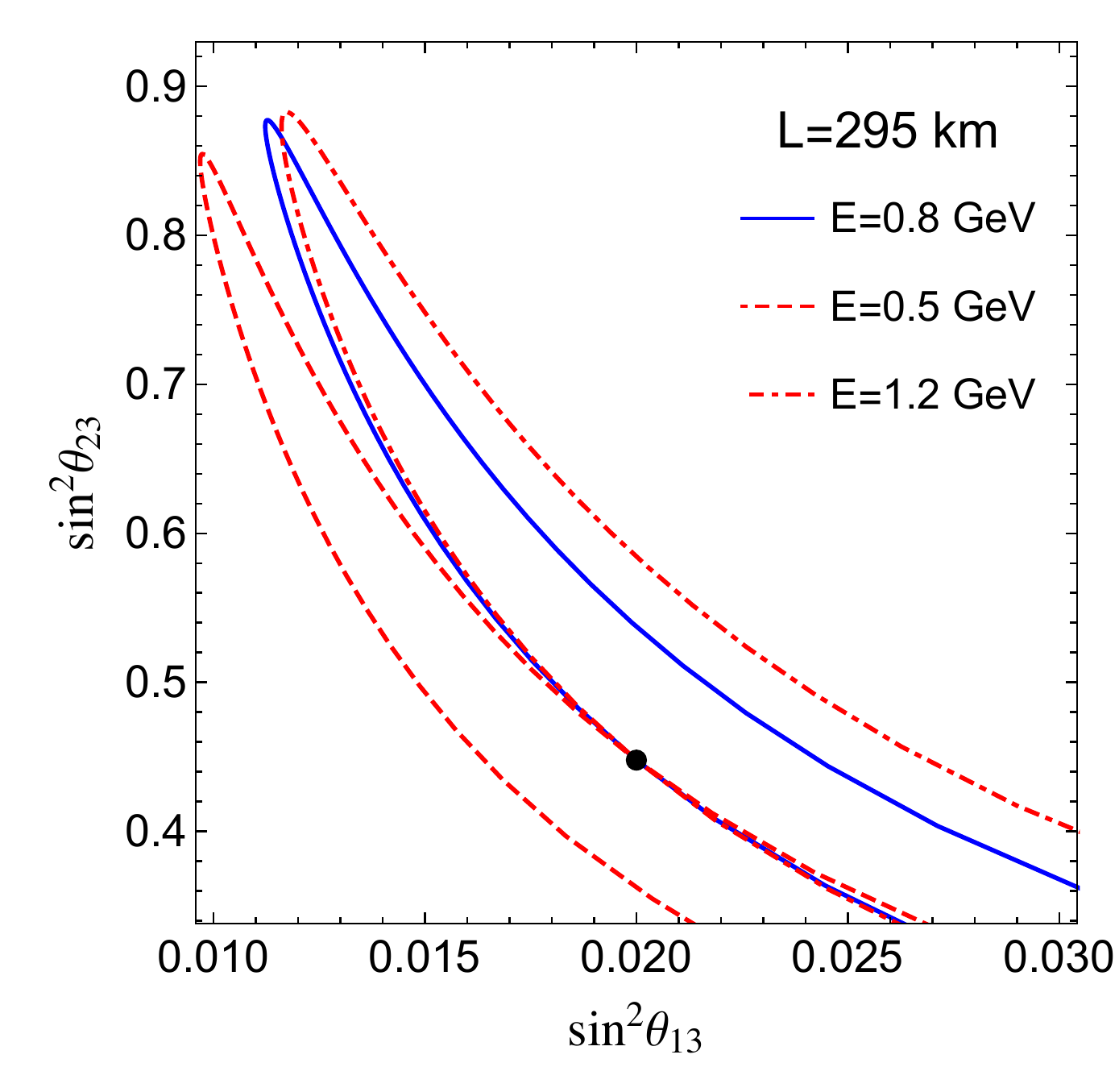}
    \caption{Illustration of the interplay between information obtained at different neutrino energies to\cut{wards the}
        resol\oscask{ve}\cut{ution} the generalized intrinsic degeneracy.
        Each curve shows the set of values of the mixing angles which are able to reproduce the same values of the appearance 
        probabilities simultaneously in the neutrino and antineutrino channels, in vacuum, and for $L=295$~km.
        The curves are obtained for different values of the neutrino energy, as indicated in the legend.
        Each point\cut{ in each curve} is obtained by varying continuously the CP phase away from its true value, which has been set
        in this example to $\dcp=30^\circ$.
        The true values of $\theta_{13}$ and $\theta_{23}$ are indicated by the black dot.
        Figure adapted from Ref.~\cite{Coloma:2014kca}, see text for details.}
    \label{fig:equiP}
\end{figure}

It is important to appreciate that measurements of neutrino oscillation probabilities typically suffer from
parametric degeneracies.
These are classified into different categories: sign degeneracies in the appearance channels involving $\dcp$ and the
sign of $\Delta m^2_{31}$~\cite{Minakata:2001qm}, octant degeneracies in the disappearance channels involving
$\theta_{23}$~\cite{Fogli:1996pv}, and intrinsic degeneracies in the appearance channels involving $\theta_{13}$, $\dcp$ and
$\theta_{23}$~\cite{BurguetCastell:2001ez} (see also Refs.~\cite{Coloma:2014kca,Minakata:2013hgk}).
On general grounds, these give rise to the so-called eightfold degeneracy problem in neutrino oscillations~\cite{Barger:2001yr}.
\oscask{In the literature, s}\cut{S}everal ways have been proposed\cut{ in the literature} to lift the degenerate solutions.
In particular, it has been shown that the intrinsic degeneracies involving $\theta_{13}$, $\theta_{23}$ and $\dcp$ can be
alleviated (or completely lifted) by combining information at different neutrino energies, or at different
baselines~\cite{BurguetCastell:2001ez}.
An example illustrating this point is shown in Fig.~\ref{fig:equiP}.
%%%%%%%%%%%%%%%%%%%%%%%%%%%%%%%%%%%%%%%
%%%%%%%%%%%%%%%%%%%%%%%%%%%%%%%%%%%%%%%
\cut{In this example, t}\oscask{T}he true input values assumed for the mixing angles are $\sin^2\theta_{13}=0.02$,
$\sin^2\theta_{23}=0.45$ and $\dcp=30^\circ$, which is indicated by the black dot.
The \cut{rest of the}\oscask{other} points in each line \cut{is}\oscask{are} obtained varying $\dcp$ continuously and \cut{imposing the
requirement that the values of}\oscask{requiring} $P$ and $\bar P$ \cut{are}\oscask{to be} constant and equal to \cut{the ones obtained
for} the\oscask{ir}\cut{ true} values \cut{of}\oscask{dictated by} the \oscask{true} oscillation parameters.
As can be seen from Fig.~\ref{fig:equiP}, \cut{multiple sets of}\oscask{many} values of ($\theta_{13},\theta_{23},\dcp$)\cut{ are able
to} recover the same oscillation probabilities\oscask{, for fixed}\cut{at a given} energy.
Thus, a measurement of the oscillation probabilities $P(\nu_\mu \to \nu_e) $ and $ P(\bar\nu_\mu \to \bar\nu_e)$ for a single value
of the neutrino energy would not suffice to determine the value of $\dcp$, unless all mixing angles are known very precisely.
However, this degeneracy is efficiently broken when the probability is measured at different neutrino energies: as can be seen from
the figure, the different lines only overlap for the point corresponding to the assumed true values for the oscillation parameters.
For a recent detailed discussion of this degeneracy at long-baseline experiments, see Ref.~\cite{Coloma:2014kca}.

On the other hand, the determination of the value of $\theta_{23}$ comes from a combination
of disappearance and appearance data.
Due to the low statistics in the appearance channels, its value is typically inferred from the observation of the 
$\nu_\mu\to\nu_\mu$ probability, which for long-baseline experiments is well-approximated by~\cite{Raut:2012dm}:

\begin{equation}
    P_{\mu\mu} = 1 - \sin^22\theta_{\mu\mu} \sin^2\left( \frac{\Delta m^2_{\mu\mu} L}{4E} \right)  , 
    \label{eq:Pmumu}
\end{equation}
where $\Delta m^2_{\mu\mu}$ is the muon neutrino weighted average of $\Delta m^2_{31}$ and
$\Delta m^2_{32}$~\cite{Nunokawa:2005nx}, and
\begin{equation}
    \sin^2 2 \theta_{\mu \mu} \equiv 4 |U_{\mu3}|^2(1- |U_{\mu3}|^2) =
        4\cos^2 \theta_{13} \sin^2 \theta_{23} (1- \cos^2 \theta_{13} \sin^2 \theta_{23}) .
\label{eq:thetamumu}
\end{equation}
Due to the large value of the atmospheric mixing angle, which is very close to maximal mixing, this probability is characterized by
a strong dip \oscask{in the event rate} at the oscillation maximum.
The measurement of the energy at which the oscillation maximum takes place determines the value of $\Delta m^2_{\mu\mu}$, while the
size of the dip itself will determine the magnitude of $\sin^22\theta_{\mu\mu}$.
The latter is directly related (and approximately equal) to $\sin^22\theta_{23}$ up to small corrections which are proportional to
$\sin^2\theta_{13}$~\cite{Raut:2012dm}.

Thus, as can be seen from Eqs.~(\ref{eq:Pmumu}) and~(\ref{eq:thetamumu}), disappearance experiments are mainly sensitive to the 
value of $\sin^22\theta_{23}$ and are unable to identify its octant.
The octant determination has to come from the combination of disappearance and appearance data: as the leading order term in the
$P_{\mu e}$ oscillation probability depends on $\sin^2\theta_{23}$ [see Eq.~(\ref{eq:Pmue})], it can potentially break this
degeneracy after combination with the constraints on $\sin^22\theta_{23}$ coming from the disappearance channels.%
\footnote{\cut{It should be n}\oscask{N}ote\cut{d} that, for values of $\theta_{23}$ very close to maximal mixing, the most precise 
measurements may come from the appearance channels instead\oscask{,}\cut{.
However, this will eventually} depend\oscask{ing eventually} on the level of systematic errors affecting this measurement and the 
statistics of the experiment; see Ref.~\cite{Coloma:2014kca}.}

A plethora of long-baseline neutrino experiments have been proposed to measure the $\nu_\mu\to\nu_e$ and $\nu_\mu\to\nu_\mu$
oscillation channels, together with their CP conjugates, and are summarized in Table~\ref{tab:expt}.
%%%%%%%%%%%%%%%%%%%%%%%
\begin{table*}[tbp]
\setlength{\tabcolsep}{5pt}
\centering
\caption{List of currently operating and future long-baseline neutrino experiments\cut{.
The listed parameters have been obtained}\oscask{, compiled} from Refs.~\cite{Adamson:2016xxw,Adamson:2016tbq,Acciarri:2015uup,HKtalkNufact}.
The flux energy range corresponds to 68\% of the total flux.
Note that most experiments receive a non-negligible flux of neutrinos with energies as high as 30~GeV.}
\label{tab:expt}
\begin{tabular}{l c c c c c c c}
\hline\hline
Experiment & Baseline & \cut{Flux peak}\oscask{Peak energy} & \cut{Flux}\oscask{Energy} range & Target & Detector  & Fiducial Mass \\
\hline 
Current: & & & & & \\
T2K   &  295~km & 0.6 GeV & 0.3--0.8 GeV & H$_2$O &         WC           & 22.5~kton \\
\nova &  810~km &   2 GeV & 1.5--2.7 GeV & CH$_2$ & Tracking+Calorimetry &   13~kton \\
\hline
Future: & & & & & \\
\HK   &  295~km & 0.6 GeV & 0.3--0.8 GeV & H$_2$O &         WC           &  520~kton \\
DUNE  & 1300~km &   2 GeV & 0.6--3.3 GeV &   Ar   & Tracking+Calorimetry &   40~kton \\
\hline\hline
\end{tabular}
\end{table*}
%%%%%%%%%%%%%%%%%%%%%%%%
Long-baseline experiments use intense neutrino (or antineutrino) beams sent through hundreds of kilometers to massive (``far'') detectors
and measure the rate of $\nu_e$ and $\nu_\mu$ interactions to infer oscillation.
The current experiments in operation are T2K and \nova.
The Tokai-to-Kamioka (T2K) experiment has a peak energy of 0.6~GeV  and \cut{neutrino travel distance 
(``}baseline\cut{'')} of 295~km to the Super-Kamiokande \oscask{water Cherenkov (WC)} detector\cut{, a water Cherenkov (WC) detector}.
T2K also has a suite of near detectors located less than a kilometer from the neutrino source.
The role of near detectors is described in more detail in Section~\ref{sec:topology}, but in the case of T2K it is notable that the
near detector technology \cut{(tracking-based) is different}\oscask{differs} from the far detector, \cut{and has both}\oscask{having}
\cut{water}\oscask{WC} and scintillator targets.
The \nova experiment will measure the same four oscillation channels as T2K but has a longer baseline, 810~km, and a higher peak
energy, 2~GeV.
\nova's detector technology \cut{is quite different and }combines tracking and calorimetric measurements.
The detector is filled with mineral oil and uses scintillation light to reconstruct the particles produced in \cut{a given}\oscask{each} event.
The \nova experiment also has a near detector, which is identical to the far detector in design and target material, but is smaller
in size: while the fiducial mass of the far detector is 14~kton, the near detector is 290~ton~\cite{Adamson:2016xxw}.
Both \cut{T2K and \nova use off-axis beams, \ie}\oscask{place} the detectors \cut{are located }at a small angle with respect to the 
beam direction (2.5$^\circ$ for T2K and $0.8^\circ$ for \nova)\cut{, which results in}\oscask{.
This technique, known as ``off axis'', yields} a narrower energy spread \cut{as compared to on-axis (or ``wide-band'') beams}\oscask{than
in a detector on the beam's axis}. Due to its longer baseline, \nova is more sensitive to matter effects, and therefore the mass ordering than T2K.
\cut{However}\oscask{That said}, the experiments provide \cut{highly }complementary information \cut{that can help}\oscask{needed to} lift
degeneracies in \cut{certain regions of the }parameter space, \cut{see}\oscask{as discussed above and in}, \eg Ref.~\cite{Abe:2014tzr}.

Two\cut{ additional} future long-baseline experiments \cut{have been pushed forward by the neutrino oscillation community}\oscask{are
being developed}: the Deep Underground Neutrino Experiment (DUNE) in the US and the Tokai-To-Hyper-Kamiokande (T2HK) experiment in Japan.
Both plan to begin operation in or around 2026.
The \HK experiment will
be very similar to T2K, operating with a similar energy spectrum and with the same detector
technology, albeit with a much larger detector of 520~kton fiducial mass~\cite{HKtalkNufact}.
DUNE will take a different approach: it will operate on-axis at higher energies, 
peaking around 3~GeV, and a baseline of $L=1300$~km.
DUNE plans to use a 40~kton liquid Argon (LAr) far detector,
which combines tracking\oscask{\;and\;}calorimeter detector, akin to \nova's approach.
Being on axis makes it possible to study a much broader range of 
energies than at off-axis experiments, although at the price of higher backgrounds.

The \cut{following}\oscask{rest of this} section\cut{s} explore\oscask{s} how neutrino interactions affect the determination of neutrino
oscillation parameters.
Experiments depend upon a\cut{ neutrino interaction (or ``cross section'')} model \oscask{of the neutrino-nucleus interaction} to
\cut{represent}\oscask{disentangle} neutrino event rates in their detectors.
The main ways in which \cut{neutrino interaction}\oscask{this} modeling \cut{may }affect\oscask{s} the oscillation physics program are
\cut{discussed in the following subsections}\oscask{are organized as follows}:
\begin{itemize}
    \item Section~\ref{sec:topology}, \textit{Event topology and experimental observables}: \cut{Multiple}\oscask{Many nucleon-level}
    processes may 
    contribute to \cut{each}\oscask{any} observ\oscask{ed}\cut{able} topology, due to the significant spread in energy of neutrino and 
    antineutrino beams\cut{ produced from meson decays}, such that signal processes are difficult to isolate. Furthermore, for each process, initial state and final state nuclear effects both play a role in the observed topology. 
    In addition, candidate selections may include processes on material other than the desired target.
    \item Section~\ref{sec:nd}, \textit{Benefits and challenges of near detectors}: The measured event rates at the near and far
    detectors differ due to \cut{the presence of }oscillations, even in the ideal case of identical near and far detectors with perfect
    efficiency.
    \cut{Practically}\oscask{In practice}, near detectors may differ from far detectors in incident source, acceptance, and/or target material.
    Near detectors may also lack precise measurements of relative difference between muon and electron (anti)neutrino interactions,
    due to the unavailability of electron (anti)neutrinos in the \cut{initial}\oscask{unoscillated} beam.
    \item Section~\ref{sec:reco-neut}, \textit{Estimation of neutrino energy}: The \cut{determination}\oscask{reconstruction} of the
    neutrino energy requires knowledge of all particles' kinematic information. However, as detection thresholds are finite and may not have the same response for all particle types, the neutrino energy may depend on the nuclear model assumed.
    \item Section~\ref{sec:efficiency}, \textit{Calculation of detection efficiency}: The efficiency used to convert the measured to
    the true event rate depends on the cross-section model\cut{ at some level}\oscask{, because the event generator (cf.\ 
    Sec.~\ref{sec:basics}) needed to determine the efficiency relies on one}.
\end{itemize}
To \cut{finalize}\oscask{conclude} our discussion,\cut{ in} Secs.~\ref{sec:impactcurrent} and~\ref{sec:impactfuture}\cut{ we will}
discuss in more detail how the\cut{ different} points raised \oscask{in Secs.~\ref{sec:topology}--\ref{sec:efficiency}}\cut{may}
impact the extraction of oscillation parameters at current and future experiments, respectively.

\subsection{Event Topology and Experimental Observables}
\label{sec:topology}

Oscillation experiments measure event rates in their far, post-oscillation detectors, which they use to extract the oscillation
probabilities discussed in Sec.~\ref{sec:basics}.
For $\nu_\alpha\to\nu_\beta$ oscillations, the event rates with a given observable topology can be naively computed as
\begin{equation}
    N^{\alpha\to\beta}_\text{FD}(\preco) = \sum_i \phi_\alpha(\Etrue) \times
        P_{\alpha\beta}(\Etrue) \times
        \sigma^i_\beta(\ptrue) \times
        \epsilon_\beta(\ptrue) \times R_i (\ptrue; \preco),
    \label{eq:fd_event_rate}
\end{equation}
where $N_\text{FD}(\preco)$ represents the event rate as a function of the reconstructed kinematic variables
$\preco\equiv(\Ereco,\pthreco)$, and $P_{\alpha\beta}(\Etrue)$ is the oscillation probability as a function of the true neutrino
energy~\Etrue.
Here, $\phi\oscask{_\alpha}$ is the neutrino flux \oscask{of flavor~$\alpha$},
$\sigma^i\oscask{_\beta}$ is the neutrino cross section for \cut{a particular type of }interaction $i$ \oscask{and flavor $\beta$},
and $\epsilon\oscask{_\beta}$ is the detector efficiency \cut{for a given set of kinematic variables}\oscask{for flavor $\beta$ as a 
function of its true four-momentum~\ptrue}\cut{ of the neutrino}.
Finally, the function $R_i(\ptrue; \preco)$ encodes the probability for the kinematic variables \ptrue to be reconstructed as \preco
due to detector smearing and nuclear effects and depends on the type of neutrino interaction~$i$.

As can be seen from Eq.~(\ref{eq:fd_event_rate}), the event sample for a given topology contains a sum over \cut{different types 
of}\oscask{several} interactions\cut{ $i$}.
This is the first way that the cross section model affects oscillation analyses\cut{-- through the degeneracy of processes in 
observable topologies}. 
Table~\ref{tab:t2k_nova_fd} shows the expected event rate predicted at the T2K and \nova experiments respectively\cut{, which are
briefly described in Sec.~\ref{sec:basics}}.
Both experiments \oscask{aim to} select \oscask{charged-current (}CC\oscask{)} \nue \cut{candidate }events.
Their most relevant backgrounds include neutral current \oscask{(NC)} \numu or \cut{charged current}\oscask{CC} \numu processes, which
mimic \nue \cut{interactions}\oscask{events}.
For example, photons from NC neutral-pion production can produce electromagnetic showers that are reconstructed as an electron from
a CC \nue interaction.
As a consequence, oscillation experiments must consider not only processes which contribute to the signal events, but also
significant (or small, but poorly understood) backgrounds which are relevant for oscillation analyses.
%%%%%%%%%%%%%%%%%%%%%%%%%%%%%
\begin{table*}[tbp]
\centering
    \caption{T2K and \nova CC \nue selection event rates at the far detector.
    The numbers of expected Monte Carlo (MC) events divided into four categories are shown after each selection criterion is applied.
    For T2K, the MC expectation is based upon three-neutrino oscillations for $\sin^{2}\theta_{23}=0.5$, $\Delta
    m^2_{32}=2.4\times10^{-3}$\evsqc, $\sin^{2}2\theta_{13}=0.1$, $\dcp=0$ and normal mass ordering (parameters chosen without
    reference to the T2K data).
    The values are reproduced from Ref.~\cite{Abe:2015awa} which correspond to a data set with an exposure of $6.60 \times 10^{20}$
    protons on target (POT).
    For \nova, the expectation is taken from Ref.~\cite{novacommunication} and corresponds to the \nova best-fit values of
    $\sin^{2}\theta_{23}=0.404$, $\Delta m^2_{32}=2.44\times10^{-3}$\evsqc, $\sin^{2}2\theta_{13}=0.085$, $\dcp=1.48\pi$,
    normal mass ordering, for full detector equivalent POT of $6.05 \times 10^{20}$.}
\label{tab:t2k_nova_fd}
\begin{tabular}{lcccccc}
\hline\hline
&                 & \ \ \numu+\numubar \ \ & \ \ \nue+\nuebar \ \ & \ \ $\nu+\bar{\nu}$ \ \ &$\nu_\tau$ & \ \ $\numu\to\nue$ \ \ \\
& \ \ MC Events\ \ & CC & CC & NC & CC \\
\hline
 T2K \nue selection   & 21.59  & 0.3\%  & 15.0\%  & 4.4\%  & -- & 80.2\%  \\
 \nova \nue selection   & 32.86  & 2.2\%      & 9.5\%    & 11.3\%   & 0.4\%    & 76.7\%   \\
\hline\hline
\end{tabular}
\end{table*}
%%%%%%%%%%%%%%%%%%%%%%%%%%%%%

\cut{Selections can be pure for a given topology, but include contributions from multiple processes.} % redundant
\oscask{Another example is \nova's CC \nue selection.}
Table~\ref{tab:breakdown} shows the \cut{\nova CC \nue }selection purity separated by process, according to their simulation.
%%%%%%%%%%%%%%%%%%%%%%%%%%%%%
\begin{table*}[tbp]
\centering
    \caption{Separation of \nova \nue CC candidate selection according to process type: quasi-elastic (QE), two-particle-two-hole
    (2p2h), resonant pion production (RES), coherent pion production (COH) and deep-inelastic scattering (DIS).
    These correspond to the \nova best-fit values of $\sin^{2}\theta_{23}=0.404$, $\Delta m^2_{32}=2.44\times10^{-3}$\evsqc,
    $\sin^{2}2\theta_{13}=0.085$, $\dcp = 1.48\pi$, normal mass ordering, and for full detector equivalent POT of
    $6.05\times10^{20}$.
    Taken from Ref.~\cite{novacommunication}.}
\label{tab:breakdown}
\begin{tabular}{lcccccc}
\hline
\hline
&                 & \ \ \numu+\numubar \ \ & \ \ \nue+\nuebar \ \ & \ \ $\nu+\bar{\nu}$ \ \ &$\nu_\tau$ & \ \ \numu $\rightarrow$ \nue \ \ \\
& \% of MC Events & CC & CC & NC &  CC & CC \\
\hline
 QE    & 28.2\%   &  0.1\%        & 10.5\%     & 0.1\%      & 0.7\%       & 88.5\%    \\
 2p2h  & 11.0\%   &  0.0\%        & 9.2\%      & 0.0\%      & 0.4\%       & 90.4\%    \\
 RES   & 39.2\%   &  1.0\%        & 10.0\%     & 6.4\%      & 0.3\%       & 82.3\%    \\
 COH   & 1.6\%    &  0.1\%        & 6.3\%      & 43.1\%     & 0.0\%       & 50.5\%    \\
 DIS   & 19.8\%   &  8.9\%        & 7.4\%      & 40.0\%     & 0.1\%       & 43.7\%    \\
\hline\hline
\end{tabular}
\end{table*}
%%%%%%%%%%%%%%%%%%%%%%%%%%%%%
At \nova the event selection is inclusive, \cut{\ie they select}\oscask{taking all} CC events with a charged lepton in the final state.
Inclusive selections such as \nova's include significant contributions from CC quasi-elastic (QE), resonant pion production (RES),
and multi-$\pi$ deep-inelastic (DIS) processes.
For T2K, the selected interactions at the far detector are charged-current event with no pions observed in the final state, denoted
CC$0\pi$.
A selection of CC0$\pi$-like interactions at the near detector contains predominantly (72.4\%) events which are CC $\nu_\mu$
interactions, with no pions exiting the nucleus and any number of nucleons in the final state.
However, achieving a pure sample according to a given process is difficult.
For example, the CC0$\pi$ topology contains contributions from CC quasi-elastic (CCQE) events, as well as from CC events with pion
production (CC$1\pi$), where the pion produced is absorbed in the nuclear medium before it can exit the nucleus.
A similar issue is present for two-particle-two-hole (2p2h) processes, where the neutrino interacts with \cut{more than one nucleon in
the system}\oscask{a correlated nucleon-nucleon pair inside the nucleus}.

There are two further complications to \oscask{predicting} the\cut{ predicted} event rate.
First, detectors are not necessarily homogeneous, so neutrino interactions on a variety of target materials may need to be simulated
in the event sample.
If the surrounding material is rock, other inactive detector material, or a magnet, the \cut{target}\oscask{struck} material may not 
match the inner detector.
For \oscask{example, in} the T2K near detector\cut{,} approximately 5\% of the event samples comes from interactions outside the
detector\oscask{, falsely} reconstructed to have happened on target material in the center of the detector.
Second, we note that Table~\ref{tab:t2k_nova_fd} is only complete assuming all relevant processes are included.
\cut{If a process is}\oscask{Any} missing\cut{, this can also affect} \oscask{process alters} the estimation of the expected event rate after 
oscillation. 

%---

\subsection{Benefits and Challenges of Near detectors}
\label{sec:nd}

Neutrino oscillation experiments often employ additional (near) detectors to measure the \emph{unoscillated} rate of interactions
\begin{equation}
    N^\alpha_\text{ND}(\preco) = \sum_i \phi_\alpha (\Etrue) \times \sigma^ i_\alpha(\ptrue) \times
        \epsilon_\alpha (\ptrue) \times R_i (\ptrue; \preco) ,
    \label{eq:nd_event_rate}
\end{equation}  
\cut{where the variables are the same as defined for}\oscask{with the same notation as in} Eq.~(\ref{eq:fd_event_rate})\cut{, but lack}%
\oscask{.
Now, however,} the oscillation probability \oscask{does not affect the rate}.
\oscask{In this way, near}\cut{Near} detectors \cut{are}\oscask{put} a powerful constraint on \cut{information also present
in}\oscask{quantities influencing} the far-detector rate.
Variations in the flux, cross section, and detection efficiency are
highly correlated between the near and far detector rates.
However, \cut{in the case of}\oscask{even when near detector data are used in} long-baseline experiments\oscask{, they do}\cut{this does}
not remove all dependence on the cross-section model\oscask{.}\cut{as they provide a measurement of the}
\oscask{Because} event rates\cut{, which are} \oscask{correspond to} a convolution of the flux and cross section\oscask{, determinations of
oscillation parameters rely on the model to relate near and far measurements to each other}.

Despite the ideal case of identical near and far detectors, in long-baseline experiments the near and far detectors typically sit
\emph{differently} in the beam and will not be identical.
First, the near detector sits in a beam from an extended source: pion decays take place along the decay 
pipe, which typically has a length of a few hundred meters.
\oscask{On the other hand, the far detector essentially sees a point source.}
\cut{The result is that the flux and}\oscask{Consequently, the} acceptance of particles is different at the near and far detector.
\cut{Moreover}\oscask{Second}, since the near detector \cut{is usually rather}\oscask{sits} close to the neutrino source, \oscask{it 
experiences} a very large number of events per beam pulse\cut{ is expected}.
This may restrict the detector technology\cut{ that can be used}, so as to ensure that data taking can be performed in a fast and
efficient manner, and that all events taking place within a given beam pulse are properly identified and recorded inside the pulse
time window.
\oscask{Third, the near and far detector may have different overburdens.}
In \nova's case, the near detector's size and particles from interactions outside the center of the detector affect acceptance
relative to the far detector; conversely, the far detector resides on the surface, and has significant backgrounds from cosmic rays
but minimal backgrounds from surrounding material.
In T2K's case, in addition to acceptance, the near and far detectors\cut{ may} have different nuclear targets\oscask{, so}\cut{and then}
extrapolation\cut{ of physics} between targets is required.

Even with differences between the near and far detectors, the cancellation of systematic uncertainty has proven to be extremely effective for oscillation experiments. At reactor oscillation
experiments, the near-far cancellation achieved impressive accuracy for the measurement of $\theta_{13}$ at Daya Bay and RENO.
Note, however, that, instead of neutrino-nucleus scattering, these analyses hinge on the inverse beta decay cross section being typically known at the 1\% level or better. For long-baseline experiments, the flux uncertainties affecting neutrino beams produced from pion decays are generally large, at the level of 10--20\%, and
present-day cross-section modeling has comparable uncertainties. Even in this case, a partial cancellation of systematic uncertainties is substantial. (See Table~\ref{tab:systematics:nsk_table_summary} in Sec.~\ref{sec:impactcurrent} for an example.)

Despite the critical role of the near detectors,  the near-far cancellation can never be complete
because of the (unknown) oscillation probability $P_{\alpha\beta}$ in Eq.~(\ref{eq:fd_event_rate}). 
The oscillated flux at the far detector is \emph{not the same} as the unoscillated flux measured at the near detector and, thus, 
the convolution of the flux and cross section will always differ among the two~\cite{Huber:2007em}. This holds even in the case of identical detector technology in a disappearance oscillation experiment ($\alpha = \beta$); overall normalization factors can cancel, but as $P_{\alpha\beta}$, $\phi$ and $\sigma$ all depend on energy, the cancellation is not complete ($\int \phi \times \sigma dE \neq \int \phi \times \sigma \times P dE$). 
%%%%%%%%%%%%%%%%%%%%%%%
\begin{figure}[tbp]
\centering
    \includegraphics[width = 0.45\textwidth]{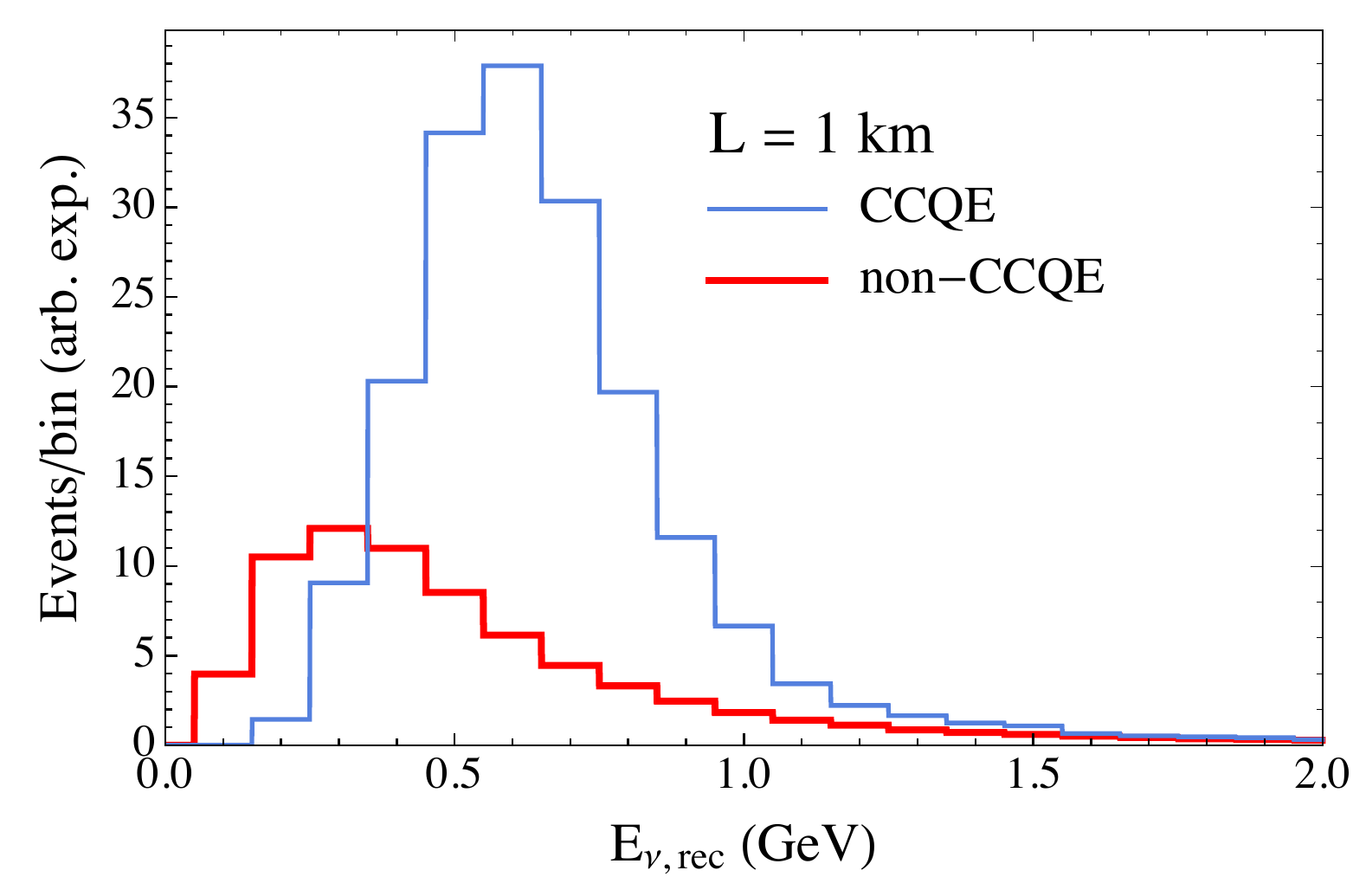}
    \includegraphics[width = 0.45\textwidth]{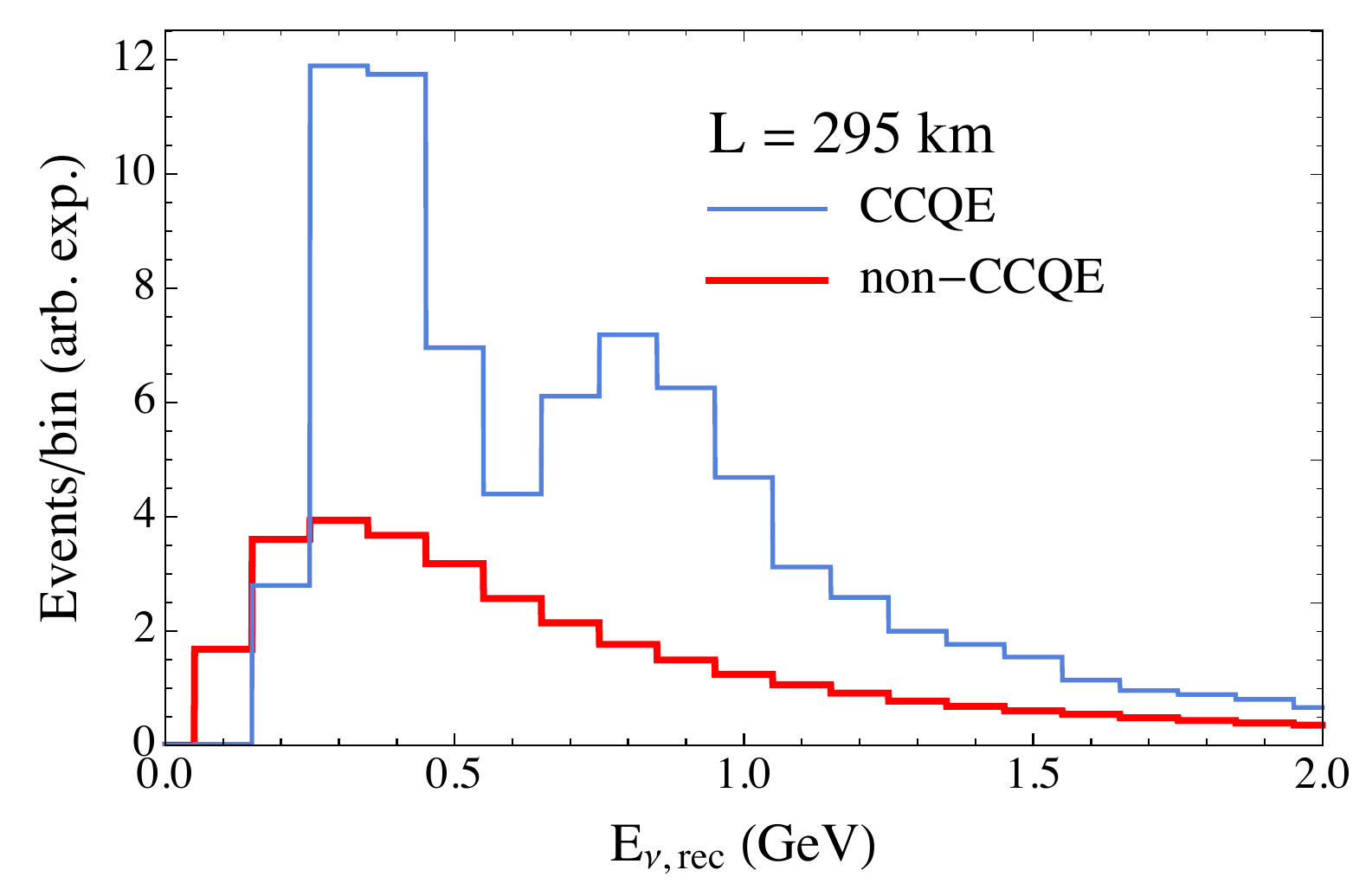}
\caption{Muon neutrino event distributions (for arbitrary exposure) as a function of the reconstructed neutrino energy for CCQE and non-CCQE interactions, for a T2K-like experimental setup. The left (right) panel show the event distributions at the near (far) detector, assuming a maximal atmospheric mixing angle. In this example, the non-CCQE events shown includes 2p2h and charged-current events with $\Delta$-production where no pion is observed in the final state. Figure adapted from Ref.~\cite{Coloma:2013rqa}. \label{fig:nearfar} }
\end{figure}
%%%%%%%%%%%%%%%%%%%%%%%
Figure~\ref{fig:nearfar} illustrates the difficulty in resolving a single cross section process using near detector data, for a $\nu_\mu \to \nu_\mu$ disappearance experiment. The event distributions are shown as a function of the reconstructed neutrino energy for CCQE and non-CCQE interactions (2p2h and events with $\Delta$-resonance production with no observed pions in the final state), for a T2K-like experimental setup, and for the near and far detectors separately. As shown in the figure, the spectrum at the near detector is quite similar for CCQE and non-CCQE. In principle, if the flux was perfectly known the peaks of the two distributions could be resolved, but the flux uncertainties are comparable to the cross section theoretical uncertainties, making this very challenging. After oscillation, the two contributions have markedly different spectra and, in particular, the contribution coming from non-CCQE events does not show an oscillating pattern with the reconstructed neutrino energy, possibly leading to a bias in the determination of the oscillation parameters.

The situation is even more challenging at appearance experiments because the final neutrino flavor is 
unavailable at the near detector.
Direct flux measurements are required in order to break these correlations.
New methods which may help address this problem are being explored.
One proposal, known as ``NuPRISM''~\cite{Bhadra:2014oma}, places a near detector at several 
different off-axis angles, which modulates the observed spectrum in a way 
designed to reproduce the oscillated far-detector flux~\cite{Bhadra:2014oma}.
This option is being further investigated and might provide an additional handle to break the flux--cross-section correlation at
experiments with narrow-band beams.
Another novel approach is to identify neutrino interactions on hydrogen, which would leverage a better known cross section to
constrain the flux~\cite{Lu:2015hea}.

\subsection{Estimation of neutrino energy}
\label{sec:reco-neut}

The second challenge for oscillation experiments is the energy estimator, which is partly based on a cross section model.
While the oscillation probability depends on the true neutrino energy, \Etrue, oscillation experiments must instead determine
the neutrino energy from the lepton's kinematic information and/or hadronic information from CC neutrino interactions.
This reconstructed \Ereco must account for unobserved energy deposition, including particles below detection threshold, 
inactive material, and escaping neutral particles.
In practice, assumptions about these effects are based on the cross-section model.  In principle, data from modern experiments on nuclear targets is certainly valuable to validate the reliability of the \Enu-\Ereco association; however, the uncertainties induced by nuclear effects and the fact that the neutrino energy is not known on an event-by-event basis make the interpretation
of such data very challenging. 
Neutrino beams have a energy distribution broader than the nuclear effects of interest, so it is not possible to isolate nuclear
effects.
It is generally not possible to measure the entire outgoing state (especially the struck nucleus) so momentum transfer in neutrino
scattering is essentially unknown.
Also, the strong-interaction physics in play alters final state particle compositions and kinematics, determination of the incident
neutrino energy, and neutrino versus antineutrino scattering. In addition, if neutrinos and antineutrinos experience different nuclear effects (as might be the case in, e.g., $^{38}$Ar due to the presence of four extra neutrons), this will directly impact our ability to definitively test for the presence of CP-violating
effects in the data. 

The determination of \Ereco depends on the detector technology used.
For example, WC detectors are only sensitive to radiation from particles above Cherenkov threshold.
Thus, protons exiting the nucleus with energy below $\sim 1$~GeV are invisible.
Low-energy mesons also may not be detected except through visible decay products, \eg via electrons
from pion or muon decay.
In the case of a single-nucleon knockout, \cut{after imposing the requirement that the nucleon is on the mass shell,}
the neutrino energy can be \cut{obtained}\oscask{estimated} as
\begin{equation}
    E_\nu^\text{kin} = \frac{2(M-\epsilon) \Ekp + M^2 - (M-\epsilon)^2 - m_\ell^2}{2(M-\epsilon - \Ekp + \n{k_\ell} \cos\theta)},
    \label{eq:kinRecEnergyW}
\end{equation}
where $M$ is the mass of the nucleon, and $\epsilon$ is known as the single-nucleon separation energy. 
Here, $m_\ell$ is the mass of the outgoing charged lepton, $E_\ell$ and $\bm{k}_\ell$ are its energy and momentum, and $\theta$ is
the angle between the outgoing lepton and the direction of the neutrino beam.
Application of the above formulas requires (i)~neglecting the unmeasured recoil momentum of the system 
and (ii)~approximating the energy of the residual nuclear system by a constant. 

This reconstruction method (dubbed ``kinematic method'' hereafter) works  well if the true nature of the event was
indeed a CCQE process, but is subject to two main limitations.
First, as discussed in Sec.~\ref{sec:topology}, \cut{multiple}\oscask{many} processes \cut{are present in}\oscask{contribute to}
a selected topology.
For non-CCQE processes \oscask{-- such as}\cut{ (\eg} CC1$\pi$ production where the pion has been absorbed in the nuclear medium, or 
\cut{2p2h}\oscask{two-nucleon knockout, with an extra neutron --}\cut{)} the energy estimator \cut{will be incorrect}\oscask{in 
Eq.~(\ref{eq:kinRecEnergyW}) is very far off~\cite{Martini:2012fa,Martini:2012uc,Nieves:2012yz,Lalakulich:2012hs}}. 
\cut{This is also true}\oscask{The same holds} for processes in the event sample where extra mesons have been produced in the final
state, but are below detection threshold or not identified by tracking software.
Second, the kinematic method assumes a fixed separation energy~$\epsilon$, while in \cut{practice there is a spread on}\oscask{reality}
the struck nucleon's momentum \cut{in}\oscask{is drawn from a distribution characteristic of} the \oscask{target} nucleus. 

Alternatively, neutrino detectors may be able to collect the majority of the calorimetric deposition in a neutrino event and be
sensitive to the hadronic part of the interaction.
Examples of detectors of this sort are liquid scintillator, magnetized iron detectors, or Liquid Argon Time Projection Chambers (LAr TPC).
Consider CC neutrino scattering off a nuclear target, resulting in the knockout of $n$ nucleons and production of $m$ mesons.
Conservation of total energy \cut{in the system leads to the equation}\oscask{implies}
\begin{equation}
    E_\nu + M_A = \Ekp + M_A - nM + E + T_{A-n} + \sum_{i=1}^n \Eppi + \sum_{j=1}^mE_{\bm{h}_j'},
    \label{eq:energyConservation}
\end{equation}
where $E_\nu$ ($\Ekp$) is the neutrino (charged lepton) energy, $\Eppi$ denotes the energy of the $i$-th knocked-out nucleon (of
momentum~$\bm{p}_j'$, $1\leq i\leq n$), $E_{\bm{h}_j'}$ stands for the energy of the $j$-th produced meson (of momentum $\bm{h}_j'$,
$1\leq j\leq m$)\cut{, and we have expressed}\oscask{.
Here,} the energy of the residual $(A-n)$ nucleon system \oscask{is expressed} in terms of the nucleon (target-nucleus)
mass $M$ ($M_A$), the recoil \oscask{kinetic} energy $T_{A-n}$, and the excitation energy~$E$.

Assuming that multinucleon effects do not introduce strong energy dependence to the cross sections, the binding energy for the
nucleons ($\epsilon_n=E+T_{A-n}$) can be treated as a constant. This simplification leads to \cut{the following expression for the neutrino energy as a function of the energies of the particles in the 
final state}
\begin{equation}
    E^\text{cal}_\nu = \Ekp + \epsilon_n + \sum_{i=1}^n(\Eppi-M) + \sum_{j=1}^mE_{\bm{h}_j'}  .
    \label{eq:calEnergy}
\end{equation}
Note that while for mesons the total energies enter the sum, for nucleons only the kinetic energies contribute.
This difference \cut{is a consequence of the fact that}\oscask{arises because} mesons are produced \cut{in}\oscask{during} the
\cut{interaction}\oscask{scattering} process, whereas nucleons \oscask{pre-exist and} are \cut{only }knocked out of the target nucleus.

This energy reconstruction procedure (dubbed ``calorimetric method'' hereafter) can in principle be applied to non-QE events
as well as to CCQE events; comparisons of the kinematic and calorimetric method are discussed in Ref~\cite{Ankowski:2015jya}.
However, this procedure is not free from \cut{relevant} systematic uncertainties affecting the determination of the incident neutrino
energy.
Each particle in the interaction must be properly identified and reconstructed, but the accurate reconstruction of hadrons poses a
formidable experimental challenge.
In particular, neutrons typically escape detection, and any undetected meson results in energy underestimation by at least the value
of the pion mass, 140~MeV.
This makes low detection and tracking thresholds a key requirement for a calorimetric detector. 
Technologies are being explored to tag neutrons (water in ANNIE~\cite{Anghel:2015xxt},
and LAr in CAPTAIN~\cite{Berns:2013usa,Liu:2015fiy}).
Further, gaseous TPC  detectors  have a lower threshold for detection than liquid detectors.
Detection alone, however, is not a panacea\cut{; the neutron acceptance will also be significantly different for near and far detectors
due to their different sizes}\oscask{, because the disparate sizes of the two detectors makes their neutron acceptance 
significantly different}.
In summary, all published literature which studies these questions in details point to the same conclusion: with the current limited
understanding of the microphysics of  neutrino-nucleus interactions the neutrino energy scale cannot be 
determined reliably in experiments like DUNE. The adverse consequences for the physics reach are profound and wide-ranging.

\subsection{Calculation of Detection Efficiency}
\label{sec:efficiency}

The third and final way the cross section model affects an oscillation analysis is subtle: through the assumed 
efficiencies \oscask{$\epsilon_\alpha (\ptrue)$} in \oscask{Eqs.~(\ref{eq:fd_event_rate}) and~(\ref{eq:nd_event_rate}), for the the 
far and near detectors}\cut{the near and far detectors}.
In principle, the detection efficiency \cut{is}\oscask{should be} independent of any underlying model -- it is merely the response of 
the detector to a particular charged particle.
In practice, \oscask{however,} detection efficiencies are calculated \cut{with}\oscask{by taking simulated particles from an event 
generator (cf.\ Sec.~\ref{generators})}, distributed \cut{in the detector geometry }according to a neutrino-interaction model.
\oscask{Uncertainties in the \ptrue dependence in each model propagate, via Eqs.~(\ref{eq:fd_event_rate}) and~(\ref{eq:nd_event_rate}),
through the whole analysis of a neutrino experiment.
Different models predict different \ptrue and particle-multiplicity distributions, so uncertainty certainly arises here.}
\cut{If the phase space is even modestly different between models, this can also affect the efficiency assumed in both
Eqs.~(\ref{eq:fd_event_rate}) and~(\ref{eq:nd_event_rate}).
Note this is a difference between the true (detector) efficiency and the calculated (interaction model) efficiency, and so again,
even in the case of identical near and far detectors, there can be a bias.}%
\oscask{And, again, any difference between the near and far detectors leads to uncertainties that do not cancel exactly.}
This concern is of particular interest in global fits to neutrino-cross-section data, \cut{where}\oscask{which must cope with}
disagreements\cut{ exist} between measurements on the same target of the same process or topology.%
\footnote{Another important concern is the ill-determined beam flux.} The disagreements between cross section measurements underpin the necessity of a more satisfactory neutrino-nucleus modeling for future experiments for use in efficiency calculations.

\subsection{Current Experimental Program}
\label{sec:impactcurrent}

The impact of cross section uncertainties on the extraction of neutrino oscillation parameters\cut{ will} generally depend\oscask{s} on
several factors, namely (1)~the type of detector being considered (which determines the reconstruction method applied for the
neutrino energy), (2)~the beam \oscask{energy} spectrum,
and (3)~the particular oscillation parameter that is being extracted from the data.
As we shall see below, these factors lead to very different problems depending on the experimental setups considered.
The conclusions will also depend on the oscillation channel being observed\cut{ (see Sec.~\ref{sec:basics} for a discussion on the
extraction of oscillation parameters from the data)}\oscask{,
extending the discussion given above in Sec.~\ref{sec:basics}}. 
Therefore, in the following we will make an explicit distinction between long-baseline experiments measuring standard neutrino
oscillation parameters, short-baseline experiments, and searches for new physics effects using neutrino oscillation experiments.

\subsubsection{Long-Baseline measurements}
\label{subsec:LBL}

At long-baseline experiments, the amplitude of the oscillation essentially determines the size of the mixing angles $\theta_{13}$
and $\theta_{23}$.
Thus, any uncertainty affecting the size of the cross section would potentially impact these measurements.
A straightforward example of relevant uncertainties for these measurements is given by the axial form factor \oscask{of the nucleon},
as the value of the effective axial mass%
\footnote{\oscask{See Eq.~(\ref{eq:FA:dipole}) in Sec.~\ref{CCQE} for a definition 
of the axial mass.
Here, it} is \oscask{just} a proxy for the axial form factor shape.}
is directly correlated with the magnitude of the interaction cross section.
Cross section uncertainties may affect the determination of other oscillation parameters in a less 
obvious manner, as follows.
For example, the current hint that $\dcp \sim -\pi/2$ comes from the combination of reactor and 
long-baseline data, where the latter is currently dominated by data taken in neutrino mode.
As $\theta_{13}$ is essentially fixed from reactor data, and long-baseline experiments measure a slightly larger number of neutrino
events than \cut{the result} expected for $\dcp = 0$, \cut{this automatically provides a}\oscask{the} hint for\cut{ the value of}
$\dcp \sim -\pi/2$ \oscask{follows automatically}.
A larger value of the axial mass would imply, however a larger cross section and, hence, more events than assumed.
In that case, the current hint might evaporate.
Thus, an improved determination of the axial form factor may affect the 
statistical significance of the current hint for CP~violation in the neutrino sector.

Several experimental techniques are used to reduce the impact of \cut{the type }systematic errors \oscask{like those} described above
on the determination of CP~violation.
At narrow-band beams, the combination of antineutrino data with neutrino data is crucial: in this case, the 
value of the CP-violating phase can be inferred from the observation of different effects in the oscillation probabilities for 
particles and antiparticles.
These measurements will be much less sensitive to those systematic uncertainties affecting both neutrino and antineutrino 
cross sections in the same manner, as in the example of the axial form factor 
mentioned above.
Instead, they will be sensitive to systematic uncertainties inducing an asymmetric behavior in neutrino vs.\ antineutrino event rates.
For example, multinucleon contributions to the cross section might be different for neutrinos and
antineutrinos~\cite{Martini:2010ex}, leading to an apparent asymmetry that could be confused with CP~violation\cut{ in oscillations};
see, \eg Ref.~\cite{Ericson:2015cva}.
A second possibility to reduce these uncertainties is exploited in wide-band beams.
In this case, the \cut{availability of a} wide energy spectrum at the far detector enables a determination of the shape of the oscillation probability, which is sensitive to the value of $\dcp$.
In this case, systematic errors affecting the determination of neutrino energy \cut{will be}\oscask{are} more relevant, as they could 
cause an apparent distortion in the shape of the probability and\oscask{, hence,} induce a bias in the determination 
of~$\dcp$~\cite{Mosel:2013fxa,Ankowski:2015kya}.

In the case of $\theta_{23}$, the strongest constraints come from $\nu_\mu\to\nu_\mu$ and $\bar\nu_\mu\to\bar\nu_\mu$
disappearance data in atmospheric and long-baseline neutrino oscillation experiments.
In fact, the main quantity that is determined from the data is the value of $\sin^22\theta_{23}$, which is extracted from the
size of the dip in the oscillation probability at the oscillation maximum (see Sec.~\ref{sec:basics}).
For maximal mixing, there should be practically no neutrino events observed in the dip region, while for nonmaximal mixing the
conversion is not complete.
However, a misreconstruction of the neutrino energy can mimic the same effect: events taking place at high energies, outside the
dip region, may end up being reconstructed in the region of the oscillation~\cite{Lalakulich:2012hs,Martini:2012uc,Coloma:2013rqa}. An example of this is shown graphically in Figure~\ref{fig:nearfar} for the T2K configuration. References~\cite{FernandezMartinez:2010dm,Lalakulich:2012hs,Meloni:2012fq,Coloma:2013rqa,Martini:2012uc,Coloma:2013tba,Jen:2014aja,Abe:2015awa,Ankowski:2016bji} have studied this effect for various models of 2p2h and observe significant effects in the determination of the mixing angle. The opposite can also hold -- if the near detector sees an excess of high-energy interactions (\eg 2p2h), 
the determination of $\sin^22\theta_{23}$ would be biased to maximal or unphysical values.
Note that the description of the cross-section uncertainties is crucial; previous efforts using an 
effective axial mass parameter could yield reasonable-quality fits to excess events seen in
near-detector data, but without the physics of 2p2h processes, the meaning of these fits is 
unclear.

If $\theta_{23}$ is not maximal, cross-section uncertainties might also affect the determination of the octant to which it belongs,
$\theta_{23}<\pi/4$ vs.\ $\theta_{23}>\pi/4$.
As explained in Sec.~\ref{sec:basics}, this measurement has to come from the combination of disappearance and appearance data.
However, cross section uncertainties affecting the $\nu_\mu \to \nu_e$ oscillation channel could affect our ability to determine the
value of $\sin^2\theta_{23}$: a larger value of $\theta_{23}$ translates into a larger appearance oscillation probability and thus a
larger number of events; however the same effect can be mimicked by a larger-than-expected interaction cross section.
Finally, it should also be kept in mind that the appearance channel will be used at the same time to determine the value of
$\dcp$, the octant of $\theta_{23}$, and the neutrino mass ordering\oscask{.
T}\cut{and, t}hus, \cut{is not free from}\oscask{these determinations are subject to} parametric degeneracies.

Table~\ref{tab:systematics:nsk_table_summary} summarizes the uncertainties in a combined analysis of \nue and \numu samples by
T2K~\cite{Abe:2015awa}.
% ND280 reaction composition for NEUT v.5.1.4.2 + tuning
%%%%%%%%%%%%%%%%%%%%%%%%%%
\begin{table}[tbp]
    \caption{Relative uncertainty (1$\sigma$) on the predicted rate of $\nu_\mu$ CC and $\nu_e$ CC candidate events in a combined 
    analysis of \nue and \numu samples by T2K~\cite{Abe:2015awa}.
    \label{tab:systematics:nsk_table_summary}}
\begin{tabular}{lcc}
    \toprule
    Source of uncertainty & \ \ $\nu_\mu$ CC \ \ &  \ \ $\nu_e$ CC \ \ \\
    \hline
    Flux and common cross sections & & \\
    (w/o near detector constraint)       &    21.7\%   &  26.0\% \\
    (w near detector constraint)      &  2.7\%   &  3.2\% \\
    \hline
    Independent cross sections    &   5.0\%   &  4.7\% \\
    \hline
    SK    &    4.0\%   &  2.7\% \\
    FSI+SI(+PN) & 3.0\% & 2.5\% \\
    \hline
    {{Total}} & & 	\\ 
    {{(w/o  near detector constraint) }}  & {{23.5\%}}   &  {{26.8\%}}  \\
    {{(w near detector constraint) }}     &   {{7.7\%}}   &  {{6.8\%}} \\
    \botrule
\end{tabular}
\end{table}
%%%%%%%%%%%%%%%%%%%%%%%%%%
Uncertainties on both appearance and disappearance channels have significant components from cross section systematic uncertainties
which did not cancel in the near/far extrapolation (5.0\% for \numu and 4.7\% for \nue).
In addition to the issues raised in Section~\ref{sec:basics}, T2K's near-detector selection predominantly includes interactions on
scintillator (carbon) which must be extrapolated \oscask{via nuclear models} to the far-detector water (oxygen) target.
Subsequent T2K analyses have used water target cross sections in the oscillation analysis, but the uncertainties~\cite{Abe:2017uxa}
due to the cross section model remain important. 
There are also \cut{significant}\oscask{important} theoretical uncertainties included in this table, notably uncertainties on the 
ratio of the \nue to \numu cross section, as the near detectors measure a yield of predominantly \numu interactions but need to 
infer the rate for \nue appearance\oscask{.
S}\cut{(s}ee, \eg Refs.~\cite{Day:2012gb,Huber:2007em,Coloma:2012ji,Akbar:2015yda,Martini:2016eec} for more information.

Cross section model uncertainties will continue to be relevant \cut{for}\oscask{until} the \cut{final}\oscask{end game of the} T2K and 
\nova \cut{programs}\oscask{experiments}.%
\cut{While T2K and \nova are just starting their experimental programs, a recent paper} 
\oscask{A~study} from T2K \oscask{positing}\cut{assuming} a \oscask{very}\cut{much} high\cut{er} statistical sample\cut{ in the oscillation 
analysis} -- $7.8\times10^{21}$~POT, approximately an order of magnitude larger than their \cut{current} data set \oscask{as of early 
2017} -- notes that 
``for the measurement of \thtt~and \dmsq, the systematic error sizes are significant compared to the statistical error''%
~\cite{Abe:2014tzr}.
Furthermore, the combined \nue appearance sample from \nova and T2K will be $\sim1,000$ events by the time the next generation of
experiments comes online.
Consequently, the ultimate measurements from T2K+\nova must confront \oscask{systematics at the}
3\%-level\cut{ systematics}~\cite{Abe:2016tii}.

\subsubsection{New physics searches}

Current and future neutrino oscillation experiments can also be used to constrain new physics models.
Like the oscillation physics program, these constraints can be severely affected by systematic uncertainties associated with
cross sections.
A relevant example is given by the measurements of neutral-current (NC) rates at the far detector in MINOS or \nova, which can be
used to put bounds on the mixing between active and sterile neutrinos~\cite{Adamson:2011ku,novaWC}.
Recent results from the \nova experiment~\cite{novaWC} show a shift in the observed event distributions towards lower values of the
calorimetric energy with respect to the Monte Carlo prediction.
Moreover, the largest contributor to the overall normalization systematic error in this channel was the NC mis-modeling
uncertainties.
Future searches using NC events will also require microscopic models for NC multinucleon interactions and their implementation into
Monte Carlo event generators, which are currently unavailable.
(The only microscopic models currently implemented into Monte Carlo event generators correspond to CC cross sections.)
NC modeling will help to reduce the overall uncertainties affecting these channels and improve the derived bounds 
from the data.

A second example is given by light dark-matter searches using oscillation experiments;
see, \eg Refs.~\cite{Batell:2009di,deNiverville:2011it,deNiverville:2012ij,Batell:2014yra,Coloma:2015pih,Dobrescu:2014ita}.
In certain models of new physics with new vector bosons, dark
matter particles could be produced at the target in neutrino oscillation experiments, either via meson decays or via direct
production in proton-nucleus collisions.
The produced particles could then lead to an observable excess of NC-like events in neutrino detectors.
The experimental signature in these models would consist of a nuclear or an electron recoil and, thus, neutrino NC interactions
constitute a sizable and irreducible background for these searches.
Therefore, a precise knowledge of neutrino NC cross sections is crucial in this case to get a strong experimental sensitivity.
Recently, the MiniBooNE collaboration performed a special run in beam-dump mode to conduct a search for
sub-GeV dark matter particles produced in this way~\cite{Aguilar-Arevalo:2017mqx}.
The initial systematic uncertainty on the NC neutrino background was determined to be at the 34\%~\cite{Aguilar-Arevalo:2017mqx}.
Moreover, the Monte Carlo simulation significantly overpredicted the NC elastic event rates at high nucleon energies; see
Refs.~\cite{Aguilar-Arevalo:2013nkf,AguilarArevalo:2010cx} for more details.

\subsubsection{Short-baseline measurements}

A host of short-baseline measurements have been planned for the near future, which use the same  neutrino interactions of interest to long-baseline experiments.
%%%%%%%%%%%%%%%%%%%%%%%%%%
% From NuINt2015 Proceedings by Anne Schukraft. Official MicroBooNE numbers.
\begin{table*}[tbp]
\centering
\caption{Composition for the expected event rate at MicroBooNE for $1\times10^{20}$ Protons on Target (POT) according to the final
state topology.
This roughly corresponds to the expected event rate from MicroBooNE's 2015 run.
No acceptance or efficiency corrections are included.
From Ref.~\cite{Schukraft:2016sxo}.}
\label{tab:microboone}
\begin{tabular}{l c }
\hline\hline
 Final State        & Events \\ 
\hline
CC inclusive & 26500 \\
CC 0$\pi$ & 17000 \\
NC elastic & 2600 \\
NC 1$\pi^0$ & 1700 \\
NC 1 $\gamma$ & 20 \\
\hline\hline
\end{tabular}
\end{table*}
%%%%%%%%%%%%%%%%%%%%%%%%%%
Here we include a short discussion of issues shared by both programs.
The MicroBooNE experiment~\cite{Chen:2007ae} will look for non-standard appearance and disappearance to search for sterile neutrinos
with a 89 ton LAr TPC.
Additional detectors are being added to the same beamline at different distances from the target~\cite{Antonello:2015lea} to
quantify the dependence of the oscillation with the distance to the source, should it be observed, and further reduce the impact of
systematics.
In addition to the issues raised for long-baseline experiments, MicroBooNE faces the use of a target material, argon,
that is significantly different from previous oscillation results using steel, carbon, or water targets.
Nuclear effects are expected to be significant for argon: for example, pion absorption is roughly twice as large in argon as it is
in carbon or oxygen~\cite{privatemicroboone}.
Moreover, the extrapolation from lighter nuclei to argon is difficult or impossible 
in many nuclear modeling frameworks.
Like long-baseline experiments, MicroBooNE will require an energy estimator based on calorimetric methods but, in principle, can also
use energy estimators based on the kinematic method described in Sec.~\ref{sec:reco-neut}.
It will have the benefit of a relatively low detection threshold but will, then, be sensitive to the kinematics and multiplicities of
final state particles through threshold effects and new reconstruction algorithms.
Unlike for long-baseline experiments aiming to perform appearance measurements, the relative importance of backgrounds is amplified
here as short-baseline appearance signals are much smaller relative to the expected intrinsic \nue and NC backgrounds.
Table~\ref{tab:microboone} shows the expected breakdown of interaction topologies at MicroBooNE for its first year of operation with
no acceptance or efficiency corrections included.
As MicroBooNE's peak energy is about 0.8~GeV, the bulk of the interactions are CC0$\pi$ topologically.

\subsection{Future Experimental Program}
\label{sec:impactfuture}

To meet the physics goals of future CP-violation searches with neutrino beams, unprecedented control of neutrino interaction
uncertainties is required.
\HK, with a design similar to T2K,
will be affected by \cut{similar }systematic uncertainties \oscask{similar to T2K}.
Table~\ref{Tab:sens-selection-nue} shows the number of \nue and \nuebar candidates expected at \HK.
Like T2K, the largest non-CC contribution is expected to be from NC interactions mis-identified as CC.
It should also be noted that, as the beams are not pure, the \cut{total flux}\oscask{beam} in the antineutrino beam configuration
\cut{(or running mode) }will also contain a significant \cut{contribution from}\oscask{component of} neutrinos.
Given the much larger cross section for neutrinos with respect to antineutrinos, the contamination in the final event sample will be
much more severe in the antineutrino running mode than in the neutrino one.
From the number of events in Table~\ref{Tab:sens-selection-nue}, it can be seen that the statistical uncertainty of the $\nu_e$
sample, $\sim 2\%$, will be comparable to the expected total systematic uncertainty for \nue (\nuebar) appearance, 3.3\% (6.2\%).
Thus, cross section systematic uncertainties will need to be controlled at the 1--2\% level for signal and background, and careful treatment is required for
the relative uncertainties between neutrinos and antineutrino interactions.
As pointed out in Sec.~\ref{subsec:LBL}, CP-violation searches with narrow-band beams (such as \HK) will be
\cut{mostly sensitive}\oscask{especially subject} to cross-section uncertainties \oscask{that} affect\cut{ing} the \oscask{neutrino and
antineutrino} event rates \cut{in an }asymmetric\oscask{ally, for instance}\cut{ manner for neutrinos and antineutrinos (\eg}
multinucleon interactions.
%%%%%%%%%%%%%%%%%%%%%%%%%%%
\begin{table}[tbp]
\centering
% Public design report, unpublished:https://lib-extopc.kek.jp/preprints/PDF/2016/1627/1627021.pdf p181
\caption{The expected number of \nue and \nuebar candidate events for the T2HK experiment.
    Normal mass \cut{hierarchy}\oscask{ordering}    with $\sin^22\theta_{13}=0.1$ and \dcp$=0$ are assumed.
    Background is categorized by the flavor before oscillation.
    Taken from Ref.~\cite{HKreport}.}
\label{Tab:sens-selection-nue}
\begin{tabular}{c@{\qquad}cc@{\qquad}ccccc@{\quad}c@{\qquad}c}
    \hline\hline
				& \multicolumn{2}{c}{\hspace*{-3em}Signal} & \multicolumn{6}{c}{Background} & \multirow{2}{*}{Total~} \\ 
				& $\numu\to\nue$	& $\numubar\to\nuebar$	&\numu CC	& \numubar CC	& \nue  CC& \nuebar CC & NC & BG Total	&  \\ \hline  
$\nu$ mode		& 2300				&	21						& 10		& 0				& 347	& 15		& 188		&	560 & 2881 \\ 
$\bar{\nu}$ mode	& 289				&	1656					& 3			& 3				& 142	& 302		& 247		&	724 & 2669 \\ \hline \hline
\end{tabular}
\end{table}
%%%%%%%%%%%%%%%%%%%%%%%%%%%

\begin{figure}[tbp]
\centering
    \includegraphics[width = 0.75\textwidth]{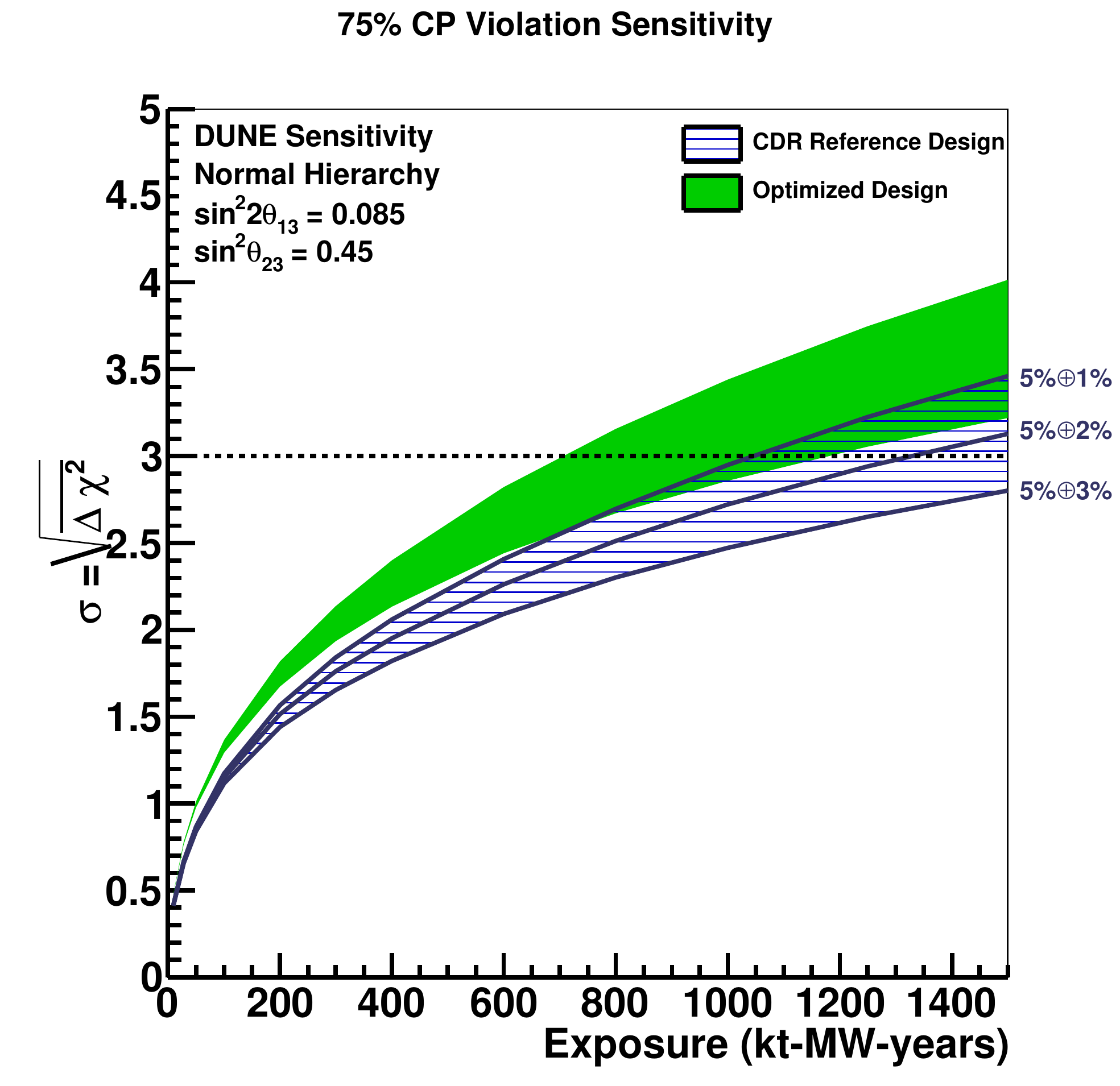}
    \caption{Effect of normalization uncertainties affecting $\nu_e$ and $\bar\nu_e$ cross sections on the sensitivity to CP 
    violation at DUNE.
    The panels show minimum significance at which the CP-conservation hypothesis can be rejected, for  75\% of values of \dcp, as a function of the total exposure in 
    $\textrm{kt}\,\textrm{MW}\,\textrm{yr}$.
    (A~priori, all possible values of $\dcp$ in the range $-\pi<\dcp \le\pi$ are assumed to be equally likely.)
    The width of the band shows the variation in sensitivity when the cross section uncertainties on the appearance sample are 
    varied between 1\% and 3\%.
    The right-axis labels X\%$\oplus$Y\% indicates that a X\% (Y\%) normalization uncertainty is assumed for the $\nu_\mu$ and 
    $\bar\nu_\mu$ ($\nu_e$ and $\bar\nu_e$) samples at the far detector, see text for details. 
    The hashed band shows the results obtained with the CDR reference beam design, while the solid band shows the results obtained 
    with the optimized beam design.
    From Ref.~\cite{Acciarri:2015uup}.}
    \label{fig:DUNE-cpv}
\end{figure}

The DUNE collaboration has set \cut{their}\oscask{a} goal \oscask{for} systematic errors at $5\%\oplus2\%$, where 5\% corresponds to the 
normalization uncertainty on the $\nu_\mu$ sample at the far detector, and 2\% is the effective uncorrelated normalization uncertainty on the
$\nu_e$ sample at the far detector, after fits to both near- and far-detector data have been performed, and all external constraints
have been included~\cite{Acciarri:2015uup}.
Figure~\ref{fig:DUNE-cpv} shows the effect of larger uncertainties on the sensitivity to CP violation at DUNE, as a function of its
total exposure~\cite{Acciarri:2015uup}.
In this figure, the signal normalization uncertainties \oscask{between neutrinos and antineutrinos}
are treated as \cut{100\%}\oscask{completely} uncorrelated\cut{ between neutrinos and antineutrinos}; for
additional details, see Ref.~\cite{Acciarri:2015uup}.
As can be seen from the figure, for exposures above $1000~\text{kt}\,\text{MW}\,\text{yr}$, the sensitivity to CP
violation obtained for 75\% of the values of \dcp  could be lowered below the 3$\sigma$ bound, if the size of the systematic
errors is increased from 2\% to 3\%.
The \cut{effect would be even larger}\oscask{degradation increases} for larger exposures, as the experiment enters the 
systematic-dominated regime.
%%%%%%%%%%%%%%%%%%%%%%%
%%%%%%%%%%%%%%%%%%%%%%%

For wide-band beams, such as the DUNE experiment, much of the information on the CP-violating phase
comes from the observation of the energy dependence of the oscillation probability over a wide range of
energies, as mentioned in Sec.~\ref{subsec:LBL}.
In this case, uncertainties affecting the neutrino reconstruction process could have a larger impact on the results.
At DUNE, hadron\cut{ic energie}s are expected to contribute more than half of the total energy deposit for many $\nu_e$ and
$\nu_\mu$ interactions in the far detector~\cite{Acciarri:2015uup}\cut{, and}\oscask{.
Thus,} the impact of pion and nucleon production through higher-energy inelastic interactions could play a \cut{significant}\oscask{key} role.
For instance, particles produced in nuclear interactions below detection threshold, or neutrons escaping detection, can lead to
a large amount of missing energy.
These effects are difficult to quantify as they rely on the predictions of a given nuclear model.
Unless they are kept under control, they will generate a bias in the determination of neutrino energy towards lower
energies, which in turn would translate into a wrong determination of the value of~\dcp.

The effect of missing energy on the measurement of \dcp is explored quantitatively in Ref.~\cite{Ankowski:2015kya} for a setup
similar to DUNE.
The authors concluded that a sizable bias would be induced in the determination of the value of \dcp if the missing energy
is underestimated by 20\% or more.
The study in Ref.~\cite{Ankowski:2015kya} assumed, however, that the reconstruction bias would be the same for neutrinos and
antineutrinos, something not expected \emph{a priori}.
The effect could be \oscask{even worse once} this assumption is relaxed.
A more detailed study is needed to determine the final impact, including a detailed simulation
of the LAr detector performance.
A further detailed study is needed of the impact of different pion absorption rates and neutron production in
argon, which both depend on the nuclear model.
Significant experience with simulation, reconstruction, and calibration of neutrino interactions in LAr TPCs is expected 
from the Intermediate Neutrino Program~\cite{Adams:2015ogl}.
In particular, Fermilab's short-baseline neutrino program~\cite{Antonello:2015lea} consists of 
three experiments with a LAr TPC: ICARUS-T600, MicroBooNE, and SBND. Moreover, an active program of detector prototypes and test-beam measurements is planned to study the reconstruction of charged and neutral particles in LAr TPC detectors, including LArIAT~\cite{Cavanna:2014iqa}, CAPTAIN~\cite{Berns:2013usa,Liu:2015fiy}, and the
CERN neutrino platform single and dual phase prototypes, also known as ProtoDUNE~\cite{Manenti:2017txy}. Finally, one should appreciate that electron scattering, with its fully defined kinematics, is 
an important testbed for \emph{any} model of neutrino-nucleus interactions, since they
necessarily must reproduce electron scattering data.
For discussions of the relevance of electron scattering to neutrino experiments, see Sec.~\ref{eA} and
Refs.~\cite{Amaro:2004bs,Pandey:2014tza,Ankowski:2014yfa,Jen:2014aja,Megias:2016lke}.

\subsection{Summary and challenges for oscillation experiments}

Several initial processes can contribute to each observable topology in our detectors due to nuclear effects and the significant energy spread of neutrino and antineutrino beams. It is clear that nuclear effects are a major issue for current and future experiments. To achieve the future program, we need a clear understanding of: 

\begin{itemize}
\item Current and future long- and short-baseline neutrino oscillation programs should evaluate and articulate what additional neutrino-nucleus interaction data or support measurements are required to meet their ambitious goals. This can be done with a combination of phenomenological and direct theoretical estimations.

\item  Near detectors are powerful in oscillation analyses, but do have fundamental and practical limitations in the near-to-far extrapolation of event rates. New experimental methods such as NuPRISM ~\cite{Bhadra:2014oma}, which enables variable neutrino energy fluxes to enter the near detector, could circumvent the problem of different fluxes at the near and far detector.

\item Of specific interest is  precise knowledge of electron/muon neutrino cross section differences, which historically has been difficult to measure in near detectors. Are there any theoretical indications of unexpected differences? What is the level that this quantity will be known by the proposed future experimental program near detectors? 

\item Neutrino energy estimators are sensitive to threshold effects and model-based particle composition and kinematics. As neutrino-antineutrino event-rate comparisons are important for \dcp measurements, the relative neutron composition of final hadronic states is key.  What are the prospects for semi-inclusive theoretical models? Experimentally, programs to detect neutrons are essential. Electron scattering data may also provide insights to the hadronic state.

\item The calculated detector efficiency often depends on the nuclear model. Whether or not current uncertainties cover this issue needs to be studied. If this is an important effect for current and future experiments, systematic errors must reflect the range of nuclear models used in calculating this efficiency. It may be that experimental and computational approaches will be necessary.
\end{itemize}
\oscask{Without these studies, the adverse consequences for neutrino-oscillation measurements are profound and wide-ranging.}

\newpage

\section{Neutrino Event Generators}
\label{generators}
% Introduction to the necessity of neutrino interaction generators
% How they work practically, power and limitations of generators and complications, recent improvements in implementing theory // global model for everyone. Consistency with nuclear models.
% Summary (and to be discussed, prioritization?) of "justified in more detail below, following issues need to be addressed."

\subsection{How do neutrino event generators work?}

Accelerator-based neutrino experiments generally feature a three-part software stack: a beam simulation with important uncertainties from hadron production sculpting the output, an event generator responsible for modeling the ``hard scattering'' process of a neutrino interacting with a nuclear target and describing the wide variety of final state interaction (FSI) processes that mask the initial process, and finally a simulation for detector response.
Note that there are important constraints implied by this factorization---the event generator must be able to consume neutrinos of definite four-momentum from the beam simulation and it must provide the full set of particles exiting the nucleus in a format compatible with the detector simulation.

All three pieces are, of course, crucial to the success of an experiment, but event generators are particularly crucial.
Because neutrinos enter the detector unobserved and reactions may proceed through neutral current channels that take away an unknown amount of energy with the outgoing neutrino, or through charged current channels that still produce large numbers of neutrons and soft particles\cut{ that may be} below detection threshold, experimenters can never measure the neutrino energy spectrum in an inference free fashion.
Neutral and charged particles have very different detector responses, and their mix is poorly constrained by experiment. 
The best we can do is build probability-weighted maps that connect the observed constellation of particles in a detector with the statistical distribution of incident energies. 
Mistakes in the weights coming from failing to understand the differential cross sections for neutrino reactions in detector observables and mistakes in the predictions for numbers and distributions of different particles produced in reactions can both lead to deadly biases.
We accommodate those possibilities with large systematic uncertainties, but those, in turn, may wash out the small effects we are searching for.

Event generators must simulate every particle that appears in the final state of an interaction on an event-by-event basis in order to accurately determine the beam energy.
At the energy frontier, generators such as MadGraph \cite{madgraph} and Pythia \cite{Sjostrand:2006za} are high-quality tools connecting theoretical predictions to experimental observables. 
This is not generally true for neutrino event generators because we lack a complete theory that can describe from first principles the neutrino interaction with a complex nuclear target and the full subsequent evolution of the reaction products.
The ideal input theory would provide internally consistent, fully-differential neutrino-nucleus cross sections in the kinematics of every final-state particle, over all reaction mechanisms, over the full energy range, for all combinations of neutrino flavor and helicity, and for every nucleus in the experiment.
However, modern theory typically provides only final state lepton kinematics, usually covering a fraction of the experimentally accessible phase space. 
Furthermore, calculations generally cover only low-multiplicity exclusive or semi-inclusive final states like quasi-elastic or single pion production modes. 
These models satisfy the requirements of the electron scattering community, but providing the extensions required for neutrino physics is far from easy.

Because we cannot wait for a complete theory to perform experiments, generators are crafted from an amalgamation of many models and prescriptions, and tuned to match data in as many kinematic variables as possible.  
To construct a generator, we assemble a good theoretical understanding of neutrino scattering from free nucleons together with measurements from charged lepton scattering that may help constrain the nuclear model, and with the best phenomenological models available.

\subsection{Cultural concerns}

There are several generator codes in use, and most of them remain actively supported as independent research projects.
Some generators are closely aligned with particular experimental efforts, but other are developed separately.
The most widely used generators today are GENIE \cite{Andreopoulos:2009rq}, NEUT \cite{Hayato:2002sd}, GiBUU \cite{Buss:2011mx}, NuWro \cite{Juszczak:2005zs,Golan:2012rfa}, and Nuance \cite{Casper:2002sd}.
The composition of the groups behind these generators varies wildly: some are composed almost entirely of experimentalists, while others are composed almost entirely of theorists.
There are numerous advantages to the variety of approaches, but one significant problem with the current state is the lack of a universal output format as well as intermediate interfaces.

The single largest problem facing event generators today is rooted in the divide between high energy and nuclear physics.
The most widely-used generators are written and maintained by high energy physics (HEP) experimenters, while the most important theory work is done by nuclear theorists.
However, it is difficult to bring these groups together in productive collaboration owing to historical issues, scientific focus, and, especially in the US, the structural funding divide between HEP and nuclear physics.
Ultimately, for neutrino event generators to serve the world's accelerator-based neutrino program, we must as a community find a way to bridge the HEP-nuclear gaps and involve the nuclear theory community in the production and maintenance of neutrino event generators more directly.

Manpower is traditionally a serious concern for generators.
A good generator is required to simultaneously contain high quality physics models and interface smoothly with modern experiments, for example, providing estimated errors for all outputs of each model.
Although previous generations of experiment managed with the work of one or two dedicated collaborators, getting sufficient accuracy in modern experiments requires larger efforts.
Ideally, experienced theorists and experimenters work together with a core of young researchers for the best product.

To date, that manpower hasn't been available despite significant effort because incentives to work on generators are not well-aligned with the research superstructure.
Theorists are not rewarded for implementation efforts, and experimentalists are not rewarded for efforts beyond the minimum required to publish a measurement.
The tradition where PhD students and postdoctoral fellows contribute to event generator development as part of their research must be significantly enhanced.

Coordination of effort between generator groups is an important topic that has not received sufficient attention and thought.
If the groups do not coordinate efforts, then scarce labor is wasted on duplication.
On the other hand, if groups coordinate too tightly, then the advantages of independent approaches to problem solving are lost.
As long as we must work with a very incomplete theory picture, this diversity of thought is very valuable.

\subsection{Theory developments}

Stitching together a global physics model is an important problem in a neutrino event generator.
Given the patchwork of phenomenonological models available, and the differing ranges of validity under which they might be reasonable approximations, it is impossible to fulfill a generator's required duty without the addition of ad-hoc extensions and blending of calculations, or without adjusting the strengths of different responses to ensure smooth, physical behavior.
Currently, each independent generator group is responsible for extending theoretical calculations to cover the full phase space seen by an experiments.

Historically, a generator's physics model was built under the assumption that everything could be decomposed as a set of free-nucleon scattering processes, with some additional modifications to account for nucleon-nucleon correlations and nuclear binding.
There is now ample experimental evidence that this approach is insufficient and a more complete model of the nucleus that includes correlations and other in-medium effects as fundamental constituents is required.

New \emph{ab initio} calculations built with this more complete description of the nucleus offer the potential to compute the fully inclusive cross section in lepton observables, although currently only for light nuclei (in the range of carbon) and in non-relativistic regimes.
We are making promising progress on extending computations to heavier nuclei and into the relativistic regime, but significant effort and resources are required.
Once available, these calculations will make it possible to tune the overall rate in a generator, and will help break degeneracies that mix the observable effects of FSI and initial state modeling.
% Here, "initial state" refers to the phase space distribution of target nucleons, correlations, and other in-medium effects that distort the hard scattering target. "FSI" is referring to the propagation of hadronic material through the nuclear medium and to the surface of the nucleus. 
It is, of course, dangerous however to view the inclusive cross section as a simple sum of various exclusive processes.
This is a recipe for internal inconsistency, double-counting, biases inherent in the set of exclusive models available, and worse.
We need more effort invested into the proper mechanisms for making use of the improvements coming in \emph{ab initio} calculations.

Various approaches exist to propagation of the daughter products out of the nucleus, ranging from sophisticated transport models (as in GiBUU) to simpler models very carefully tuned to hadron scattering data (as in GENIE and NEUT). 
Transport models likely contain better physics, but they are often prohibitively slow and effort may be required to improve their performance.

In general, neutrino event generators need to do more to better position themselves to take advantage of improvements in parallel hardware.
Because event generator groups tend to be small, and code may be old or based on unfashionable programming languages, it is difficult to organize the effort to properly parallelize code, or make it thread safe in such a way that the generator is truly able to use multicore machines.
As computational physicists continue to drive improvements in Monte Carlo integration techniques that leverage multiple cores, the neutrino event generator community needs to find ways to take advantage of those efforts.

\subsection{Interplay with experiments}

Recent work focused on understanding the relationship between different experimental measurements of nominally comparable quantities has highlighted the need for very careful presentation of experimental results.
Recall that the cross sections measured by a neutrino experiment are integrated over that experiment's specific flux and carry the influence of the specific mix of nuclear targets in the detector.
As such, direct comparison of two results is often impossible and interpretation between the pair requires an event generator, which, itself carries many highly relevant model biases.

As such, extreme care must be taken to minimize the impact of model dependence in experimental results.
Cross sections should be presented in terms of experimental observables first and foremost.
Cross sections in quantities that require the interpretation of a model (such as neutrino energy and four momentum transfer) should be accompanied whenever possible by the model-independent constituent inputs (e.g., final state lepton variables).
Quantities that are highly dependent on the detector model, or on the generator inputs such as hadronic energy should be presented in terms of the nominal particle content whenever possible.

Results should also be presented in observable phase space.
If certain particle angles or energies in the final state topology are not visible to an experiment, it is important to at least present results in that restricted phase space.
When reporting full phase space results, experiments should be careful to note the configuration of the generator used to integrate into the unobserved regions of phase space.
Numerous subtle traps await in this process.
For example, when reporting one dimensional differential cross sections, care must be taken with all the input variables to produce cross sections based on observables.

In the end, of course, generators cannot be more accurate than the measurements that inform and constrain them.
Improvements in the input theory models is crucial, but it is equally crucial for experiments to publish as many good cross section results as possible.

\subsection{Top challenges}

Further details about event generators are covered in the following sections, focusing on the specifics at hand.
While each aspect of neutrino physics brings specific challenges to event generators, several general issues emerge:
\begin{itemize}
\item How do we coordinate efforts between generator groups in such a way as to reduce duplication of effort while still preserving the advantages of independent approaches and ideas?
\item How do we fully engage the nuclear theory community in the design and implementation of event generators?
\item How do we take best advantage of new and upcoming \emph{ab initio} calculations of the nucleus?
\item When tuning generators, how do we reconcile tensions in existing datasets? How aggressively should we pursue one model meant to work across all energies and targets, an idea known as a ``universal tune?''
\item We may do well to follow the lead of the energy-frontier generator community, by organizing meetings explicitly devoted to common data formats and interfaces between the various stages of neutrino-nucleus event generation.
\end{itemize}

\newpage

\section{Electron-nucleus Scattering as input to neutrino scattering}
\label{eA}

\subsection{Introduction}

Many high-quality electron-nucleus scattering data exist, covering a wide energy range corresponding to different reaction mechanisms: from quasi-elastic (QE) scattering to the region of the $\Delta$ resonance and the complete inelastic spectrum -- resonant, non-resonant and deep inelastic scattering (DIS).
Any nuclear model used to describe neutrino-nucleus scattering should first be validated against these data. Since the vector part of the weak response is related to the electromagnetic response through CVC, such a test is necessary, but not sufficient, to ensure the validity of a model for given kinematics, namely given values of the transferred energy $\omega$ (= $\nu$ for neutrinos) and momentum $q$.
Valuable information on the axial response could in principle be extracted from parity-violating (PV) electron scattering off complex nuclei \cite{Musolf:1992xm,Barbaro:1993jg}, where however few data exist and are  mostly limited to the elastic part of the spectrum. 
In particular, from measurements of the PV asymmetry at backward scattering angles in the QE regime good knowledge of the radiative corrections entering in the isovector axial-vector sector could be gained~\cite{Gonzalez-Jimenez:2015tla}.

\subsection{Experimental input}

The cross section for neutrino scattering from nuclei is sensitive to the same underlying structure determined by QCD, and as probed with pure electromagnetic processes, such as charged lepton scattering from nucleons and nuclei.  As such, there are a number of ways that electron scattering data inform $\nu-A$ cross section modeling, as well as 
providing a test-bed for model validation.  In contrast to past and current neutrino beams, charged lepton scattering has 
the distinct advantage of nearly monochromatic beams with well determined energies, allowing for a significantly cleaner 
kinematic separation of the various production mechanisms in inclusive scattering, such as resonance production and 
nucleon elastic scattering.      
In addition to providing important experimental input such as nucleon isovector elastic form factors and resonance transition 
form factors, 
%the ratio of longitudinal to transverse inelastic structure functions ($R = {F_L \over 2x F_1}$), 
electron scattering data provide critical information on the distributions of initial state momentum and energy for nucleons 
in nuclei, the importance of 2-body currents and final state interaction effects.  In this section we will give a brief 
overview of the experimental input provided by electron scattering data.   

%Additionally, the 
%  Electron scattering experiments have studied the energy and momentum distributions of nucleons in nucleons, as determined 
% by the nuclear wavefunction.  These initial state distributions 

% Study of the breakdown of the IA

%In this section we will briefly review the status of the electron data, including 

%In addition, the final state hadrons are sensitive to   

At beam energies of a few GeV and below the $\nu-A$ cross section is dominated by nucleon elastic scattering and resonance 
production, as well as contributions from 2-body currents.  In elastic scattering the cross section is sensitive to isovector, 
axial, and pseudoscalar nucleon form factors, with the isovector form factors determined from the nucleon electromagnetic form 
factors through Conservation of Vector Current (CVC) and the pseudoscaler determined from Partial Conservation of Axial Current 
(PCAC).  Therefore, extractions of the weak-axial form factor from neutrino deuteron (and anti-neutrino hydrogen) data from 
spark and bubble chambers depend critically on the values of the electric and magnetic form factors utilized to construct 
the isovector form factors.  This was highlighted in a re-extraction of the axial form factor and mass parameter, $M_A$, 
utilizing an updated fit to the electromagnetic form factors~\cite{bodek2008vector}, which included low $Q^2$ data on the 
proton electric to magnetic form factor ratio from Jefferson Lab and the BLAST experiment~\cite{gegm-blast} as recent as 2007 
(see ~\cite{perdrisat2007ff-review} for a review of the experimental status of nucleon form factors prior to 2008).  
This study showed a variation of several percent in $M_A$ relative to the original extractions. 

Since this study was published several new data sets on both proton and neutron elastic form factor have been become available.  
These include measurements 
of the neutron electric to magnetic form factor ratio ($G_E^n / G_M^n$) from the BLAST~\cite{blast-gen} experiment in the $Q^2$ 
range of 0.14 to 0.55 $\rm (GeV / c)^2$ and from Jefferon Lab Hall A~\cite{riordan-gen} in the range 1.72 to 3.41 $\rm (GeV / c)^2$ 
using double polarization observables with polarized deuteron and $^3$He targets, respectively, as well as measurements of 
the neutron magnetic form factor by the CLAS collaboration in Hall B utilizing a deuterium target in the range
 1~$<$~$Q^2$~$<$~4.75~$\rm (GeV/c)^2$.  In regards to modeling the neutrino quasielastic cross section, vector form factor 
parameterizations should be utilized that include the most recent, high precision, data.  

However, an open question remains in regards to the observed discrepancy in the ratio of the proton electric to magnetic form factors determined from polarization transfer measurements and Rosenbluth separations of the cross section for $Q^2$~$>$~1$\rm (GeV/c)^2$.  While two photon exchange contributions beyond the standard radiative corrections remains the most likely explanation for the difference between the ratios extracted from the two techniques, the current experimental evidence for such effects has not been definitive.  This leaves open the question of which value for the electric form factor should be utilized for constructing the isovector form factors for input in neutrino scattering.  

\begin{figure}
\begin{center}
\includegraphics[width=7.5cm]{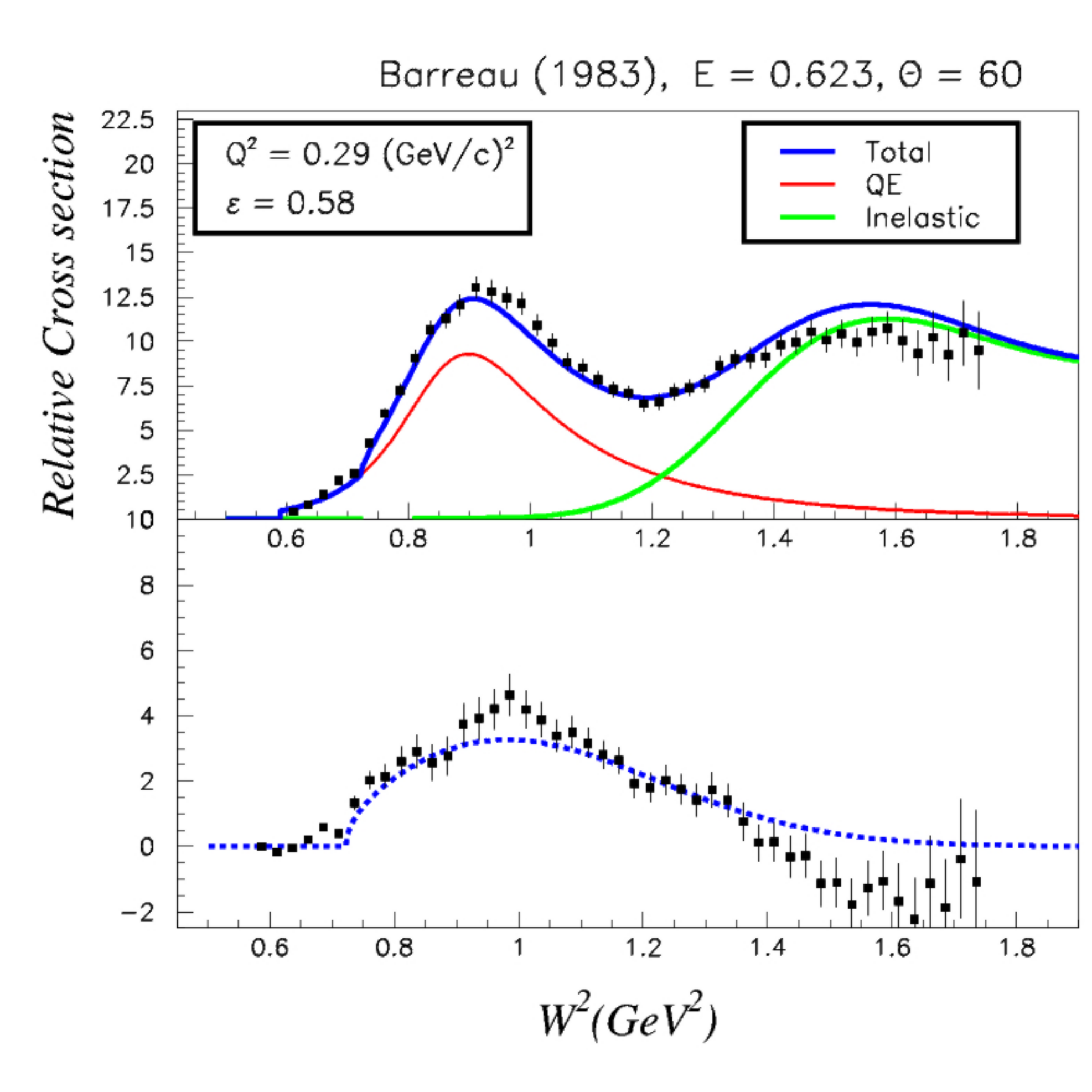}
\includegraphics[width=7.5cm]{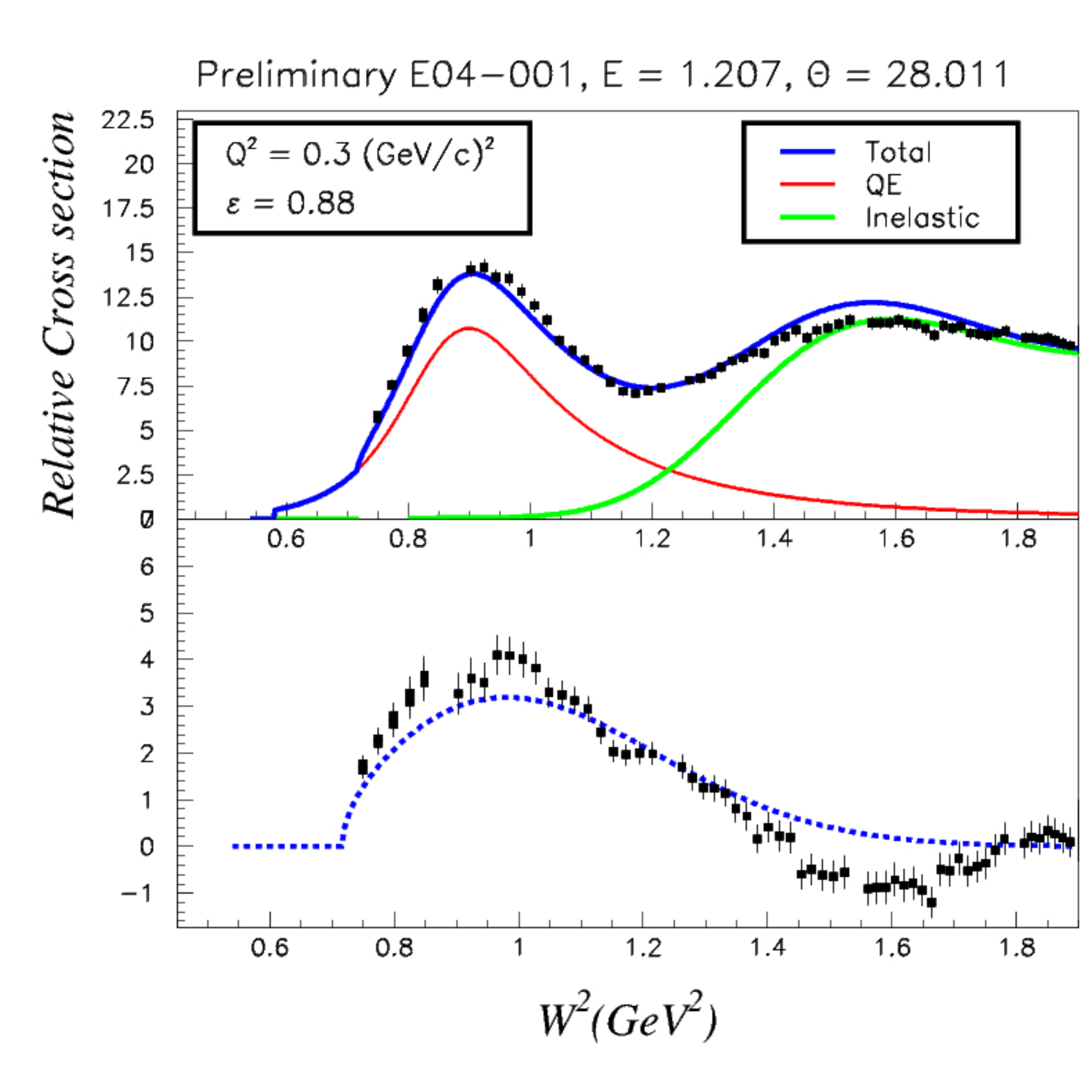}
\end{center}

\caption{\label{fig:12C-inclusive} (Color online) [Note: want to replace figure]  Inclusive reduced cross sections for electron scattering from $\rm ^{12}C$ from SLAC (Left Panel) and preliminary data from JLab Hall C experiment E04-001 (Right Panel) at a similar 
$Q^2 \approx 0.3$~$\rm (GeV/c)^2$.  The bottom panels show the data residuals after subtracting the quasielastic and inelastic contributions from global fits with the dashed curve representing the 2-body contributions determined from the fit.}
\end{figure}

As noted above, many different reaction mechanisms contribute to the inclusive cross section for scattering of electrons from nuclei 
depending on the invariant hadronic mass $W$ of the final state.  In the energy and $Q^2$ range of current oscillation experiments, an 
abundance of electron scattering cross section data currently exists for a large array of target nuclei and with many data sets having 
relatively high precision.  The quasielastic database of Benhar, Day, and Sick~\cite{Benhar:2006er} provides a valuable resource for 
data covering the quasielastic and the delta resonance region.  A number of these data sets were utilized for separation of the cross 
section for longitudinally and transversely polarized photons (so-called L/T separations), which typically require uncertainties 
point-to-point in the photon polarization parameter $\epsilon$ of 2-3\% or better.  
In addition, new high precision inclusive data 
have, or will shortly become, available from Jefferson Lab on nucleons and nuclei which can be used both to determine the vector 
contributions to inclusive structure functions, as well as to test the modeling of the nuclear medium.  These data include Jefferson Lab 
Hall C experiments E99-118 (targets p, $^2\text{H}$)~\cite{Tvaskis:2006tv,Tvaskis:f2}, E94-110 (p)~\cite{Liang:2004tj}, 
%Tvaskis:R,
%,e99118-crosssections
E02-019 (p,$\rm ^2H$)~\cite{fomin-e02019}, E03-103 ($\rm ^2H$, $\rm ^3He$, $\rm ^4He$, $\rm ^9Be$ and 
$\rm ^{12}C$)~\cite{seely:2009}, 
E02-109 (p, $\rm ^2H$) , E06-009 ($\rm ^2H$)~\cite{Albayrak:dissertation}, and
E04-001 ($\rm ^{12}C$, $\rm ^{27}Al$, $\rm ^{56}Fe$, $\rm ^{64}Cu$)~\cite{Mamyan:dissertation} from the JUPITER collaboration.  

Figure~\ref{fig:12C-inclusive} shows examples of reduced inclusive cross section data as a function of $W$ from the quasielastic region 
through the $\Delta(1232)$ resonance region from SLAC (Barreau {\it et.al}~\cite{barreau1983}) and preliminary data from E04-001 at 
$Q^2 \approx 0.3$~$\rm (GeV/c)^2$.  The data at all $W$ have been centered to the common $Q^2$ utilizing a global fit to the available 
cross sections as described in ~\cite{bodek2013further}.  
Also shown are the contributions from quasielastic scattering utilizing the superscaling formalism with updated form factor 
parameterizations, as well as the inelastic contribution based on a gaussian smearing of fits to the proton~\cite{Christy-bosted:2010} and 
neutron~\cite{Bosted-christy:2008} cross sections and a medium modification factor.  The remaining strength is assumed to be due to 2-body 
currents in scattering from quasi-deuterons, such as meson-exchange.  This additional strength required beyond the independent nucleon 
impulse approximation is consistent with previous observations that this enhancement is only in the transverse cross section, as indicated 
by the independence of the enhancement on $\epsilon$.  The new JLab E04-001 data further underscores this observation and is 
expected to have final point-to-point uncertainties of 2\% or better, allowing for a separation of the longitudinal and transverse cross 
sections and structure functions for a range of nuclei from the quasielastic region through the resonance region to 
$W~\approx$~4.5~$\rm GeV^2$ and for $0.3 < Q^2 < 4.5$~$\rm (GeV/c)^2$.  This experiment ran in parallel with experiments E02-109 and E06-009 
on deuterium targets.  First publications on the L/T separation results are in preparation and will feature the deuteron $R_d$ ($F_L^d$) 
and the modification of $R$ ($F_L$) in the nuclear medium from $R_A - R_d$.  

The latter is of interest to low energy neutrino scattering experiments as statistics and kinematic limitations will not allow a separation of all three structure functions and $R$ must be taken from other data sets, such as electron scattering data.  The problem here is that due to the $Q^2$ behavior of $F_L$ for neutrinos, $R_{\nu}$ is different than $R_{em}$. However, while $R$ from electron scattering on a proton target was well measured by E94-110, $R$ on nuclear targets in this kinematic region has not previously been well measured prior to the Jupiter experiments with rough estimates of the impact of the uncertainty on the input $R$ on the predicted neutrino cross section in this region in excess of several percent~\cite{jupiter-prop}.  The results from these soon to finalized Hall C experiments are expected to reduce the uncertainty of $R_A$ by at least a factor of four or better.

In the quasielastic region, Hall A experiment E05-110 ($^4$He, $^{12}$C, $^{56}$Fe, and $^{208}$Pb) is completing precision L/T separations of 
response functions in the 3-momentum transfer range of 0.55 to 0.9~GeV/c, which is expected to resolve long standing discrepancies in the 
integral of the longitudinal response function extracted from different experiments (for a review of inclusive quasielastic 
electron-nucleus scattering and the Coulomb Sum Rule prior to 2009 see ~\cite{RevModPhys.80.189}).

A critical component to modeling the quasielastic and resonance region inclusive cross sections for electron scattering from nuclear targets is the distribution of momentum and binding energy for the nucleon on which the scattering occurs, and which is encoded in the spectral functions.  Electron scattering data over the last several decades or more have provided experimental access to proton spectral functions through the $(e,e^{\prime}p)$ reaction from light nuclei such as $^3$He through heavy nuclei such as $^{208}Pb$ with energy resolutions of several hundred KeV or better, which is necessary to isolate individual shell model states.  For $\nu-A$ interactions the spectral function is important for the prediction of the energy and momenta of final state nucleons, which is often used as a cut parameter to isolate the quasielastic process from inelastic processes.   Currently Argon, a nucleus important to the experimental neutrino oscillation community,  is one of the nuclei for which the spectral function has not been well studied.  An experiment to remedy this situation was recently approved~\cite{Benhar:2014nca} and will soon be underway in Jefferson Lab Hall A.

To further complicate the matter of isolating quasielastic events, the presence of final-state multi-nucleons in scattering from correlated  nucleon pairs obfuscates the quasielastic experimental signature.  Significant experimental progress has been made in recent years to determine the fraction and type of correlated pairs ($p-n$ versus $p-p$).  One of the significant challenges in $\nu-A$ event generators is properly sampling the momenta and energy distributions for both single nucleon and correlated initial states.  The data discussed is critical for confronting {\it ab initio} calculations and testing the adequacy of models utilized in generators.

The identification of particular scattering processes in neutrino scattering experiments, such as resonance production, relies on the identification of one or more hadrons.  Furthermore, reconstruction of the neutrino energy depends on the observed energies of these hadrons.  It is therefore imperative to model as accurately as possible the effects of hadron formation and propagation in the nucleus.  Data from semi-inclusive electron scattering provides experimental observables which can directly confront models of hadronization and propagation utilized for neutrino scattering.  Such studies with electron beams were first performed at SLAC~\cite{Osborne:1978ai} and subsequently measured with muons by at CERN~\cite{Ashman:1991cj}, and more recently with 27.6~GeV electron and positrons at HERMES\cite{Airapetian:2000ks,Airapetian:2007vu}, and at significantly lower energies around 5~GeV with CLAS at Jefferson Lab~\cite{Hafidi:2006ig}.  The data from both HERMES and CLAS provide the kinematic dependence of the ratio of hadron multiplicities on nuclei to those from the deuteron ($R_A^h$) for nuclei with different A and a range of hadrons (such as $\pi^+$, $\pi^-$, $\pi^0$, $K^+$, $K^-$).  This kinematic dependence of this ratio is sensitive to hadronic formation times and mechanisms.  
                   
%[PT broadening:  The broadening of the distribution of hadron momentum transverse to the photon ($p_T$).]

\subsection{Modeling}  

An important distinction in studying various classes of lepton-nucleus reaction should be made clear:
one should distinguish {\em inclusive} reactions, where only the scattered lepton is presumed to be detected, from more {\em exclusive} reactions where, in addition to the final-state lepton, additional particles are presumed to be detected. Examples of the former are $(e,e^{\prime})$ and $(\nu_{\mu}, \mu^-)$ reactions, while examples of the latter are the so-called semi-inclusive reactions $(e,e^{\prime}p)$ and $(\nu_{\mu}, \mu^- p)$. 
This separation into inclusive and more exclusive reactions is of considerable importance for the nuclear theory being employed. For instance, {\it ab initio} non-relativistic approaches are designed to work for inclusive reactions: by suitable manipulation it becomes possible to insert a complete set of final nuclear states and thereby {\em implicitly} include all classes of final-state interactions. However, the final states are not treated {\em explicitly}, and thus this approach is not directly applicable for the more exclusive reactions. The result is that very sophisticated non-relativistic studies are possible for inclusive reactions, and that these must be extended (typically by making approximations such as factorization, employing spectral functions, {\it etc.}) when the goals of the measurements require more exclusive modeling. Note that one cannot obtain a semi-inclusive cross section from an inclusive one, whereas the reverse is possible by integrating over all open channels.

% success and difficulties of electron scattering models
Different theoretical approaches used to model inclusive $(e,e^\prime)$ scattering in the quasielastic regime and beyond have recently been extended to the study of neutrino reactions.
In some cases, such as the simple and commonly-used relativistic Fermi gas model (RFG), models fail to reproduce both inclusive electron scattering in the quasielastic regime as well as recent measurements of QE neutrino and antineutrino scattering cross sections. In particular models based on the impulse approximation (IA) usually fail to reproduce the existing L/T separated data, which point to a transverse (with respect to ${\bf q}$)  response larger that the longitudinal one. Furthermore, both the shape and size of the  responses are different from the experimental ones, due to the simplified description of the reaction mechanism and of the nuclear dynamics.
Hence a proper evaluation of the effects introduced by final-state interactions (FSI) and mechanisms beyond the IA, such as nuclear correlations and two-particle two-hole excitations, are needed. 

% brief summary of ``ab-initio'' and phenomenological approaches
Ab initio approaches describe electron-nucleus scattering processes starting from a realistic nuclear Hamiltonians. Among these is the model based on the Green's Function Monte Carlo (GFMC) algorithm, which has been used to calculate the inclusive electromagnetic QE response functions of $^4$He and $^{12}$C in the regime of moderate momentum transfer, 
including nuclear correlations and consistent 2-body meson-exchange currents \cite{Lovato:2015qka,Lovato:2016gkq}. 
The main drawbacks of this method are its computational cost and the severe difficulties involved in its extension to include relativistic kinematics and resonance production. The Pavia Relativistic Green's Function (RGF) approach accounts for final state interactions (FSI) in a relativistic framework using a technique which allows one to conserve the total flux \cite{Capuzzi:1991qd,Meucci:2003uy}.
The formalism based on spectral function (SF) and factorization of the nuclear
transition matrix elements has been used by Benhar \textit{et al.} \cite{Rocco:2015cil} to model the QE peak \cite{Ankowski:2014yfa} and has been recently extended to include two-particle–two-hole final states \cite{Benhar:2015ula}. The model accounts for the effects of FSI by means of a folding function which contains a real optical potential and a nuclear transparency factor extracted from $(e,e'p)$ data.
The Valencia  group used a local Fermi gas model with RPA correlations based on phenomenological Landau-Migdal parmeters \cite{Gil:1997bm}. The model, which accounts for medium effects through the use of particle and hole spectral functions, is used to describe the QE and $\Delta$ peaks and includes 2p2h excitations.
The Ghent HF-CRPA model starts from a Skyrme-based Hartree-Fock mean field and adds long-range correlations through a continuum RPA approach using the same Skyrme residual interaction \cite{Pandey:2014tza}; the model is particuarly suited to study low-energy excitations.
The approach of the Sofia group is based on Coherent Density Fluctuation Model (CDFM) which accounts for the high-momentum tail of the nucleon momentum distribution arising from short-range NN correlations \cite{Antonov:2006md,Antonov:2009qg}.
The Giessen group uses the GiBUU implementation of quantum-kinetic transport theory to describe the QE and $\Delta$ regions \cite{Leitner:2008ue}. The key ingredient of the model in the cross section computation is a momentum dependent potential translated into an effective nucleon mass. The model has been recently complemented with a phenomenological fit of the 2p2h response 
\cite{Gallmeister:2016dnq}. 
 More details on the different treatments of two-body currents are given in Section VI.C.

\begin{figure}[tbh]
\centering
\includegraphics[scale=0.25, angle=0]{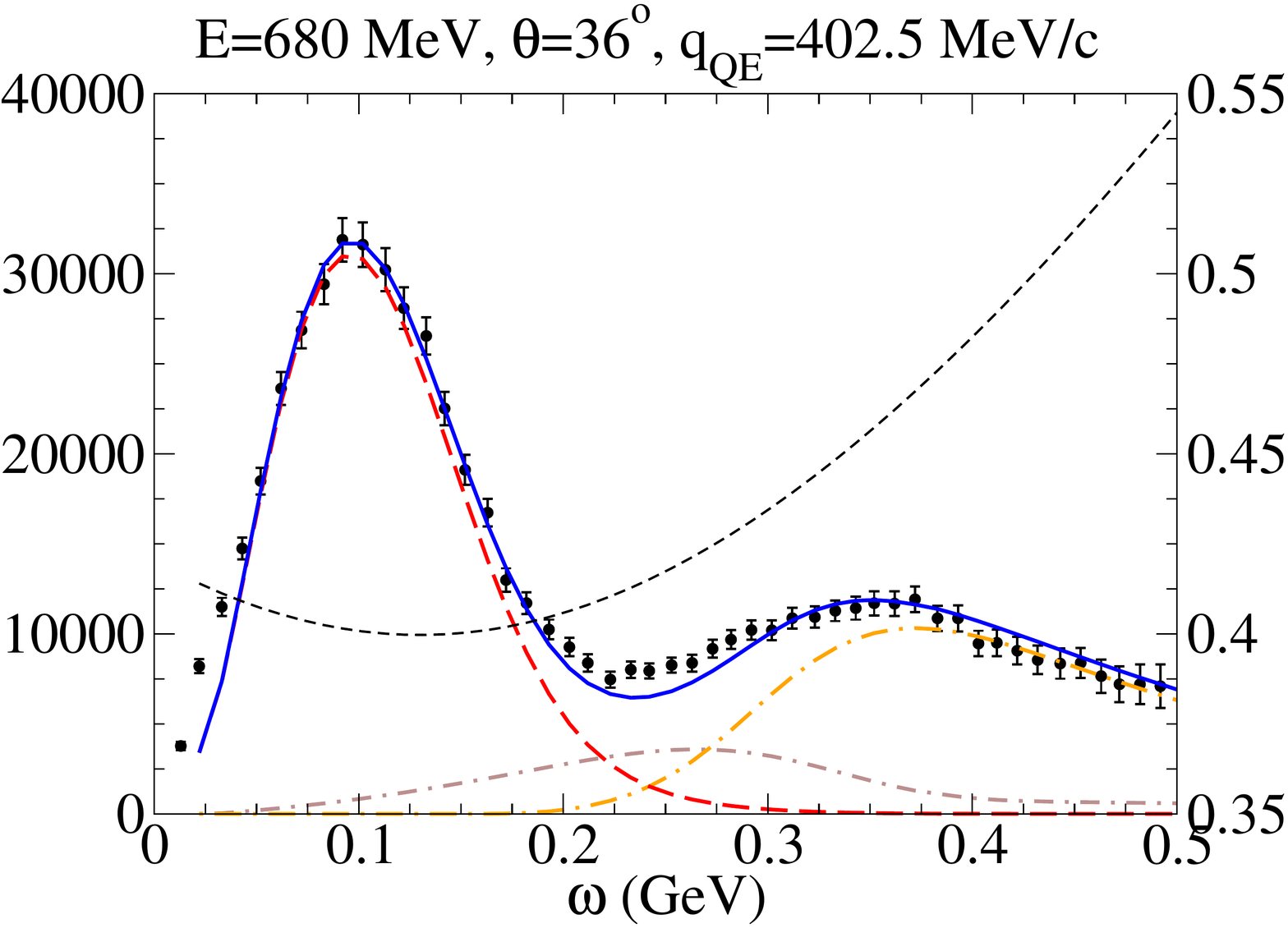}
\includegraphics[scale=0.25, angle=0]{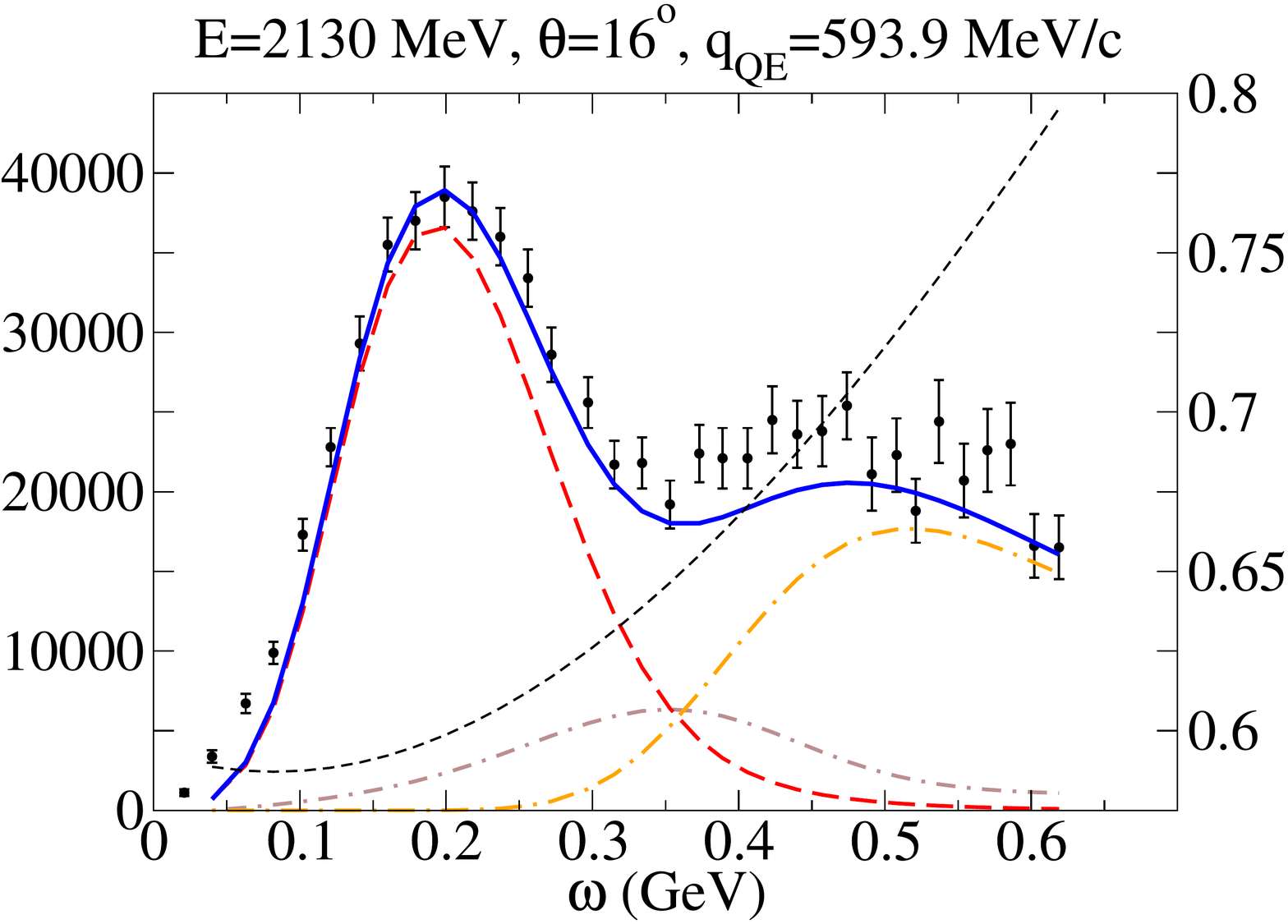}
\caption{Comparison of inclusive $^{12}$C$(e,e')$ cross sections and predictions of the QE-SuSAv2 model (long-dashed red line), 2p-2h MEC model (dot-dashed brown line) and
inelastic-SuSAv2 model (long dot-dashed orange line). The sum of the three contributions is represented with a solid blue line. The $q$-dependence upon $\omega$ is also shown (short-dashed black line). The y-axis on the left represents $d^2\sigma/d\Omega/d\omega$ in nb/GeV/sr whereas the one on the right represents the $q$ value 
in GeV/c. Figure from Ref.\cite{Megias:2016lke}.}
\label{fig:eepsusa}
\end{figure}

% superscaling
In the SuperScaling approach  \cite{Amaro:2004bs} (denoted as SuSA), instead of starting from a microscopic Hamiltonian, the scaling and superscaling properties of electron-nucleus interactions 
\cite{Day:1990mf,Donnelly:1998xg,Donnelly:1999sw,Alberico:1988bv,Maieron:2001it}
have been used to construct a semi-phenomenological model for lepton-nucleus scattering. A similar approach is also taken in the Transverse Enhancement Model (TEM) of Ref.~\cite{Bodek:2011ps}.
The SuSA model assumes the existence of universal scaling functions for electromagnetic and weak interactions. The general procedure adopted in this analysis consists of dividing the inclusive $(e,e')$ experimental cross section by an appropriate single-nucleon one to obtain a reduced cross section.
When this is plotted as a function of the ``scaling variable'', itself a function of $\omega$ and $q$, some particular properties emerge:
at energy transfers below the QE peak, the reduced cross section is largely independent of the momentum transfer, which is called scaling of first kind, and of the nuclear target, which is defined as scaling of second kind. The simultaneous occurrence of scaling of both kinds is denoted as superscaling. At higher energies, above the QE peak, both kinds of scaling are shown to be violated as a consequence of the contributions introduced by effects beyond the IA, such as meson-exchange currents (MEC) and inelastic scattering.
The scaling formalism, originally introduced to describe the QE domain, has been extended to the region of the $\Delta$ resonance \cite{Maieron:2009an} and the complete inelastic spectrum \cite{Barbaro:2003ie}. 
Recently an improved version of the superscaling model has been developed, called SuSAv2 \cite{Gonzalez-Jimenez:2014eqa}, that incorporates relativistic
mean field (RMF) effects in the longitudinal and transverse nuclear
responses, as well as in the isovector and isoscalar channels independently.
Within the RMF model the bound and scattered nucleon wave functions are solutions of the Dirac-Hartree equation in the presence of energy-independent real scalar (attractive) and vector (repulsive) potentials. Because the same relativistic
potential is used to describe the initial and final nucleon states, the model preserves the continuity equation.
An important result is that the model reproduces surprisingly well the magnitude and shape of the experimental longitudinal superscaling function. On the other hand, it predicts a larger transverse scaling function, an effect due to the distortion of the lower components of the outgoing nucleon Dirac wave function by
the FSI which agrees with the available separated L/T data (see 
Refs.\cite{Caballero:2006wi,Caballero:2005sj,Caballero:2007tz,Martinez:2008ve,Ivanov:2008ng} for details of the model and its predictions on electron and 
neutrino reactions).
2p-2h MEC effects, which play an important role in the “dip” region between the
QE and the $\Delta$ peaks, are included in the SuSAv2 model following the work of De Pace \textit{et al.} \cite{DePace:2003xu,DePace:2004cr}, who performed the first fully relativistic calculation of the electromagnetic two-body currents contribution to inclusive electron scattering. Detailed comparison of the SuSAv2 predictions with electron scattering data on $^{12}$C at many different kinematics can be found in Ref.\cite{Megias:2016lke}, showing a very satisfactory agreement of the model with inclusive data. Two illustrative examples are shown in Fig. \ref{fig:eepsusa}.

%relativity
It is important to notice that the regime of interest in present and future neutrino experiments is high-energy and relativistic aspects of the problem are critical. What exists in modeling this regime is either very limited or requires making approximations. For instance, it is possible to treat the deuteron relativistically with sophisticated treatment of both initial and final (NN) states. 
However, for heavy nuclei only the non-relativistic {\it ab initio} approach can claim a high level of consistency when treating the nuclear many-body problem for inclusive reactions. 
Making relativistic extensions to this approach without also making approximations cannot be anticipated for the foreseeable future. Accordingly, it is essential to continue to pursue modeling where reasonable approximations are made, but where relativistic quantum mechanics of one form or other is incorporated. 
For heavy nuclei this means, for instance, employing simple models such as the RFG to get some insights into the significance of relativistic effects, or more sophisticated approaches such as RMF to explore how some aspects of relativistic dynamics play a role. One goal for the near future should be to inter-compare the results of the {\it ab initio} non-relativistic studies with those of the approximate, but relativistic modeling. Additionally, it is important to make contact with SuperScaling Analyses of electron scattering data, which provide a stringent test of nuclear models. The continued validation of the last is important, since scaling analyses allow one to get some insight into the roles played by the various contributions to the inclusive cross section (quasielastic, MEC, inelastic).

%coherent scattering and PV
In passing let us also comment on coherent scattering from nuclei.  
In elastic scattering of either electrons or neutrinos from nuclei one has all multipoles allowed by conservation of angular momentum, parity and time reversal invariance.  
For instance, in elastic electron scattering from a spin-5/2 nucleus one has C0, C2, C4, M1, M3 and M5 multipoles.  Of these, the monopole is coherent 
(involves all of the nucleons in the nuclear ground state), whereas the others, while contributing to elastic scattering, are not coherent.  
For electron scattering this implies that the C0 multipole is proportional to Z, which is large for a heavy nucleus, while the others are proportional to quantities of order 
unity and thus are typically much smaller.  Accordingly, at modest momentum transfers one can expect that the coherent monopole contribution is dominant.  
In Section IX the problem of coherent neutral-current neutrino scattering at low energies is discussed.  Here we only note that the ground-state neutral current matrix element 
can also be probed using parity-violating electron scattering.  Indeed, the neutrino scattering cross section in leading-order is equal to the product of the parity-conserving 
electron scattering cross section times the square of the parity-violating asymmetry:
\begin{equation}
\left[ \frac{d\sigma}{d\Omega} \right]_{neutrinos} = \left[ \frac{d\sigma}{d\Omega} \right]_{electrons, PC} (A_{PV})^2 .
\end{equation}
Any deviation from equality can in principle be used to explore physics beyond the Standard Model.  
The projection of present experimental opportunities suggests that the right-hand side of this equation could be determined at the sub one percent level, 
although getting the left-hand side to this level will be a challenge.  For details on these issues and for a brief discussion of corrections to this equation see \cite{Moreno:2015bta}.

\subsection{Challenges}

Summarizing, we list what we view as the main challenges in connecting
electron and neutrino reactions:

\begin{itemize}
\item matching models used to predict neutrino-nucleus observables to
electron scattering data
\item expanding theory to include more semi-inclusive predictions
\item provide semi-inclusive neutron, proton and pion data sets with as
broad an angular range as possible
 \end{itemize}

%\bibliography{wpelectron}

\newpage
\section{Quasi-elastic, quasi-elastic-like scattering}
\label{CCQE} 
Quasi-elastic (QE) scattering is the main interaction mechanism for neutrinos with energies up to about 1~GeV.
As this region is at the core of the neutrino energy distribution for many neutrino experiments, quasi-elastic scattering is key to the understanding of neutrinos and their interactions with nuclei.
In quasi-elastic scattering, the incoming neutrino scatters off a nucleon, bound by the nuclear potential.  Instead of a sharp peak in the excitation spectrum as found in true {\it elastic} scattering, the  scattering off the nucleons moving in the nuclear medium, gives rise to a broad peak in the excitation spectrum, centered around the quasi-elastic value $\omega=\frac{Q^2}{2M}-S$,  with $Q^2$ the four-momentum transfer, $M$ the nucleon mass and $S$ a shift correcting for the binding of the nucleons in the potential. 
Central parameters in the dynamical behavior of the cross section are  energy and momentum transfer ; incoming energy mainly affects kinematic aspects of the cross section.  The influence of the nuclear medium and nuclear correlations on the scattering
process and on the ejectiles make this processes far more challenging to model than could naively be expected.  

It is noteworthy that different definitions of 'quasi-elastic' are used.  In general, the term might refer to events that are close to the quasi-elastic peak in lepton kinematics.  In experimental situations, events are usually classified as being quasi-elastic when their final state obeys certain restrictions, 'a lepton, no pions' being the most common one.   
In a theoretical context, interactions are dubbed quasi-elastic when the scattering occurs elastically off a single bound nucleon. 
The confrontation with the experimental practice then leads to the identification of 'QE-like' events, involving more complicated scattering mechanisms such as short-range correlations or meson-exchange currents, but with a final state matching the experimental QE constraints.  To avoid the confusion of the signal definition,
it becomes increasingly  common to present the data in terms of the final-state particles,
such as '1 muon and 0 pions, with any number of protons'.  
This corresponds to the charged-current CCQE-like data without subtracting any intrinsic backgrounds (except detector related effects) and is dubbed {\bf CC0$\pi$}.

Experimental results are typically evaluated with strong biases from theoretical predictions. 
Improvement on accelerator intensities provide intense neutrino fluxes and systematic errors are becoming  relevant in oscillation neutrino experiments. Both experiment and theory must improve in parallel to  help upcoming and future neutrino oscillation analyses.  Quasi-elastic scattering is still the reference cross-section for low energy oscillation experiments like T2K/TH2K and the Fermilab Booster neutrino beam experiments and it is still very relevant for higher energy neutrinos used by the NOvA and DUNE experiments. Uncertainties in the modeling of the cross-section impact the neutrino energy reconstruction used by these neutrino oscillation experiments. Running and future experiments require better sensitivities to improve on discerning between different models but they also require more solid theoretical predictions for both  quasi-elastic and background events.\\

\subsection{QE scattering on the nucleon}
Whereas the fundamental interaction at play in QE scattering is the neutrino communicating with quarks through the exchange of a Z-boson for neutral-current interactions or W-boson for charged-current processes, at the energy scales at play in QE interactions, protons and neutrons are efficiently used as effective degrees of freedom.  Cross sections are then calculated using form factors for the nucleon, parameterizing our lack of knowledge about QCD at low energies and effectively  taking into account the internal structure of the baryons and their coupling to the lepton current.

At the nucleon level, (quasi)elastic $\nu$A scattering refers to the charged-current processes $\nu_\ell n\to\ell^-p$
and $\bar\nu_\ell p\to\ell^+n$, for lepton flavor $\ell$, as well as the neutral-current process $\nu
N\to\nu N$, where $N$ can be the neutron $n$ or proton~$p$, and $\nu$  can be a neutrino or
antineutrino.
The neutral-current process is actually elastic, but it is convenient to group them together. It is
imperative to understand these relatively simple hadronic transitions, since uncertainties at the level of elementary amplitudes limit the achievable precision of any nuclear cross section.
This desire also holds for more complicated hadronic transitions, $N\to\Delta$ or $N\to N\pi$, discussed below.

\subsubsection{Invariant form factors} 
Working at leading order in electroweak couplings, quark-level interactions with neutrinos are described by
the Lagrangian
\begin{align}
  {\cal L}_{\rm eff} &=
  - \frac{G_F}{\sqrt{2}} \left[ J^{+ \mu} J^-_{\mu} +  J^{0 \mu} J^0_\mu \right]
\end{align}
after integrating out the $W$ and $Z$ bosons.
Here $J^\pm$ and $J^0$ are charged and neutral currents, $J^-_\mu={J^+_\mu}^\dagger$,
\begin{align}
    J^{+}_\mu &= \sum_\ell \bar{\nu}_\ell \gamma_\mu(1-\gamma_5) \ell
      + \sum_{ij} V_{ij} \bar{U}_i \gamma_\mu (1-\gamma_5) D_j ,
    \label{eq:current+} \\
    J^{0}_\mu &= \sum_f \left[ g_L^f \bar{f} \gamma_\mu (1-\gamma_5) f
      + g_R^f \bar{f} \gamma_\mu (1 + \gamma_5) f \right] ,
    \label{eq:current0}
\end{align}
and $g_{L,R}^f = I_3 (f_{L,R} ) - Q(f) \sin^2\theta_W$, with $I_3$ the third component of isospin, $Q$ the
electric charge in units of proton charge, and $\theta_W$ the weak mixing angle.
Here $V_{ij}$ is the CKM matrix element relating the electroweak gauge eigenstate basis for quarks
to the mass eigenstate basis.

The hadronic matrix elements $\langle p|J^+_\mu|n\rangle$, $\langle n|J^-_\mu|p\rangle$, and
$\langle N|J^0_\mu|N\rangle$ of the
currents (\ref{eq:current+}) and (\ref{eq:current0}) are decomposed into Lorentz-covariant forms of the
nucleon four-momenta, multiplied by  functions of $q^2$ known as ``form factors.''
For example, 
\begin{multline}
    \langle p(p') | J^{+}_{\mu} | n(p) \rangle = \bar{u}^{(p)}(p') \left\{
        \gamma_\mu F_1^\text{CC}(q^2) + \frac{i}{2m_N} \sigma_{\mu\nu} q^\nu F_2^\text{CC}(q^2) +
        \gamma_\mu \gamma_5 F_A^{CC}(q^2) 
    \right. \\ \left.
      + \frac{1}{m_N} q_\mu \gamma_5 F_P^\text{CC}(q^2) \right\} u^{(n)}(p),
  \label{eq:amp}
\end{multline}
and corresponding expressions with form factors $F_i^{\text{NC, }p}$ and $F_i^{\text{NC, }n}$ for
neutral-current scattering matrix elements $\langle p | J^{0}_{\mu} | p \rangle$ and
$\langle n|J^{0}_{\mu} | n \rangle$.
For the vector case, these are the Dirac and Pauli form factors, $F_1(q^2)$ and $F_2(q^2)$, respectively,
which are often expressed in terms of the electric and magnetic form factors, $G_E(q^2)=F_1(q^2)+q^2
F_2(q^2)/(4m_N^2)$ and $G_M(q^2)=F_1(q^2)+F_2(q^2)$.
For the axial-vector case, one has  two more form factors, $F_A(q^2)$ and $F_P(q^2)$, known as the axial
and pseudoscalar form factors.
In the cross section, the contribution of the pseudoscalar form factor, $F_P(q^2)$, is suppressed by a
factor $m_\ell^2$ (for free nucleons), so it is less
important than~$F_A(q^2)$.
The discrete symmetries $C$, $P$, $T$ respected by QCD imply that the basis (\ref{eq:amp})
is complete.

The form factors are of two types: vector, and axial (the latter including pseudoscalar).
Here we summarize current knowledge of the two types of form factors from a range of experimental and
theoretical constraints.
Constraints may be divided into three categories: form factor normalization at $q^2=0$,
form factor slopes at $q^2=0$; and general $q^2$ dependence.  As discussed below, for 
the $q^2$ range of interest, and with the appropriate choice of variable, the form factors
become approximately linear.  We describe here the experimental constraints on normalization,
slope and residual shape parameters.

\subsubsection{Electromagnetic form factors}
Form factor normalizations are defined by electric charges (in units of the positron charge)
and magnetic moments of
the nucleons:
\begin{align}
    G_E^N(0)= Q_N \,, \quad G_M^N(0)=\mu_N ,
\end{align}
where $Q_p=1$, $Q_n=0$, $\mu_p=2.79$ and $\mu_n=-1.91$. 
Form factor slopes are conventionally defined as charge and magnetic radii
\begin{align}
    {d G_E^{N} \over dq^2}\bigg|_{q^2=0} &= \frac16 (r_E^N)^2 , \quad
    {1 \over G_M^N(0)} { d G_M^{N} \over dq^2}\bigg|_{q^2=0} = \frac16 (r_M^N)^2 . 
\end{align}
The most precise determination of the neutron charge radius is from low energy neutron
scattering on the electrons of heavy nuclei, $(r_E^n)^2=-0.1161\pm0.0022~\text{fm}^2$.
For the proton charge radius, the recent development of muonic hydrogen spectroscopy
has provided the most precise determination, $r_E^p=0.84087(26)(29)\, {\rm fm}$~\cite{Antognini:1900ns}
from the muonic hydrogen Lamb shift.  There is a $5.6\sigma$ discrepancy,
representing a $\sim 8\%$ discrepancy in the value of the slope $(r_E^p)^2$
between this value and previous determinations based on regular hydrogen spectroscopy
and electron scattering, $r_E^p=0.8751(61)\,{\rm fm}$~\cite{Mohr:2015ccw}.  
This discrepancy has become known as the proton radius puzzle and remains controversial.
The magnetic radii are primarily determined by electron scattering measurements,
$r_M^p = 0.776(34)(17)\,{\rm fm}$~\cite{Lee:2015jqa} and $r_M^n = 0.864^{+0.009}_{-0.008}\,{\rm fm}$~\cite{Olive:2016xmw,Belushkin:2006qa,Epstein:2014zua}.  
The general $q^2$ dependence of the vector form factors is
constrained by electron-proton scattering, and
from electron scattering on light nuclear targets, interpreted as  
electron-neutron scattering after correcting for nuclear effects.

\subsubsection{Charged current vector form factors}
The relevant hadronic matrix element for charged current process involves 
the isovector quark current.  
Neglecting isospin violations from up- and down-quark mass terms and higher-order
electroweak effects, the isovector electroweak form factors are given
by the difference of proton and neutron electromagnetic form factors.
Many current neutrino scattering analyses employ the
BBBA2005 parameterization~\cite{Bradford:2006yz} for the isovector
nucleon form factors.
The global data for nucleon electromagnetic form factors
has been more recently analyzed using the $z$ expansion in Refs.~\cite{Lee:2015jqa,global}.   

\subsubsection{Neutral current vector form factors}
The neutral current vector form factors, restricting to 3-flavor QCD,
consist of linear combinations of $u$, $d$ and $s$ quark currents, and
are thus not fully determined by the electromagnetic form factors 
for proton and neutron.  
Many current neutrino scattering analyses neglect strange- and other
heavy-quark contributions, and assume a common dipole $Q^2$ dependence for
the remaining isoscalar and isovector combinations~\cite{Ahrens:1986xe,Andreopoulos:2015wxa}.  
It may be necessary to revisit these approximations with future precision
neutral-current neutrino data.
A discussion and further references for the vector form factor
normalization and slopes within 3-flavor QCD is found in Sec.~4.1 and
Appendix~B of Ref.~\cite{Hill:2014yxa}.  

\subsubsection{Axial form factors: charged current}
Constraints may again be divided into three categories: form factor normalization at $q^2=0$,
form factor slopes at $q^2=0$; and general $q^2$ dependence. 
Form factor normalizations are determined by neutron beta decay, 
\begin{align}
  F_A(0) &= g_A \,, 
\end{align}
with~\cite{Olive:2016xmw} $g_A=-1.2723(23)$.
The axial radius is defined analogously to the vector radii,
\begin{align}
  {1\over F_A(0)} {d F_A \over dq^2}\bigg|_{q^2=0} &= \frac16 (r_A)^2 \,. 
\end{align}
The axial radius, and general $q^2$ dependence of $F_A$, is
constrained by several processes.
Neutrino-deuteron scattering, interpreted as
neutrino-neutron scattering after correcting for nuclear effects,
provides the most direct access
to $F_A$ over a broad $q^2$ range.  A recent analysis~\cite{Meyer:2016oeg}
using the $z$ expansion obtains an axial radius $r_A^2=0.46(22)\,{\rm fm}^2$
from existing bubble chamber data~\cite{Mann:1973pr,Barish:1977qk,Miller:1982qi,Baker:1981su,
  Kitagaki:1983px,Kitagaki:1990vs}.
No future high-statistics measurements of neutrino scattering on
hydrogen or deuterium are presently foreseen. 
Existing constraints on $F_A(q^2)$ inferred from charged pion
electroproduction~\cite{Amaldi:1972vf,Brauel:1973cw,
  DelGuerra:1975uiy,DelGuerra:1976uj,Esaulov:1978ed,Amaldi:1970tg,Bloom:1973fn,Joos:1976ng,Choi:1993vt,
  Liesenfeld:1999mv,Friscic:2016tbx}
have similar statistical power~\cite{Bhattacharya:2011ah} but suffer from
model-dependent corrections to the chiral limit~\cite{Bernard:2001rs}.
The muon capture process $\mu^- p \to \nu_\mu n$ from the
muonic hydrogen ground state probes a combination of
$F_A(q_0^2)$ and $F_P(q_0^2)$,
where $q_0^2=-0.88\,m_\mu^2$~\cite{Andreev:2012fj}.

\subsubsection{Axial form factors: neutral current}
The neutral current axial-vector form factors, restricting to 3-flavor QCD,
consist of linear combinations of $u$, $d$ and $s$ quark currents.
Many current neutrino scattering analyses account for strange- and other
heavy-quark contributions by rescaling the normalization at $q^2=0$
that would be obtained from the purely isovector case: 
\begin{align}
 F_A^{\rm NC}(0) = F_A^{\rm CC}(0)\big( 1 + \eta \big) \,,
\end{align}
with default value $\eta = 0.12$, and assuming a common dipole $Q^2$
dependence~\cite{Ahrens:1986xe,Andreopoulos:2015wxa}.
It may be necessary to revisit these approximations with future precision
neutral-current neutrino data.
A discussion and further references for the axial-vector form factor
normalization and slopes within 3-flavor QCD is found in Sec.~4.2 and
Appendix~B of Ref.~\cite{Hill:2014yxa}.  

\subsubsection{Form factor parameterizations}
A range of parameterizations has been used for the
form factors appearing in neutrino scattering analyses.
Historical benchmarks include the dipole ansatz \ask{for the axial form factor}~\cite{LlewellynSmith:1971uhs},
\begin{equation}
    F_A(q^2) = \frac{g_A}{(1+q^2/M_A^2)^2},
    \label{eq:FA:dipole}
\end{equation}
and ratios of polynomials for
vector form factors~\cite{Olsson:1978dw}.  A variety of other forms
have been used more recently~\cite{Sick:2003gm,Bodek:2007ym,Bernauer:2013tpr,Amaro:2015lga}. 
The so-called $z$ expansion provides a model independent
description of form factor shape and quantification of shape uncertainty. 
The formalism for the $z$ expansion and nucleon form factors is
  described in  Refs.~\cite{Bhattacharya:2011ah,Hill:2010yb}, and
  several applications are  found in
  Refs.~\cite{Lorenz:2014vha,Epstein:2014zua,Lee:2015jqa,Bhattacharya:2015mpa}.
  Related formalism and applications may
  be found in~\cite{Hill:2006ub, Bourrely:1980gp,
    Boyd:1994tt,Boyd:1995sq,Lellouch:1995yv,Caprini:1997mu,Arnesen:2005ez,
    Becher:2005bg,Hill:2006bq,Bourrely:2008za,Bharucha:2010im,Amhis:2014hma,Bouchard:2013pna,Bailey:2015dka,Horgan:2013hoa,Lattice:2015tia,Detmold:2015aaa}.
The underlying analytic structure of
the form factor implies that a change of variable from $q^2$ to $z$,
\begin{align}
  z(q^2) &= \dfrac{ \sqrt{t_{\rm cut} - q^2} - \sqrt{t_{\rm cut} - t_0} }{
    \sqrt{t_{\rm cut} - q^2} + \sqrt{t_{\rm cut} - t_0} } \,,
\end{align}
maps the form factor shape onto a convergent Taylor expansion throughout
the entire spacelike scattering region: (for generic form factor $F$) 
\begin{align}
 F(q^2) &= \sum_{k=0}^\infty a_k [z(q^2)]^k \,. 
\end{align}
Here $a_k$ are dimensionless numbers encoding nucleon structure, 
$t_{\rm cut}$ is the mass
of the lightest state that can be produced by the current under consideration,
and $t_0$ is a free parameter chosen for convenience.   The number
of relevant parameters is determined a priori by the kinematic
range and precision of data.   For example, in the case of the
axial form factor, for $0<-q^2<1\,{\rm GeV}^2$, we can choose $t_0$
so that $|z|<0.23$, and it can be readily seen that quadratic,
cubic and quartic terms enter at the level of $5\%$, $1\%$ and $0.3\%$.

\subsubsection{Practical prospects for lattice QCD} 
The matrix elements in (\ref{eq:current+}) and (\ref{eq:current0}) can be computed directly from the QCD
Lagrangian using lattice gauge theory.
It is worth noting that similar calculations of $B$-meson form factors are used, together with the $z$
expansion to determine the CKM matrix elements $|V_{ub}|$~\cite{Bailey:2008wp,Flynn:2015mha,Lattice:2015tia}
and $|V_{cb}|$~\cite{Lattice:2015rga,Na:2015kha}.
More recently, this approach has been extended to $\Lambda_b$ decays, using $\Lambda_b\to p\ell\nu$ and
$\Lambda_b\to\Lambda_c\ell\nu$ to determine $|V_{ub}|/|V_{cb}|$~\cite{Detmold:2015aaa}.
These are examples of a wide range of successful calculations over the past decade~\cite{Kronfeld:2012uk},
including several predicitions of quantities that had not been measured well.

Lattice-QCD calculations of nucleon properties suffer from a larger signal-to-noise ratio than the
corresponding meson quantities, for well-understood reasons~\cite{Lepage:1989hd}.
In practice, there are ways to circumvent this problem~\cite{Beane:2010em}.
In the baryon sector, notable achievements are the mass spectrum (see, e.g., a summary plot
in~\cite{Kronfeld:2012uk}, papers cited therein, and newer work in~\cite{Alexandrou:2014sha}) and the
neutron-proton mass difference~\cite{Borsanyi:2014jba,Horsley:2015eaa}.

The vector and axial form factors have been calculated by many groups interested in nucleon structure.
A special focus has been on $g_A$, because it is precisely known from neutron $\beta$~decay.
This quantity seems to be susceptible to every technical challenge in lattice QCD: contamination from 
excited states, finite-volume effects, and unphysically heavy up and down quarks.
The most recent calculations~\cite{Bhattacharya:2016zcn} find agreement with experiment, with 3\% errors.
This work profited from ensembles of gauge-field configurations with essentially physical up and down 
quarks (isospin averaged), from~\cite{Bazavov:2012xda}.
A recent paper~\cite{Green:2014xba} similarly uses physical-quark-mass ensembles (from~\cite{Durr:2010aw}) 
to compute the $q^2$ dependence of the nucleon electromagnetic form factors.
This work uses dipole fits to extract the charge radii, which are found to be in agreement with experiment.
In the future, as discussed above, it will be preferable to treat the shape in~$q^2$ with Ansatz-free 
functional forms, such as the $z$~expansion~\cite{Meyer:2016kwb}.

For the NC processes, an additional challenge arises for lattice QCD.
The flavor-singlet part of the neutral current can be mediated through a virtual quark loop, which 
communicates via nonperturbative QCD interactions with the valence quarks.
It is then necessary to compute the quark propagator for the loop for all lattice sites to all other 
lattice sites.
The computational demand is prohibitive unless stochastic methods are employed to estimate these effects.

In summary,  lattice-QCD calculations of the axial and also electromagnetic (as a 
cross-check) form factors are a promising prospect.
Even so, the situation is similar to that for meson form factors 10--15 years ago: a lot of work has been 
carried out, and the main obstacles and their workarounds are understood.
We should begin to see calculations with full error budgets, suitable for incorporating into 
nuclear-physics calculations.
With suitable support, several such calculations will exist, and they can be scrutinized and (again as in 
meson physics) averaged.\\
%{\bf Contributors : Richard Hill, Andreas Kronfeld}

\subsection{QE on the nucleus : 1p1h processes}\label{QE:1p1h}
The usual analysis in the region of the quasielastic (QE) peak assumes that the dominant process is elastic scattering from nucleons in the nuclear ground
state, followed by quasifree ejection of the nucleons from the nucleus. This is known as the Impulse Approximation (IA) and corresponds to one-particle-one-hole (1p1h) excitations. In spite of the simplicity of the elementary reaction mechanism, this is a complicated many-body problem, which involves the proper treatment of nuclear correlations and of interactions of the knocked-out nucleon in both the initial and final state.

The basic ingredients needed to describe this process are the vector and axial elastic form factors of the nucleon, discussed in the previous section, and a model to describe the nuclear dynamics. The latter must take into account the nuclear mean field and nucleon-nucleon (NN) short- and long-range correlations in the ground state, as well as the final-state interactions (FSI) of the outgoing nucleon with the residual nucleus.
 
Reliable theoretical models are required to describe charged-current (CC) reactions, where the incident neutrino (or antineutrino) is converted into a charged lepton, and neutral-current (NC) ones, where the outgoing lepton is an unobserved (anti)neutrino. The two processes imply different kinematics and a model capable of describing one of them is not necessarily optimal  for the other.
In the former case one must also take into account the fact that the charged lepton in the final state is not a plane wave but is influenced by the Coulomb potential generated by the nucleus.

The kinematics involved in ongoing and future neutrino experiments typically lie in a domain where relativistic effects are important, with typical energies of the order of or larger than the nucleon mass; not only should the reaction mechanism incorporate relativity, but also the nuclear dynamics must be described in a relativistic framework. Some models, such as the relativistic mean field (RMF) are fully relativistic in nature, most of the other descriptions are based on a non-relativistic reduction of the nuclear dynamics that is subsequently relativized by means of an effective scheme.

The simplest nuclear model used to describe the QE region is the Relativistic Fermi Gas (RFG), where nucleons in the nuclear ground state are free, moving (Fermi motion) particles, correlated only by the Pauli principle. The RFG, used in most MC generators, is clearly inadequate to describe electron scattering data and therefore should not be expected to give reliable predictions for neutrino scattering observables. 
A wide variety of more sophisticated models, in most cases originally developed for electron-scattering studies, have been applied in recent years to the case of charged-current quasi elastic (CCQE) and neutral-current elastic (NCE) scattering. 
 These models rely on quite different hypotheses and approximations and utilize different theoretical frameworks.
    
The Giessen group uses the GiBUU transport model to describe various processes, including QE neutrino-scattering reactions \cite{Mosel:2016cwa}.
The modeling by Benhar~\textit{et al.} is based on a spectral-function approach \cite{Benhar:2006nr,Benhar:2009wi,Ankowski:2014yfa}.
Amaro~\textit{et al.} considered the relativistic super-scaling approach (SuSA)~\cite{Amaro:2004bs} based on
the super-scaling behavior exhibited by electron scattering data. 
It was extended to the SuSAv2 model by Gonzalez-Jimenez~\textit{et al.}~\cite{Gonzalez-Jimenez:2014eqa}
in order to take into account the different behavior of the longitudinal and transverse nuclear responses due to relativistic mean field effects.  It incorporates effects \cite{Gonzalez-Jimenez:2014eqa} stemming from an RMF description of the nucleus.  These scaling ideas were also explored within the Coherent Density Fluctuation Model (CDFM)~\cite{Antonov:2009qg}, which accounts for NN correlations giving rise to high-momentum components of the nucleon momentum distribution.
The models used in \cite{Meucci:2011vd,Meucci:2015bea} are based on a relativistic Green's function approach.
Other authors have reported on models based on relativistic distorted wave impulse approximation (RDWIA) and relativistic multiple-scattering Glauber approach to describe final-state interactions in QE scattering processes \cite{Martinez:2005xe}.
In the correlated-basis approaches as the one of Lovato~\textit{et al.}~\cite{Lovato:2014eva,Lovato:2015qka,Lovato:2016rhk},  NN correlations are included in the description of the nuclear wave functions within a Green's function Monte Carlo approach. 
 The models of  Nieves~\textit{et al.} and Martini~\textit{et al.} are similar :
they start from a local Fermi gas picture of the nucleus and consider medium polarization and collective effects through the random phase approximation (RPA) 
including $\Delta$-hole degrees of freedom, $\pi$ and $\rho$ meson exchange and $g'$ Landau-Migdal parameters in the effective $p-h$ interaction.  
The Ghent Hartree-Fock Continuum RPA model (HF-CRPA) \cite{Pandey:2014tza} starts from a Skyrme-based Hartree-Fock description of the nuclear mean field and adds long-range correlations through a continuum RPA approach using the same Skyrme parameterization as the residual interaction. 
The latter models are based on a non-relativistic description of the nucleon current and are relativized using an effective scheme.
It is noteworthy that even with typical neutrino energies of the order of 1 GeV, reactions with low energy transfers play a non-negligible role, especially for forward lepton scattering \cite{Pandey:2016jju,Amaro:2013yna}, making a detailed microscopic modeling of these processes that are very sensitive to nuclear structure details not addressed in Fermi-gas based models, important in these kinematic regions.

Without entering into the details of each calculation, it is important to point out that, despite the sometimes very different approaches, all of these models seem to provide rather similar predictions for $(\nu_l,l)$ CCQE cross sections. However, comparison with neutrino scattering data is not sufficient to test the validity of the various models. First, the experimental error bars are too large to discriminate between various calculations. Second, the specific experimental conditions, namely the fact that the neutrino energy is not exactly known, do not allow to select true QE events, which are usually mixed with different elementary processes.  For this reason it is important to emphasize the needed comparison of these model predictions to the much more accurate $(e,e')$ scattering data with known incoming lepton kinematics.  It is worth mentioning that two-body currents can also excite 1p1h states \cite{VanCuyck:2016fab,Amaro:2003yd,Umino:1996cz}. This channel has often been disregarded in the literature.

Most of the theoretical work performed up to now has been focusing on CCQE inclusive reactions, where only the outgoing lepton (muon or electron) is detected. When more exclusive reactions become the focus, the problem changes. For instance, if the semi-inclusive reaction $(\nu_{\mu},\mu^- p)$ must be modeled, the theoretical models need to be extended.
First, the elementary reaction, even on single nucleons, becomes more
complicated : instead of the seven familiar 'standard' nuclear responses,
eleven  more response functions that cancel in the angular integration for inclusive reactions,
enter for semi-inclusive reactions \cite{Moreno:2016sht}. And as the responses become dependent on the direction and momentum of the final hadron, they  are functions of four kinematical variables (e.g. $Q^2$, $\omega$, missing momentum $p_m$ and missing energy $E_m$), rather than simply $Q^2$ and $\omega$ as in
the inclusive case.
Moreover, it is  necessary to model the final state for the specific
channel and kinematics being considered. 
Clearly, given that experimental neutrino oscillation
measurements require the treatment of more exclusive processes, an
important goal will be to explore the model dependence that inevitably
arises in this more complicated case. 

Further studies on this subject include the extension of the present models, mainly designed for $^{12}$C and $^{16}$O nuclei, to heavier nuclei, in particular $^{40}$Ar, which will play a crucial role in future experiments. To this scope it is important to explore the density dependence of the various  nuclear effects mentioned above.\\

\subsection{Multinucleon processes and 2-nucleon knockout}
In the discussion of CCQE-like cross sections, the MiniBooNE measurement played a central role in revealing the presence of additional nuclear processes impacting neutrino scattering.  This measurement was obtained using a high-statistics sample of $\nu_\mu$ CCQE events on $^{12}$C and showed considerable discrepancies with simple RFG-based predictions ~\cite{Katori:2009du, AguilarArevalo:2010zc}. These discrepancies revealed the fact that  experimental signatures measuring one lepton and no pion final states will include more involved reaction mechanisms than the pure 1p1h  QE channel. At the time, alternative solutions were adopted to fit the data,  e.g. higher values of the axial mass, but it can now be appreciated that in fact the models were missing significant cross section contributions (NN correlations, MEC, large range correlations, etc).  

Prior to the release of the MiniBooNE data, larger possible neutrino scattering cross sections were suggested by Martini \textit{et al.}~\cite{Martini:2009uj}, drawing the attention to the existence of additional mechanisms beyond the interaction of the neutrino with a single nucleon in the nucleus.  

In addition  to the absorption of  the $W$ boson by a single nucleon which is knocked out leading to a 1p1h excitation,  coupling to nucleons belonging to correlated pairs (short-range NN correlations) and to two-nucleon currents arising from meson exchange (MEC) must also be considered. 
This leads to the excitation of multinucleon or np-nh excitations. 
The addition of the np-nh excitations 
to the genuine quasielastic (1p-1h) contribution leads to an agreement with the MiniBooNE data without any increase of the axial mass.  It is noteworthy that  multinucleon mechanisms also influence processes with only one nucleon in the final state \cite{VanCuyck:2016fab,VanCuyck:2017wfn}.

Models agree on the crucial role of  multinucleon processes in the modeling of the 
MiniBooNE, T2K and MINERvA cross-section data. 
Nevertheless there are some differences in the results obtained for this np-nh channel by the different theoretical approaches. 
In the following, we will review the current theoretical status on this subject.

The np-nh channel is taken into account in a phenomenological approach by Lalakulich, Mosel~\textit{et al.} 
\cite{Lalakulich:2012ac,Lalakulich:2012hs,Mosel:2014lja,Gallmeister:2016dnq} 
in GiBUU and by Bodek~\textit{et al.}~\cite{Bodek:2011ps} in the so called Transverse Enhancement Model (TEM). Other recent
 interesting calculations discussing the 2p-2h effects in connection with the neutrino scattering include \cite{Lovato:2014eva,Lovato:2015qka,Benhar:2015ula,Rocco:2015cil,VanCuyck:2016fab}. 
The most complete  theoretical calculations of np-nh excitations contributions to neutrino-nucleus cross sections have been  performed by 
Martini~\textit{et al.}~\cite{Martini:2009uj,Martini:2010ex,Martini:2011wp,Martini:2012fa,Martini:2012uc,Martini:2013sha,Martini:2014dqa,Ericson:2015cva,Martini:2016eec}, 
 Nieves~\textit{et al.}~\cite{Nieves:2011pp,Nieves:2011yp,Nieves:2012yz,Nieves:2013fr,Gran:2013kda}
and  Amaro~\textit{et al.}~\cite{Amaro:2010sd,Amaro:2011qb,Amaro:2011aa,Simo:2014wka,Simo:2014esa,Megias:2014qva,Ivanov:2015aya,Simo:2016ikv,RuizSimo:2016ikw,Megias:2016fjk}; these authors include comparisons with flux-folded data.

All  models for multinucleon processes build on an underlying description of the pure QE process, and add additional mechanisms.  It is important to realize that model-dependencies and a scheme-dependent separation between genuine 1p1h and np-nh effects are  nearly impossible to avoid and  difficult to discern.

In the np-nh sector, several contributions to two-body currents are active
\cite{Alberico:1983zg,DePace:2003xu,Nieves:2011pp,Simo:2016ikv}. 
In the electromagnetic case, the pion-in-flight term $J^{\mu}_{\pi}$,
the contact term $J^{\mu}_{\textrm{contact}}$ and the $\Delta$-intermediate state or $\Delta$-MEC term $J^{\mu}_{\Delta}$ contribute. In the weak-interaction case, in addition the pion-pole term $J^{\mu}_{\textrm{pole}}$ appears. 
It has only an axial component and is therefore  absent in the electromagnetic case.\\
If  in the 1p-1h sector a basis of uncorrelated independent nucleons is used, as in the Fermi-gas or, to some extent, also in mean-field based models,
one needs to consider  the nucleon-nucleon (NN) correlation contributions
since the protons and the neutrons in the nucleus are correlated, and 
 correlated pairs act as a unique entity in the nuclear response to an external field.
In independent-particle models,
 NN correlations are included by considering an additional two-body current,
the correlation current $J^{\mu}_{\textrm{NN-corr}}$. 
Detailed calculations and results for these NN correlation current contributions are given for example 
in Refs.~\cite{Alberico:1983zg,Alberico:1990fc,Amaro:2010iu,VanCuyck:2016fab}.
On the other hand, in approaches as in the one of Lovato~\textit{et al.}~\cite{Lovato:2014eva,Lovato:2015qka} 
the NN correlations are included in the description of the nuclear wave functions. 
With the introduction of the NN correlation contributions,
also the NN correlations-MEC interference contribution to the 2p-2h excitations naturally appears.
In the correlated-basis approach, these contributions are referred as one nucleon-two nucleon currents interference.

It is important to stress that even in this simple model exact  calculations are difficult, for several reasons. The first difficulty is that one needs to perform 7-dimensional integrals for a huge number of 2p-2h response Feynman diagrams. 
Second, divergences in the NN correlations sector and in the angular distribution of the ejected nucleons
may appear and need to be regularized.
Furthermore, the neutrino cross section calculations should
be performed for all the kinematics compatible with the experimental neutrino flux. For these reasons an exact  calculation is computationally very demanding,
and as a consequence different approximation schemes are employed by the different groups. 
The selection of the subset of diagrams and terms that are calculated also presents important differences. 
In this connection Amaro \textit{et al.} only explicitly add the MEC contributions but not the NN correlations-MEC interference terms
(these last terms were analyzed for electron scattering in Ref.\cite{Amaro:2010iu})
to the genuine quasi-elastic interaction.  
MEC contributions, NN correlations and NN correlations-MEC interference are present both in Martini \textit{et al.} and Nieves \textit{et al.} 
Martini~\textit{et al.} only consider the $\Delta$-MEC as this constitutes  the dominant contribution.
The treatment of Amaro~\textit{et al.} is fully relativistic as well as the one of Nieves~\textit{et al.} 
(even if the non pionic $\Delta$ decay contribution of $\Delta$-MEC are taken from
the non-relativistic work~\cite{Oset:1987re}, as in the case of Martini~\textit{et al.}) 
while the results of Martini~\textit{et al.} are based on a non-relativistic reduction of the two-body currents.
With the recent results of Refs.~\cite{Simo:2016ikv,Megias:2016fjk}, one of the major differences between the results of Amaro~\textit{et al.} on one hand and Martini~\textit{et al.} and Nieves~\textit{et al.}
on the other hand, related to the presence of 2p-2h contributions in the axial and  vector-axial interference term (and as a consequence,
on the relative role of 2p-2h contributions for neutrinos and antineutrinos) has disappeared. 
The major differences that still remain, 
are related to the treatment of  NN correlations and various interferences as NN correlations-MEC interference terms and direct, exchange and direct-exchange interferences.
\\

\subsection{Experimental situation: results on nucleon, MINERvA, MiniBooNE, T2K}
    Neutrino  interactions on free nucleons have been studied at the first bubble experiments in the seventies \cite{Barish:1977qk,Kitagaki:1990vs}. To date, these experiments provide  the only single-nucleon experimental data available. The statistical power of these experiments is very limited and the data cannot  easily be reanalyzed using modern techniques and interaction models. 
The difficulties to design and operate hydrogen and deuterium target detectors might compromise our understanding of neutrino CCQE scattering in the future. Recent studies \cite{Lu:2015hea} propose the use of transverse variables to isolate single-nucleon interactions in neutrino and antineutrino interactions on nuclei, but the ability to control heavy-nucleus backgrounds needs to be addressed. 

 Current CCQE experiments use heavy nuclei (usually hydrocarbonate, water or argon) as target material. First experimental analyses with heavy nuclei seemed to point to  $M_A$ values that were larger than the ones obtained with the hydrogen and deuterium interactions around the expected value 1.0~GeV. Further comparison with more advanced theoretical models however  showed that the addition of nuclear correlations was able to explain some of the higher $M_A$ results.

Even the most recent experiments such as T2K, \minerva and \nova have difficulties to  provide definitive answers to the modeling issues for  CCQE reactions. 
The data-simulation disagreement that have been reported by a number of
collaborations~\cite{Gran:2006jn,AlcarazAunion:2009ku,Adamson:2014pgc,AguilarArevalo:2007ab,AguilarArevalo:2010zc,AguilarArevalo:2010cx,AguilarArevalo:2013hm,Aguilar-Arevalo:2013nkf,Abe:2014iza,Abe:2015oar,Abe:2016tmq,Fields:2013zhk,Fiorentini:2013ezn} show three main features: data are suppressed at low $Q^2$, data seem enhanced at high $Q^2$,
and the overall data normalization tends to be higher than simulations. 
The community agrees that nuclear correlations and
 meson-exchange currents
are essential to explain the neutrino QE data, and 
state-of-the-art theoretical models including correlations and multinucleon mechanisms can qualitatively explain
lepton kinematics of QE-like data from MINERvA, MiniBooNE, and T2K.

However, at this moment the community is not successful in verifying proposed models in a quantitative sense.
Most notably, the global fit performed by T2K shows very poor results~\cite{Wilkinson:2016wmz}.
This is mainly due to two reasons: MiniBooNE data do not come with a full covariant matrix,
and there are no realistic systematic errors implemented in the proposed new models.
A successful global fit is  essential for the validation of new models and understanding of the data,
but it requires agreement on more standardized procedures within the community, such as  uniform data-reporting methods by experimentalists,
and inclusion of systematic errors in models by theorists. 

\begin{figure}[tbp]
 \begin{center}
\includegraphics[scale=0.75]{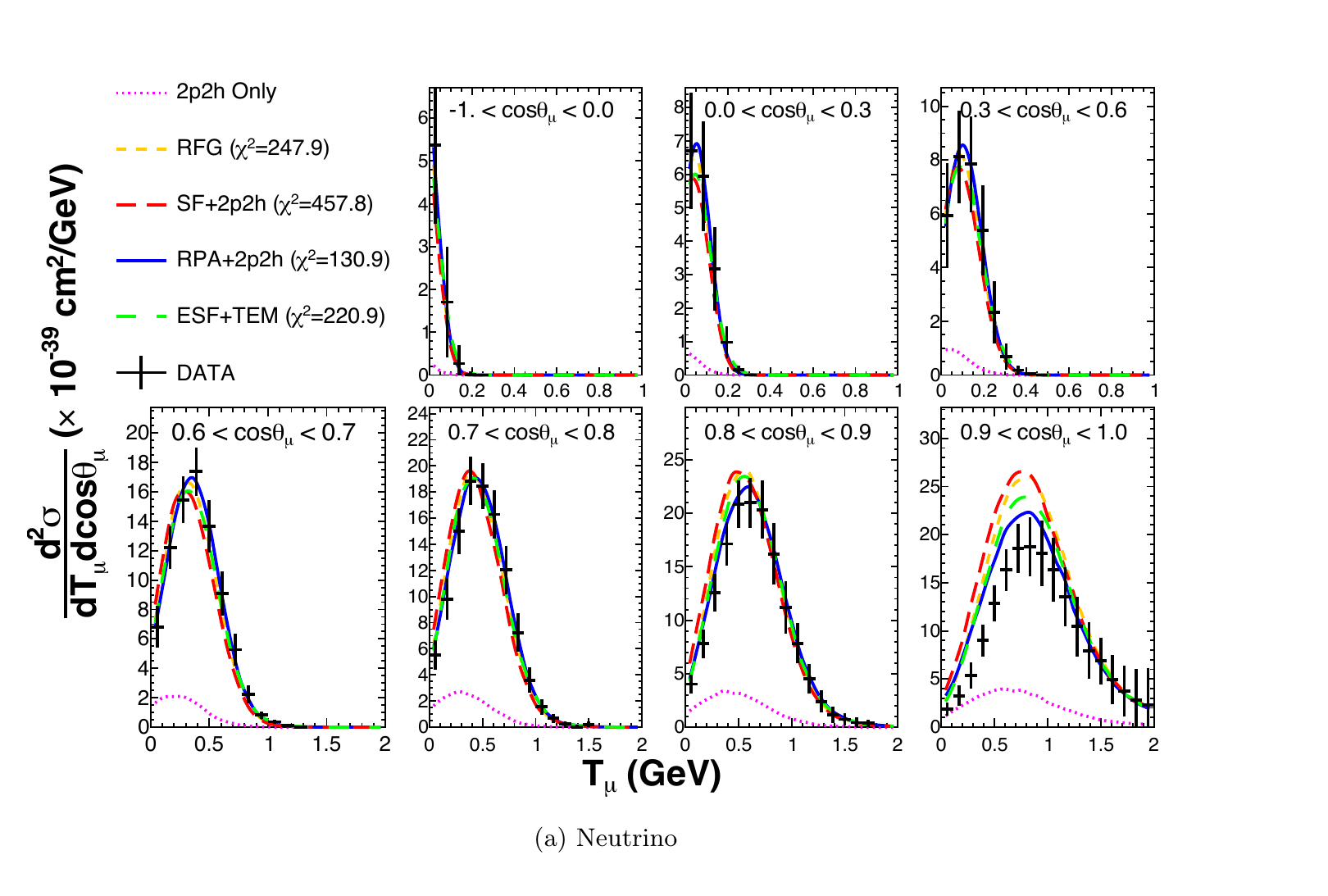}
\end{center}
\caption{\label{fig:MiniBooNE} Neutrino charged current quasi-elastic double differential cross section as a function of muon angle and kinetic energy measured by MiniBooNE\cite{AguilarArevalo:2013hm} Results are compared to several models with different ingredients \cite{Wilkinson:2016wmz}}
\end{figure}

\begin{figure}[tbp]
\begin{center}
\includegraphics[scale=0.07]{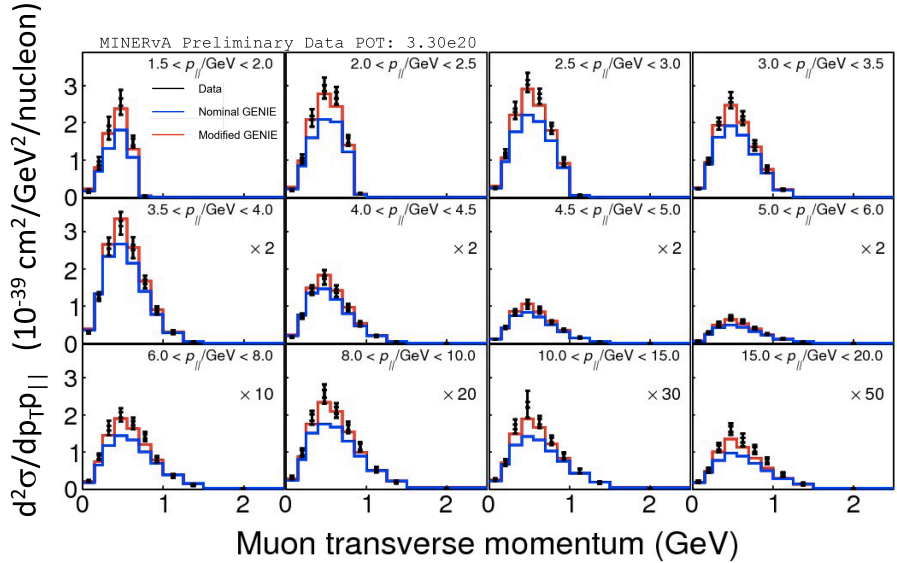}
\end{center}
\caption{\label{fig:Minerva} Double differential cross section as a function of transverse $p_T$ and longitudinal $p_L$ muon momentum for neutrinos measured by MINERvA \cite{Minervaptpl}. The predictions from nominal GENIE MC (blue) without RPA and 2p2h and a modified version (red) with RPA and an enhanced 2p2h based on data from the MINERvA are shown.}
\end{figure}

Modern experiments with high statistics have measured double differential cross sections; MiniBooNE experiment measured for the first time neutrino charged current quasi-elastic double differential cross section as a function of muon angle and kinetic energy. Fig.\ref{fig:MiniBooNE} shows the comparison of MiniBooNE data and predictions that includes 2p2h and RPA, MiniBooNE data agrees with predictions\cite{Wilkinson:2016wmz}. MINERvA has measured double differential cross section as a function of transverse $p_T$ and longitudinal $p_L$ muon momentum for neutrinos \cite{Minervaptpl}, where $p_T$ is correlated with four momentum transfer $Q^2$ and $p_L$ is correlated with the neutrino energy $E_{\nu}$. Fig.\ref{fig:Minerva} shows the $d^2\sigma /d p_T d p_L$ cross section of neutrino charged current quasi-elastic for muons with a momenta of $1.5 < p_L < 20.0$~GeV and $0 < p_T < 2.5$~GeV for muons with an angle less than 20 degrees. MINERvA data agrees with simulations that include multinuclear process.  T2K measures the double differential cross-section  as function of the muon momentum and angle ($d\sigma / d p_{\mu} d \cos{\theta_{\mu}}$) for charge current events with no pions in the final state \cite{Abe:2016tmq} and compare the data to Martini \cite{Martini:2009uj} and Nieves \cite{Nieves:2012yz} models.

\begin{figure}[tbp]
\begin{center}
\includegraphics[scale=0.75]{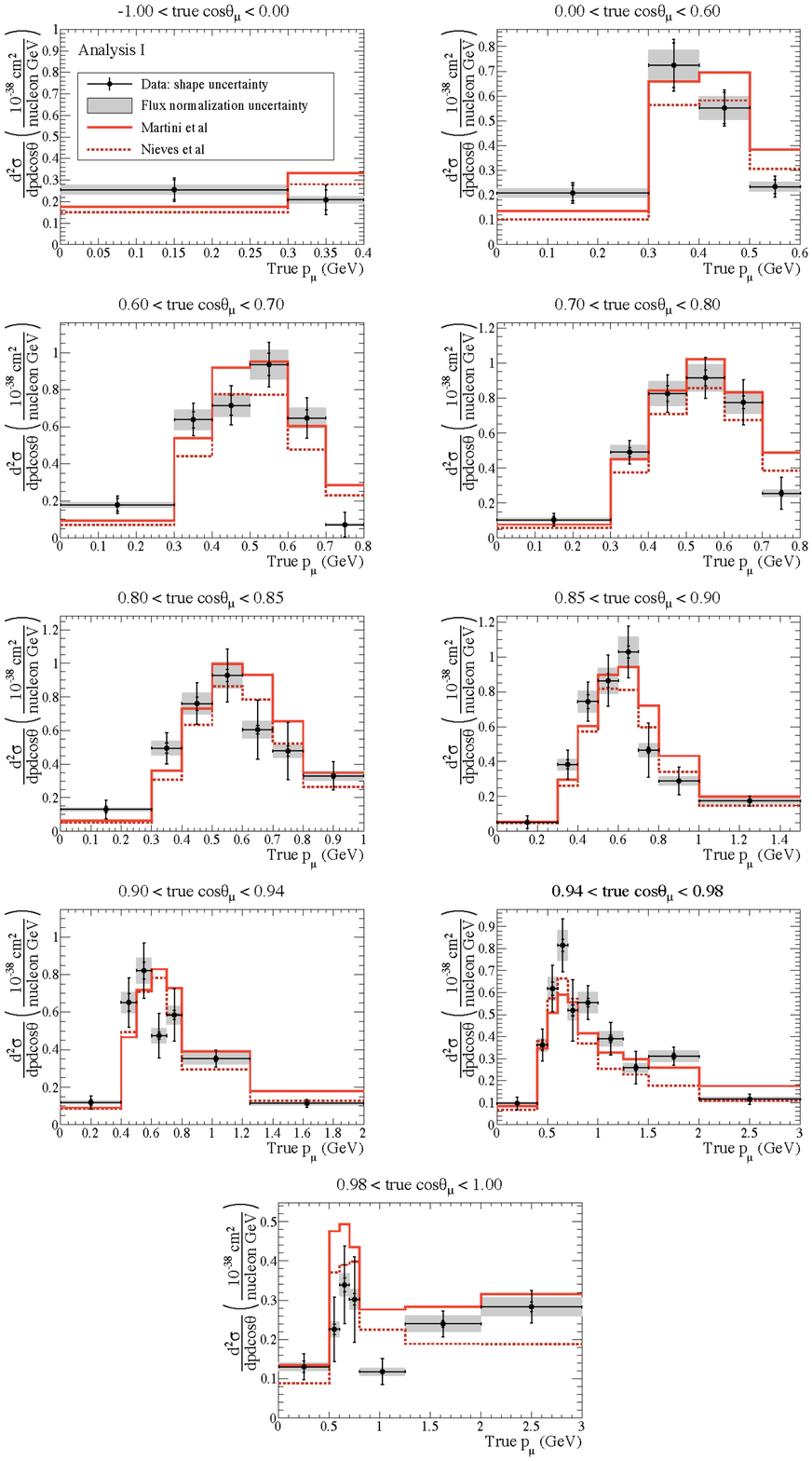}
\end{center}
 \caption{\label{fig:T2K} Measured double differential crosssection Charge Current with no pions as a function of $\cos\theta_{\mu}$ and longitudinal $p_{\mu}$ muon momentum for neutrinos measured by T2K\cite{Abe:2016tmq}. The predictions are from the Martini \textit{et al.} cite{Martini:2009uj} and Nieves \textit{et al.} models \cite{Nieves:2012yz}.}
\end{figure}

Near future data on the same distributions are unlikely to improve on the model selection,
unless new data dramatically reduce systematic errors, which is unlikely
with our current knowledge of the neutrino flux and background models. 
Therefore, there is a growing interest in  kinematics measurements of the hadronic system.
In the past, muon-proton kinematics were used to select the QE sample in experiments, and  
details of the hadronic information  were not published in a format useful for comparison with models~\cite{Gran:2006jn,Lyubushkin:2008pe}.
T2K published CCQE total cross sections for the one- and two-track sample separately~\cite{Abe:2015oar}.
The disagreements between them may be  key to the understanding of  the hadronic system. 
The majority of hadronic studies is provided by the MINERvA collaboration.
MINERvA utilized  vertex activity to identify extra hadronic energy deposits, which indicates
the presence of extra protons stemming from multinucleon interactions~\cite{Fields:2013zhk,Fiorentini:2013ezn}.
For the first time MINERvA also tested $Q^2$ reconstruction using lepton versus hadron kinematics~\cite{Walton:2014esl}.

\begin{figure}[tbp]
\begin{center}
\includegraphics[scale=0.75]{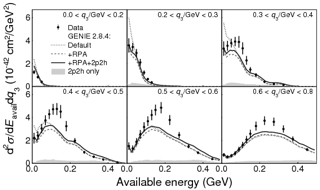}
\end{center}
\caption{\label{fig:MinervaAvailableEnergy}  Double-differential cross section $d\sigma / d E_{available} dq_3$ in six regions of $q_3$ as a function of available energy measured by MINERvA \cite{Rodrigues:2015hik}.}
\end{figure}

Recently, MINERvA performed the once-thought-impossible energy-momentum transfer ($\omega$ and $|{\bf q}|$)
reconstruction by using the measurement of the  hadronic energy deposit~\cite{Rodrigues:2015hik},
which shows that current 2p-2h models are  not able to describe the MINERvA data. 
Figure \ref{fig:MinervaAvailableEnergy} shows the double-differential cross section $d\sigma / d E dq_3$ in six regions of $q_3$ as a function of available energy. The available energy
quantity is a metric for the visible energy in the MINERvA detector and is the sum of kinetic energies of proton and charged pions, and total energy of photons and elections. 
ArgoNeuT performed the first two-proton final-state CC measurement~\cite{Acciarri:2014gev}.
This is  analogous to the triple coincidence measurement at JLab's Hall A~\cite{Shneor:2007tu},
which identifies short-range correlations (SRC) by reconstructing back-to-back protons in the initial state.
 ArgoNeuT's low statistics do not make it possible to arrive at any final conclusions,
but  high statistics data from new LArTPC experiments, such as Fermilab's short baseline SBN program~\cite{Antonello:2015lea} are expected. 
These measurements are all interesting because the hadron system provides such a rich source of information.  However, extracting information of the primary weak interaction is complicated by the presence of final stat interactions (FSI), which also contribute significantly to observed final state particles and their kinematics

Another type of constraint is provided by electron kinematics from $\nu_e$CC interactions.
These measurements are essential to reduce uncertainties in the knowledge of $\nu_e/\nu_{\mu}$ ratio  error,
as most  cross-section model studies are done with muon neutrinos.
 Still, systematic errors on $\nu_e$CC cross sections need to be reduced for  $\delta_{CP}$ studies. 
At this moment, MINERvA is the only experiment that has published  $\nu_e$CCQE-like cross section data~\cite{Wolcott:2015hda}. 
QE-like interactions dominate $\nu_e$CC inclusive samples
from T2K and agree with simulations including 2p-2h contributions~\cite{Abe:2014agb}. 
However, the errors on the data  ares too large to make any conclusions. 

These new experimental approaches address the issues mentioned above from two independent directions. On the one hand, the construction of experiments sensitive to the low-energy hadrons produced in the interactions, this is done with high resolution tracker detectors or Time Projection Chambers. Recent results from ArgoNeuT \cite{Weinstein:2016inx} show the potential of this detector concept. On the other hand,  increasing the statistics and improving on the hermeticity of the detector \cite{Abe:2016tii} and the particle detection acceptance,  reduces our dependence on limited Monte Carlo models. All these developments will profit from improved models and more accurate Monte Carlo simulations. Using various target nuclei, including hydrogen and deuterium, will help to factorize nucleon cross-sections and nuclear uncertainties such as Fermi momentum or final state interactions. Adding the capability to change the neutrino energy-distributions will help to reduce the degeneracies caused by the convolution of the neutrino flux with the cross-sections. \\
%{\bf Contributors : Teppei Katori, Federico Sanchez}

 \subsection{Relation theory-experiment} 
At this moment, the experimental and theoretical communities agree to meet
at the `flux-integrated differential cross section'~\cite{AguilarArevalo:2010zc}. 
\begin{equation}
\left(\frac{d\sigma}{dX}\right)_i=\frac{\sum_jU_{ij}(d_j-b_j)}{\Phi\cdot T\cdot\epsilon_i\cdot\Delta X_i}~.
\label{eq:dexp}
\end{equation}
Here $d_j$ is the data vector as a function of a direct observable $X$,
$b_j$ represents the background to be subtracted,
$U_{ij}$ represents the unsmearing matrix, $\epsilon_i$ is the efficiency, $\Phi$ the total integrated flux,
$T$ the total target number, and $\Delta X_i$  the bin width of the $i^{th}$ bin.
 Eq.~(\ref{eq:dexp}) is  symbolic rather than exact,
because there are several ways to remove backgrounds and unsmear the distributions, 
but it describes all elements we need to calculate the flux-integrated differential cross section.  
In the neutrino scattering community, experimentalists measure them, and theorists calculate them to allow for  comparison \cite{Katori:2016yel}.

This constitutes  a major difference with the situation in  electron scattering,
where the beam energy is precisely known and all kinematics ($\omega$ and $|{\bf q}|$) can be fixed for a given interaction. 
In the neutrino interaction physics community, these choices had to  be made : in neutrino experiments with their broad incoming neutrino-energy distribution, the
incoming neutrino energy is not known, and this prevents full kinematics reconstruction. 
Most notably, neutrino energy reconstruction, and hence flux unfolding is possible only in model-dependent ways.  Moreover,
  the primary neutrino-nucleon interaction can be determined  only in a model-dependent way : data samples need to be defined by topology, such as '1 muon + 0 pion + N protons' (CC0$\pi$), and
not by the primary neutrino interactions mechanism, as e.g. CCQE. 

This hampers theorists, because models for CCQE are not directly comparable with CC0$\pi$,
unless they include all processes which contribute to the CC0$\pi$ topology (2p2h, pion production and reabsorption, ...).
Second, the differential cross sections are function of direct observables,
such as muon kinematics, hadron energy, etc., and not inferred variables,
such as neutrino energy, energy and momentum transfer, etc.
This is a further complication for theoretical models, because 
direct observables need to be determined by integrating over the neutrino flux and including all relevant processes.

\subsection{ What can be obtained from electron scattering ?} 
Valuable information can be extracted from many high quality electron scattering data, which should be used to validate models before these are applied to the study of neutrino scattering. 
Experiments on inclusive electron scattering off a wide variety of nuclei, from $^3$He to $^{208}$Pb, were performed at several facilities, including Bates, JLab, Saclay and SLAC.
Some data are also available on the separated longitudinal and transverse response functions, $R_L$ and $R_T$, obtained through the Rosenbluth separation. 

In $(e,e')$ electron scattering experiments the beam energy, unlike in the case of neutrino scattering, is precisely known. This allows one to determine the energy transferred to the nucleus from the kinematics of the outgoing electron, and hence helps to identify the corresponding reaction mechanisms. In particular, $\omega=\sqrt{q^2+m_N^2}-m_N$ corresponds roughly to the center of the quasielastic peak, $\omega=\sqrt{q^2+m_\Delta^2}-m_N$ to the $\Delta$-resonce peak, and the region between the two peaks to two body excitations. When the beam energy is not too high these regions are clearly separated in the data, therefore allowing for a test of theoretical models for each specific process. At high energy different regions tend to overlap: in this case only the comparison with a complete model, including all different mechanisms from QE to DIS, is meaningful.

In the quasielastic regime, the scaling properties displayed by electron scattering data can be used to constrain nuclear models. 
Scaling consists in the fact that the reduced inclusive cross section, which a priori is a function of two independent variables ({\it e.g.}, $q$ and $\omega$), actually depends on a single variable $y(q,\omega)$ or, alternatively, $\psi(q,\omega)$. 
This property is known as scaling of the first kind.
A second kind of scaling concerns the independence of the scaling function on the specific nucleus. The simultaneous occurrence of both kinds of scaling is called Superscaling.
At sufficiently high energies good first-kind scaling is observed at excitation energies below the QE peak, namely, in the so-called scaling region. 
At energies above the peak, where nucleon resonances (especially the $\Delta$) are important, scaling is broken for the total reduced cross section. On
the other hand, from longitudinal/transverse separated data, it is known that these scaling violations reside in the transverse response, but not in the longitudinal. The latter appears to superscale. 
Scaling violations in the transverse channel are due in part to the contribution of reaction mechanisms different from one-nucleon knockout, such as the excitation of nucleon resonances, as well as non-impulsive mechanisms, such as two-body excitation induced by two-body meson-exchange currents. Therefore, the scaling violations observed in $(e,e')$ data carry important information on how the dynamics go beyond the simple IA.

The phenomenological longitudinal scaling function extracted from the data displays two main features, which represent strong constraints for theoretical calculations.
First, its value at the maximum ($\sim$0.6) is much lower than the one predicted by the RFG (0.75). Second, it exhibits an asymmetric shape, with a sizeable high energy tail, not present in the RFG model nor in most models based on the IA. This region has been shown to be particularly sensitive to the treatment of FSI and to relativistic effects. A simple test against these two properties can give insight on the reliability of theoretical models.
Alternatively, the phenomenological scaling function can be multiplied by the elementary weak vertex in order to predict neutrino-nucleus cross sections. This is the basic idea underlying the superscaling (SuSA) approach.
Although it has been designed for the QE region, the scaling analysis can be extended to the resonance region and to the full inelastic spectrum, provided the corresponding elementary form factors are sufficiently well known.
It should be stressed that scaling arguments hold for not too low momentum and energy transfer (roughly, $q>$400 MeV/c and $\omega>$50 MeV). At lower energies the SuSA approach is bound to fail, as any other model based on the IA. In this region collective nuclear effects become important and alternative approaches, such as RPA, are expected to be more reliable.

Useful information can also be extracted from the analysis of the longitudinal/transverse separated electromagnetic responses. Unfortunately not many data of this type exist and some of them are still controversial. However, they are crucially important for the present purposes, since the balance between longitudinal and transverse responses is different in electron and neutrino scattering. 
The data indicate that the transverse scaling function is higher than the longitudinal one, a property not fulfilled by most IA models and certainly violated by the RFG, which predicts $f_L=f_T$. One of the few models able to reproduce this feature is the Relativistic Mean Field, where the enhancement of the transverse response emerges as a genuine relativistic effect.

In Section V the issues of inclusive versus semi-inclusive electroweak processes is briefly addressed. Much remains to be done on the theoretical side in modeling the latter, especially at the high energies where relativistic effects must be included. Initial studies indicate that detecting, for instance, nucleons in the final state (in addition to detecting the final-state charged lepton) might help in determining the energy of the incident neutrino in CC$\nu$ reactions. In particular, detecting a muon and one proton (semi-inclusive), while placing new demands on theory beyond those that are required in treating inclusive scattering (see Section V), holds promise in this regard. Detecting more than two hadrons ({\it i.e.,} being even more exclusive) may be too demanding for any detailed interpretation beyond basic calorimetry. In this vein, one possibility was discussed during the 2016 INT neutrino workshop: namely, perhaps one could ``mine'' data from the CLAS detector at JLab to test how well one might reconstruct the beam energy using only the final-state charged lepton (an electron in that case) together with final-state nucleons, but not using the incident electron's energy.

Finally we note that unpolarized electron scattering data only provide information on the vector response, although we remark in passing that the VA interference could in principle be tested through comparison with parity-violating electron scattering in the QE region and beyond.\\

\subsection{Generator Status}
       The experiments' need to simulate all particles involved in the neutrino-nucleus interaction has driven the implementation of  different models in the Monte Carlo generators which are discussed in more detail in Sec.~\ref{generators}. The typical model starts from the impulse approximation and then adds various nuclear effects.  The neutrino-nucleon interaction  follows a modified Lewellyn Smith \cite{LlewellynSmith:1971uhs} approximation with the usual banning of the scalar and tensor terms, the PCAC and CVC approaches, and the use of dipole form factors \cite{Smith:1972xh}. More elaborate Monte Carlos models replace the vector dipole by form factors extracted from electron scattering \cite{Bradford:2006yz}. The new form factors alter the high $Q^2$ region of the interaction. Under these assumptions, only the axial mass ($M_A$) related to the axial form factor is kept as a free parameter of the model. There are potential contributions to the single-nucleon final state such as those induced by two-body currents that are not taken into account in current models and Monte Carlos. 

The impulse approximation selects a target nucleon  in the nuclear potential with a Fermi momentum following some of the three traditional nuclear descriptions: Relativistic Fermi Gas (RFG), Local Fermi Gas (LFG) or Spectral Function (SF). These three models differ in the nucleon dispersion relation and  its variations depending on the position in the nucleus. Binding  is considered in two different ways : as a transition energy from initial to final nuclear states or as a change in the dispersion relation by associating a modified nucleon energy.  Both approaches are equivalent to the level of precision that was needed until now, and is driven by the needs of the model: Spectral functions provide the dispersion relation, LFG requires nuclear mass and RFG is 'a priori' independent of the chosen approach. In addition, all Monte Carlos include Pauli blocking by excluding interactions with final-state nucleons below the Fermi momentum.

Recent developments implement  long-range nuclear correlations as a quenching factor depending on $q^2$, derived from the RPA implementation of \cite{Nieves:2004wx,Gran:2013kda}. The $q^2$ shape of the quenching factor has been shown to be independent of the neutrino energy. The RPA model is based on a phenomenological nuclear potential that is being validated for vanishing $q^2$ values, but the model provides free parameters to adjust it to  neutrino-nucleus experimental results. 

 Due to the complexity of two-particle models, there are several implementations at the nucleon level: 

\begin{enumerate}

\item Modified 1p1h cross-section with modified Sachs magnetic form factor ($G_M$,p/n), to emphasize the transverse nature of two-particle production. \cite{Bodek:2011ps} .

\item Generated $(p_{\mu},\theta_{\mu},E_{\nu})$ lookup tables from the Valencia model \cite{Nieves:2011yp} 

\item Generated from the $(q_0,q_3)$ hadron tensors pre-computed by the Nieves model \cite{Gran:2013kda}. The advantage of this model is the compactness and the possibility of adding any model provided its hadron tensors can be computed. This and the previous model have limited validity for values of $q_3$ above ~2 GeV/c which will limit their precision for neutrino beams with high energy tails. 

\end{enumerate}

The first of these models wrongly assumes the 1p1h kinematics. This assumption alters the energy reconstruction based on lepton kinematics, but it might be a good approximation for calorimetric energy reconstruction. The second and third model do not have a prediction for the final state kinematics. This is normally circumvented by generating final states under certain reasonable assumptions \cite{Golan:2012wx}: back-to-back initial state nucleons with momentum generated up to the Fermi level and no correlation between initial and final state hadron directions. In all cases, the Pauli blocking algorithm is applied to all final-state interactions.
There are more effects, such as initial and final state bremsstrahlung and the Coulomb potential that might have a large impact on the cross-section modeling and are critical for  $\nu_e$ scattering. These effects are currently ignored in available Monte Carlo models. 

Particles generated during the interaction are propagated through the nucleus following the different generator implementations of final-state interactions. 
Many of these models have been derived from electron scattering data, but the  implementation departed from the original theoretical models. The need of a MC implementation able to describe both electron and neutrino scattering has been acknowledged by the various Monte Carlo generator teams, but we are still far from being able to perform detailed comparisons. \\

\subsection{Challenges and open questions}
 After the acceptance of two-body currents as  relevant contribution to the CCQE cross section,  several issues  still remain. The most urgent one is that of agreement between different models, and  between models and experiments. 
Theoretical results need to be compared in a systematic way to all available data, and validated against electron-scattering data. The various assumptions and differences in models that lead to discrepancies need to be understood.  
This  would be of great help in assessing the range of validity of each approach and facilitate the incorporation of more detailed models in generators. 
\begin{itemize}
 \item
 The complete implementation of multinucleon phenomena in the generators is still pending. The adoption of the hadron-tensor approach simplifies both the numerical calculation and the adoption of several models in the same generator. These models integrate the hadron production that needs to be estimated later during the event generation, losing potential correlations between the lepton and the hadron currents. This must be carried out in parallel with experiments able to resolve different final states and with the proper theoretical developments. These developments might be more critical for calorimetric experiments such as \nova or DUNE but the understanding of the energy-scale biases between a pure calorimetric and a kinematic reconstruction typical of Cerenkov detectors, needs to be worked out. 
 \item
 The multinucleon discussion has also uncovered several deficits in the description of  1p1h interactions in the nucleus. Long-range correlations, Fermi momentum description, binding energy corrections, etc... need to be revisited in a manner free of parameters, and uncertainties need to be identified. One of the main problems faced by experimentalists is the rigidity of the theoretical models. The identification of  parameter uncertainties and the errors  they bring along, might help to improve our understanding of CCQE interactions. 
 \item
 There is  an ongoing discussion about the superposition of  microscopic two-body  models and the presence of initial nucleon correlations that appear in spectral functions. 
 \item
  The validity of the Smith-Moniz parameterization for the single-nucleon interaction and the modeling of the axial vector form factor as a simple dipole needs to be revisited both theoretically and experimentally. 
\item
  The development of a 'universal' model able to cover all experimental needs from 200 MeV to 10s of GeVs is an open issue. None of the  theories currently in use cover this vast energy region, models to match and fill the gaps between different predictions need to be developed.
\item
 A better quantitative evaluation of the differences in cross-section between muon and electron neutrinos will be very relevant for CP violation measurements in future experiments.  Some initial studies are available but are just starting.  Experimentalists  need to identify facilities and techniques to measure these cross-sections over a broad energy range. 
\item
  Experiments need to identify  measurements able to identify and reduce  theoretical uncertainties. Several examples based on transverse variables have been proposed, but other measurements based on very forward scattering ($q^2 \approx 0$) might better control certain nuclear uncertainties.
\item
From a purely theoretical view, the modeling of outgoing hadrons and hadronic final-state interactions 
is an issue that needs increased efforts.
\item
Moreover, the understanding of theoretical predictions and discrepancies among them would benefit from a more careful treatment of  interferences between various nuclear effects and a meticulous study of double counting hazards.  It is important  to identify model-dependences and basis-dependent separations between different approaches.
\end{itemize}
   
\newpage

\newpage

%\section{Resonance Model}

\section{Resonance Model}
%- Topic Contributors: Mohammad Sajjad Athar, {\bf Steve Dytman}, Juan Nieves, David Richards, {\bf Toru Sato}, Jan Sobczyk, Geralyn Zeller}
\label{1pi}

\subsection{Introduction and Motivation}
%    1)  Introduction and motivation for studying the channel
%         Include impact on Oscillation Physics (briefly and referring to the dedicated section of the white paper)

The resonance　(RES) region is typically characterized by invariant masses $W < 2$ GeV with broad nucleon excitations on a smooth background.  Here, the effective degrees of freedom are chosen to be baryons and mesons; a nucleon has a transition to an excited state (N* or $\Delta$) and its main decay mode is emission of one or more pions.
%pions, kaons, and photons are emitted when it quickly decays.  
At higher $W$ and momentum transfers $Q^2>1$ (GeV/c)$^2$, the reaction is described by the interaction with partons described in the next section on SIS and DIS interactions.  
%These latter methods are also used to calculate an average cross section at lower $W$.  
Data from all probes at low $W$ has a strong contribution from the lowest resonant state $\Delta_{33}(1232)$ with a smaller contribution from nonresonant processes; these data are widely described as a resonance dominant process.  For the single pion production, nonresonant Born terms remain important up to $W\approx$1.6 GeV.

Resonance excitation is a large part of the response for neutrinos of energy in the range 0.5-3 GeV.  For neutrinos in the lower part of this range, pions are mainly produced through the $\Delta_{33}(1232)$ resonance, especially for CC $\pi^+$; this process is an important background to the QE process when the pion is not detected.  At higher neutrino energies, higher mass resonances $P_{11}(1440), S_{11}(1535)$ and $D_{13}(1520)$ in the second resonance region become important and resonance production is the dominant process.  Although decay to single pion is most important, resonances also decay with emission of multiple pions, kaons, and photons.  Resonances are excited by many probes - electromagnetic, hadronic, and weak.  Since data for weak probes contain larger error bars, progress to date has leaned heavily on electromagnetic and hadronic work.  
% Resonances decay by emitting a nucleon with pions, kaons, photons; in addition, there are competing processes for all final states that come from nonresonant mechanisms.

Lack of quality data for pion production by weak probes from nucleon targets remains a critical hindrance for development.  Two older experiments (BNL~\cite{Kitagaki:1986ct} and ANL~\cite{Radecky:1981fn}) from 1980's bubble chambers have generated a lot of controversy due to normalization differences of $\sim$ 30\% and a new experiment is the most proper way to settle this controversy. The results of all these experiments are based on hundreds of events.  This makes any separation into individual resonances and nonresonant
amplitudes problematic. 

Two neutrino experiments (\mb~\cite{AguilarArevalo:2010bm} and \minerva~\cite{Eberly:2014mra,McGivern:2016bwh}) have published high quality data for pion production from light nuclear targets - $CH_2$ and $CH$, respectively.  Although the best theoretical calculations have been unable to reproduce the \mb data, the models in event generators have more success.  All calculations are based on nucleon pion production and pion final state interaction (FSI) models based on previous data.  Newer \minerva data have features similar to the \mb data, but event generators are unable to  reproduce simultaneously the magnitude of both data sets.

All these issues with pion production must be handled in any oscillation experiment.  Typically, systematic errors are increased to account for any discrepancies and the problems cited above will have significant contributions.  Complications are magnified because oscillation experiments must use nuclei as detector materials, e.g $CH$, $H_2O$, and $Ar$, and nuclear models are then required.

Data sets for kaon production are much smaller than for pion production and theoretical models are less well developed.  Kaon production is important for proton decay experiments.  There is no data for single photon production from nucleon or nuclear targets.  Calculations are based on diagrammatic approaches~\cite{Wang:2013wva}.  This turns out to be a major source of uncertainty for ($\nu_\mu \rightarrow \nu_e$) oscillation experiments \cut{such as \mb}.

\subsection{Resonance production from the nucleon - Theory}
\label{se:piN}
 
Resonance processes are characterized by form factors which describe internal dynamics.  Both vector (related by isospin symmetry to the electromagnetic interaction) and axial vector form factors are important.  As a probe of the  nucleon axial vector response, the neutrino reaction is unique in hadron physics.  In RES interactions, the interesting quantities are the nucleon to resonance axial transition form factors, e.g. $N \rightarrow N^*,\Delta$. 
The dominant axial transition couplings at low $Q^2$ can be reliably estimated thanks to the PCAC symmetry.  Deviations from  PCAC predictions~\cite{Gorringe:2002xx,bernard1995chiral}, are studied using chiral perturbation theory (ChPT).  While the original theory was a low energy Taylor expansion, modern calculations extend it into the resonance region and are applied to the vector and axial vector excitations of $\Delta_{33}(1232)$.  The axial $N\Delta$ transition form factors are calculated with relativistic ChPT at one-loop  using the $\delta$ expansion scheme\cite{Geng:2008bm} and heavy baryon ChPT\cite{Procura:2008ze}.  The theoretical challenge is to describe the weak pion production amplitude within ChPT going beyond the transition form factor approach.
%, e.g. lattice QCD calculations.  

\subsubsection{reaction models of pion production for nucleons}
\paragraph{Diagrammatic approaches}
To describe the meson production reaction, the original work of Rein and Sehgal~\cite{Rein:1980wg} guides the field.  It defines the amplitudes in a helicity basis and the resonances have the Breit-Wigner form.  Event generators typically use it with updated parameters to get reasonable agreement with most data sets.  In ~\cite{Lalakulich:2006sw}, the Breit-Wigner resonance amplitude uses the MAID analysis of pion electroproduction~\cite{Drechsel:2007if} (vector) and a modified dipole form (axial vector) for the $Q^2$ dependence of transition form factors.  
Other reaction models have been recently developed~\cite{Alvarez-Ruso:2014bla,Nakamura:2016cnn}.
Tree diagrams based on the chiral Lagrangian are included as non-resonant mechanisms in Refs.~\cite{Hernandez:2007qq,Hernandez:2010jf,Leitner:2008ue}.  Models must account for resonant and nonresonant processes which interfere.  These approaches provide a good description of existing data within a robust theoretical approach.  

Another scheme is the dynamical approach, where
the hadronic rescattering processes are taken into account by solving a
coupled channel equation for the $\Delta_{33}(1232)$ and higher resonances~\cite{Sato:2003rq,Nakamura:2015rta}.  In this approach, a unified treatment of all resonance production processes satisfying unitarity is provided.
Another method partially restores unitarity via Watson's theorem~\cite{Alvarez-Ruso:2015eva}, fitting  data in the $\Delta_{33}(1232)$ region.
%In the $\Delta(1232)$ region, unitarity requires that the
%phase of the weak pion production amplitude is given by that of pion-nucleon elastic scattering.
With a detailed comparison between the dynamical approach and Olsson's implementation of unitarity it  will be interesting to see the consequences of unitarity on the axial vector coupling constant.  In the resonance region above the $\Delta_{33}(1232)$, it becomes non-trivial to satisfy unitarity because new two-body and three-body meson-baryon channels are open.  

Implementing three-body $\pi\pi N$ unitarity is technically difficult.  This is achieved in Ref. \cite{Nakamura:2015rta} by solving coupled channel equations.  Since this is the only model describing neutrino reactions in the higher resonance region, other approaches (e.g. J\"{u}lich-Bonn model~\cite{Ronchen:2015vfa,Ronchen:2012eg}) should also be applied. Since the dynamical model is numerically demanding, the work must now be transferred to the larger community for further development.

\paragraph{Lattice calculations}
Lattice calculations are performed in the Euclidean space, which naively might suggest that the calculation of scattering amplitudes is precluded.  However, lattice calculations are also performed in boxes of finite volume, thereby constraining particles to interact and indeed infinite-volume momentum-dependent scattering amplitudes can be directly computed in lattice QCD calculations performed at finite volume. The earliest formulation was applicable to elastic scattering~\cite{Luscher:1990ux}, but recently the formalism has been extended both to inelastic channels, and to three-body final states~\cite{Guo:2012hv,He:2005ey,Hansen:2012tf,Briceno:2012rv}, both relevant for pion production.  Thus far, most of the applications have been to the meson sector, or in an idealized regime where the $\Delta_{33}(1232)$ and $N \pi$ are comparable in mass~\cite{McNeile:2002fh}, but the first application of the method to pion production in the positive-parity Roper channel has now appeared~\cite{Lang:2016hnn}.  With the availability of increasingly powerful computational resources, and our increasing refinements of the method, further applications to meson production are within reach.

Calculations of the $N \Delta$ transition form factors in lattice QCD, and of the axial and vector form factors of the $\Delta_{33}(1232)$ have been an important theme of lattice calculations.  However, all the existing calculations treat the $\Delta_{33}(1232)$ as a stable, single-hadron state.  Recently, the formalism to rigorously compute the single to two-particle transition amplitudes has been developed where the transition is mediated through an external current, and in particular the electromagnetic and axial vector current. For example, the implementation to $N\pi$ production is illustrated\cite{Agadjanov:2014kha}.  The method has been demonstrated in the meson sector, notably for the $P$-wave $\pi\pi \rightarrow \pi \gamma^*$ transition revealing the enhancement due to the $\rho$ resonance, and the application to pion production from the nucleon is now computationally within reach.  

\subsubsection{testing reaction models}
\begin{figure}
\begin{center}
\includegraphics[width=7.5cm]{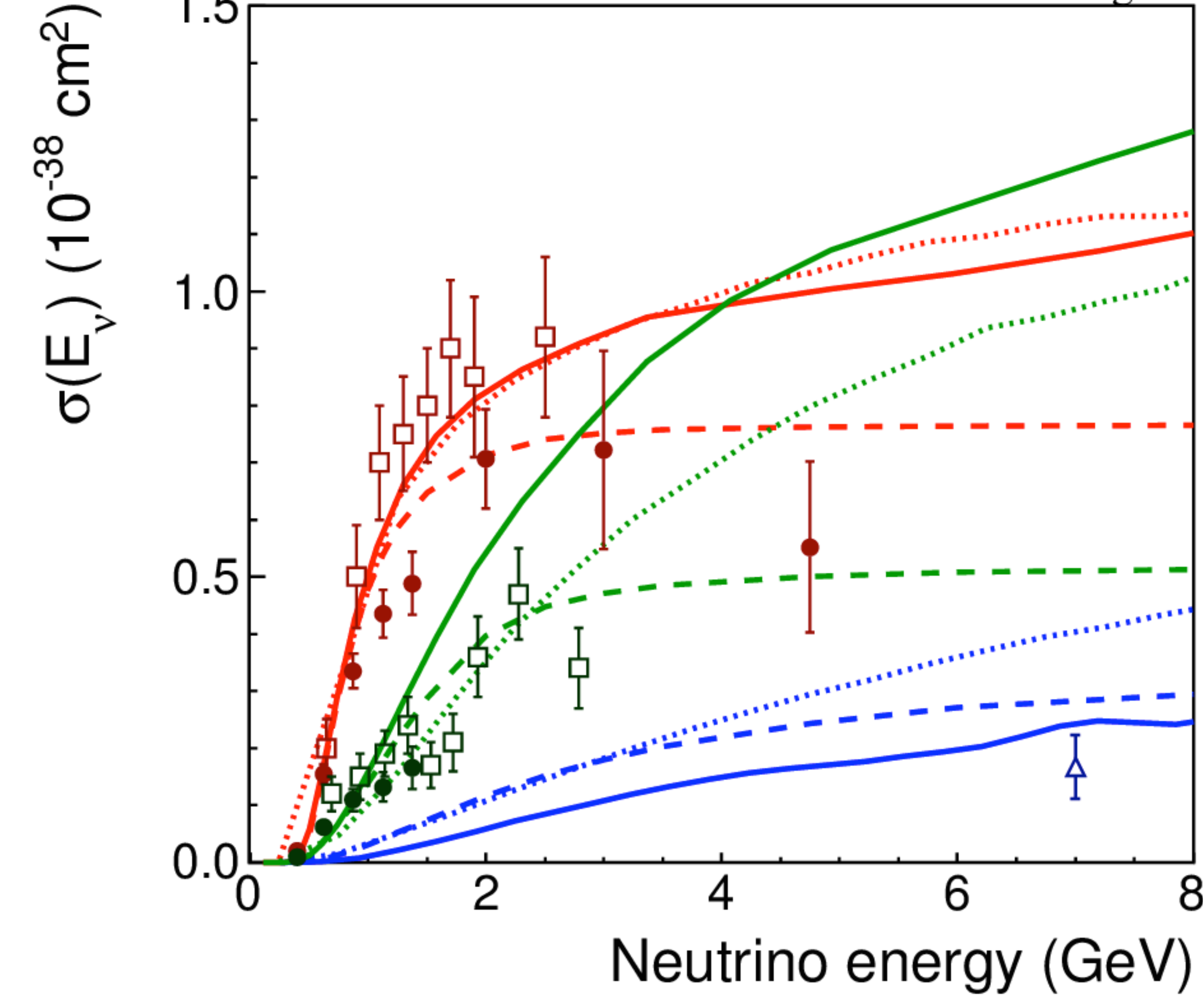}
\includegraphics[width=7.5cm]{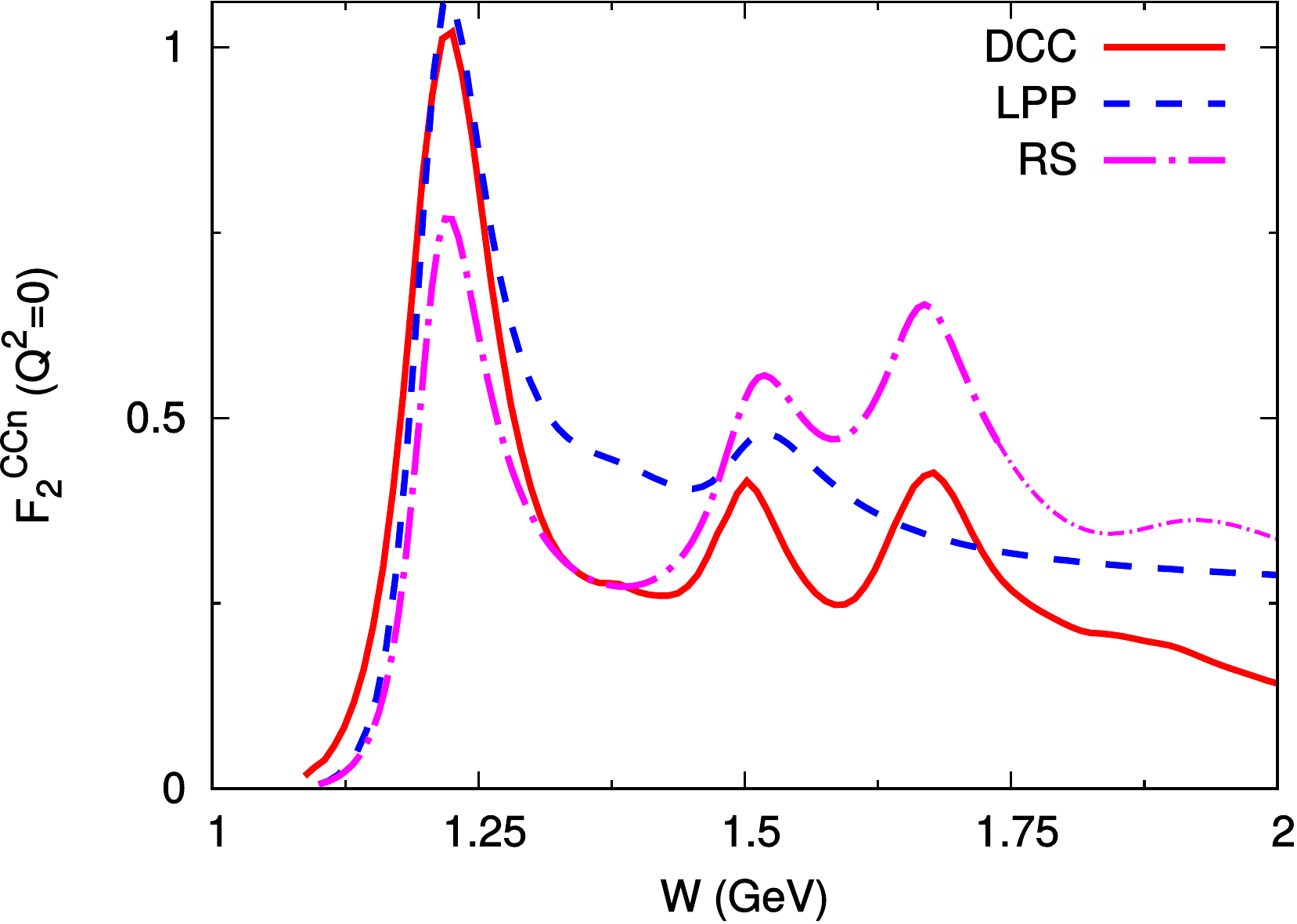}
\end{center}
\caption{\label{fig:pion-nucleon} (left) Various data with generator fits as of 2015 (P. Rodrigues, private communication). In general, $\nu_\mu p \rightarrow\mu^- \pi^+ p$ is red, $\nu_\mu n \rightarrow \mu^- \pi^+ n$ is green, and $\overline{\nu}_\mu p \rightarrow \mu^+ \pi^0 n$ is blue.  Data are shown as open squares for BNL~\cite{Kitagaki:1986ct}, closed circles for ANL~\cite{Radecky:1981fn}, and open triangle for SKAT~\cite{Grabosch:1988gw}.  Calculations are shown as solid lines (GENIE), dashed lines (NEUT), and dotted lines (NuWRO). (right) Comparison of theoretical calculations for $\nu_\mu n$ CC $1\pi$~\cite{Nakamura:2015rta}. RS is Rein-Sehgal~\cite{Rein:1980wg}; LPP is an isobar-model calculation~\cite{Lalakulich:2006sw}, and DCC is full coupled channel calculation~\cite{Nakamura:2015rta}.}
\end{figure}

Theoretical reaction models have to be confronted with  data.
A large data set for electron and pion induced reactions for resonance production is available. 
Most theoretical work is based on analyses of these data giving a description of the vector current induced meson production amplitudes.
At the limit of zero momentum transfer, one can relate the axial vector induced meson production
and the elastic pion  cross sections.  For neutrino reactions, most analyses use the ANL~\cite{Radecky:1981fn} and 
BNL~\cite{Kitagaki:1986ct} deuterium bubble chamber data with large accompanying systematic errors for $E_\nu <$ 8 GeV.  Fig.~\ref{fig:pion-nucleon} shows cross section data for three of the pion production channels for nucleon targets. 
Updated differential cross section data. e.g. $d\sigma/dW$ or $d\sigma/dQ^2$, for proton and neutron targets are strongly desired.  At the same time, theoretical work to extract pion production cross sections on the nucleon from the deuterium data by including nuclear effects~\cite{Wu:2014rga,Moreno:2015nsa} is underway.

\subsection{Resonance production from the nucleus - Theory}
%2)  Summary of the theoretical studies of this channel
%             Theoretical studies of this channel off the nucleon. 

%         Theoretical studies of this channel off the nucleus comparing various models with heavy             
%         referencing for details
Pion production in nuclei depends on models for the initial nuclear state,
the production of pions on a bound nucleon, and the interaction of the
pions and nucleons in the residual nucleus.  The 
initial nuclear state is usually approximated by a Fermi gas (FG)
of non-interacting nucleons, in its global~\cite{Kim:1996bt} and
local~\cite{Singh:1998ha} versions. More precise descriptions
based on realistic spectral functions~\cite{Benhar:2005dj},
bound-state wave functions~\cite{Praet:2008yn}, or RPA calculations~\cite{Martini:2009uj} have been also
developed. It is noteworthy that the integrated cross sections obtained with FG models are
very similar to those from sophisticated approaches. This is because
at the higher energy transfers present in resonance production, the
details of nuclear structure can be less relevant. 

Models for meson production in nuclei need a model for the free nucleon from  Sect.~\ref{se:piN}.  Since all those models are based on the same bubble chamber data, there is some natural deviation among them given the choices made.  At energy transfers above approximately 200 MeV, inelastic excitations of the nucleon connected with pion production become possible. Most of the nucleon resonances have spin 1/2 or spin 3/2. Pion
production is through the weak excitation of the $\Delta(1232)$ resonance and its subsequent decay into $N\pi$ is dominant. Thus, the in-medium modification of the $\Delta(1232)$ properties represents the most important nuclear effect, as already stressed in the early work of Refs.~\cite{Kim:1996bt,Singh:1998ha}.
Current assumptions are that higher energy resonances have small excitation cross sections and don't play
a strong role in the interpretation of any existing measurement.  However, this  is indeed only an assumption and needs confirmation for higher energy experiments such as DUNE.

On the other hand, FSI takes into account
that pions can be absorbed in their way out of the nucleus, and can
also suffer different quasielastic collisions that modify their
energy, angle, and charge when they come off the nucleus.  For
instance, in the case of NC $\pi^0$ production, signal events
originate mostly from a NC1$\pi^0$ primary interaction with a $\pi^0$
not being affected by FSI, but also from a NC1$\pi^+$ primary
interaction with the $\pi^+$ being transformed into $\pi^0$ in a
charge exchange FSI reaction. In this particular case, an additional
difficulty in interpreting the NC$\pi^0$ production comes from the
presence of a coherent contribution. FSI definitely alters the
signature of the event and thus the correct simulation of pion
production requires a model not only able to describe the elementary
reactions, but also the final state interactions.

To compute the incoherent pion production on a nucleus, one should sum
the nucleon cross section over all nucleons in the nucleus.  For
instance within the local density approximation (LDA) and for a neutrino CC process  one gets for initial pion production (prior to any pion FSI)
\begin{eqnarray}
 \frac{d\sigma}{d|\vec{k}\,|4\pi  r^2\,dr\,d\cos\theta_\pi\, dE_\pi} &=&
 \nonumber \\
\Phi(|\vec k |)\,
 \hspace{-.15cm}
  \sum_{N=n,p}%\hspace{-.15cm}
 2\int \frac{d^3p_N}{(2\pi)^3}\
&\theta&(E_F^N(r)-E_N) \,\theta(E_N+q^0-E_\pi-E_F^{N'}(r))
\frac{d\hat\sigma(\nu N\to l^-N'\pi)}{d\cos\theta_\pi dE_\pi}
\label{eq:eq1}
\end{eqnarray}
with $E_F^N(r)=\sqrt{M^2+(p_F^N(r))^2}$, given in terms of the local
Fermi momentum $p_F^{n,p}(r)=[3\pi^2 \rho_{n,p}(r)]^\frac13$. The
step functions implement Pauli blocking and $\Phi(|\vec k|)$ is
the neutrino flux with incoming-neutrino energy $E_\nu
\equiv |\vec k|$. In addition, $\hat\sigma(\nu N\to l^-N'\pi)$ is the cross
section at the nucleon level modified by medium effects, where the
modification of the $\Delta(1232)$ spectral function is the most
relevant one. The $\Delta$ properties are strongly modified in the
nuclear medium~\cite{Oset:1987re}, and since the direct
$\Delta$-mechanism is dominant, a correct treatment is needed for
$\pi$ production inside a nucleus. This is achieved by 
using a realistic spreading potential ($\Delta-$selfenergy). In the
nuclear medium, on one hand, the width is reduced due to Pauli
blocking, but on the other hand, it is increased by the collisions
inside the nucleus.  For example, via the processes $\Delta N \to NN$
and $\Delta NN \to NNN$, the $\Delta$ can disappear without producing
a pion. Secondary pion production is also possible, namely via the
process $\Delta N \to \pi NN$. These processes contribute to the in
medium $\Delta$ width that generally becomes larger than in the free
space.

FSI effects must use effective models because of the difficulty
of describing the interaction of hadrons in the nuclear environment.
Often, FSI effects are implemented by means of a semiclassical intranuclear
cascade, including different nuclear corrections. There is a long history
of success for these models in describing hadron beam data.
The in medium differential cross section of Eq.~(\ref{eq:eq1}) is used
in the simulation code to generate, at a given point $\vec r$ inside the
nucleus and by neutrinos of a given energy, on-shell pions with a
certain momentum. These pions are followed through their path across
the nucleus.  One should bear in
mind that the $\pi N$ interaction is also dominated by the $\Delta$
resonance excitation, modified in the nuclear medium in the same way
as it was modified in $\hat\sigma(\nu N\to l^-N'\pi)$. The different
contributions to the imaginary part of its self-energy account for
pion, two- and three-nucleon absorption and quasielastic
processes. One solution to this problem was given
by Salcedo and Oset~\cite{Salcedo:1987md} which uses the
$\Delta-$selfenergy calculated in \cite{Oset:1987re}.
This approach is used in the Valencia calculation~\cite{Salcedo:1987md}.

A different approach to account for FSI effects is based on the
Giessen Boltzmann-Uehling-Uhlenbeck (GiBUU) model~\cite{Buss:2011mx}. It is a transport
model where FSI are implemented by solving the semi-classical
Boltzmann-Uehling-Uhlenbeck equation.  It describes the dynamical
evolution of the phase space density for each particle species under
the influence of the mean field potential, introduced in the
description of the initial nucleus state.  Equations for various
particle species are coupled through this mean field and also through
the collision term. GiBUU provides a unified framework for nucleon--,
nucleus--, pion--, electron-- and neutrino interactions with nuclei,
from around a hundred MeV to tens of GeV, where medium effects like
the $\Delta-$spreading potential can be taken into account.  

Coherent contributions, when relevant, need also to be
evaluated.  In the coherent processes, the final nucleus is
left in its ground state.  Here, FSI can be described with
multiple iterations of an optical potential between the outgoing pion and the nucleus in the ground state, giving a new view of the problem. For example, the coherent channel can provide a clear insight into the details on modifications of the $\Delta$-propagation in a nuclear environment. This is because non-resonant background contributions are  suppressed~\cite{Amaro:2008hd,Hernandez:2010jf}.

\subsection{Generator status}
% 3)  Status of the generator picture of this channel.
As discussed elsewhere in this document, event generators are at the interface between experiment and theory.  There are several Monte Carlo codes in use, GENIE~\cite{Andreopoulos:2009rq}, NuWro~\cite{Golan:2012wx},
Neut~\cite{Hayato:2009zz}, Nuance~\cite{Casper:2002sd}.  Each must make
choices similar to the theoretical calculations described above.  Although they try to include modern theoretical treatments, their first responsibility
is to provide a fast model that has robust tools for interfacing to experiments and predicts experimental results in a large variety of circumstances.  Unfortunately, these needs are sometimes in conflict.  Although event generators can only include partial versions of most theoretical models, they are able to make calculations for any observable measured in an experiment.

Event generator builders make different decisions about the very definition of the resonant region.  On one extreme side one can define it as $W<1.6$~GeV (i.e. mostly $\Delta(1232)$ region) while others use an upper limit of up to 2.0~GeV. If $W$ is extended to large values generators rely on a Rein-Sehgal model~\cite{Rein:1980wg} which is easily implemented and covers a large fraction of the phase space. By adjusting a value of the axial mass parameter $M_A^{res}$ one can reproduce typical data sets~\cite{Radecky:1981fn,Kitagaki:1986ct,Grabosch:1988gw}.  

In the original Rein-Sehgal model~\cite{Rein:1980wg} non-resonant contributions are approximated by an artificial extra resonance.  This approach can be improved using computations done by Rein~\cite{Rein:1987cb} who proposed a model based on three Born diagrams.  The Rein-Sehgal model included interference terms; unfortunately all generators disregard them. Furthermore there are no models for non-resonant multi-pion production employed in any generators.  Similarly, one can include information about angular distribution of pions resulting from resonance decays. In the simplest approach the distribution is uniform. More realistic implementations include e.g. density matrix measurements done in ANL and BNL experiments.

Event generators include explicit contributions from heavier resonances (the focus is on the second resonance region) using expressions for resonance excitation matrix elements. This allows calculation of events at all kinematics.  Often, the outdated Rein-Sehgal parameterizations are updated to modern values.  While the vector part of those elements are known~\cite{Drechsel:2007if}, there is practically no information on the axial part from data and one must rely on educated guesses. Similarly, ad hoc ansatze are presently used for the non resonant background. A possible procedure is to extrapolate fits done to electron pion production data or to use ChPT models.

As for nuclear effects, generators typically describe target nucleons in terms of the local Fermi gas model. Medium corrections to pion production are sometimes included as the $\Delta$ self-energy, but on very different levels of sophistication. The GiBUU cross section 
formula includes both nucleon and $\Delta$ spectral functions. NEUT assumes a fixed fraction of pionless $\Delta$ decays, using results of Singh at al. NuWro takes the fraction to be neutrino energy dependent.  GENIE presently has none of these effects.

For final state interactions the event generators NEUT and NuWro use the cascade code developed in \cite{Salcedo:1987md}, while GENIE uses an effective model assuming pion absorption cross section to be a fixed fraction of the pion reaction cross section.  The recent GENIE release v2.12.0 has a variety of FSI models that can be substituted for the default model, allowing interesting comparisons.

Event generators performance cannot exceed the data precision. In the resonance region it is rather difficult to take decisions how to improve their performance. Typically, the generators reproduce either \mb or \minerva carbon target pion production data quite well.  Old bubble chamber ANL and BNL deuterium pion production data are not very difficult to reproduce with reasonable precision. Thus generators need more precise experimental data to justify more ambitious upgrades. 

Electro- and photoproduction data provide an important test of nuclear models and FSI used in the description of resonance excitations.  Unfortunately, options for these processes are presently only included in GENIE.

\subsection{Existing Experimental Results}
%4)  Summary of the experimental studies of this channel
%             Experimental results off the nucleon
%             Experimental results off the nucleus
%             Importance of e-nucleus (and/or other pertinent non-neutrino) results for this channel
\subsubsection{pion production from the nucleon with neutrinos}
These data all come from bubble chamber experiments from the 1980's.
At neutrino energies less than about 2 GeV, the ANL~\cite{Radecky:1981fn} and BNL~\cite{Kitagaki:1986ct} experiments are the primary source.  Each has low statistics (few hundred events per energy), excellent reconstruction, and uncertain normalization.
Estimated uncertainties are dominated by statistical and absolute normalization errors.  Nevertheless, there are systematical differences between the 2 measurements well outside estimated uncertainties (30-40\%).
A recent paper~\cite{Wilkinson:2014yfa} suggests that the QE measurements be used as a benchmark.  They base the normalization on the ratio of QE to pion production data and QE theoretical calculations and reevaluate systematic errors of each experiment.  The result is that both experiments are in better agreement at approximately the cross section level of the published ANL data.

Two pion production data~\cite{Day:1984nf} are of very low quality.  At higher energies, the SKAT data~\cite{Allen:1985ti,Grabosch:1988gw} are very important for 3-10 GeV beam energy.

For antineutrinos, the data is of considerable lower quality.  For example, the only 
data for $\pi^0$ production with $\overline{\nu}_\mu$ beam was at a 
single average energy (7 GeV) for a freon ($CF_3Br$) target~\cite{Grabosch:1988gw}.  Confidence (misplaced) in their nuclear model led them to quote results for a free proton.

\subsubsection{pion production from nuclei with neutrinos}

\begin{figure}
\begin{center}
\includegraphics[width=7.5cm]{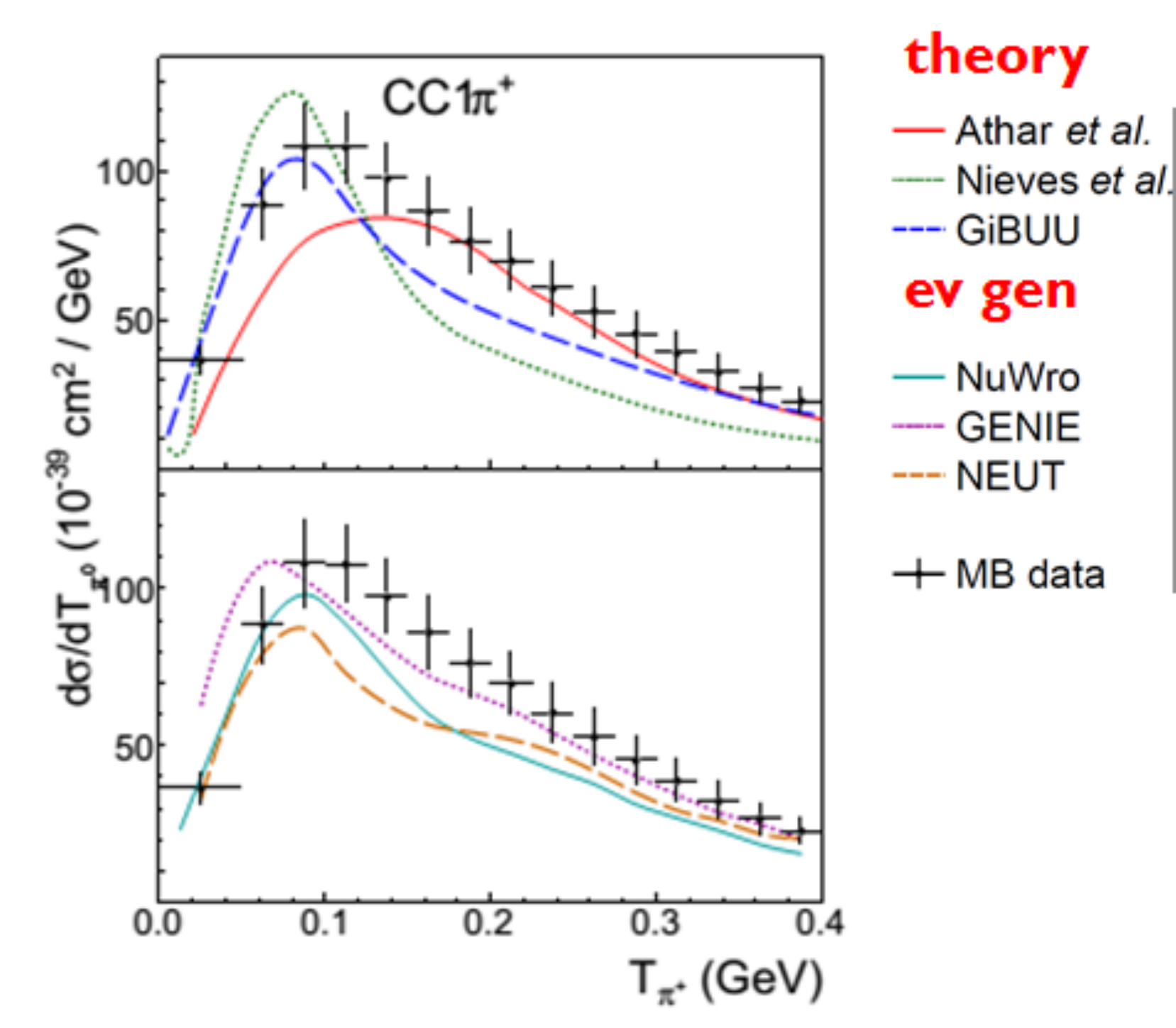}
\includegraphics[width=7.5cm]{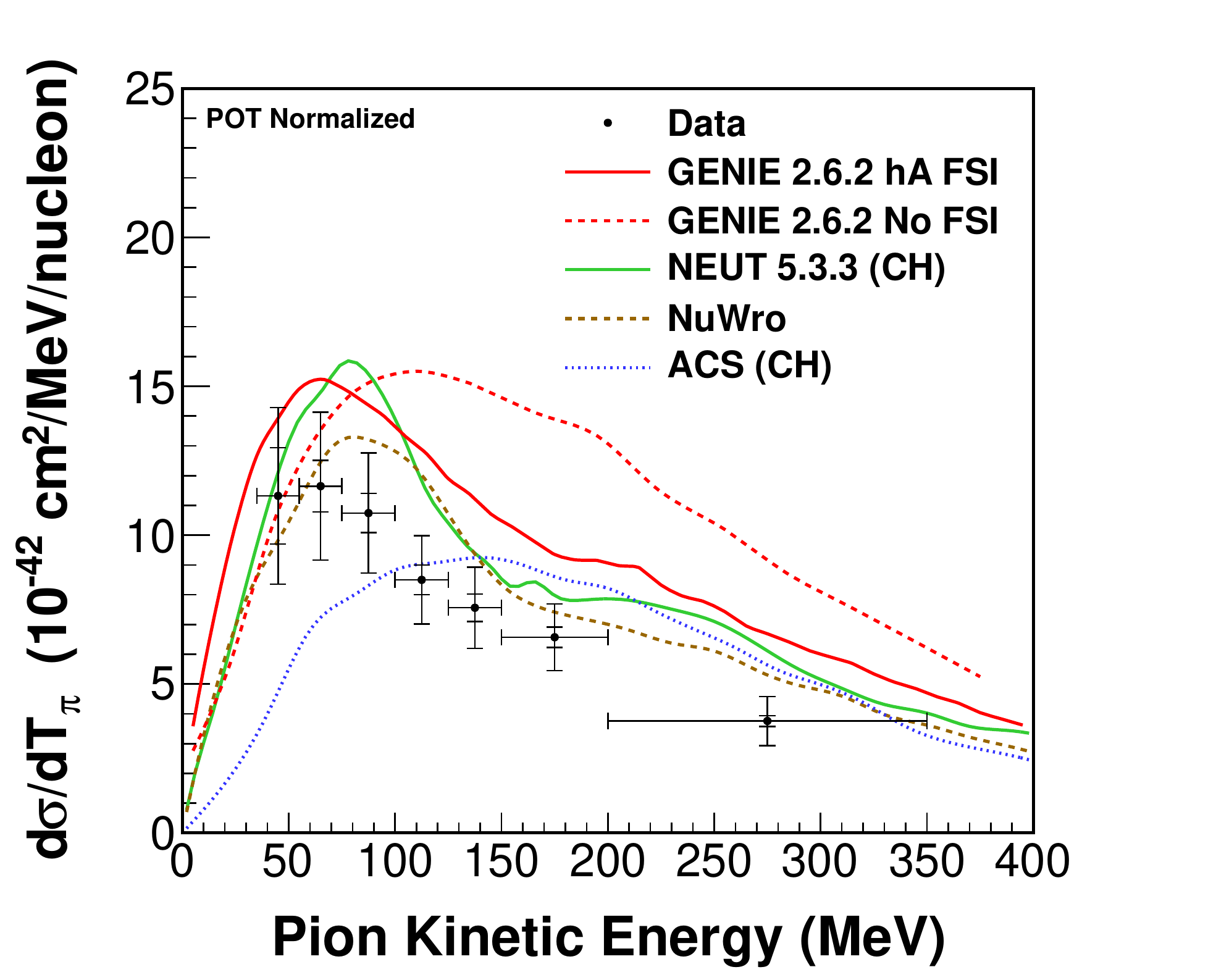}
\end{center}
\caption{\label{fig:pion-data-calc} (left) Comparison of theoretical and event generator calculations available in 2014 with \mb $\nu_\mu CH_2$ CC $\pi^+$ production data~\cite{Rodrigues:2014jfa} (right) Comparison of theoretical and event generator calculations with \minerva $\nu_\mu CH$ CC $\pi^+$ data~\cite{Eberly:2014mra}.   }
\end{figure}

Recent publications from \mb ($\nu_\mu CH_2$ CC)~\cite{AguilarArevalo:2010bm} and \minerva 
($\nu_\mu CH$ CC)~\cite{Eberly:2014mra} have provided two separate pictures of the low energy $E_\nu \sim$ 1 GeV and $E_\nu \sim 4$ GeV regions.  There are both consistencies and at the same time confusing features.  The \mb data was first, providing muon and pion cross sections in a comprehensive
data set.  Primary focus has been on the pion kinetic energy distribution where sensitivity depends strongly on the pion production model for the free nucleon and FSI.  For a variety of nucleon production models, calculations have had trouble matching the pion kinetic energy distribution, see Fig.~\ref{fig:pion-data-calc} from Ref.~\cite{Rodrigues:2014jfa}.  The best theoretical calculations~\cite{Hernandez:2013jka,Lalakulich:2012cj} have a dip in both $\pi^+$ and $\pi^0$ spectra at the energy where the pion interacts most strongly with nuclei ($T_\pi \sim$ 160 MeV).   At the same time, the event generator model predictions see a much more shallow dip using a variety of FSI models (including a FSI model identical to one of the theoretical calculations). 

\minerva pion production data was first published in 2014 and they see no
dip for both $\nu_\mu \pi^+$~\cite{Eberly:2014mra} and $\overline{\nu}_\mu \pi^0$~\cite{Aliaga:2015wva} production.  Shapes for \minerva and \mb are similar but not identical.  Generator simulations~\cite{Sobczyk:2014xza} find almost identical shape for $\pi^+$ independent of beam energy,
perhaps because the pion kinetic energy distributions for the primary
$\pi^+$ process are very similar.  No theoretical calculations are yet available for these data.

The magnitude of these two data sets has proved more problematic.  NuWro and
GENIE simulations both find the ratio between the average cross section ratio
(\minerva / \mb) about 30\% larger than seen in the data; the energy
dependence is determined by the $\pi^+$ primary production process.
The first \minerva $\pi^+$ paper attempted to reproduce the conditions
of the \mb experiment. That wasn't totally possible because both experiments
used $W$ cuts in different ways. \mb cut out events with $W>$1.35 GeV because the signal process was ambiguous, then added the higher $W$ response
back in ($\sim$ 25\% effect) using the NUANCE generator.  \minerva used a
cut $W<$1.4 GeV to eliminate the contribution of higher energy resonances,
but used a model dependent definition of $W$.  They also provided a separate
analysis with $W<$ 1.8 GeV and found very similar results and conclusions.

\begin{figure}
\begin{center}
\includegraphics[width=7.5cm]{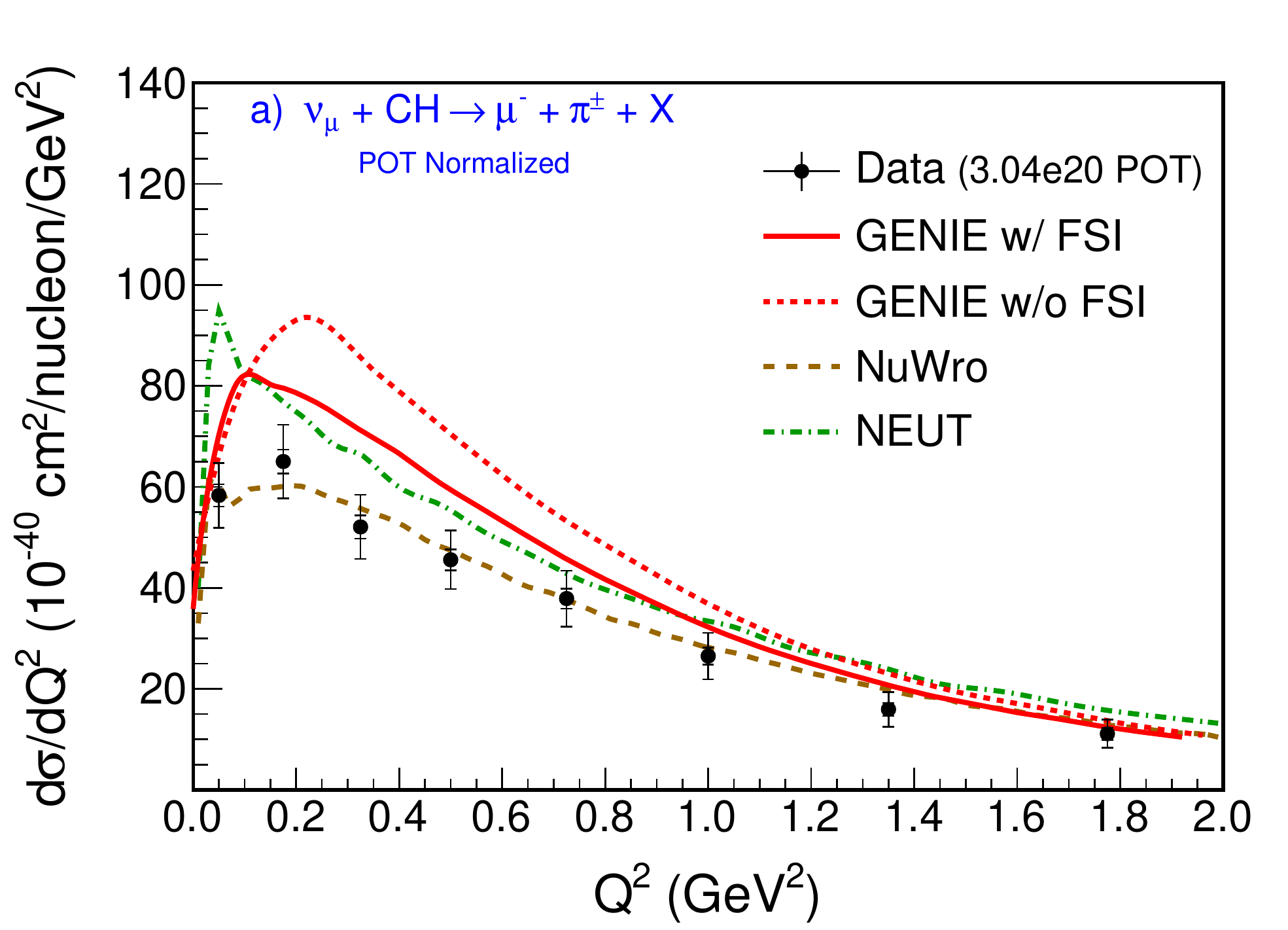}
\includegraphics[width=7.5cm]{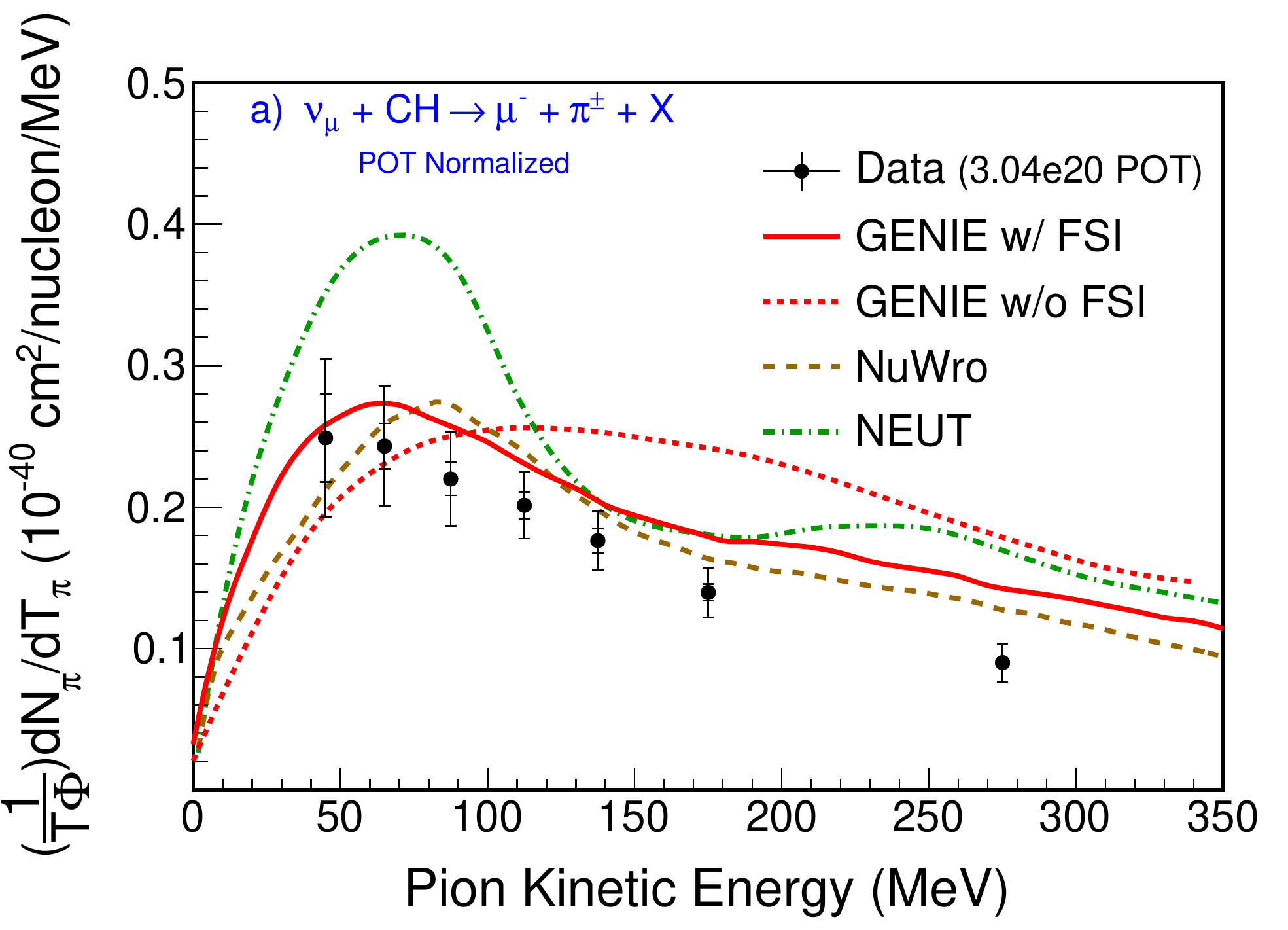}
\end{center}
\caption{\label{fig:pion-data-calc-2016} Comparisons of event generator calculations with \minerva $\nu_\mu CH$ CC $\pi^+$ data~\cite{McGivern:2016bwh} (left) $Q^2$ and (right) kinetic energy.  Both results include resonances at $W <$ 1.8 GeV.}
\end{figure}

More recent \minerva data~\cite{McGivern:2016bwh} has broadened the discussion significantly.  Both $\nu_\mu \pi^+$ and $\overline{\nu}_\mu \pi^0$ data are presented in parallel analyses.  They use a cut of $W<$ 1.8 GeV, thereby including higher energy resonances.  They also use a definition of $W$ based on experimental observables, removing much of the model dependence in the first result.  A major change from earlier papers is due to an updated flux calculation~\cite{Aliaga:2016oaz}; the result is an average 13\% (12\%) increase in the \numu (\numubar) cross sections.  The focus of this paper is on the muon and other associated variables, i.e. $E_\mu$, $E_\nu$, and $Q^2$, where the latter two quantities involve model dependent reconstruction.
Sample $\pi^+$ kinetic energy and $Q^2$ distributions are shown in Fig.~\ref{fig:pion-data-calc-2016}, the $Q^2$ distribution is largely dependent on nuclear structure models.  Both \mb and \minerva data have rapid falloffs above $Q^2> 0.5 (GeV/c)^2$.  At lower values, each data set has a hint of suppression that would be due to long range $NN$ correlations. At low $Q^2$, differences in the coherent cross section models used cause large disagreements.  Otherwise, the generator simulations are similar despite different nuclear structure conditions.

Most recently, $T2K$ data (\numu CC $\pi^+$) has been shown in preliminary form.  This will be an important check of the \mb data since the $\nu$ energy range is very similar.

\subsubsection{pion production with electromagnetic beams}

Using electromagnetic beams numerous inclusive pion production measurements are available for proton and deuteron targets and fits for inclusive response~\cite{Christy:2007ve} and resonance couplings~\cite{Drechsel:2007if} are available.  These results have been the basis for the
vector response of each theoretical model.  However, there are very few experiments with nuclear targets.  The pion photoproduction data from Mainz~\cite{Krusche:2004uw} are notable. Reanalysis of older JLab data~\cite{Manly:2015fva} for single charged pion production with
5 GeV electrons and C, Fe, and Pb targets has appeared in preliminary form.
The published results are anxiously awaited as they will have important
repercussions on all calculations.

\subsection{Challenges and Open Questions}
%5)  Challenges and open questions for this channel
%Open challenges in the theoretical and experimental treatment of this channel
%                  Differences in various theoretical models and their sources of uncertainty
%                  New experimental observations required for a better understanding and modeling 
%                  of this channel and to meet the accuracy goals 
\subsubsection{Theory}
Full calculations of any of the observables discussed here must involve approximations because the many body problem has no well-established solution.  Correct descriptions of experiments need accurate descriptions of nuclear structure (momentum distributions and effects of NN correlations) and medium corrections.  This work has greatly advanced in the last decade as theorists extend successful descriptions of electromagnetic and hadronic interactions into weak interactions~\cite{Alvarez-Ruso:2015eva,Buss:2011mx}.  The models use effective interactions with form factors, including amplitudes for both resonant and nonresonant meson production.  One example is Valencia~\cite{Hernandez:2013jka} which is sophisticated but simple enough to use in event generators.  GiBUU~\cite{Buss:2011mx} is a more sophisticated model but more difficult to include in event generators.  At present, neither GiBUU nor the Valencia nonresonant model is in any
of the event generators.

Problems in developing a rigorous model for QE interactions are further amplified for meson production interactions.  Unlike electromagnetic experiments, a complete picture of the final state is required to properly simulate the event in experiments.  This means solving two difficult problems, nuclear structure above pion production threshold and final state interactions.  To date, the effects of NN correlations are seldom included in pion production models.  Pauli blocking is sometimes included in simple ways.  Final state interactions are even more complicated because many channels are open.  

The basis for every model for the nucleus is meson production on a single nucleon.
Excellent models for nucleon targets are available, but lack of quality data
prevents additional progress.  Medium effects should be compatible with the FSI model used.

The problems are then many-sided.  Descriptions of neutrino interactions experiments need more sophisticated models than those that have been used previously for
electromagnetic or hadronic interactions.  Although we surely will end up with reasonable descriptions using effective degrees of freedom, we have a variety of models which are likely incomplete.  Furthermore, the most sophisticated models are not always useful in event generators which depend on simple algorithms for simulation speed and an efficient evaluation of systematic errors.  What is the optimal compromise between best theory and best applicability to experiments?

\subsubsection{Experiment}

The ability to do quality experiments has also grown dramatically in the last decade.  Oscillation experiments have realized that programs to understand the models describing neutrino interactions couple into their systematic errors in important ways.  Advances in theory and applications to event generators will lead to corresponding advances in oscillation results.  

Neutrino experiments with nuclei are complex.  Separate models are needed to calculate the neutrino flux (can't be measured yet) over a broad energy range.  These models also require nuclear calculations of pion and kaon production.  One consequence is that the signal definition is complicated and experiment-dependent.  Better and more uniform signal  definitions are required to enable direct comparisons of experimental results.  

New measurements addressing some of the issues discussed here have been reported at conferences.  The first T2K pion production measurement (CC $\pi^+$) will be at neutrino energies similar to \mb.  The $\pi$ electroproduction results from CLAS~\cite{Manly:2015fva} will provide the vector response for a range of kinematics and a variety of targets. Both \minerva and T2K are still taking data and new results are in progress.  \minerva $\pi^+$ data for C, Fe, and Pb will be especially interesting for exploring the FSI medium dependence. Finally, new results from the NOMAD experiment are expected for $E_\nu>$ 5 GeV.

Liquid $Ar$ detectors promise a new generation of experiments with heightened ability to measure low energy particles.  These techniques must be perfected in experiments now running.  A liquid $Ar$ cross section experiment with a neutrino flux similar to DUNE would be valuable as there are minimal pion production measurements for nuclei with $A>$20.  For the future, new experimental results on $H$ or deuterium targets are essential.  This will require new technological solutions.  However, these experiments will produce results which decrease systematic errors in oscillation experiments.  DUNE plans propose new measurements with a hydrogen target.

\newpage

\section{Shallow and Deep Inelastic Scattering}
\label{sisdis}

\subsection{Introduction} 

The definition of deep-inelastic scattering (DIS) is based upon the kinematics of the 
interaction products and there is no precise way to distinguish the onset of the DIS region 
from the resonance region. 
Usually, the region $ W \ge 2.0 $ GeV and $Q^2 \ge 1.0$ GeV$^2$ is considered to be 
the safe DIS region, beyond the resonance contributions.
 
In the (anti)neutrino-nucleon DIS process, the (anti)neutrino interacts with a quark in a
nucleon ($N$), producing a lepton ($l$) and a jet of hadrons ($X$) in the final state: 
\begin{equation} 	\label{reaction}
\nu_l/\bar \nu_l(k) + N(p) \rightarrow l^{\mp}(k^\prime) + X(p^\prime),~l=~e,~\mu,
\end{equation}
 where the quantities in parenthesis represent the four momenta of the corresponding particles.
 If the nucleon $N$ is bound inside a nucleus $A$ its structure is influenced by a number of 
 nuclear effects including Fermi motion, binding energy, off mass shell, nucleon-nucleon 
 correlations, as well as by non-nucleonic degrees of freedom like meson exchange currents, 
 quark clusters and nuclear shadowing. Experimental and theoretical studies of DIS 
 using charged leptons and (anti)neutrinos off various nuclear targets show ample 
 evidence that these nuclear effects result in a modification of the bound nucleon. 
 
The general expression of the double differential cross section for the (anti)neutrino induced DIS 
off a nucleon/nucleus can be written as:
\begin{equation} 	\label{ch2:dif_cross_nucleus}
\frac{d^2 \sigma_{\nu,\bar\nu}^A}{d \Omega' d E'} 
= \frac{G_F^2}{(2\pi)^2} \; \frac{|{\vec k}^\prime|}{|{\vec k}|} \;
\left(\frac{m_W^2}{q^2-m_W^2}\right)^2
L^{\alpha \beta}_{\nu, \bar\nu}
\; W_{\alpha \beta}^i\,,
\end{equation}
where $q=k-k'$ is the four momentum transferred, $m_W$ is the mass of the $W$-boson, 
and $L^{\alpha \beta}_{\nu, \bar\nu}$ is the leptonic tensor. The quantity $W_{\alpha \beta}^i$ 
represents the nucleonic tensor for $i=N$ and the nuclear hadronic tensor for $i=A$, respectively.  
The leptonic tensor $L^{\alpha \beta}_{\nu, \bar\nu}$ is given by:
\begin{equation} 	\label{ch2:dif_cross2}
L^{\alpha \beta}_{\nu, \bar\nu}=k^{\alpha}k'^{\beta}+k^{\beta}k'^{\alpha}
-k.k^\prime g^{\alpha \beta} \pm i \epsilon^{\alpha \beta \rho \sigma} k_{\rho} 
k'_{\sigma}\,,
\end{equation}
where the plus sign is for antineutrino and the minus sign for neutrino. 
In the limit $m_l \to 0$, the hadronic tensor $W_{\alpha \beta}^i$ in Eq. (\ref{ch2:dif_cross_nucleus}) 
can be expressed in terms of structure functions $W^i_{1-3}(x,Q^2)$ as: 
\begin{eqnarray}\label{ch2:had_ten}
W^i_{\alpha \beta} &=&
\left( \frac{q_{\alpha} q_{\beta}}{q^2} - g_{\alpha \beta} \right) \;
W_1^i
+ \frac{1}{M_i^2}\left( p_{\alpha} - \frac{p . q}{q^2} \; q_{\alpha} \right)
\left( p_{\beta} - \frac{p . q}{q^2} \; q_{\beta} \right)
W_2^i \nonumber\\
&-& \frac{i}{2 M_A^i} \epsilon_{\alpha \beta \rho \sigma} p^{\rho} q^{\sigma} W_3^i,
\end{eqnarray}
where $M_i=M_N$ is the mass of the nucleon and $M_i=M_A$ is the mass of the nucleus. Usually the functions 
$W^i_{1-3}(x,Q^2)$ are redefined in terms of the dimensionless structure functions 
$F^i_{1-3}(x,Q^2)$ through the relations:  
\begin{eqnarray}\label{ch2:relation}
M_i W_1^i(\nu, Q^2)&=&F_1^i(x, Q^2),~~~\nu W_2^i(\nu, Q^2)=F_2^i(x, Q^2),~~~\nu W_3^i(\nu, Q^2)=F_3^i(x, Q^2),~~~~~
\end{eqnarray}
where $Q^2=-q^2$, $\nu=p\cdot{q}/M_N$ is the energy transfer to the nucleon, $x=Q^2/(p\cdot{q})$ is the momentum fraction carried by the partons in the nucleon, and $x=Q^2/(p_A\cdot{q})$ is the momentum fraction carried by the partons in the nucleus. These structure functions can be associated in quantum chromodynamics (QCD) to the 
partonic structure of nucleons and can be expressed in terms of parton distribution functions (PDFs), representing 
the momentum distribution of the partons within the hadron target. 

For electromagnetic interactions, the DIS cross section depends only on the two structure functions 
$F_{1,2}^i(x, Q^2)$. However, for the charged current (CC) (anti)neutrino-nucleus DIS process, 
three structure functions $F_{1,2,3}^i(x, Q^2)$ are required. 
While the first two can be measured both in 
charged lepton and (anti)neutrino scattering, the third one, $F_3$, can only be accessed by 
parity-violating processes like weak interactions. The goal of future DIS experiments is to 
independently determine these structure functions in neutrino and antineutrino scattering 
from nuclear targets. At leading order (LO) in perturbative QCD and assuming four parton flavors 
(up, down, strange, and charm quarks), they can be defined as: 
\begin{eqnarray}\label{ch2:f2f3}
F^{\nu p}_{2}& = & 2 x [d(x) + s(x) + \bar{u}(x) +\bar{c}(x)], \,\;\;\;\;\;\;  F^{\bar{\nu} p}_{2} =  2 x [u(x) + c(x) + \bar{d}(x) +\bar{s}(x)], \nonumber\\ 
F^{\nu n}_{2} & = & 2 x [u(x) + s(x) + \bar{d}(x) +\bar{c}(x)], \,\;\;\;\;\;\; F^{\bar{\nu} n}_{2} =  2 x [d(x) + c(x) + \bar{u}(x) +\bar{s}(x)], \nonumber\\
xF^{\nu p}_{3}&=& 2x[d(x) + s(x) -{\bar u}(x) - {\bar c}(x)], \;\;\;\;\; xF^{{\bar \nu} p}_{3} = 2x[u(x) + c(x) -{\bar d}(x) - {\bar s}(x)], \nonumber\\
xF^{\nu n}_{3} &=&  2x[u(x) + s(x) -{\bar d}(x) - {\bar c}(x)], \;\;\;\;\; xF^{{\bar \nu} n}_{3} = 2x[d(x) + c(x) -{\bar u}(x) - {\bar s}(x)]. 
\end{eqnarray}
 where $q(x)$ is the probability of finding a quark or an anti-quark carrying a fraction $x$ of the nucleon momentum.
These structure functions are related by the Callan-Gross relation $ 2 x F_{1}(x) = F_{2}(x)$. 
 
Alternatively, the DIS cross-section can be also described in terms of transverse $F_T$ and longitudinal $F_L$ structure functions defined as
\begin{equation}
 F_T (x,Q^2)= 2xF_1(x,Q^2), \;\;\;\;\;\;\; F_L(x,Q^2)=\gamma^2 F_2(x,Q^2)-F_T(x,Q^2), 
 \label{eq:f1f2sig}
\end{equation}
where $\gamma=1+4x^2M_N^2/Q^2$. Due to the different behavior of the transverse and longitudinal 
components, the ratio $R=F_L/F_T$ provides an interesting observable in 
(anti)neutrino DIS.

\subsection{Inelastic Scattering off Nucleons} 

\subsubsection{Quark-Hadron Duality}
\label{sec:DIS:QHduality} 

At low energy, the inclusive cross sections describing the 
scattering processes induced by charged leptons and (anti)neutrinos on nucleons and nuclei 
can be expressed in terms of structure functions (form factors) corresponding to the excitation of various 
discrete resonances like $\Delta,~N^\ast$, etc., characterized by increasing values of the 
 CM energy $W$ of the final hadrons. 
At high energy and $Q^2$, the inclusive cross sections are usually expressed in
terms of the structure functions corresponding to the continuum DIS processes. 
The description of inclusive lepton scattering in the transition between the resonance excitation 
and the DIS, occurring in the intermediate energy region, is still a subject of continuing study. 
This region is also known as shallow inelastic scattering (SIS). 
The quark-hadron (QH) duality, first introduced by Bloom and Gilman~\cite{Bloom:1970xb, Bloom:1971ye} to explain electron-proton scattering, 
states that the resonance structure functions 
in the low $Q^2$ region, suitably averaged over an energy interval, provides the same 
result as the corresponding DIS structure functions at high $Q^2$, in the same energy interval.
This phenomenon can thus provide a connection between 
quark-gluon description of the DIS formalism at high $Q^2$, and the pion-nucleon description 
in the region of resonance excitation at low $Q^2$. 
The QH duality seems to be valid individually in each resonance region, 
as well as over the entire resonance region, if the structure functions are summed over the higher resonances. 
This phenomenon is called local duality. When the local QH duality is observed for 
higher moments of structure functions, it is called global duality. 

In the weak sector, the QH duality has been shown to work in neutral current (NC) 
interactions for polarized electron-nucleon scattering, as observed from the 
parity violation (PV) asymmetry of electrons from proton and deuteron targets.
From isospin symmetry arguments, it can be argued that in the case of (anti)neutrino scattering 
the QH duality does not hold for proton and neutron targets separately, but rather, 
with a limited accuracy, for an average isoscalar target. 
A similar picture is expected in (anti)neutrino interactions with bound nucleons 
in nuclear targets~\cite{Lalakulich:2009zza}.
A verification of the validity of QH duality in the charged current (CC) and NC sectors of 
weak interactions can provide a way to describe the (anti)neutrino-nucleon and
(anti)neutrino-nucleus scattering cross sections in the transition region, in which 
the use of either the effective Lagrangian or the quark-parton description is not adequate. 
 Further studies are necessary to understand the concept of QH duality in weak interactions 
 on non-isoscalar nuclei.

Different approaches to the modeling of the SIS region are used in modern event generators. While NuWro~\cite{Graczyk:2005uv} has a smooth transition from resonance to DIS region attempting to imitate the QH duality, both GENIE~\cite{Andreopoulos:2009rq} and NEUT~\cite{Hayato:2009zz} have discontinuities of cross sections in the SIS region as a function of $W$. The transition from the resonance to the DIS formalism occurs abruptly at $W=1.7$ GeV in GENIE and at $W=2.0$ GeV in NEUT. In all generators the DIS models are also extended into the resonance region in order to simulate non-resonant pion-production backgrounds.

\subsubsection{Perturbative and Electroweak Corrections} 
\label{sec:DIS:pQCD} 
 
\begin{figure}[tbp]
\begin{center}
\includegraphics[scale=0.90]{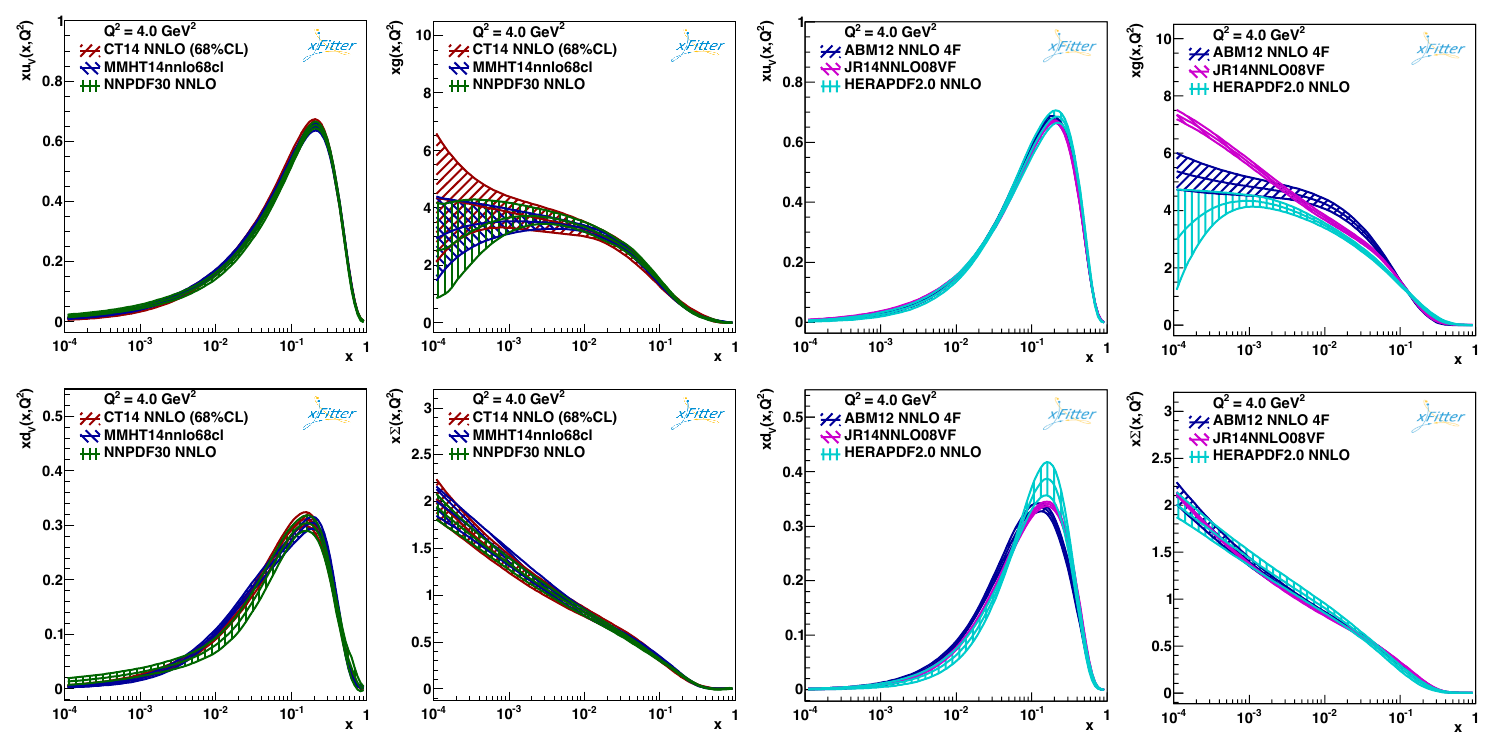}
\end{center}
\caption{Example of parton distribution functions for the proton and their uncertainties at $Q^2=4$ GeV$^2$. 
Figure adapted from Ref.~\cite{Accardi:2016ndt}. 
  }
\label{fig:PDFs} 
\end{figure}

At high momentum transfer $Q$ the lepton-nucleon cross sections are 
well described in terms of PDFs, whose $Q^2$ 
evolution is well-understood in perturbative quantum chromodynamics (QCD). 
The PDF content of the nucleon is extracted from global fits
\cite{Alekhin:2017kpj,Accardi:2016qay,Harland-Lang:2014zoa,Ball:2014uwa,Dulat:2015mca,Abramowicz:2015mha,Jimenez-Delgado:2014twa}
to experimental data at large momentum transfer, including lepton DIS,
lepton-pair production (Drell-Yan process), jet production, and
$W$ and $Z$ boson production in hadron collisions. As an example, Fig.~\ref{fig:PDFs} 
illustrates the PDFs obtained by various groups as a function of
$x$ for $Q^2$ = 4 $GeV^2$.

%%%%%%%%%%%%%%%%%%%%%%%%%%%%%%%%%%%%%%%
%%%%%%%%%%%%%%%%%%%%%%%%%%%%%%%%%%%%%%%

The Wilson coefficients entering the massless DIS structure functions are known at the 
NNLO (next-to-next-to-leading-order)~\cite{vanNeerven:1991nn,Zijlstra:1991qc,Zijlstra:1992qd,Zijlstra:1992kj,Moch:1999eb,Moch:2004xu,Vermaseren:2005qc} 
or at the N$^3$LO (next-to-next-to-next-to-leading-order)~\cite{Vermaseren:2005qc,Moch:2008fj}.   
The heavy quark Wilson coefficients entering the DIS structure functions for charm production 
are known exactly only to the next-to-leading-order (NLO)~\cite{Gottschalk:1980rv,Gluck:1996ve}.
It is worth noting that exclusive charm production in CC (anti)neutrino DIS provides a direct probe of the strange quark content of the nucleon and of the charm quark mass~\cite{Alekhin:2014sya}. 

In the analysis of experimental data, and in comparisons between measurements and theoretical models, electroweak radiative corrections beyond the Born approximation 
must be applied to structure functions and cross-sections. 
One-loop calculations for the elementary partonic processes are 
available~\cite{Arbuzov:2004zr,Diener:2005me} including 
virtual corrections, hard and soft photon radiation, quark and muon mass singularities. 
The dominant correction in CC interactions is related to hard photon radiation. It is worth noting 
that electroweak corrections depend upon the inelasticity $y(=\frac{\nu}{E})$, and are significant in the 
low-$x$ and large-$x$ regions, where they can be of comparable size with respect to 
the nuclear corrections. 

In the very low-$x$ region, one needs to take into account 
saturation~\cite{Arguelles:2015wba,Block:2014kza}, which may be relevant for the detection of ultrahigh-energy astrophysical 
 neutrinos with $E_\nu\simeq 1$ EeV.
 
\subsubsection{High Twist Contributions} 
\label{sec:DIS:HT} 

For lower values of $Q$, a few GeV or less, non-perturbative 
phenomena become important for a precise modeling of cross sections, 
in addition to high-order QCD corrections~\cite{Alekhin:2007fh}. 
In the formalism of the operator product expansion (OPE), unpolarized structure 
functions can be expressed in terms of powers of $1/Q^2$ (power corrections):  
\begin{equation}
F_{i}(x,Q^2) = F_{i}^{\tau = 2}(x,Q^2)
+ {H_{i}^{\tau = 4}(x) \over Q^2} 
+ {H_{i}^{\tau = 6}(x) \over Q^4} + .....   \;\;\; i=1,2,3, 
\label{eqn:ht}
\end{equation}
where the first term ($\tau=2$) is known as the twist-two or leading twist (LT) term, 
and corresponds to the scattering off a free quark. 
This term is expressed in terms of PDFs and is responsible for the evolution 
of structure functions via perturbative QCD $\alpha_s(Q^2)$ corrections. 
The higher twist (HT) terms with $\tau = 4,6$,\ldots
reflect the strength of multi-parton correlations ($qq$ and $qg$). 
The HT corrections spoil the QCD factorization, so one 
has to consider their impact on the PDFs extracted in the analysis of low-$Q$ data.  
Due to their nonperturbative origin, current models can only provide a 
qualitative description for such contributions, which are usually determined via
reasonable assumptions from data~\cite{Alekhin:2013nda,Accardi:2016qay}.   

%%%%%%%%%%%%%%%%%%%%%%%%%%%%%%%%%%%%%%%
\begin{figure}[tbp]
\begin{center}
\includegraphics[scale=1.00]{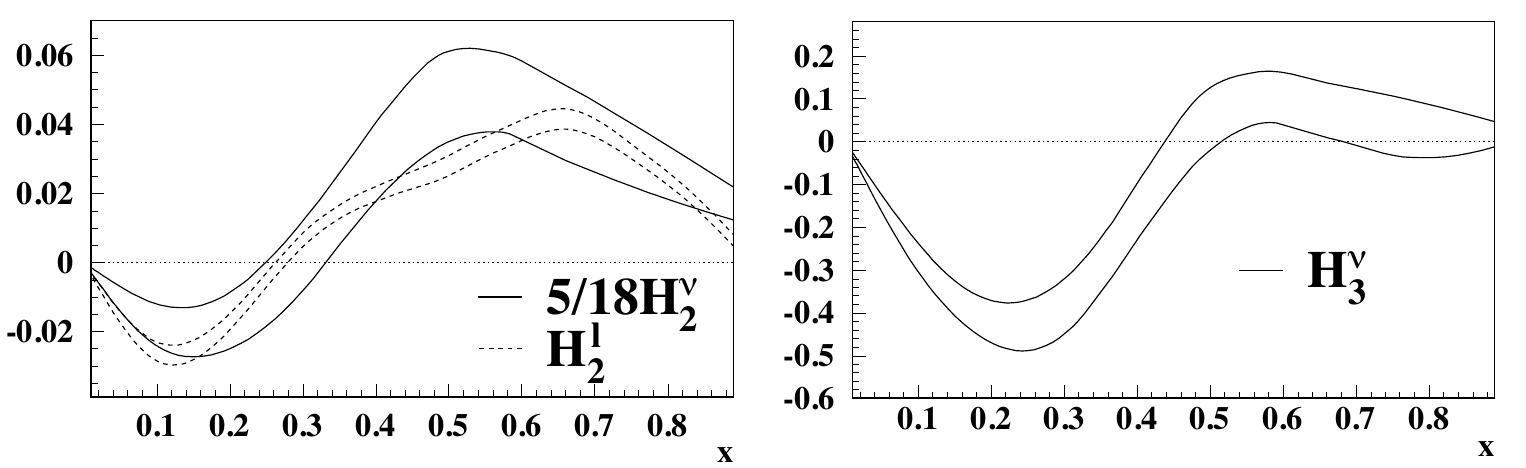}
\end{center}
\caption{High twist terms $H_2^{\tau=2}(x)$ determined from global QCD fits to charged-lepton and (anti)neutrino DIS data. Figure adapted from Ref.~\cite{Alekhin:2007fh}.
  }
\label{fig:HTs}
\end{figure}
%%%%%%%%%%%%%%%%%%%%%%%%%%%%%%%%%%%%%%%

In addition to the dynamical HT terms defined in Eq. (\ref{eqn:ht}), we also have 
kinematic HT contributions associated with the finite mass of the target nucleon $M_N$, 
which are mostly relevant when $x^2M_N^2/Q^2$ is large. The corresponding target mass 
corrections (TMC) involving powers of $1/Q^2$ are usually incorporated into the LT 
term following the prescription of Refs.~\cite{Nachtmann:1973mr,Georgi:1976ve}. For a discussion of the impact of TMC see also Ref.~\cite{Steffens:2012jx}. 

Existing information about dynamical HT terms in lepton-nucleon structure functions is 
scarce and somewhat controversial. Early analyses~\cite{Miramontes:1989ni,Whitlow:1990gk} 
suggested a significant HT contribution to the longitudinal SF $F_L$. The subsequent 
studies with both charged leptons~\cite{Virchaux:1991jc,Yang:1999xg,Alekhin:2002fv} and 
neutrinos~\cite{Kataev:1999bp} raised the question 
of a possible dependence on the order of QCD calculation used for the leading twist.  
More recent HT studies~\cite{Alekhin:2007fh} including both charged lepton and neutrino/antineutrino 
DIS data indicated that dynamic HT corrections affect the region of $Q^2 < 10$ GeV$^2$ and are 
largely independent from the order of the QCD calculation. 
Most notably, as shown in Fig.~\ref{fig:HTs}, the HT corrections to the $F_2$ and $F_T$ structure functions in neutrino/antineutrino DIS are consistent with the 
ones extracted from charged lepton DIS after a charge rescaling~\cite{Alekhin:2007fh}. 

An empirical approach to take into account the effects of both kinematic and dynamical HT 
corrections on structure functions~\cite{Bodek:2003wd} is often implemented in MC generators. 
This method is based upon LO structure functions (using GRV98 PDFs) in which the 
Bjorken variable $x$ is replaced by an adhoc scaling variable $\xi_w$ and all PDFs are 
modified by $Q$-dependent $K$ factors. The free parameters in the $\xi_w$ variable and 
in the $K$ factors are fitted to existing data. 

An extrapolation of the HT terms on DIS structure functions to the transition 
and resonance region results in sizable corrections at low invariant masses 
$W<1.9$ GeV. However, the verification of QH duality (Sec.~\ref{sec:DIS:QHduality}) 
at JLab implies a suppression of additional HTs with respect to the average DIS behavior, 
down to low $Q^2\sim 1$ GeV$^2$~\cite{Niculescu:2000tk}. 

It is worth noting that the transition from the high $Q^2$ behavior of structure functions, 
well described in terms of perturbative QCD at leading twist, to the asymptotic limit for $Q^2\to 0$ defined by current conservation arguments, is largely controlled by the HT contributions. 
In this respect (anti)neutrino interactions are different with respect to charged leptons, 
due to the presence of an axial-vector current dominating the cross sections at low $Q^2$. 
The effect of the Partially Conserved Axial Current (PCAC)~\cite{Adler:1964yx,Kopeliovich:2004px} 
in this transition region can be formally considered as an additional HT contribution 
and can be described with phenomenological form factors~\cite{Kulagin:2007ju}. 
In the limit $Q^2 \to 0$ for both charged leptons and neutrino scattering $F_T \propto Q^2$, while $F_L \propto Q^4$ in the electromagnetic current and is dominated by the finite PCAC contribution in the weak current. As a result, the ratio $R=F_L/F_T$ has a very different behavior in neutrino scattering at small $Q^2$ values~\cite{Kulagin:2007ju} and this fact must be considered in the extraction of (anti)neutrino structure functions from the measured differential cross-sections.

\subsubsection{Hadronization} 
\label{sec:DIS:hadr} 

The formation of hadrons in inelastic interactions is characterized by nonperturbative fragmentation 
functions (FF), which in an infinite momentum frame can be interpreted as probability distributions 
to produce a specific hadron of type $h$ with a fraction $z$ of the longitudinal momentum 
of the scattered parton. These universal fragmentation functions can not be easily calculated 
but can be determined phenomenologically from the analysis of high-energy scattering data. 
A recent study of $\pi$ and $K$ FF in $e^+e^-$ collisions can be found in Ref.~\cite{Sato:2016wqj}. 
The FF for charmed hadrons ($D,D_s,\Lambda_c$) in neutrino DIS interactions was 
studied in Ref.~\cite{Samoylov:2013xoa}. 

Modern event generators often use the LUND string fragmentation model, as implemented 
in the PYTHIA/JETSET packages, to describe the hadronization process. 
This model results in a chain like production of hadrons with local compensation of 
quantum numbers. The original partons are associated with the endpoints of a massless relativistic string to approximate a linearly confining color flux tube, while gluons are associated with 
energy and momentum carrying kinks on the string. 
The production rate of the created 
$q\bar q$ pairs leads to a Gaussian spectrum of the transverse momentum $p^2_\perp$ 
for the produced hadron, while an associated FF provides the probability that a 
given ratio $z$ between the hadron energy and the energy transfer is selected. 
The PYTHIA/JETSET implementation of this model is controlled by many free parameters, 
which can be tuned to describe the data. A detailed study of the PYTHIA fragmentation 
parameters with neutrino data~\cite{Kuzmin:2013tza} 
from proton and deuterium targets was performed in Ref.~\cite{Katori:2014fxa}. 
In particular, the various parameter sets determined by the HERMES experiment were 
used within the GENIE event generator obtaining predictions in agreement with the 
measured hadron multiplicities. An independent tuning of the JETSET fragmentation 
parameters was performed in Ref.~\cite{Chukanov16} with NOMAD data from 
exclusive strange hadron production and inclusive momentum and angular distributions 
in neutrino-carbon DIS interactions. It must be noted that in neutrino-nucleus interactions 
the hadrons originated from the primary vertex can re-interact inside the nucleus. 
Final state interactions must be therefore taken into account in the determination 
of the effective fragmentation parameters from the observed final state hadrons. 

At lower values of the invariant mass $W<3$ $GeV$ the LUND hadronization model 
deteriorates. A better description of the data can be achieved with  a phenomenological 
description of the hadronization process in which the average hadron multiplicities are 
parameterized as linear functions of $\log W$ for each channel. The Koba-Nielsen-Olesen (KNO) 
scaling law~\cite{Koba:1972ng} can then be used to relate the 
dispersion of the hadron multiplicities at different invariant masses with a 
universal scaling function parameterized in terms of the Levy function. 
Both the averaged hadron multiplicities and the KNO functions are usually tuned from 
neutrino bubble chamber data. 

The GENIE~\cite{Andreopoulos:2009rq} generator uses the hybrid AGKY 
approach~\cite{Yang:2009zx}, which has a gradual transition from the KNO hadronization 
model to PYTHIA in the region $2.3 \leq W \leq 3.0$ GeV and allows the average multiplicities to be continuous as a function of $W$. However, since PYTHIA underestimates the 
dispersions at low $W$ with respect to bubble chamber data, the AGKY model 
is characterized by some discontinuities of the topological cross-sections 
in the hadronization transition region. 
The NuWro~\cite{Juszczak:2005zs} generator tuned both the average multiplicities and the 
corresponding dispersions to the available bubble chamber data in order to achieve 
continuous topological cross-sections. 
The NEUT~\cite{Hayato:2009zz} generator has a more abrupt transition for the hadronization process, 
using KNO for $W<2$ GeV and PYTHIA for $W>2$ GeV. Similarly to GENIE and NuWro, 
the average hadron multiplicities and dispersions are tuned from bubble chamber data. 

It must be noted that all generators effectively use the hadronization models within the DIS 
formalism to produce non-resonant mesons in the resonance region. This mechanism provides 
the main contribution for multi-meson production in the resonance region, while resonance models focus on single meson production. 

\subsection{Inelastic Scattering off Nuclei} 

\subsubsection{Nuclear Modifications of Structure Functions} 
\label{sec:DIS:NPDFs} 

In order to collect high statistics samples, neutrino experiments typically use massive nuclear targets, 
which are particularly critical in long-baseline oscillation experiments because of the reduced
flux at the far detector. This fact requires an understanding of the structure and interactions of hadrons inside the nuclear targets, in which nuclear medium effects like 
 Fermi motion, Pauli blocking, strong nucleon-nucleon interactions, meson cloud contributions, 
 final state interactions, etc.\ play important roles in different regions of the 
 Bjorken scaling variable $x$ and momentum transfer square $Q^2$.  
 
While several microscopic models for the dynamics of nucleons in the nuclear medium 
have been applied to electromagnetic interactions, only a few studies are available 
for weak interactions. The Kulagin-Petti (KP) model~\cite{Kulagin:2004ie,Kulagin:2007ju,Kulagin:2010gd,Kulagin:2014vsa,Ru:2016wfx,Alekhin:2016giu} 
incorporates several mechanisms of nuclear modifications of structure functions and 
parton distributions functions, including smearing with the spectral function describing 
the energy-momentum distribution of bound nucleons (Fermi motion and binding), 
an off-shell correction for bound nucleons, contributions from meson exchange 
currents and the coherent propagation of the hadronic component
of the virtual intermediate boson in the nuclear environment (nuclear shadowing).
The model of Refs.~\cite{SajjadAthar:2009cr,Haider:2011qs,Haider:2012nf,Haider:2012ic,Haider:2016zrk,Haider:2014iia} includes the nuclear effects related to the 
Fermi motion and binding, the meson exchange currents, and the coherent processes 
responsible of the nuclear shadowing. 

\begin{figure}[tbp]
\begin{center}
\includegraphics[scale=1.15]{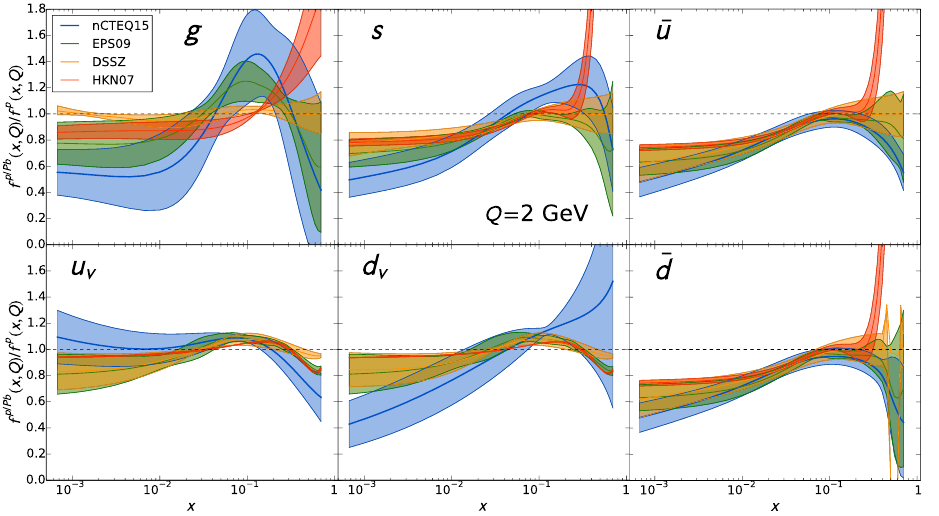}
\end{center}
\caption{Nuclear modification factors defined as the ratios of NPDFs with respect to the corresponding proton PDFs.
The uncertainty bands obtained from different global analyses are shown for lead at $Q^2=4$ GeV$^2$. 
Figure adapted from Ref.~\cite{Kovarik:2015cma}. 
  }
\label{fig:NPDFs} 
\end{figure}

A phenomenological approach is often used to parameterize the 
nuclear modifications of structure functions in terms of nuclear parton 
distributions functions (NPDFs), which are conventionally extracted from global 
QCD fits to nuclear data including DIS, Drell-Yan (DY) production, and 
heavy ion collisions at colliders. To this end, two different procedures are available 
in literature. The first one assumes a given set of free proton PDFs as input and 
introduces separate nuclear correction factors $R_i^A(x,Q_0)$ for each proton 
PDF of flavor $i=u,d,s,c,..$ in the nucleus $A$. These factors are parameterized and 
determined from the global QCD fits. This approach is followed by groups like 
HKN~\cite{Hirai:2007sx}, EPS~\cite{Eskola:2009uj}, DSSZ\cite{deFlorian:2011fp}, 
KA~\cite{Khanpour:2016pph}, etc. A second approach is followed by the nCTEQ 
group~\cite{Kovarik:2015cma}, which is performing a native QCD fit for 
nPDFs without assuming fixed proton PDFs as input. 
Figure~\ref{fig:NPDFs} illustrates the nuclear modification factors and their 
uncertainties obtained by different groups.   
It must be noted that the nuclear structure functions for the (anti)neutrino-nucleus DIS 
are not simply a combination of NPDFs, as discussed in the following. Furthermore, 
the unresolved discrepancies reported between charged lepton and (anti)neutrino scattering 
data limit the applicability of NPDFs to the latter.  

%%%%%%%%%%%%%%%%%%%%%%%%%%%%%%%%%%%%%%%
%%%%%%%%%%%%%%%%%%%%%%%%%%%%%%%%%%%%%%%

As a result of the higher complexity of weak interactions with respect to 
electromagnetic ones, mainly due to the presence of the axial-vector current,  
significant differences are expected in nuclear effects for charged lepton 
and (anti)neutrino DIS. In general, nuclear modifications of structure functions depend 
on the isospin of the target and on the $C$-parity and can therefore differ 
for neutrino and antineutrino interactions. It is also worth noting that at the typical 
$Q^2$ values accessible in (anti)neutrino inelastic scattering, high twist contributions 
play an important role, both at the nucleon and at the nuclear level. The kinematic regions 
mostly affected are low $x<0.1$ with the nuclear shadowing and large $x>0.5$ 
with the combination of TMC with the nuclear binding and off-shell effects. 
Further theoretical and experimental studies are needed for both HT contributions 
(Sec.~\ref{sec:DIS:HT} ) and nuclear effects.

\subsubsection{Final State Interactions} 
\label{sec:FSI} 

In experiments with nuclear targets the hadrons originated from the primary interaction vertex 
may rescatter inside the target nucleus producing additional hadrons, 
knocking-out other nucleons, or even be absorbed inside the nucleus.
These final state interactions (FSI) can occur uniformly throughout the nuclear volume 
and can mask the primary neutrino interaction, smearing the visible hadron 
multiplicities and neutrino energy. This makes the determination of fragmentation functions 
from neutrino-nucleus interactions challenging and dependent on the nuclear models. 

A few different approaches are available to describe the effect of FSI in neutrino 
interactions~\cite{Dytman:2009zza}. They typically use a intranuclear cascade (INC) model 
based upon the assumption that the interactions in the nuclear medium can be 
described by the corresponding hadron-nucleon cross sections. 
Hadron interactions are located inside the nuclear volume according 
to probability distributions and the outgoing particles produced in each interactions are 
then propagated through the nucleus within the same framework. An alternative approach 
available in GENIE~\cite{Andreopoulos:2009rq} uses hadron-nucleus 
cross sections with selected final state particles, like taking meson formation lengths into account, 
thus avoiding a complete nuclear cascade. The FLUKA~\cite{Battistoni:2006da} 
and DPMJET~\cite{Ranft:1994fd,Ranft:1999qe}  
generators use a sophisticated INC model taking into account quantum mechanical effects 
like the coherence time, the effects of the nuclear potential between scatterings, etc. 
The development of the intranuclear cascade is usually controlled by a formation zone 
for the hadrons inside the nucleus, which can be determined from neutrino data of 
different types of interactions. 

The GiBUU framework~\cite{Buss:2011mx} takes into account FSI by solving 
the Boltzmann-Uehling-Uhlenbeck (BUU) 
equation, which provides a semiclassical description of the particle propagation through 
the nuclear medium. It describes the dynamical evolution of the phase space density 
for each particle species under the influence of the mean field potential for the initial 
nucleus state. The GiBUU transport model has similar assumptions for the hadron-nucleon 
cross sections as the INC models but takes into account nuclear medium effects.

\subsection{Experimental Measurements} 

Experimental measurements in the DIS kinematic region require (anti)neutrino beams 
of relatively high energies, several GeV and higher. Although historically the study of this 
region was one of the primary goals of early experiments, the focus on the measure of 
 neutrino-oscillation parameters in modern experiments tends to emphasize 
lower neutrino energies. For this reason, the opportunities to explore the DIS region in 
current and planned experiments are somewhat limited. It is worth noting that the 
understanding of the inelastic region is important for long-baseline oscillation experiments. 
For instance, in the future DUNE experiment~\cite{Acciarri:2015uup} more than 30\% of the 
interactions will be in the DIS region and more than 40\% in the resonance and transition region.  

The use of proton and deuterium targets in combination with both neutrino and 
antineutrino beams offers an ideal tool to probe electroweak interactions and the structure 
of the nucleon. The flavor separation offered by the weak charged current allows a direct 
access to different structure functions and parton distributions inside the nucleon. 
However, the only available data from 
(anti)neutrino DIS off proton and deuterium still comes from the early bubble chamber 
experiments ANL~\cite{Barish:1978pj}, BNL~\cite{Baker:1982ty}, BEBC~\cite{Colley:1979rt,Allasia:1985hw}, and FNAL. In spite of the excellent experimental resolution of 
these bubble chamber measurements, the overall statistics is rather limited and 
totally insufficient for modern needs (e.g. only about 9,000 $\bar \nu$ and 5,000 $\nu$ events 
were collected by BEBC on hydrogen~\cite{Allasia:1985hw}). 
There is a growing voice for new high-statistics measurements of (anti)neutrino interactions 
off hydrogen and deuterium within the community.

Measurements from heavy nuclear targets are more abundant but are often limited by the 
experimental granularity and resolution. Some of the existing higher statistics measurements 
also provide somewhat conflicting results. Early bubble 
chamber measurements (ANL, BNL, BEBC, and FNAL) also took data with heavy nuclei 
like neon, propane and freon. The first high statistics measurements (${\cal{O}}(10^7)$ events) 
were performed by relatively coarse detectors like CDHS (iron)~\cite{Berge:1987zw,Berge:1989hr} 
and CHARM/CHARM II (marble/glass)~\cite{DeWinter:1989zg} mostly based upon large 
passive nuclear targets. The CCFR~\cite{Seligman:1997fe,Bodek:2000pj} and 
NuTeV~\cite{Tzanov:2005kr,Goncharov:2001qe} 
experiments (iron) are based upon the same technique and can be considered 
the first modern experiments. The E531~\cite{Ushida:1988rt} and CHORUS~\cite{KayisTopaksu:2011mx} experiments performed high 
resolution measurements of neutrino interactions (most notably charm production) in 
nuclear emulsions with $\langle A \rangle \sim 80$. The CHORUS experiment also 
performed cross section measurements using the lead calorimeter as a target~\cite{Onengut:2005kv}. 
The NOMAD experiment provides high resolution measurements from carbon and iron 
targets~\cite{Wu:2007ab,Samoylov:2013xoa}. The MINOS experiment performed cross section measurements in iron~\cite{Adamson:2009ju}, albeit with 
somewhat limited experimental resolution. More recently, the MINER$\nu$A experiment 
has measured CC induced $\nu$-A DIS cross sections on polystyrene, graphite, 
iron and lead targets\cite{DeVan:2016rkm,Mousseau:2016sn}. 

Most of the experimental measurements from heavy targets are related to inclusive 
$\nu$ and $\bar \nu$ cross sections or to exclusive studies of particle production 
and multiplicities. Very limited information is currently available on nuclear modifications 
of cross sections and structure functions in (anti)neutrino inelastic interactions. 
The first measurement of nuclear effects was performed by BEBC from the ratio 
of neon and deuterium targets~\cite{Allport:1989vf}, providing evidence of nuclear shadowing at small 
$Q^2$ values. The MINER$\nu$A experiment has recently presented the results of the differential scattering cross section in the form of ratios
 $\frac{d\sigma^i}{dx}/\frac{d\sigma^{CH}}{dx}$,~i=C, Fe, and Pb~\cite{Mousseau:2016sn}.

\subsection{Comparisons between Models and Measurements} 

Experimental measurements of inelastic cross sections are limited and somewhat 
contradictory. The total cross section $\sigma(E)$ was measured with good accuracy by 
CDHS~\cite{Berge:1987zw}, CCFR~\cite{Seligman:1997fe}, and NuTeV~\cite{Tzanov:2005kr} at high energies, resulting in a combined normalization 
uncertainty of 2.1\% on $\sigma(E)/E$ for $E>40$ GeV. The recent measurements 
by NOMAD~\cite{Wu:2007ab}, MINOS~\cite{Adamson:2009ju} and MINER$\nu$A~\cite{DeVan:2016rkm} achieved good precisions down to $E\sim 4$ GeV. However, 
for $E<4$ GeV large uncertainties are still present, especially for anti-neutrino 
scattering, which has being plagued by scarce measurements. Available models tend to 
describe well the total cross sections. We note that partial cancellations of nuclear effects 
on the total cross sections are expected as a result of DIS sum rules.  

%%%%%%%%%%%%%%%%%%%%%%%%%%%%%%%%%%%%%%%
\begin{figure}[tbp]
\begin{center}
%\hspace*{-0.60cm}\includegraphics[scale=0.95]{NuTeV-CCFR-xsec.pdf}\includegraphics[scale=0.95]{F2nu-vs-F2cl.pdf}
\hspace*{-0.60cm}\includegraphics[scale=1.00]{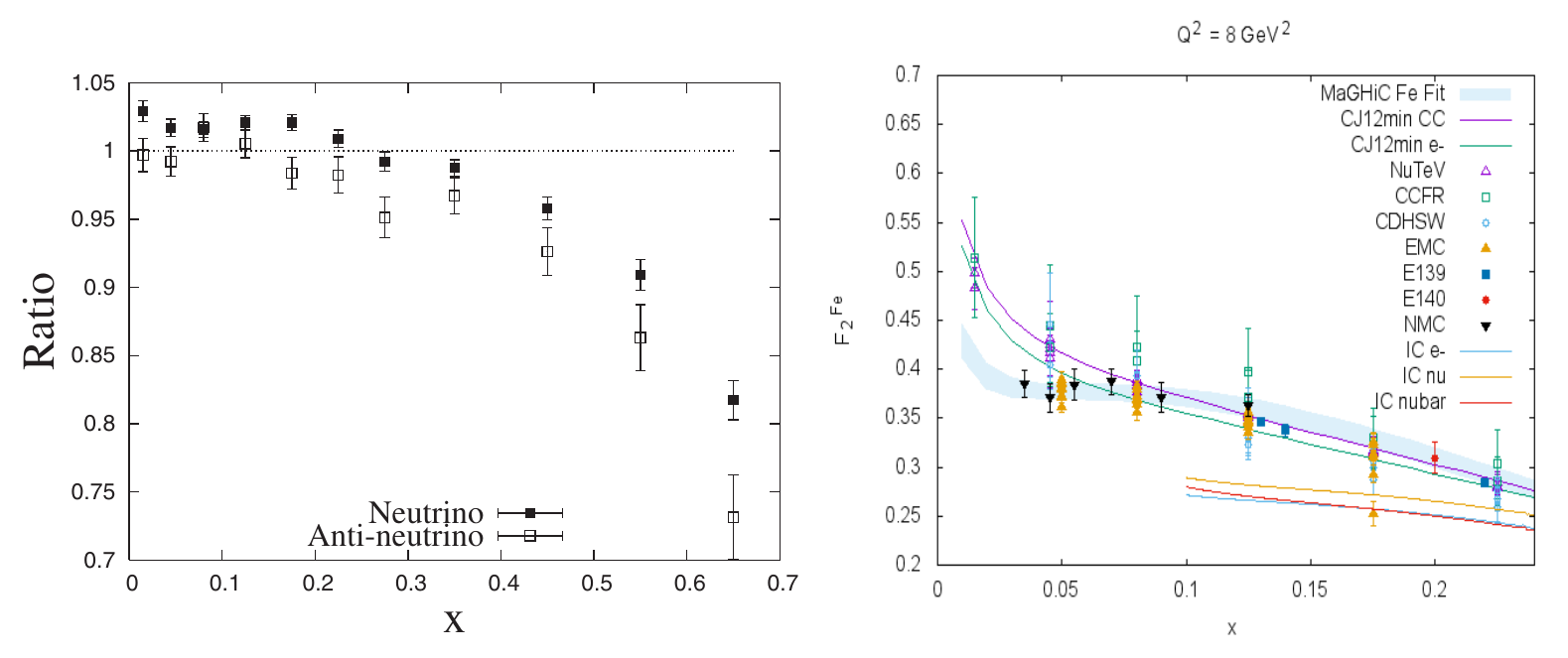} 
\end{center}
\caption{Left panel: Ratio between the NuTeV and CCFR measurements of the (anti)neutrino differential cross-sections on Fe target. Each $x$ point is averaged over all available measurements in different bins of $E$ and $y$. Figure from Ref.~\cite{Tzanov:2005kr}. Right panel: Comparison between the values of the structure function $F_2$ determined from (anti)neutrino and charged lepton DIS on 
an Fe target. The neutrino data are scaled by 5/18 to account for the quark charges. 
Figure from Ref.~\cite{Kalantarians:2017mkj}. 
} 
\label{fig:R-FE-nu} 
\end{figure}
%%%%%%%%%%%%%%%%%%%%%%%%%%%%%%%%%%%%%%%

The current understanding of the double differential cross sections $d \sigma / dxdy$ is less clear. 
The most recent measurements from CCFR (Fe)~\cite{Bodek:2000pj}, NuTeV (Fe)~\cite{Tzanov:2005kr} and CHORUS (Pb)~\cite{Onengut:2005kv} indicate 
tensions among different data sets, albeit the latter experiment uses a different nuclear target. 
In particular, while the NuTeV and CCFR measurements agree for $x~\le~0.4$, for $x>0.5$ the NuTeV data show an excess up to 20\% above the CCFR results (Fig.~\ref{fig:R-FE-nu} ). 
Available models are roughly in agreement with CCFR and CHORUS at large $x$ values,  
but can not fully explain the excess observed in NuTeV data~\cite{Kulagin:2007ju,Haider:2011qs}. 
In addition, the data sets from all available experiments consistently suggest that in 
the small $x<0.05$ region (anti)neutrino cross sections are significantly higher than 
predictions obtained by a simple re-scaling of the charged lepton cross sections.
The analysis by the nCTEQ~\cite{Kovarik:2010uv} group showed that the existing $\nu A$ and $l^\pm A$ DIS data prefer different nuclear correction factors. Possible explanations include unexpectedly large HT effects, or even non-universal nuclear effects~\cite{Kovarik:2010uv}. This result has implications for the extraction of both nuclear and proton PDFs using combined (anti)neutrino and charged-lepton data. 
The HKN~\cite{Nakamura:2016cnn} group also finds some inconsistencies between 
(anti)neutrino and charged-lepton data. 
The analysis performed by the EPS group ~\cite{Paukkunen:2013grz} using different statistical methods suggest that 
the $\nu A$ and $l^\pm A$ DIS data can be statistically consistent and relates the 
discrepancies to possible 
energy-dependent fluctuations. 
Similar results are obtained by the 
DSSZ group~\cite{deFlorian:2011fp}.
The available measurements of the $F_2$ and $xF_3$ structure functions from 
CCFR, NuTeV, and CHORUS are characterized by the same issues observed in the 
differential cross sections. However, since only cross-sections are directly observable 
experimentally, the structure function measurements require some model-dependent 
assumptions. Figure~\ref{fig:R-FE-nu} illustrate the differences observed between (anti)neutrino and charged lepton scattering for the structure function $F_2$ in an Fe target~\cite{Kalantarians:2017mkj}. 
 
\begin{figure}[tbp]
\begin{center}
\includegraphics[scale=0.45]{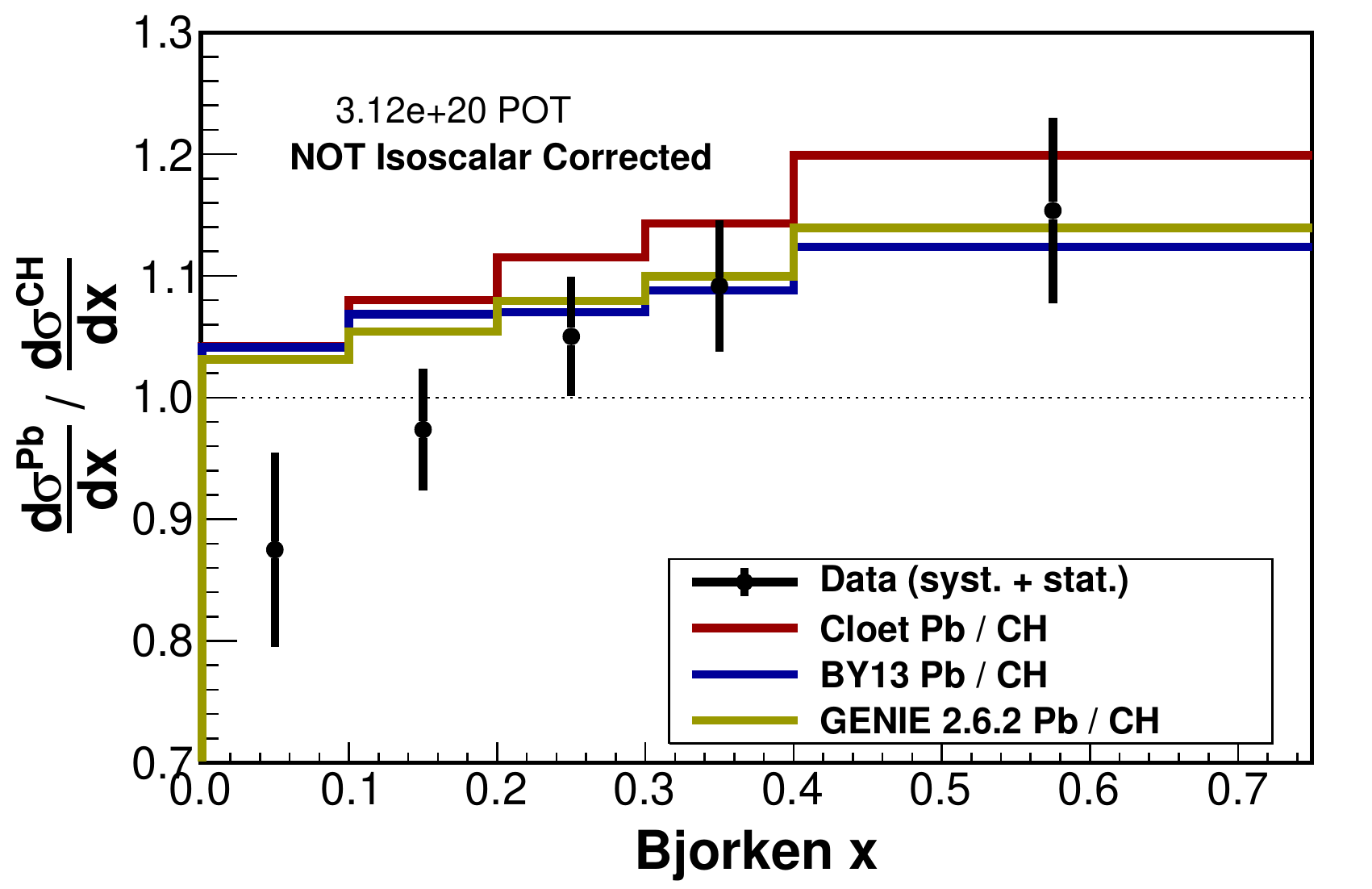}\hspace*{-0.30cm}\includegraphics[scale=0.45]{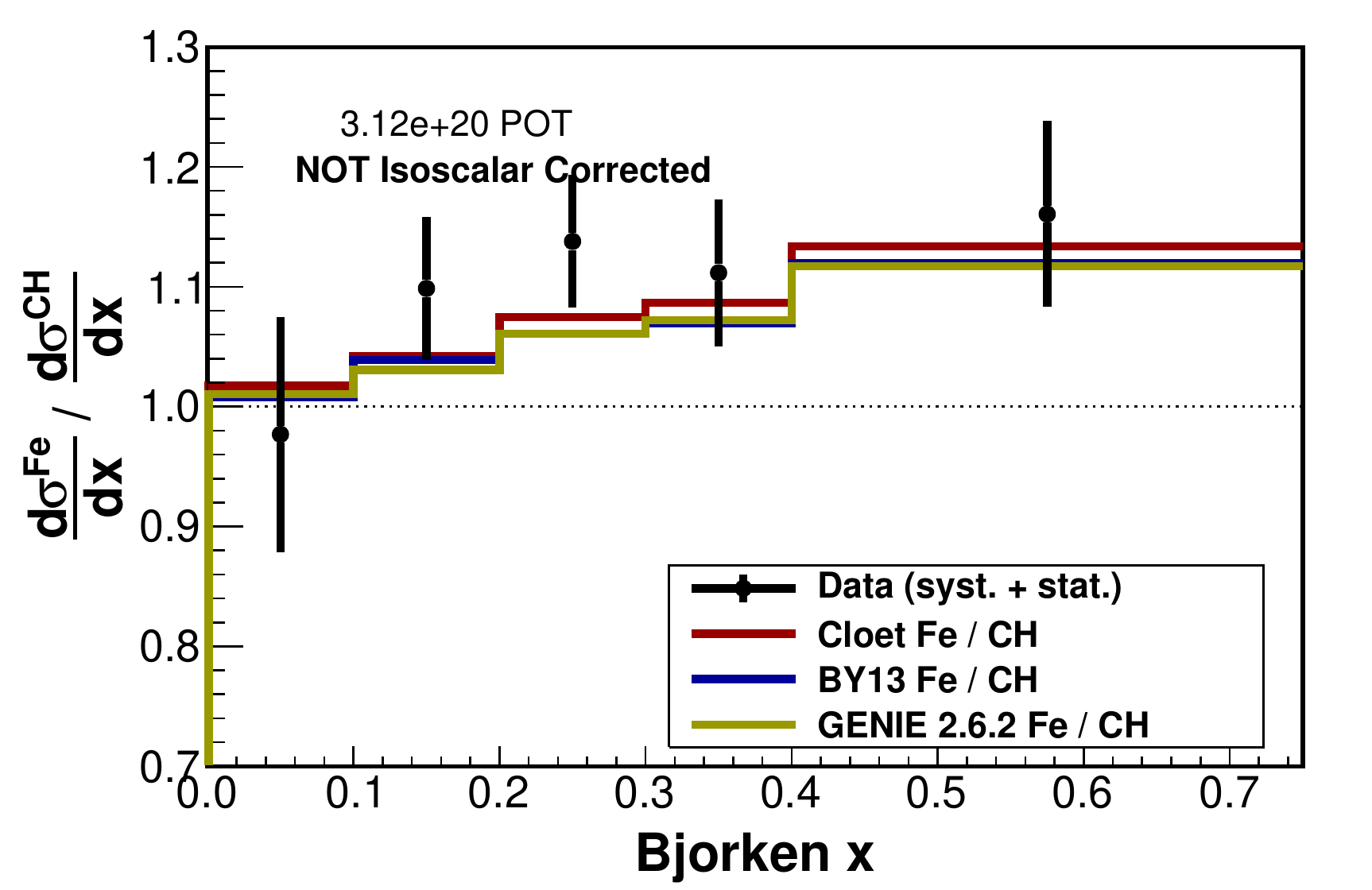}
\end{center}
\caption{Ratio of neutrino cross-sections $\frac{d\sigma^i}{dx}/\frac{d\sigma^{CH}}{dx}$ with $i=$Pb and Fe, measured by the MINER$\nu$A experiment, compared to different models and Monte Carlo simulations. Figure adapted from Ref.~\cite{Mousseau:2016sn}.
  }
\label{fig:PbCH-Minerva}
\end{figure}

The direct measurements of nuclear effects in neutrino inelastic scattering
from the BEBC and MINER$\nu$A experiments provide inconsistent results.
The BEBC data show evidence~\cite{Allport:1989vf} for the presence of nuclear shadowing at small
$x$ values, which is roughly in accord with the expectations. However, the excess
observed at small $x$ in the differential cross sections measured by NuTeV~\cite{Tzanov:2005kr}
and CHORUS~\cite{Onengut:2005kv} may indicate a somewhat
reduced shadowing correction with respect to charged leptons (Fig.~\ref{fig:R-FE-nu}). 
The MINER$\nu$A measurements~\cite{Mousseau:2016sn} of cross section
ratios off different nuclear targets instead suggest a more pronounced shadowing
in the lead target (Fig.~\ref{fig:PbCH-Minerva}). The results from MINER$\nu$A are not consistent with the GENIE
MC generator, based upon the Bodek-Yang model, but are consistent with the hypothesis
that the coherence length of the axial-vector current is different than the vector current~\cite{Kopeliovich:2012kw}.
In order to clarify the existing discrepancies higher precision measurements are needed.

In general, inelastic cross sections are much better understood at high $Q^2$ 
than at relatively low $Q^2$ and $W$. This latter region is characterized by an interplay 
between HT and nuclear corrections. Existing data are scarce and, if available, contradictory. 
Since current and future neutrino oscillation experiments are predominately in this 
low $Q^2$ and $W$ region, more experimental and theoretical studies of this region are needed. 

%%%%%%%%%%%%%%%%%%%%%%%%%%%%%%%%%%%%%%%
%%%%%%%%%%%%%%%%%%%%%%%%%%%%%%%%%%%%%%%

\subsection{Challenges} 

\subsubsection{Modeling Issues} 
\begin{itemize}
\item Optimize the description of the transition region from DIS to resonance production and define the kinematic limits of applicability of the DIS formalism for structure functions and cross-sections. 
\item Study the QH duality for neutrino and antineutrino interactions as a function of the 
isospin of the target. 
\item Study the impact of radiative corrections and their applicability in the transition region close to the kinematic limits of the parton treatment. 
\item Study outgoing charged-lepton mass terms and cross-section for $\nu_\tau$ CC interactions. 
\item Study the impact of the structure functions $F_L$ and $R=F_L/F_T$ on (anti)neutrino cross-sections and violations of the Callan-Gross relation. 
\item Study the role of the PCAC contributions to structure functions and cross-sections at low and moderate $Q^2$. 
\item Quantify the HT contributions to the different structure functions $F_2, xF_3, F_T, F_L$ and 
comparisons with the corresponding HT terms in electromagnetic interactions. 
\item Improve hadronization models in modern generators in order to describe exclusive hadron production at all W values. 
 \item Study the interplay of the various nuclear effects (Fermi motion and nuclear binding, meson exchange currents, nuclear shadowing, off-shell effect, etc.) in different regions of the Bjorken $x$ and $Q^2$ for neutrino and 
 antineutrino interactions off bound protons and neutrons in nuclear targets. 
 \item Understand the differences in the nuclear effects for electromagnetic and weak DIS structure functions and cross-sections (e.g. for coherent nuclear effects at small $x$). 
 \item Study nuclear effects for different structure functions $F_2, xF_3, F_T, F_L$ and the role of nuclear HT contributions. 
 \item Understand the role of the nuclear medium on structure functions and parton distributions 
 (e.g., collective mode effects due to mesons and other particles, off-shell modifications of 
 bound nucleons, etc.). 
 \item Improve the description of the FSI and nuclear transport in nuclei. 
 \item Study of DIS sum rules and normalization constraints for different nuclear targets.
 \item Update SIS/DIS description in generators following the recent parton distributions and models available.
\end{itemize}

 \subsubsection{Experimental Issues} 

\begin{itemize}

 \item Need measurements of cross-sections (both total and differential) with neutrino and antineutrino beams on free proton and deuteron targets, since the earlier results from bubble chambers are limited by statistics. 
 
 \item Need new precise measurements of both neutrino and anti-neutrino differential and total 
 cross-sections off various nuclear targets. These measurements should have a wide 
 $x$ and $Q^2$ coverage, like the ones performed at Jefferson Laboratory using 
 charged lepton beams, in order to compare the structure functions $F_2$ and $xF_3$, 
 as well as the weak $F_2^{\nu(\bar \nu)}$ and the electromagnetic $F_2^{l^\pm}$.
 
\item Need model-independent measurements of nuclear effects on (anti)neutrino structure 
functions and cross-sections by comparing, within the same experiment, results from heavy 
nuclear targets with proton and deuteron targets in different regions $x$ and $Q^2$. 
  
 \item Need to perform detailed measurements in the transition region from DIS to resonance production ($1.5~\le~W~\le~2$GeV) at moderate and low $Q^2$ to clarify QH duality and the 
 role of HT contributions. 
 
 \item Need to perform detailed exclusive measurements of hadron production, multiplicities, angular and momentum distributions in order to constrain hadronization and FSI models. 
  
\item Need to clarify the inconsistent results from existing measurements 
(BEBC, MINER$\nu$A, NuTeV, CHORUS) of nuclear effects 
at small $x$ values and in particular differences between (anti)neutrinos and charged leptons 
for the nuclear shadowing effect. 

\item Need to clarify the discrepancies among existing measurements and between 
(anti)neu\-tri\-nos and charged leptons at large Bjorken $x$ values (e.g. NuTeV cross-sections). 

\item Measure $\nu_e$ CC vs. $\nu_\mu$ CC vs. $\nu_\tau$ CC and test of lepton universality. 

\end{itemize}

Since precise DIS measurements typically require medium to high energy (anti)neutrino beams, it is worth noting that the only opportunity to study experimentally this region in the near future is offered by the Fermilab neutrino program using the Main Injector. The MINER$\nu$A experiment is expected to perform measurements in the DIS and transition region off various nuclear targets addressing several topics listed above. The planned DUNE Near Detector complex can potentially cover most of the required measurements with unprecedented precision~\cite{Acciarri:2015uup}. In particular, measurements of neutrino and anti-neutrino scattering off free protons are planned. 

\newpage

\section{Coherent and Diffractive Scattering}
\label{cohdiff}

Coherent scattering refers to processes in which the final-state nucleus is left in its ground state, rather than in an excited one. The simplest example is coherent elastic neutrino-nucleus scattering (CE$\nu$NS), $\nu A\to \nu A$, which could be sensitive to non-standard neutrino interactions and is an irreducible background for many direct dark matter searches. Several experiments~\cite{Akimov:2015nza,Dutta:2015vwa,Aguilar-Arevalo:2016qen} have been proposed to observe for the first time and study this reaction. Lacking any direct impact on oscillation measurements, we do not consider this process any further. 

As discussed below, we focus on coherent production of mesons and photons, because they can mimic signal events for neutrino oscillations. The same applies to diffractive scattering, which has similar kinematics as the coherent one, but arise from forward scattering off a nucleon, with associated meson or photon emission.

\subsection{Basics of Coherent and Diffractive Processes}
\label{subsec:def}
 
\begin{figure}[b]
\centering
    \includegraphics[width=0.4\textwidth]{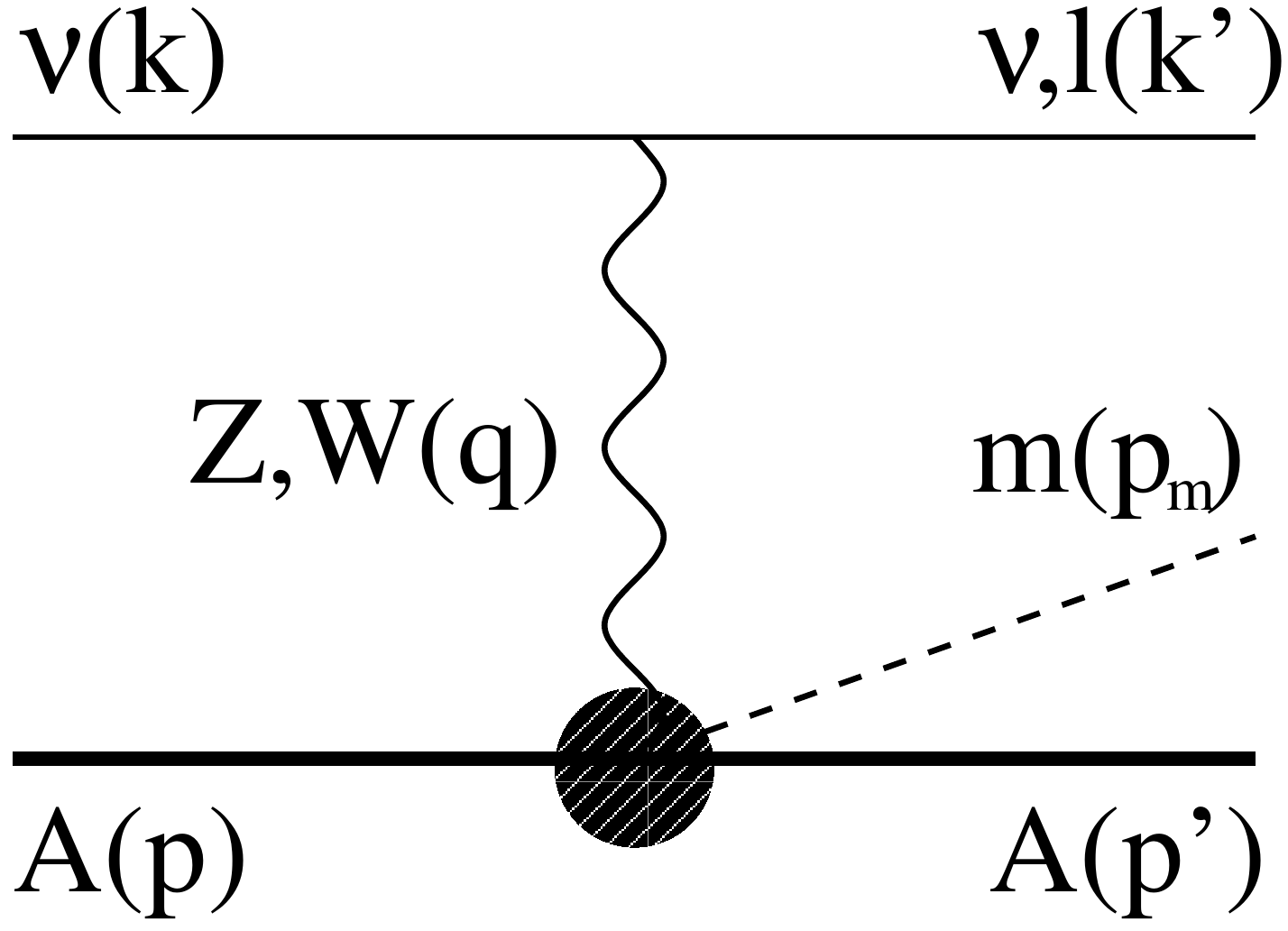} \hspace{0.1\textwidth}
    \includegraphics[width=0.3\columnwidth]{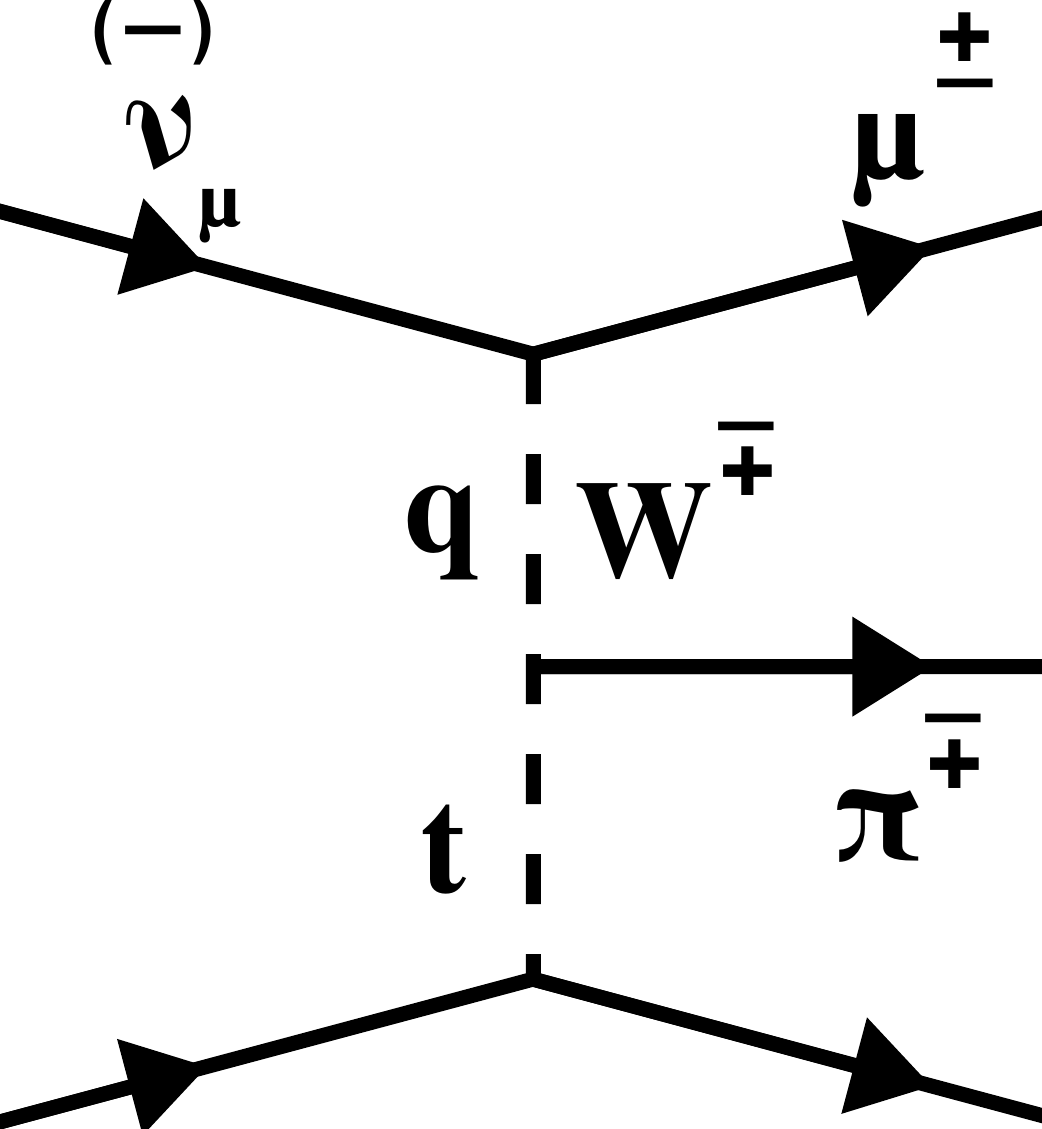} 
    \caption{Left panel: generic diagram for coherent particle production by neutrino-nucleus scattering.
    Four-momenta of the incoming neutrino~($k$) and nucleus~($p$), outgoing lepton~($k'$) and nucleus~($p'$), \ask{coherently} 
    produced particle ($p_m$), and the four-momentum transferred by the lepton~($q$) are indicated.  
    Right panel: diagram for coherent CC pion production highlighting $t\ask{=(p'-p)^2}$ as the square of the 4-momentum 
    transferred to the nucleus.} 
\label{fig:CohDiag}
\end{figure}

In CC interactions, charged mesons can be coherently produced 
\begin{align}
    \nu_l A &\to l^- m^+ A, \\ 
    \bar{\nu}_l A &\to l^+ m^- A ,
\end{align}
with $m^\pm = \pi^\pm$, $K^\pm$, $\rho^\pm, \ldots$, while in the NC case, one has
\begin{align}
    \nu_l A &\to  \nu_l m^0 A ,\\
    \bar{\nu}_l A &\to  \bar{\nu}_l m^0 A,
\end{align}
with $m^0 = \gamma$, $\pi^0$, $\rho^0,\ldots$.
The absence of tree-level flavor-changing NC suppresses production of neutral (anti)kaons or any other 
strange particles below any observable rate.  These processes and the kinematic variables associates with them are shown in Fig.~\ref{fig:CohDiag}.
The 4-momentum transferred by the leptons is denoted $q = k - k'$ as usual, while the one transfered to the nucleus is $p'- p = q - p_m$, so that $t=(p'-p)^2=(q-p_m)^2 = -2 M_A T_A$, where $M_A$ is the mass of the nucleus and $T_A$ its final-state kinetic energy in the laboratory frame.

Coherent reactions have smaller cross sections and are clearly more forwardly peaked than corresponding incoherent ones, where the final nucleus goes to any allowed excited state. Indeed, at large absolute values of $t$, the cross
section is significantly reduced by the nuclear form factor. Because small $|t|$ corresponds to negligible $T_A$, the energy of the outgoing particle, $p^0_{m}$, nearly coincides with the lepton transferred energy~$q^0$. Therefore, $t \approx -(\bm{q} - \bm{p}_{m})^2$. Taking into account that $|\bm{p}_{m}| \approx \sqrt{q_0^2 - m^2}$, one finds that small $|t|$ occur when (i) $\bm{q}$ and $\bm{p}_{m}$ are nearly parallel, (ii) $q_0\approx|\bm{q}|$, implying forward scattering with $q^2 \approx 0$, and (iii) the produced particle mass $m$ is small.
In fact, at low energies the ratio of coherent to incoherent kaon production cross sections is much smaller than the corresponding one for pions because of the relatively larger kaon mass~\cite{AlvarezRuso:2012fc}. The opposite holds for photon emission where the mass is zero. However, in this case, the amplitude squared cancels exactly at $q^2=0$ because of symmetry reasons, so that the largest differential cross section are found away from this optimal kinematics~\cite{Wang:2013wva}.

Particle production in the kinematic conditions described above can also take place in neutrino-nucleon scattering.
Since $t = (p' - p)^2 = - 2 m_N T_N$, with $m_N$ denoting the nucleon mass, low $t$ implies small kinetic energies for the outgoing nucleon, $T_N$. Nevertheless, these $T_N$ are larger than the corresponding ones in coherent particle production in nuclei due to the target mass difference. For this reason, outgoing protons can be experimentally detected. This scenario is often called diffractive scattering. In targets containing both hydrogen and heavier nuclei, such as water or scintillator materials, particle production by coherent scattering on nuclei and by diffractive scattering on protons coexist in the same kinematic regime.

\subsection{Relevance for oscillation experiments}
\label{subsec:osc}

A proper understanding of the coherent and diffractive processes is very important to the analysis of
neutrino oscillation experiments. In particular, such NC $\pi^0$ and $\gamma$ production can be important $\nu_\mu$-induced backgrounds to $\nu_\mu\to\nu_e$ oscillations because for some detection techniques, the electromagnetic shower of $\gamma$ or $\pi^0$  events can mimic the final-state electron in $\nu_e$ signal events. 

In the case of $\pi^0$ production, the misidentification can occur when the two photons from the $\pi^0\to \gamma\gamma$ decay are collinear or one of them is not detected. This might happen when the missed photon exits the detector before showering or does not have enough energy to initiate a shower.

The $\pi^0$ background to $\nu_\mu\to\nu_e$ can be significantly reduced with dedicated reconstruction algorithms~\cite{Abe:2013hdq}, While the smaller single-$\gamma$ background can also be greatly reduced in scintillator and LAr detectors (see, for example, Ref.~\cite{Wolcott:2015hda}) it remains irreducible in Cherenkov detectors. 
The number and distributions of coherent NC$\gamma$ events at the Super Kamiokande detector in the T2K experiment was calculated in Ref.~\cite{Wang:2015ivq}.
Coherent photon production driven by axial-anomaly-induced $Z\gamma\omega$ interactions was also suggested as an explanation for the MiniBooNE excess of events at low reconstructed neutrino energies, although subsequent theoretical work showed that it is not really so~\cite{Hill:2010zy,Zhang:2012xi,Rosner:2015fwa}.

Furthermore, in many experiments, single showers induced by coherent NC $\gamma$ emission can hardly be distinguished from those
coming from neutrino-electron elastic scattering, which is a reference process in neutrino physics.
On the other hand, in view of its relative simplicity, the use of coherent pion production (Coh$\pi$) as a \emph{standard candle} has
been considered to help constrain neutrino fluxes and neutrino-energy reconstruction for oscillation analyses.
However, such an ambitious goal requires consensus on the correct theoretical description of coherent scattering with an acceptable fit to experimental data.
We are not there yet.

Finally, the misidentification of CC coherent $\pi^+$s as protons distorts the reconstructed $E_{\nu}$ distribution in $\nu_\mu$ disappearance searches.
For laboratory pion energies $100~\text{MeV}\lesssim E_\pi \lesssim 500$~MeV, coherent CC $\pi^\pm$ production is largely dominated by $\Delta(1232)$ excitation.
Accurate data may then provide better constraints in the leading $N$-$\Delta$ axial transition coupling [known as $C_{5}^A(0)$ in the notation of Ref.~\cite{LlewellynSmith:1971zm}] and in-medium effects in $\Delta(1232)$ production which are crucial ingredients of pion production models as described in Sec.~\ref{1pi}.

\subsection{Theoretical status}
\label{subsec:theory}

Models for weak coherent scattering are usually labeled as either \emph{PCAC} or \emph{microscopic}. More details are given below.

\subsubsection{PCAC models of coherent particle production}
\label{subsubsec:PCAC}

Models of Coh$\pi$ based on the partial conservation of the axial current (PCAC)
take advantage of the fact that, at $q^2=0$, the Coh$\pi$ cross section can be related to pion-nucleus elastic scattering by a soft-pion theorem~\cite{Adler:1964yx}:
\begin{equation}
    \left.\frac{d\sigma}{dq^2dydt}\right|_{q^2=0} = r \frac{G_F^2 f_\pi^2}{2\pi^2} \frac{1-y}{y}
        \left.\frac{d\sigma}{dt}(\pi A \to \pi A\cut{_{gs}})\right|_{q^2=0,\,\omega_\pi=q^0},
\label{PCAC}
\end{equation}
where $y=q^0/E_\nu$,  $r=2 \,|V_{ud}|^2$ (1) for CC (NC) and $f_\pi = 92.4$~MeV is the pion decay constant. The above result in the CC case also neglects the final lepton mass, which is important at $E_\nu \lesssim 1$~GeV. Corrections to Eq.~(\ref{PCAC}) for nonzero lepton mass have been derived~\cite{Kopeliovich:1992ym,Rein:2006di,Paschos:2005km}.

Using this equation, Rein and Sehgal (RS) built a simple and elegant Coh$\pi^0 $ model using empirical information about pion-nucleon elastic and inelastic scattering~\cite{Rein:1982pf}.
A common issue of PCAC models is that the $q^2 =0$ approximation neglects terms in the cross section that vanish in this limit but not at finite $q^2$.
This leads to pion angular distributions that are too wide~\cite{Amaro:2008hd,Hernandez:2009vm}.
Nevertheless, the main problem of the RS model resides in its poor description of pion-nucleus elastic scattering (see Fig.~2 of Ref.~\cite{Hernandez:2009vm}). The work of Refs.~\cite{Berger:2008xs,Paschos:2009ag} offers a remedy by directly using experimental pion-nucleus elastic cross sections. Then, however, the off-shell dependence of the pion-nucleus amplitude is neglected: in Coh$\pi$ $q^2\lesssim0$, unlike  $m_\pi^2$ for real pions. The impact of this correction is not yet understood.

PCAC has also been applied to relate the axial-vector contribution to coherent NC$\gamma$ at $q^2 = 0$ to the $\pi^0 \,A \to \gamma
\, A$ differential cross section~\cite{Rein:1981ys}.
This however amounts only to the rather small longitudinal contribution (see Fig.~2 of Ref.~\cite{Rein:1981ys}). The majority of the cross section has to be calculated using model assumptions that are critically reviewed in Ref.~\cite{Hill:2010zy}.

\subsubsection{Microscopic models of coherent particle production}
\label{subsubsec:micro}

Microscopic models have been extensively developed for pion
production~\cite{Kelkar:1996iv,Singh:2006bm,AlvarezRuso:2007tt,AlvarezRuso:2007it,Amaro:2008hd,Leitner:2009ph,Nakamura:2009iq,Hernandez:2009vm,Zhang:2012xi}
and have recently become also available for photon~\cite{Zhang:2012xi,Wang:2013wva} and (anti)kaon emission~\cite{AlvarezRuso:2012fc}. These approaches start from particle production models on nucleons and perform a coherent sum over all nucleonic currents.
Modifications of the elementary amplitudes in the nuclear medium are also taken into account when pertinent. They are very important for the $\Delta(1232)$ resonance in pion and photon emission. In addition, pion and (anti)kaon outgoing wave functions are distorted inside the nuclei. This distortion is particularly strong in the case of few-hundred MeV pions, owing to the $\Delta(1232)$ presence in the pion-nucleus optical potential, and rather mild for kaons due to the absence of $K N$ resonances. A quantum treatment of the meson distortion is usually applied via the Klein-Gordon~\cite{AlvarezRuso:2007tt,Amaro:2008hd} or the
Lippmann-Schwinger~\cite{Nakamura:2009iq} equations although the semiclassical eikonal approximation has also been employed~\cite{Singh:2006bm,Zhang:2012xi}.
The nonlocality in the $\Delta$ propagation is neglected in most models although it might have a sizable impact on the cross section~\cite{Leitner:2009ph}. It has been partially implemented in Ref.~\cite{Nakamura:2009iq} for the $\Delta$ kinetic term. Although the mismatch between the non-local recoil effects and the local approximation might be minimized if the $\Delta$ selfenergy parameters are adjusted to describe pion-nucleus scattering data, the problem of nonlocality calls for further investigation.

These models comply with PCAC but do not critically rely on it. This feature makes validation with coherent pion photo and electroproduction data possible~\cite{Nakamura:2009iq}. The main challenge for microscopic models developed so far is that they are restricted to low energy transfers (where weak particle production models and meson optical potentials are mostly available). In the case of $\pi$ and $\gamma$, this is the region where the excitation of the $\Delta(1232)$ is dominant. In the case of $K^\pm$, the validity is in principle restricted to the threshold region, although for $K^+$, the absence of baryon resonances makes the extrapolation of the threshold model more reliable.

\subsubsection{Diffractive contribution to meson production}
\label{subsubsec:diff}

When small momentum is transferred to the nucleon in a neutrino-nucleon collision, the wavelength is large enough to see the nucleon as a whole. Such a kinematic scenario, which closely resembles coherent pion production on nuclei, is called diffractive or peripheral meson
production. Unlike the Coh$\pi$ case, the relatively small nucleon mass makes the outgoing nucleon experimentally detectable as it has been the case in MINER$\nu$A (see Sec.
\ref{subsec:diffpi}). Diffractive meson production is present for all available invariant masses of the final meson-nucleon system, $W_{\pi N}$. Actually, for pion production at threshold $W_{\pi N} = m_N + m_\pi$, the amplitude is fully determined by chiral symmetry (see
Sec.~\ref{1pi}). However, for $W_{\pi N} < 2$~GeV, the diffractive contribution will be masked by the dominant resonance excitation so it is more
easily identifiable at high $W_{\pi N}$. In the $q^2 \rightarrow 0$ limit, the nucleon version of the soft-pion theorem [Eq.~(\ref{PCAC})] can be used to relate diffractive meson production to meson-nucleon elastic scattering~\cite{Rein:1986cd}.

\subsection{Coherent and diffractive scattering in event generators}
\label{subsec:gen}

\begin{figure}[tb]
    \centering
    \includegraphics[width=0.7\textwidth]{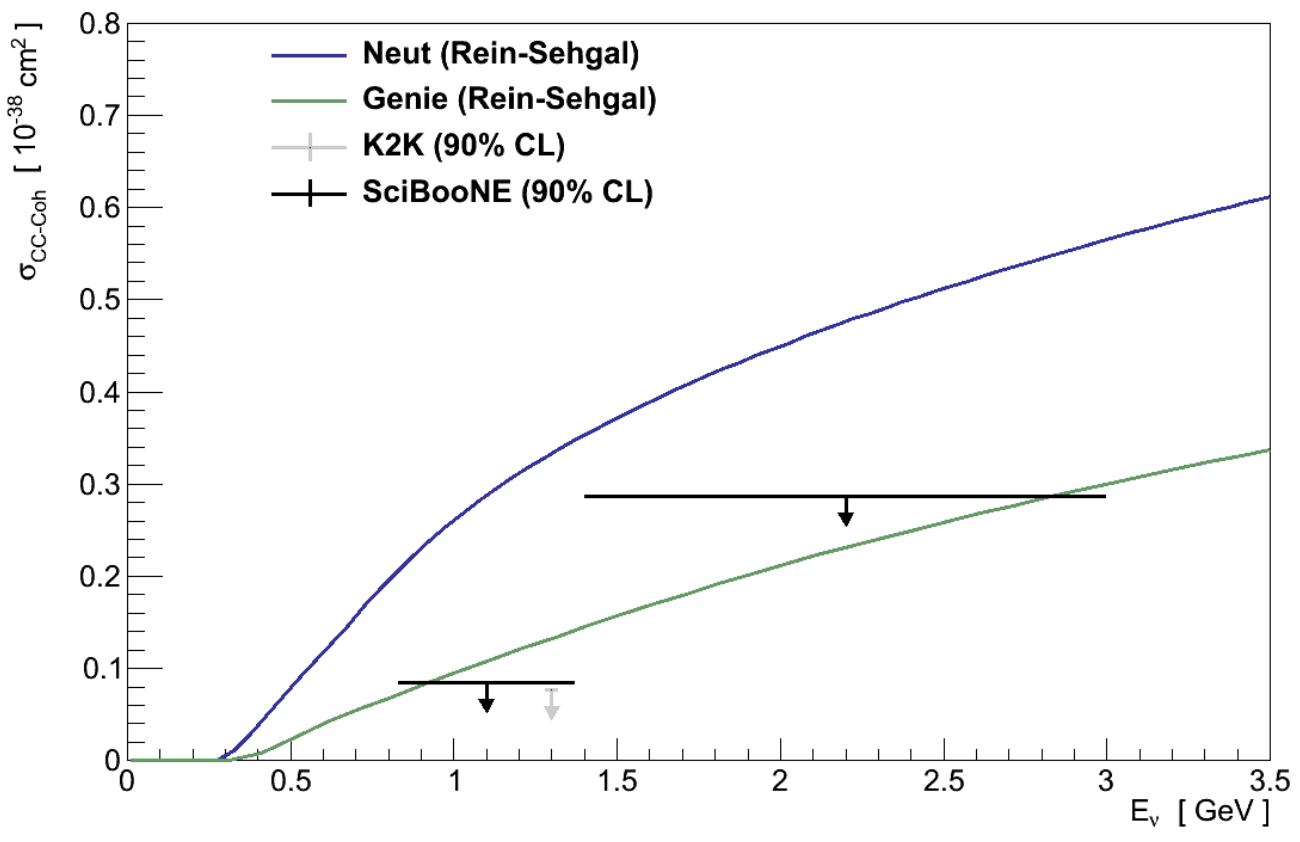}
    \caption{Differences in the cross section predictions of the Coh$\pi$ RS model within the NEUT and GENIE simulation programs as 
    a function of the neutrino energy (figure from Ref.~\cite{Scully:2013zba}).
    The predictions are compared to the 90\% CL upper limits 
    set for CC coherent pion production on $^{12}$C by the K2K~\cite{Hasegawa:2005td,Sanchez:2006hp,Tanaka:2006zm} and
    SciBooNE~\cite{Hiraide:2008eu,Hiraide:2009zz} experiments.}
    \label{fig_NvsG}
\end{figure}

At present, neutrino event generators simulate only coherent pion
production but not other coherent and diffractive processes. Owing to its simplicity, generators have implemented the RS model. Comparisons reveal, however, that the RS model has been interpreted differently within the particular neutrino event generators used
by different experiments. Figure~\ref{fig_NvsG} shows results from the GENIE and NEUT generators, together with upper limits from K2K~\cite{Hasegawa:2005td,Sanchez:2006hp,Tanaka:2006zm} and SciBooNE~\cite{Hiraide:2008eu,Hiraide:2009zz}.
As a way to remove this ambiguity, the model of Berger-Sehgal~\cite{Berger:2008xs} using pion-nucleus elastic scattering data has been implemented in GENIE and NuWRO.
However, such data are not available for all the targets of interest, particularly argon.
The nuclear target dependence of the Coh$\pi$ cross section is presently not well understood.
Albeit slow, a version of the microscopic model of Ref.~\cite{AlvarezRuso:2007tt}
has recently become available in GENIE.%
\footnote{For simplicity, the pion wave function is obtained in the eikonal approximation rather than by solving the Klein-Gordon 
equation as in the original paper.}
Such implementation has been used by T2K to compare to their Coh$\pi$ data~\cite{Abe:2016fic}.
A more complete comparison of experimental results with various theoretical descriptions of Coh$\pi$ is presented below in Sec.~\ref{subsec:compare}.

\subsection{Experimental Status: Coherent and Diffractive Meson Production}
\label{subsec:expt}

\subsubsection{Early experiments on CC and NC coherent pion production}
\label{subsubsec:earlyexpt}

Coherent pion production was first observed in early 1983 by the Aachen-Padova spark-chamber experiment~\cite{Faissner:1983ng} while studying isolated $\pi^0$s produced in their \numu and \numubar exposures.
This discovery was confirmed with a study performed by the Aachen Gargamelle group~\cite{Isiksal:1984vh} that isolated a sample of coherent NC $\pi^0$ events in the Gargamelle heavy Freon exposure.

Following these early discoveries, there were several \numu and \numubar experiments, CHARM~\cite{Bergsma:1985qy, Vilain:1993sf} and SKAT
\cite{Grabosch:1985mt, Nahnhauer:1986xh}, that observed NC Coh$\pi$ across a wide-range
of neutrino energies, nuclear targets and detection techniques.
The first observation of CC Coh$\pi$ was with $\langle E_{\nu}\rangle\approx7$~GeV by the SKAT experiment, followed by a series of measurements studying CC Coh$\pi$ including BEBC~\cite{Marage:1986cy, Allport:1988cq}, CHARM II~\cite{Vilain:1993sf} and
FNAL-E632~\cite{Aderholz:1988cs, Willocq:1992fv}, all of them with neutrino beams of $\langle E_{\nu} \rangle\geq 7$~GeV.

\subsubsection{More Recent Coherent Pion Production Experimental Results}
\label{subsubsec:modexpt}

The experimental search for \numu and \numubar Coh$\pi$ then lapsed for over a decade until the discovery of neutrino oscillations revitalized neutrino physics.
It is important to note that recent and current accelerator-based neutrino oscillation experiments require a low-energy neutrino beam with $E_{\nu} \lesssim 2.5$~GeV; it has been with these low-energy beams that the experimental study of Coh$\pi$ has continued.

This new generation of experiments started with the K2K~\cite{Hasegawa:2005td,Sanchez:2006hp,Tanaka:2006zm} search for CC Coh$\pi$
at $\langle E_{\nu} \rangle$ of $\simeq$ 1.3 GeV.
K2K found no evidence for CC Coh$\pi$ and could only set an upper limit on the cross section.
This surprising lack of CC Coh$\pi$ was later confirmed by the SciBooNE experiment~\cite{Hiraide:2008eu, Hiraide:2009zz}, which also
set upper cross-section limits with two different $\langle E_{\nu} \rangle\approx1.1$~GeV and 2.2~GeV; see Fig.~\ref{fig_NvsG}.

Searches for NC coherent pion production in the same energy range at SciBooNE~\cite{Kurimoto:2010rc} and MiniBooNE~\cite{AguilarArevalo:2008xs} experiments found evidence for this process. In addition, the NOMAD collaboration~\cite{Kullenberg:2009pu} provided a higher-energy (25 GeV) measurement of the NC Coh$\pi$ cross section.

%\subsubsection{Latest Coherent Pion Production Results ArgoNeuT, MINERvA and T2K}
There have been three more recent studies of CC Coh$\pi$ by MINERvA~\cite{Higuera:2014azj},
ArgoNeut~\cite{Acciarri:2014eit}, and T2K~\cite{Abe:2016fic}.
As opposed to the earlier K2K and SciBooNE analyses, these three experiments attempted to employ kinematical constraints coming directly from the dynamics of coherent pion production.
The ArgoNeut result in the NuMI beam at Fermilab, although with limited statistics of the order of 10 events each for \numu
($\langle E_{\nu} \rangle$ $\simeq$ 9.6 GeV) and \numubar ($\langle E_{\nu} \rangle$ $\simeq$ 3.6 GeV) was the first experiment to
detect Coh$\pi$ in a LAr TPC. The T2K experiment using a neutrino beam with $\langle E_{\nu} \rangle$ $\simeq$ 1.5 GeV is the first experiment to yield a signal
for this process at low $E_{\nu}$, in contrast to the null result previously obtained by T2K and SciBooNE. The MINERvA experiment used the NuMI wide-band neutrino beam and measured the energy dependent cross section from $E_{\nu}=1.5$--20~GeV, as well as the $\pi$ energy and angular distributions for both \numu and \numubar. The MINERvA experiment will be used to further illustrate the experimental technique for isolating the coherent signal.

\subsubsection{Experimental Isolation of the Coherent Pion Production Signal}
\label{subsubsec:expttech}

It is important to note that coherent pion production is only a small fraction of the total \numu and \numubar pion production cross section, which is dominated by resonant production  (Sec.~\ref{1pi}). To isolate the coherent signal, the two main kinematic characteristics that distinguish coherent from other pion production processe are used.
The nucleus remains intact so that there is no indication of nuclear breakup measured at the interaction point. To further ensure that the nucleus does not break up, the 4-momentum transfer to the nucleus $|t|$ must be small, $\tcoh\lesssim \hbar^2/R^2$, where $R$ is the nuclear radius.
In terms of experimentally measured variables $|t|$  is given by ($q^0 \approx E_\pi$)
\begin{equation}
    \mid t \mid = -(q - p_\pi)^2 =
        - q^2 + 2 (E_\pi^2 - E_\nu |\bm{p}_\pi|\cos{\theta_\pi} + |\bm{k}'| |\bm{p}_\pi| \cos{\theta_{\pi\mu}}) - m_\pi^2,
\end{equation}
where $\theta_\pi$ is the pion angle with respect to the neutrino beam direction and $\theta_{\pi\mu}$ is the opening angle between the muon and the pion track. 

\begin{figure}[tbp]
\centering 

\includegraphics[width=0.49\columnwidth]{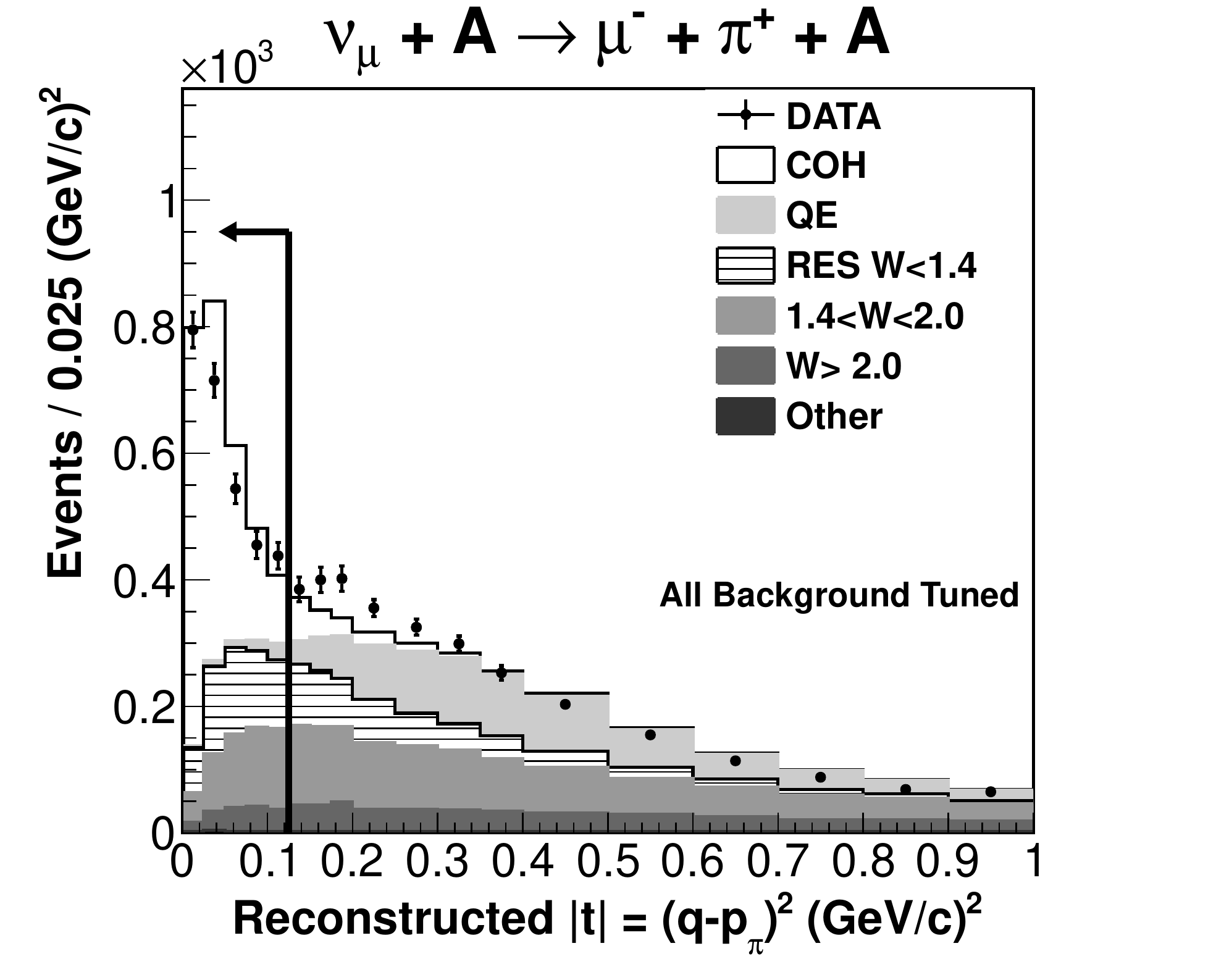}

\caption{An example of an experimental $\tcoh$ distribution from Ref.~\cite{Higuera:2014azj} showing the signal distribution peaking near zero and the relative size of the (GENIE) predicted background.}
%    \vspace{-10pt}
\label{fig:tdists}
\end{figure}

The \minerva experiment identified coherent $\pi^\pm$ candidates from \numu and \numubar beams on a scintillator (primarily CH) target by reconstructing the final state $\mu^\mp$ and $\pi^\pm$, requiring minimal additional energy near the
neutrino interaction vertex and small \tcoh\ as a signature of the coherent reaction. As an example of the strength these criteria, the reconstructed \tcoh\ distribution presented in  
Fig.~\ref{fig:tdists} displays a significant excess of low \tcoh\ events over the background  
after employing the vertex energy constrain. A further cut of \tcoh\ $\leq 0.12$~(GeV$/c)^2$  provided an enriched sample of coherent pion production candidates.

\subsubsection{Charged-Current Coherent Kaon Production }
\label{subsubsec:Kexpt}

Neutrino-induced CC coherent kaon production, $\nu_{\mu}A\rightarrow\mu^{-}K^{+}A$, is a process
yielding a $\mu^-$ with a single $K^+$ and no other (observable) detector activity around the interaction vertex.  In comparison to coherent pion production, this process has a much lower rate due to both Cabibbo suppression and a kinematic suppression caused by the larger kaon mass.

The MINERvA experiment isolated this rare channel~\cite{Wang:2016pww} by using the minimal vertex energy requirement and the kinematics of the $\mu^-$ and $K^+$ to reconstruct \tcoh\ that was required to be small. After background subtraction, the the evidence for this signal is of $3.0\, \sigma$ significance.

\subsubsection{Coherent Photon Production}
\label{subsubsec:cohgam}

As mentioned in Sec.~\ref{subsec:osc}, this process is a background for $\nu_e$ appearance experiments. It has also been mentioned that this process can become a background for the study of $\nu_{\mu}e \rightarrow\nu_{\mu}e$. Since $\nu_{\mu}e$ scattering is being proposed to constrain the energy-dependent neutrino flux, neglecting this background could result in a false flux constraint. The current study of $\nu_{\mu}e$ scattering in the medium energy configuration of the MINERvA experiment is addressing this coherent reaction. Although there has been no explicit search for coherent gamma production in recent experiments, it is worth noting that the high energy, $\langle E_\nu \rangle \approx 25$~GeV, NOMAD experiment found no significant single-$\gamma$ signal in the forward direction, setting an upper limit of
\begin{equation}
    \frac{\sigma(\mathrm{NC}\gamma,\mathrm{forward})}{\sigma(\nu_\mu A \rightarrow \mu^- X)} < 1.6 \times 10^{-4}
\end{equation}
at 90\% confidence level~\cite{Kullenberg:2011rd}.

\subsection{Diffractive Pion Production off a Nucleon}
\label{subsec:diffpi}

During the extraction of the MINERvA CC coherent pion signal, it was noted that the scintillator target (CH) has as many free protons as $^{12}$C nuclei. Diffractive production of pions from these protons could then also produce events at low \tcoh. The theoretical treatment of neutrino-induced diffractive pion production off nucleons, for example~\cite{Rein:1986cd}, does not apply in the lower $W_{\pi N}$ region mainly covered by the MINERvA coherent pion study; there was no process in GENIE for this channel.
Following discussions with Boris Kopeliovich~\cite{Kopeliovich-private} and recognizing that for a nucleon target the recoiling proton could cause the event to be rejected by a vertex energy cut,  a rough estimate of the event rate of diffractive pion production off protons in the MINERvA data was found to be equivalent to order 5\% of the GENIE prediction for the Coh$\pi$ cross section on $^{12}$C.

Subsequently, the MINERvA experiment detected a signal that could be interpreted as diffractive pion production~\cite{Wolcott:2016hws} while extracting the signal for charged current quasi-elastic scattering of \nue in the NuMI beam~\cite{Wolcott:2015hda}.
Indeed, an unexpectedly large number of events with electromagnetic showers likely caused by photon conversions was observed. The features of the excess events were consistent with those expected from NC diffractive $\pi^{0}$ production from hydrogen in the CH target.
The measured cross section for this process for $E_{\pi} \geq 3$~GeV, and integrated over the \minerva\ flux, is $0.26\pm0.02(\text{stat})\pm0.08(\text{sys})\times10^{-39}\text{cm}^{2}/\text{CH}$, comparable to that for NC coherent $\pi^{0}$ production from carbon.
This process can be important for the background studies of oscillation experiments, which emphasizes the need for models of diffractive pion production covering also the lower $W_{\pi N}$ kinematic region.

\subsection{Comparisons between theory and experiment: Open questions}
\label{subsec:compare}

As noted, the renewed attention to CC Coh$\pi$ was initiated by the K2K experiment's surprising result of no extractable signal, which was then confirmed by SciBooNE. To put these upper limits in perspective, one should recall that, as stated in Sec.~\ref{subsec:gen}, there are different implementations of the RS model. K2K and SciBooNE both used the NEUT~\cite{Hayato:2002sd} simulation program; their resulting experimental limits were well below the level predicted by the  version of the RS model implemented in NEUT, as shown in Fig.~\ref{fig_NvsG}.  

The negative K2K result spurred a careful re-examination of both the experimental results and the original RS model. As described in the theoretical summary above, Sec.~\ref{subsec:theory}, this model did not include the outgoing finite lepton (in this case, muon) mass in the calculations. This effect was small for the $E_\nu \geq 7$~GeV neutrino beams employed in the successful early searches for CC coherent pion production. However, it is particularly significant for the low neutrino energies employed by both K2K and SciBooNE.
It was further established that the approximation for the pion-nucleus cross sections employed in the PCAC expression, Eq.~(\ref{PCAC}), for coherent pion production in the original RS model was not consistent with current experimental results.

\begin{figure}[tbp]
    \centering
    \includegraphics[width=0.41\textwidth]{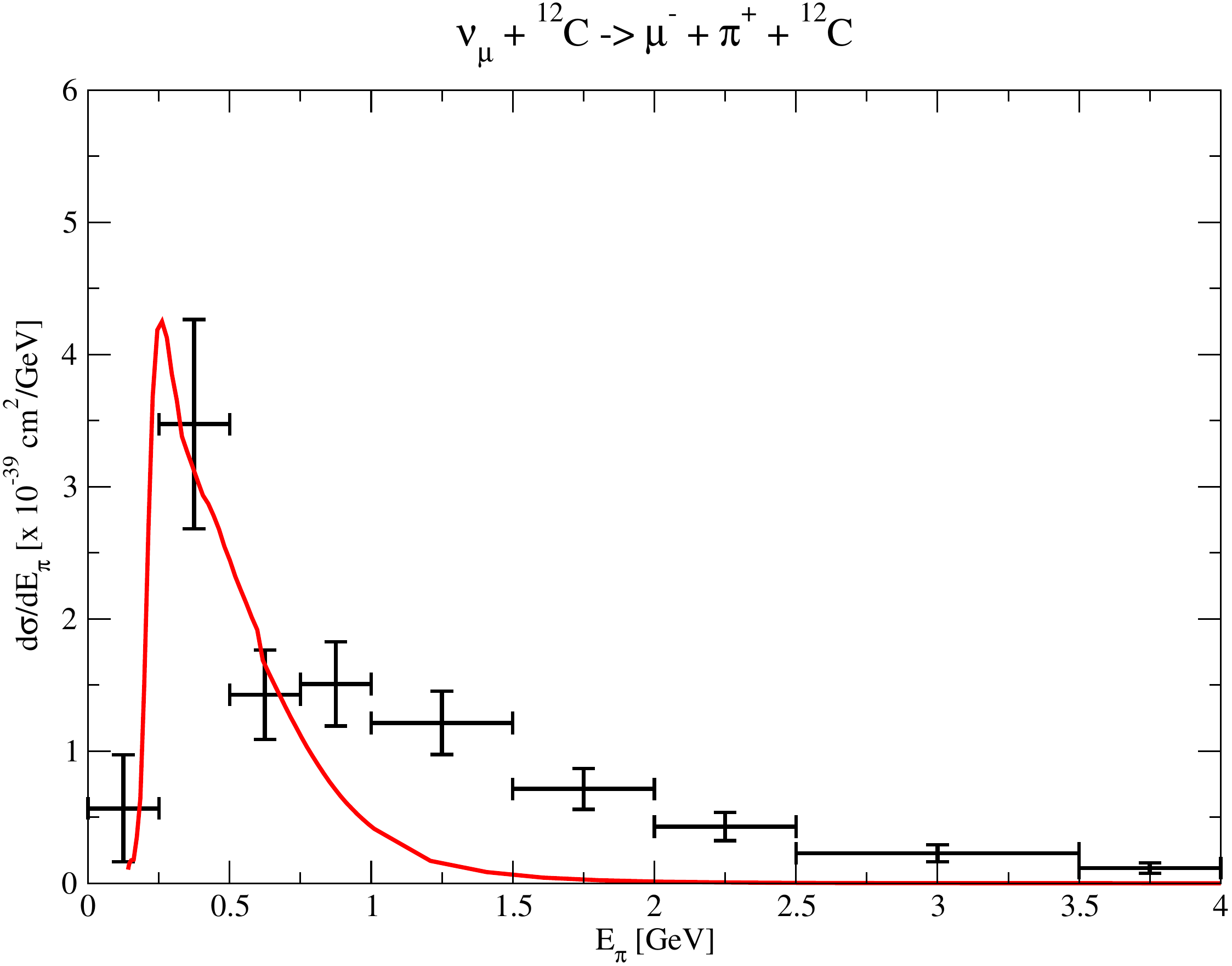}\hfill
    \includegraphics[width=0.41\textwidth]{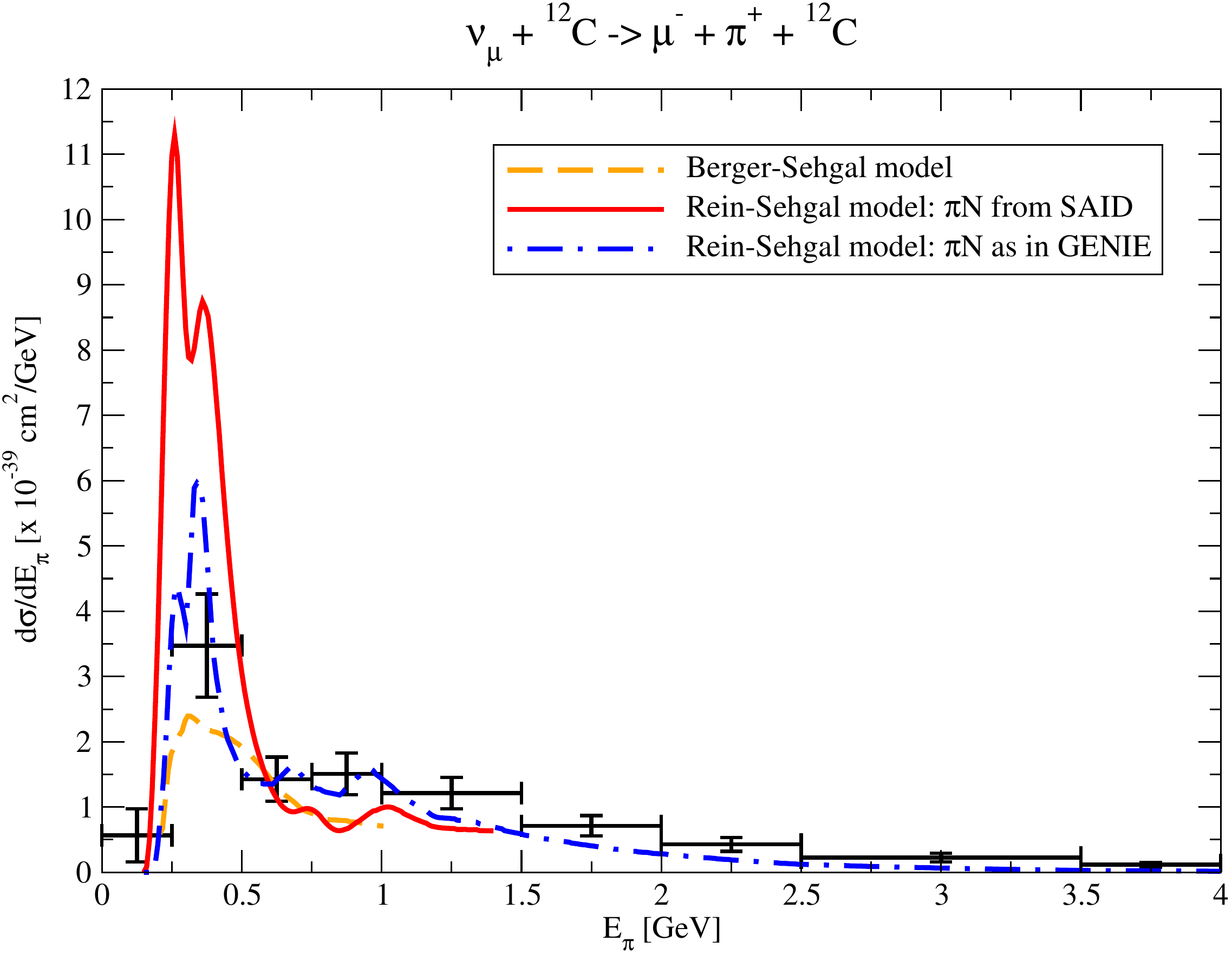}
    \caption{Coh$\pi^+$ MINERvA data~\cite{Higuera:2014azj} confronted with different theoretical models (also in
        Ref.~\cite{Alvarez-Ruso:2016ikb}).
        Left: microscopic model of Ref.~\cite{AlvarezRuso:2007tt}.
        Right: implementations of Rein-Sehgal~\cite{Rein:1982pf} and Berger-Sehgal~\cite{Berger:2008xs} models.
        For the Rein-Sehgal model, input as in GENIE~\cite{Perdue} and from SAID~\cite{SAID} have been used.}
    \label{fig:CohPi}
\end{figure}

From the MINERvA experiment, we now have detailed information about the energy (Fig.~\ref{fig:CohPi}) and angular distributions (Fig. 4 of~\cite{Higuera:2014azj} ) of pions produced in (anti)neutrino interactions on nuclei, where the target remains in the ground state~\cite{Higuera:2014azj}. Although the data fit the indicated version of the GENIE prediction better than the version of NEUT there are still significant disagreements between data and GENIE.  A proper understanding of these coherent pion production data is a new challenge for reaction model builders.

In Fig.~\ref{fig:CohPi} (right) we compare the implementation by the authors of Ref.~\cite{Alvarez-Ruso:2016ikb} of the RS~\cite{Rein:1982pf} and the Berger-Sehgal (BS)~\cite{Berger:2008xs} approaches to the MINERvA data.
Within the RS model, the $\pi N$ parametrizations as implemented in GENIE~\cite{Perdue} are considered, as well as the state-of-the-art ones from SAID~\cite{SAID}.
The plot shows that the RS cross section is very sensitive to this input. An improvement in the parametrizations does actually cause a worse agreement with data. From this perspective, the good agreement obtained by the GENIE implementation, particularly above $E_\pi = 500$~MeV, can be
regarded as accidental; see also Fig.~4 of Ref.~\cite{Higuera:2014azj}. The prediction from the BS model is better but not entirely satisfactory as it underestimates both the low-energy peak and the region of $E_\pi = 0.6$--1~GeV.

In the left panel of Fig~\ref{fig:CohPi}, the prediction of the model of Ref.~\cite{AlvarezRuso:2007tt} for the differential cross
section as a function of the pion energy averaged over the MINERvA flux is compared to the data of Ref.~\cite{Higuera:2014azj}. A good description is found at low pion energies, where the model is applicable, while the high energy tail is missed. Coh$\pi$ is dominated by low $q^2$.
In this limit the predicted cross section strongly depends on the value of the leading $N\Delta$ axial coupling, denoted $C_5^A(0)$. The results in Fig.~\ref{fig:CohPi} (left) are obtained using $C_5^A(0)=1.2$, which is consistent with pion-nucleon scattering via
the off-diagonal Goldberger-Treiman relation; see, for instance, Sec.~2.1 of Ref.~\cite{Alvarez-Ruso:2016ikb}.

Other comparisons have been performed by the T2K collaboration in Ref.~\cite{Abe:2016fic}.
Using the GENIE 2.6.4 implementations of the RS model and the one of Alvarez-Ruso \emph{et al.}~\cite{AlvarezRuso:2007tt} for selection efficiency and to extrapolate to the full phase space, T2K finds a Coh$\pi^+$ flux averaged cross section of
$3.9\pm1.0(\text{stat})^{+1.5}_{-1.4}(\text{sys})\times10^{-40}~\text{cm}^2$ and
$3.3\pm0.8(\text{stat})^{+1.3}_{-1.2}(\text{sys})\times10^{-40}~\text{cm}^2$, respectively.
These results should be compared to the predictions of $6.4\times10^{-40}~\text{cm}^2$ and $5.3\times10^{-40}~\text{cm}^2$ by the
correspondent model implementations~\cite{Abe:2016fic}.
In addition it is worth stressing that the standard untuned NEUT predicts a much larger value of
$15.3\times10^{-40}~\text{cm}^2$~\cite{Abe:2016fic}.

Although coherent and diffractive meson production have been addressed by several recent experiments, there are still several outstanding open experimental questions that need to be addressed by the community. The most inclusive recent experimental investigation of this topic has been performed by MINERvA in the so-called ``low-energy'' configuration of the NuMI beam.
MINERvA is addressing these channels again in the current ``medium-energy'' NuMI configuration that will allow a study with significantly increased statistics over a wider range of neutrino energies. Most importantly, this new configuration will also provide sufficient statistics to study the variety of MINERvA nuclear targets (C, CH, Fe and Pb) and provide a measurement of the A-dependence of CC Coh$\pi$. Such data will permit tests of the different theoretical predictions available in the literature for the A dependence. To reliably predict a possible background to upcoming LAr oscillation experiments, we need an experimental measurement of the A~dependence of these processes in order to extrapolate to $^{40}$Ar. Finally, it is always preferable to have a second experimental method to check the available results. However, there is currently no second experiment to check MINERvA measurements over a comparable neutrino energy range.

From the theoretical perspective, microscopic models, which can and should be validated with other coherent reactions, need to be extended to higher energies to cover the kinematic range probed by MINERvA. Although the microscopic model of Ref.~\cite{AlvarezRuso:2007tt} is available in GENIE, more efficient implementations of this and other microscopic models are needed.
Regarding PCAC pion production models, in spite of the limitations spelled out in 
Sec.~\ref{subsubsec:PCAC}, their simplicity makes them valuable. Indeed, in the case of meson production, it is important to understand if the accuracy goals justify the need for models better than the simple and fast PCAC based ones. The presence of multiple and inconsistent implementations of a given model is however harmful and should be avoided: pion-nucleus elastic scattering data might be used for validation purposes. From this perspective, the more phenomenological approach  of Refs.~\cite{Berger:2008xs,Paschos:2009ag} that rely on pion-nucleus scattering data may be preferable but it should be understood how to reliably estimate its errors and extrapolate results to different  nuclear targets.

\newpage
\section{Acknowledgements}
Support for participating scientists was provided by: 
Spanish Ministerio de Econom\'ia y Competitividad and the European Regional Development Fund, under contracts FIS2014-51948-C2-1-P, FPA2016-77347-C2-2 and SEV-2014-0398; Generalitat Valenciana under contract PROMETEOII/2014/0068;
Universit\`a degli Studi di Torino under Projects BARM-RILO-15-02 and BARM-RILO-17-01; Istituto Nazionale di Fisica Nucleare under Project MANYBODY;
Fermi National Accelerator Laboratory, operated by Fermi Research Alliance, LLC under Contract No. DE-AC02-07CH11359 with the United States Department of Energy; University of Pittsburgh, U.S. Department of Energy award DE-SC0007914, DE-SC0015903, DE-SC0010005, DE-SC0009973 and DE-SC0010073; Colorado State University, DE-FG02-93ER40788; Michigan State University, DE-SC0015903; Virginia Tech, DE-SC0009973; Jefferson Laboratory, U.S. Department of Energy, Office of Science, Office of Nuclear Physics under contract DE-AC05-06OR23177;
Alfred P. Sloan Foundation; MEXT KAKENHI, Japan, Grant No. JP25105010;
the European Union's Horizon 2020 research and innovation programme under the Marie Sklodowska-Curie grant agreement No. 674896; Polish NCN Grant No. UMO-2014/14/M/ST2/0085; the Research Foundation Flanders  (FWO-Flanders);  the  Interuniversity  Attraction Poles Programme P7/12 initiated by the Belgian Science Policy Office. The authors supported during the INT-16-63W workshop thank the Institute for Nuclear Theory at the University of Washington. 

\newpage
\bibliography{NuSTEC,resref1,wpqe,disref,Coherent,wpelectron}
%\bibliography{resref1}
%resref1,
\end{document}